%% file: main.tex
\documentclass[12pt]{book}

\usepackage[numbers,sort&compress]{natbib}

\usepackage{a4wide}
\usepackage{fancyhdr}
\usepackage{graphicx}
\usepackage{wrapfig}
\usepackage{comment}
\usepackage{amsmath}
\usepackage{amssymb}
\usepackage{epsfig}
\usepackage{graphicx}
\usepackage{caption}
\usepackage{subcaption}
\usepackage{color}
\usepackage[colorlinks=true,
            urlcolor=teal,
            anchorcolor=teal,
            citecolor=teal,
            filecolor=teal,
            linkcolor=teal,
            menucolor=teal,
            linktocpage=true,
            pdfproducer=medialab,
            pdfa=true,
            unicode,
            psdextra]{hyperref}
\hypersetup{
pdfpagemode = {UseNone},
pdftitle = {Spectral distortion and polarization of the cosmic microwave background: Measurement, challenges and perspectives},
pdfauthor = {Vyoma Muralidhara},
pdflang = {en-GB}
}
\newcommand\pdfmath[1]{\texorpdfstring{$#1$}{#1}}
\usepackage{tikz}
\usepackage{tikz-3dplot}
\usetikzlibrary{calc,shadings}
\usepackage{enumitem}
\usepackage{mathtools}
\usepackage{longtable}
\usepackage{multirow, bigdelim, makecell, booktabs} 

\usepackage{array}
\usepackage{colortbl}
\usepackage{MnSymbol}
\usepackage{braket}
\usepackage{tensor}

\usepackage{listings}

\newcommand{\plk}{{\it Planck~}}

\tikzset{%
>=latex,
inner sep=0pt,%
outer sep=2pt,%
mark coordinate/.style={inner sep=0pt,outer sep=0pt,minimum size=3pt,
fill=black,circle}%
}

\usetikzlibrary{arrows}
\usepackage{pgfplots}
\usetikzlibrary{calc,fadings,decorations.pathreplacing,positioning}
\pgfplotsset{compat=1.14}

\AtBeginDocument{\numberwithin{lstlisting}{section}}  

\definecolor{background}{gray}{.98}                 
\definecolor{comments}{RGB}{51,102,0}               
\definecolor{keywords}{RGB}{0,0,120}                
\definecolor{keywords2}{RGB}{204,0,102}             
\definecolor{numbers}{RGB}{127, 0, 127}             
\definecolor{Maroon}{RGB}{128, 0, 0}

\makeatletter
\newenvironment{btHighlight}[1][]
{\begingroup\tikzset{bt@Highlight@par/.style={#1}}\begin{lrbox}{\@tempboxa}}
{\end{lrbox}\bt@HL@box[bt@Highlight@par]{\@tempboxa}\endgroup}

\newcommand\btHL[1][]{%
  \begin{btHighlight}[#1]\bgroup\aftergroup\bt@HL@endenv%
}
\def\bt@HL@endenv{%
  \end{btHighlight}%
  \egroup
}
\newcommand{\bt@HL@box}[2][]{%
  \tikz[#1]{%
    \pgfpathrectangle{\pgfpoint{1pt}{0pt}}{\pgfpoint{\wd #2}{\ht #2}}%
    \pgfusepath{use as bounding box}%
    \node[anchor=base west, fill=orange!30,outer sep=0pt,inner xsep=1pt, inner ysep=0pt, rounded corners=3pt, minimum height=\ht\strutbox+1pt,#1]{\raisebox{1pt}{\strut}\strut\usebox{#2}};
  }%
}
\makeatother

\usepackage[font=small]{caption}
\usepackage{float}

\usepackage{pdflscape}

\usepackage{scalerel}
\newlength\bshft
\def\fakebold#1{\ThisStyle{\ooalign{$\SavedStyle#1$\cr%
  \kern-\bshft$\SavedStyle#1$\cr%
  \kern\bshft$\SavedStyle#1$}}}

\DeclareMathAlphabet{\mathbbmsl}{U}{bbm}{m}{sl}

\renewcommand{\chaptermark}[1]%
         {\markboth{\thechapter.\ #1}{}}
\renewcommand{\sectionmark}[1]%
         {\markright{\thesection\ #1}}
\newcommand{\LMUTitle}[9]{
  \thispagestyle{empty}
  \vspace*{\stretch{1}}
  {\parindent0cm
   \rule{\linewidth}{.7ex}}
  \begin{flushright}

    \vspace*{\stretch{1}}
    \sffamily\bfseries\Huge
    #1\\
    \vspace*{\stretch{1}}
    \sffamily\bfseries\large
    #2
    \vspace*{\stretch{1}}
  \end{flushright}
  \rule{\linewidth}{.7ex}
  \vspace*{\stretch{5}}
  \begin{center}
    \includegraphics[width=2in]{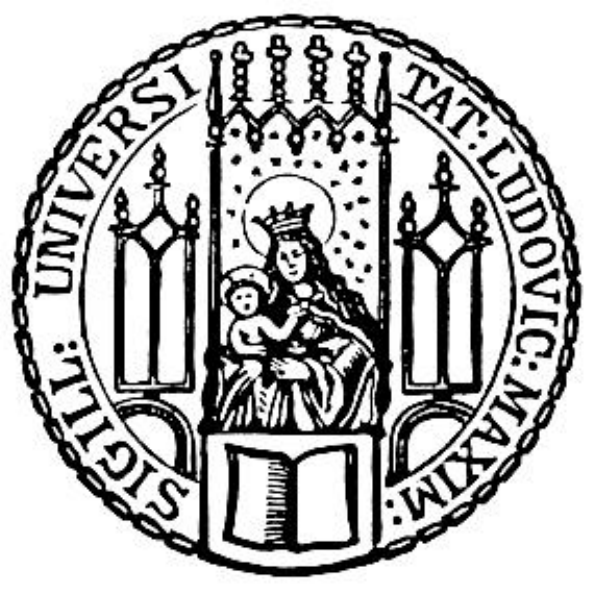}
  \end{center}
  \vspace*{\stretch{1}}
  \begin{center}\sffamily\LARGE{#5}\end{center}
  \newpage
  \thispagestyle{empty}

  \cleardoublepage
  \thispagestyle{empty}

  \vspace*{\stretch{1}}
  {\parindent0cm
  \rule{\linewidth}{.7ex}}
  \begin{flushright}
    \vspace*{\stretch{1}}
    \sffamily\bfseries\Huge
    #1\\
    \vspace*{\stretch{1}}
    \sffamily\bfseries\large
    #2
    \vspace*{\stretch{1}}
  \end{flushright}
  \rule{\linewidth}{.7ex}

  \vspace*{\stretch{3}}
  \begin{center}
    \Large Dissertation\\
    \Large an der #4\\
    \Large der Ludwig--Maximilians--Universit\"at\\
    \Large M\"unchen\\
    \vspace*{\stretch{1}}
    \Large vorgelegt von\\
    \Large #2\\
    \Large aus #3\\
    \vspace*{\stretch{2}}
    \Large M\"unchen, den #6
  \end{center}

  \newpage
  \thispagestyle{empty}

  \vspace*{\stretch{1}}

  \begin{flushleft}
    \large Erstgutachter:  #7 \\[1mm]
    \large Zweitgutachter: #8 \\[1mm]
    \large Tag der m\"undlichen Pr\"ufung: #9\\
  \end{flushleft}

  \cleardoublepage
}

\usepackage{pgfplots,pgfplotstable}
\usepgfplotslibrary{colormaps}
\pgfplotsset{compat=newest}

\definecolor{temp}{HTML}{74355d}

\newcommand{\npp}{\texttt{NPIPE~}}
\newcommand{\toast}{\texttt{TOAST~}}

\usepackage[scale=3, color=gray]{draftwatermark}

\usepackage{csquotes}

\begin{document}

  \SetWatermarkText{}

  \frontmatter

  \LMUTitle
      {Spectral distortion and polarization of the cosmic microwave background: Measurement, challenges and perspectives}               
      {Vyoma Muralidhara}                       
      {Bengaluru, Indien}                             
      {Fakult\"at f\"ur Physik}                         
      {M\"unchen 2024}                          
      {23.09.2024}                            
      {Prof.\ Dr.\ Eiichiro Komatsu}                          
      {Prof. Dr. Joseph Mohr}                         
      {05.11.2024}                         

  \tableofcontents
  \markboth{CONTENTS}{CONTENTS}

  \listoffigures
  \markboth{List of figures}{List of figures}

  \listoftables
  \markboth{List of tables}{List of tables}
  \cleardoublepage

  \markboth{Abstract}{Abstract}
  \include{abstract}

  \mainmatter\setcounter{page}{1}
  \include{chapters/intro}
  \include{chapters/chp01}
  \include{chapters/chp02}
  \include{chapters/chp03}

  \include{chapters/chp04}

  \include{chapters/chp05}
  \include{chapters/chp06}
  \include{chapters/chp07}

  \include{chapters/chp08}

  \begin{appendix}
    \cleardoublepage
    \addcontentsline{toc}{chapter}{\protect Appendices}
    \part*{Appendices}
    \cleardoublepage
    \include{chapters/app01}
    \include{chapters/app02}
  \end{appendix}

  \backmatter
  \bibliographystyle{bibliography}
  \addcontentsline{toc}{chapter}{\protect Bibliography}
  \bibliography{bibliography.bib}
  \markboth{}{}

  \include{danksagung}

\end{document}

%% file: abstract.tex
\chapter*{Zusammenfassung}\addcontentsline{toc}{chapter}{\protect Zusammenfassung}
Der kosmische Mikrowellenhintergrund (CMB) ist f\"{u}r die moderne Kosmologie von grundlegender Bedeutung. Die Polarisierung des CMB liefert entscheidende Hinweise auf neue physikalische Ph\"{a}nomene wie Parit\"{a}tsverletzung und Inflation liefert.  Abweichungen vom Schwarzk\"{o}rperspektrum des CMBs testen das Intracluster-Medium (ICM) von Galaxienhaufen und erlauben Rückschlüsse auf kosmologische Parameter. Die Genauigkeit, mit der CMB Polarisierung und Spektrum gemessen werden können, ist limitiert durch Vordergrundemission, durch instrumentelle Systematiken und durch die Empfindlichkeit der Instrumente.

Zun\"{a}chst betrachten wir Abweichungen vom Schwarzk\"{o}rperspektrum des CMBs, das durch die Inverse-Compton-Streuung von CMB-Photonen an nicht-thermischen Elektronen im ICM von Galaxienhaufen verursacht wird. Dieser Effekt, genannt nicht-thermischer Sunyaev-Zeldovich-Effekt (ntSZ), h\"{a}ngt von der nicht-thermischen Energiedichte in Galaxienhaufen ab. Die Amplitude des ntSZ-Effekts in Verbindung mit Messungen der Synchrotronemission bei 1,4 GHz liefert Hinweise auf die volumen gemittelte Magnetfeldst\"{a}rke innerhalb des ICM. Wir identifizieren Galaxienhaufen die Radio Halos beherbergen und Synchrotronstrahlung emittieren in den All-Sky-Karten des Planck Satelliten um den ntSZ-Effekt zu messen und die Magnetfeldst\"{a}rke einzuschr\"{a}nken. Des Weiteren untersuchen wir, wie sich die Messungen durch kommende bodengest\"{u}tzte CMB Experimente verbessern lassen.

Parit\"{a}tsverletzung kann eine kosmische Doppelbrechung erzeugen, die in der CMB-Polarisation detektiert werden kann. In j\"{u}ngster Zeit wurden mehrere Methoden zur Messung der kosmischen Doppelbrechung anhand von \plk- und \textit{WMAP}-Daten eingesetzt. Eine große Herausforderung bei der Messung ist die Fehlkalibrierung der Polarisationsrichtung der Detektoren, die im Mikrowellenbereich beobachtet werden. Diese Herausforderung kann durch die  absolute Winkel Kalibrierung der Instrumente mit einer hellen, bekannten astrophysikalischen Quelle gel\"{o}st werden. In dieser Arbeit versuchen wir eine absolute Polarisationswinkel Kalibrierung der \plk Detektoren mit Beobachtungen des Crab Nebula im Mikrowellen- und R\"{o}ntgenbereich.

Eine weitere Herausforderung für genaue Messungen des CMB ist das räumlich und zeitlich korrelierte Detektor- und atmosph\"{a}rische Rauschen, das die Rekonstruktion von CMB-Karten aus zeitlich-geordneten Daten (TOD) erschwert. Bestimmte Strategien, wie ein Teleskop den Himmel abtastet und wie die Rohdaten reduziert werden, k\"{o}nnen eingesetzt werden, um die Auswirkungen dieser Rauschkomponenten zu mildern. Um diese Effekte und die verwendeten Strategien zu untersuchen, haben wir eine Simulationspipeline basierend auf dem Software-Frameworks \texttt{TOAST} entwickelt, die es uns erlaubt TOD Daten eines zuk\"{u}nftigen bodengebundenen Sub-mm-Teleskops zu erzeugen. Wir verwenden diese Simulationen, um die st\"{o}renden Effekte aufzuzeigen, die durch diese Rauschkomponenten auf der Kartenebene entstehen, und er\"{o}rtern mehrere Methoden zur Kartenerstellung, um diese zu minimieren.

In dieser Arbeit stellen wir fest, dass die kombinierte Empfindlichkeit von kommenden CMB-Experimenten und eine breite Frequenzabdeckung im sub-mm-Bereich zu einer besseren Charakterisierung von Vordergr\"{u}nden sowie einer verbesserten Einschr\"{a}nkungen von Abweichungen vom Schwarzk\"{o}rperspektrum des CMBs f\"{u}hrt. W\"{a}hrend \plk Beobachtungen des Crab Nebula zu einer verzerrten Sch\"{a}tzungen des Positionswinkels der linearen Polarisation ergeben, unterstreicht dieses Ergebnis die Notwendigkeit spezieller bodengest\"{u}tzter Messungen. Diese erm\"{o}glichen unabh\"{a}ngige Methoden zur Kalibrierung von CMB-Instrumenten und der genauen Messung der Polarisation des CMB.

\chapter*{Abstract}\addcontentsline{toc}{chapter}{\protect Abstract}
The Cosmic Microwave Background (CMB) is an essential observational probe to modern cosmology. The linear polarization of the CMB provides a crucial observational tool for exploring new physics, including the inflationary paradigm and parity-violating phenomena. The spectral distortion of the CMB can be used as a probe of the intracluster medium (ICM) of galaxy clusters and to infer cosmological parameters. However, accurate measurements of the spectral distortion and polarization of the CMB are limited by the characterization of the foreground emission and systematic effects introduced by the instruments of CMB experiments. 
\paragraph{}
We first study the spectral distortion of the CMB introduced by Inverse-Compton scattering of CMB photons by non-thermal electrons in the ICM of galaxy clusters, known as the non-thermal Sunyaev-Zeldovich (ntSZ) effect, which serves as a probe of the non-thermal energy budget within galaxy clusters. The amplitude of the ntSZ effect, which, when combined with measurements of the synchrotron emission at 1.4 GHz, provides constraints on the volume-averaged magnetic field strength within the ICM. Observations of galaxy clusters that host radio halos and emit diffuse synchrotron radiation are selected from \plk all-sky maps to measure the ntSZ effect and subsequently the magnetic field strength. We also study the improvement in the uncertainties of these constraints when considering the sensitivities of upcoming ground-based CMB experiments. 
\paragraph{}
The imprints of parity-violating physics on the polarization of the CMB are predicted theoretically and recently, several methods have been used to measure the effect of cosmic birefringence from \plk and \textit{WMAP} data. A major challenge in the detecting the cosmic birefringence angle has been the miscalibration of the orientation of polarization sensitivity of the detectors that observe in the microwave regime. One of the methods to overcome this challenge is the absolute angle calibration of the instruments with a bright, well-known astrophysical source. In this work, we attempt a relative and absolute polarization angle calibration of the \plk detectors with microwave and X-ray observations of the Crab nebula.
\paragraph{}
Another challenge to accurate measurements of the CMB is the spatially and temporally correlated detector and atmospheric noise which complicates the reconstruction of CMB maps from the raw data. Certain strategies can be employed in terms of how a telescope scans the sky and how the raw data are reduced to mitigate the effects of these noise components. To study these effects and the strategies employed, we developed a simulation pipeline to generate time-ordered data (TOD) from an upcoming ground-based sub-mm telescope using the \texttt{TOAST} software framework. We use these simulations to show the spurious effects that are introduced by these noise components at the map level and discuss several map reconstruction methods employed to minimize them.
\paragraph{}
In this work, we find that the combined sensitivity of upcoming CMB experiments and a broad observing frequency coverage in the sub-mm regime leads to better characterization of foregrounds and improved constraints on the spectral distortion of the CMB. While \plk observations of the Crab nebula result in biased estimates of the position angle of linear polarization due to systematic effects, it highlights the need for dedicated ground-based measurements of the object that enable independent methods of calibration of CMB instruments to accurately measure the polarization of the CMB.

%% file: chapters/intro.tex
\chapter{Introduction}
\label{ch:introduction}
\begin{center}
    \begin{minipage}{0.75\textwidth}
        \textbf{Summary:} This chapter establishes the context of this thesis. We start with a brief description of how certain astrophysical phenomena and new physics leave their imprint in the Cosmic Microwave Background radiation and how it serves as an observational probe of these phenomena (of astrophysical and cosmological significance) in synergy with other observations. This is followed by a description of specific observational challenges to these studies. We then describe the contents of this thesis.\\[3mm]
    \end{minipage}
\end{center}
The radiation density in the current Universe is dominated by the Cosmic Microwave Background radiation (CMB) which has been essential for empirical Cosmology. The CMB as a relic of the Big Bang was first theorized by \cite{Alpher:1948srz, Alpher:1949sef, Gamow:1948pob} in which they studied the physics of the early Universe and Big Bang Nucleosynthesis. It was detected for the first time 20 years later by \cite{1965PenziasWilson, 1965DickePeebles}. Since its discovery, the CMB has been the cornerstone in establishing an observational window into the theories of the Big Bang model, $\Lambda$CDM model and the inflationary paradigm that remain the basis for Modern Cosmology.
\paragraph{}
Roughly 380,000 years after the Big Bang, the CMB photons decoupled from baryonic matter during the epoch of recombination. Before the epoch of recombination, photons and baryons formed a coupled fluid that would undergo oscillations due to any perturbations, such as by gravitational potentials. These baryonic acoustic oscillations (BAOs) cause small temperature anisotropies that are referred to as \textit{primary} anisotropies with typical amplitudes of the order of $10^{-5}$. After the epoch of recombination, the CMB photons have traversed the Universe from what is referred to as the last scattering surface (LSS). The primary anisotropies thus form an imprint of the matter distribution on the CMB at $z\approx 1080$. After the epoch of recombination, a variety of processes lead to additional temperature anisotropies of the CMB that are referred to as \textit{secondary} anisotropies. Thomson scattering after the reionization of the Universe, integrated Sachs-Wolfe effect, gravitational lensing and the Sunyaev-Zeldovich (SZ) effect are some of the \textit{secondary} anisotropies.
\paragraph{}
The SZ effect, first predicted by \cite{Sunyaev:1970er, Sunyaev:1972eq}, is the Inverse-Compton scattering of CMB photons by energetic electrons leading to a characteristic spectral distortion of the CMB towards galaxy clusters. The SZ effect is actually a composite of distortions due to different electron populations within the intracluster medium (ICM) of galaxy clusters. While the thermal populations cause a distortion known as the thermal SZ (tSZ) effect, non-thermal populations of electrons cause the non-thermal SZ (ntSZ) effect. The same non-thermal electrons lose energy by emitting synchrotron radiation in the radio waves, providing the first hints of their presence beyond the \textit{localised} regions of Active Galactic Nuclei (AGN) and of a magnetic field that extends to the outskirts of galaxy clusters. Thus, a measurement of the spectral distortion of the CMB due to the ntSZ effect in conjunction with observations of diffuse emission of synchrotron radiation gives a measure of the non-thermal electron energy budget within galaxy clusters and a volume-averaged magnetic field strength that permeates the ICM. This information aids in understanding scenarios of the origin of cosmic rays, mergers of galaxy clusters and the dynamic processes within the ICM.
\paragraph{}
Apart from determining physical processes that leave imprints on the Spectral Energy Density (SED) or the temperature angular auto-power spectrum of the CMB, theoretical predictions indicate the effects of parity-violating physics on CMB polarization. One such phenomenon is cosmic birefringence which is when the plane of linear polarization of the CMB is rotated over the course of its propagation by a time-dependent parity violating pseudoscalar. An ultra-light axion-like field coupled to electromagnetism through a Chern-Simons interaction \cite{Carroll:1989vb, Carroll:1991zs} is one of the proposed mechanisms that introduces this effect. The phenomenon results in a non-zero $E-B$ cross angular power spectra ($C_\ell^{EB}$). Recent works indicate hints of a non-zero $C_\ell^{EB}$ produced by an isotropic cosmic birefringence angle of $\beta \approx 0.3^\circ$ from the analysis of \textit{WMAP} and \plk data \cite{Minami:2019ruj, Minami:2020fin, Diego-Palazuelos:2022dsq, Eskilt:2022cff}.
\paragraph{}
A major challenge in this endeavour, however, has been the calibration of the instruments that measure the CMB in the microwave wavelength regime. A miscalibration of the orientation of polarization sensitivity of the instruments introduces a similar effect as that of the cosmic birefringence angle and several self-calibration techniques are unable to disentangle $\beta$ from the miscalibration angles. A potential solution is to use a well-known, bright, astrophysical source as an absolute calibrator. The Crab nebula is one of the brightest extended sources in the sky whose observations by the CMB instruments of interest and observations from other instruments (and/or at other wavelengths, such as in the X-rays where the systematic uncertainties are lower) offer an independent method of absolute polarization angle calibration.
\paragraph{}
Systematic uncertainties due to miscalibration of orientation of polarization sensitivity, however, is only one of the many challenges encountered in the accurate measurement of the polarization of CMB. Despite major advances in technology of polarization-sensitive detectors that operate in the microwave regime, the temporally correlated noise introduced by the instruments and atmosphere (in the case of ground-based telescopes) pose a challenge. While atmosphere is not polarized, beam asymmetries of the instrument and pair-differencing of orthogonally oriented detectors cause a leakage from intensity-to-polarization that result in spurious correlations. Several strategies are employed in terms of how a telescope scans the sky and how the observed data are reduced to mitigate the effects of these systematic uncertainties. Simulations of raw telescope observations (time-ordered data, TOD), with the incorporation of parameters of observation such as telescope location, weather conditions and field of observation, the effects of systematics and the methods employed to mitigate them can be studied under the context of the telescope's observing strategy.
\subsubsection{Structure of the thesis}
The scientific goals addressed by this thesis are
\begin{itemize}
    \item[-] Constraints on the non-thermal electron energy budget and the volume-averaged magnetic field strength within galaxy clusters hosting radio halos using the ntSZ effect are obtained from observations in the microwave regime by \plk satellite and measurements of synchrotron emission from at 1.4 GHz. We also make forecasts of the same by considering the sensitivities of upcoming (ground-based) CMB experiments.
    \item[-] The absolute and relative calibration of \plk detectors with observations of the Crab nebula in the microwave and X-ray regimes.
    \item[-] A study of the effects of temporally correlated detector and atmospheric noise on timestream data and maps for a given observing strategy of an upcoming ground-based telescope called the Fred Young Submillimeter Telescope (FYST) is performed. The CMB is considered the signal of interest here.
\end{itemize}
We begin with a discussion of radio halos and the Crab nebula, which are two astrophysical objects of interest in Chapter \ref{ch:astro-objects}. In this thesis, we have worked with multi-wavelength observations from the microwave to X-rays that involved three observatories which are introduced in Chapter \ref{ch:observatories}.
\paragraph{} Physical processes that occur due to non-thermal particles (in objects like radio halos and the Crab nebula) are presented in Chapter \ref{ch:physical-processes}. The phenomenon of cosmic birefringence and its effect on the observed CMB polarization are also discussed in Chapter \ref{ch:physical-processes}. 
\paragraph{} In Chapter \ref{ch:ntSZ}, the work on the study of ntSZ effect in galaxy clusters and application of the measurement on understanding the non-thermal components within the ICM is presented. The simulation of timestreams of a ground-based sub-mm telescope, components that are involved in such a simulation and map-making techniques are presented in Chapter \ref{ch:TOAST-TOD}. 
\paragraph{} Chapters \ref{ch:tauA-pol-angle} and \ref{ch:x-ray} present the analysis of the observations of the Crab nebula in the microwave and X-ray regimes, respectively, with a study of the measurement of its polarization properties in the microwave for relative and absolute calibration of orientation of polarization of \plk detectors. Finally the conclusions and outlook are presented in Chapter \ref{ch:conclusion}. 
\paragraph{}
Chapter \ref{ch:ntSZ} has been submitted as a paper and is currently under peer review:
\begin{itemize}
    \item[\cite{Muralidhara:2024ipg}] \textbf{V.\ Muralidhara} and K.\ Basu: \emph{``Constraining the average magnetic field in galaxy clusters with current and upcoming CMB surveys''}. Preprint: \href{https://arxiv.org/pdf/2402.17445.pdf}{\texttt{2402.17445}}. Submitted to JCAP in February 2024.
\end{itemize}
Some results from that study are published in Section 8.4 of \cite{CCAT-Prime2021} as part of presenting the scientific goals of the CCAT collaboration using FYST.

%% file: chapters/chp01.tex
\chapter{Astrophysical objects of interest}
\label{ch:astro-objects}
\begin{center}
    \begin{minipage}{0.75\textwidth}
        \textbf{Summary:} The Crab nebula is one of the brightest objects in the radio, microwave and X-ray sky. Synchrotron emission is the dominant emission mechanism attributed to this brightness. Another object of interest that also emits synchrotron radiation is a radio halo. Both of these objects are discussed in subsequent chapters of this thesis and we describe them in this chapter.\\[3mm]
    \end{minipage}
\end{center}
The non-thermal Universe is complex and intriguing. It is a complex interplay between magnetic fields, turbulent media and highly energetic particles. These particles gain enormous energies from processes that result in power-law spectra that extend over many orders of magnitude in energy. Sources of such non-thermal processes result from a number of events such as stellar collapse, rotational energy of a neutron star, accretion of matter onto a black hole or even from mergers of galaxy clusters.
\paragraph{}
The ``cosmic rays" (CR) are considered messengers from these highly-energetic processes that meander through our Galaxy and the Universe. The particles can be atomic nuclei, electrons and their anti-particles. Measurements of the spectra of these particles (either directly in case of electrons and neutrinos or indirectly with $\gamma-$rays emitted by proton-proton interactions) are often used to probe a number of questions ranging from stellar evolution, description of acceleration mechanisms, origin of cosmic rays and magnetic fields, to the search for annihilation of dark matter or other particles.
\paragraph{}
We consider in this work, radio halos and a supernova remnant (the Crab Nebula) as sources of high-energy processes, and use electromagnetic radiation from these sources in the radio, microwave and X-ray regimes as probes of non-thermal electrons and the underlying magnetic field strength. In this chapter, we introduce the them as the astrophysical objects of interest. In Chapter \ref{ch:physical-processes}, we describe the emission mechanisms resulting from these non-thermal electrons.
\section{Radio Halos}
\begin{figure}[ht]
    \centering
    \includegraphics[width=0.8\textwidth]{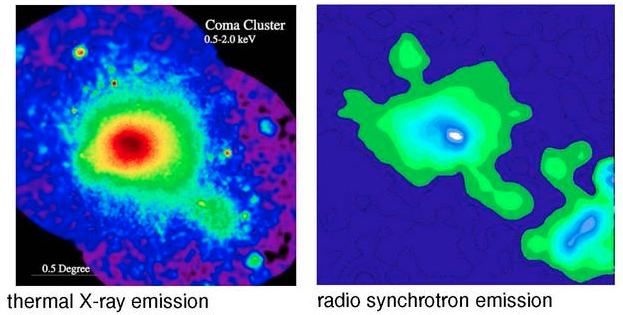}
    \caption{\textit{Left}: Thermal X-ray Bremsstrahlung emission in the $0.5-2.0$ keV energy range. \textit{Right}: Synchrotron emission from the Coma cluster in the radio wavelengths. The diffuse nature of the emission is apparent. The spatial profiles of the emissions in the X-ray and radio wavelengths are similar which can indicate that the non-thermal electrons can be spatially distributed in a similar manner to the thermal electrons. Image credit: Christoph Pfrommer, KITP Online Talks}
    \label{fig:coma-radiohalo}
\end{figure}

Radio halos are large-scale diffuse radio emission observed in galaxy clusters, and are among the largest gravitationally bound structures in the universe. These halos are characterized by their relatively low surface brightness and their emission remains unpolarized. They provide crucial insights into the physical processes occurring in galaxy clusters and the intracluster medium (ICM). Synchrotron emission is attributed to radio halos with emitting regions of size of the order of a few hundred kpc. The Coma cluster, for example, is known to host a radio halo. The emission from this cluster observed in the soft X-ray and radio wavelengths is shown in Figure \ref{fig:coma-radiohalo}. The right panel of the figure shows the characteristic feature of radio halos which is a diffuse radio emission that fills the volume of the cluster. 
\paragraph{}
Synchrotron emission indicates the presence of non-thermal electrons with $\sim$ GeV energies and the observed flux is a product of the non-thermal electron number density and the magnetic field strength within the ICM (more information on synchrotron emission is given in Section \ref{subsec:ch3-nonthermal-emission-theory}). The origin of radio halos remains a mystery as one must speculate for the presence of non-thermal electron populations that permeate such large volumes within the ICM.

\paragraph{}
{\it Localized creation} of non-thermal electrons due to AGN activity at the cluster cores, for example, cannot explain the presence of the CR electrons within the global cluster volume. {\it In-situ} creation mechanisms such as passage of merger shocks, acceleration driven by turbulence following major mergers, or the collision and decay of CR protons which are themselves energized by the merger process have been proposed  \cite{vanweeren2019}. These models can be classified into two classes:
\begin{enumerate}
    \item Primary models: The re-acceleration of seed electrons due to highly turbulent states of the ICM, often following major mergers. It is considered that turbulence in the ICM re-accelerates relativistic electrons resulting from hadronic decay after having experienced cooling over a short time span. 
    \item Secondary models: CR protons (whose origin remains unknown) interact with protons in the ICM, leading to the production of pions. Charged pions further decay into electrons and positrons, and neutral pions decay into $\gamma-$ray photons.
\end{enumerate}
\cite{Liang2002} present simulations which examine the influence of turbulent states on the non-thermal electrons. \cite{Pinzke2010, Donnert2013} provide a comprehensive discussion on the mechanisms of the secondary models and the resulting spectral energy profiles of the non-thermal electrons.
\paragraph{}
In Chapter \ref{ch:ntSZ}, we measure the ntSZ effect from a stack of radio halos to constrain the non-thermal energy budget within galaxy clusters. This information is further used to constrain volume-averaged magnetic field strength within the ICM by combining synchrotron emission measurements at 1.4 GHz.

\section{Crab Nebula}
The Crab Nebula is a supernova remnant (SNR) located at a distance of 2 kpc. Historical Chinese and Japanese records indicate a \textit{guest star} that appeared in the night sky ca. July 1054-1056. Charles Messier observed the same object and recorded it as his first non-cometary nebulous object in 1758 (hence the identifier M1, \cite{Hester2008}). Irish astronomer Lord Rosse drew his observation of the nebula observed through a 36-inch telescope and suggested that it resembles a crab. Even though subsequent observations revealed to him that the object does not resemble a crab, the name has stuck ever since.
\paragraph{} 
For the first time, \cite{Duyvendak:1942, Mayall:1942} verified that the Crab nebula was indeed the remnant of SN1054 that occurred on the 4th of July 1054. SNRs such as the Crab should display shells of emission. However, in the case of Crab, a lack of such shell-like emission and a deduced estimate of a low-energy explosion, leads to the proposition that it may have originated from electron-capture or Type IIn supernova \cite{Miyaji:1980, Moriya:2014uaa}.
\paragraph{}
Massive stars with masses (in the Zero Age Main Sequence phase) $>10\,M_\odot$ fuse heavy elements in their cores such as Nickel (\textsuperscript{56}Ni) and Iron (\textsuperscript{56}Fe). However, massive stars with slightly lower mass can still evolve to have electron degenerate O+Ne+Mg cores that then trigger core collapse through electron-capture reactions \cite{Miyaji:1980}. When energy production from nuclear fusion reaches its limit, the star's core can no longer counteract its own gravity with radiation pressure, leading to a gravitational collapse. This collapse triggers an outward-propagating shockwave which heats the core's shell and propels it outward. This explosion, or a  supernova, releases energy of the order of $\sim 10^{53}$ erg. While most of this energy is emitted as neutrinos, the remaining energy is dissipated in the form of kinetic energy in the shell and electromagnetic radiation. This phenomenon is referred to as a core-collapse or Type-II supernova.
\paragraph{}
The gravitational collapse of a star can lead to the formation of a neutron star depending on the initial mass. The electron-capture or inverse $\beta$ decay process ($e^- + p \rightarrow n+\nu_e$) create a nuclear composition that is more neutron-rich with increasing core density and heavier nuclei, and the neutrinos generated by the process also carry away energy, further cooling the core \cite{Langanke:2014rya}. The Crab Pulsar, which is a neutron star, lies in the center of the Crab Nebula.
\subsection{Components of the Crab Nebula}
The Crab nebula is a \textit{plerion-}type or \textit{plerionic} SNR \cite{Weiler:1978, Bamba:2022spk}. Young neutron stars formed from core-collapse supernovae often create pulsar wind nebulae. These nebulae emit bright nonthermal radiation across a wide spectrum, from radio waves to very high energy $\gamma-$rays, through synchrotron and inverse Compton processes (described in Sections \ref{subsec:ch3-nonthermal-emission-theory} and \ref{subsec:ch3-IC-theory}). In the subsequent sections, the various components that constitute the Crab nebula are discussed.
\subsubsection{Supernova Remnant}
Most of the non-thermal emission observed from Crab are in the UV and X-ray regimes which require high energy particles with lifetimes shorter than the age of the SNR. This prompted the consideration that the central pulsar was injecting relativistic particles into the medium \cite{Rees:1974}. It is assumed that the ultrarelativistic wind from the pulsar terminates in a standing shock in the transition phase from fast wind from the pulsar to the nebula. Positrons and electrons are accelerated at the shock which then move downstream of the shock (undergoing adiabatic losses) and further lose energy through synchrotron emission \cite{Kennel:1984}. The SNR thus appears as a volume-filled nebula. The shock fronts display fine structures called \textit{wisps} \cite{Rees:1974} that display a time variability of the order of days.

\subsubsection{Pulsar}
A pulsar (PSR B0531+21) is located in the center of the nebula. It was first discovered in the optical regime by \cite{Cocke:1969} and found to have a rotational period of $33\,$ms. It is considered to be the most powerful pulsar in our Galaxy \cite{Kuiper:2001ev, MAGIC:2008jib} with a spin-down luminosity of $4.6\times 10^{38}\,\mathrm{erg\,s}^{-1}$. The emission profile of the pulsar is characterised by three components: two pulsed emissions separated by a pulse phase\footnote{The instantaneous pulse phase $\phi$ is described as $\frac{d\phi}{dt}=\frac{1}{P}$, where $P$ is the Pulse period. Since $\phi$ is measured in \textit{turns} of $2\pi$ radians, $0<\phi<1$.} of 0.4 (measured in \textit{turns} of $2\pi$ radians) and a Bridge emission that occurs between the two pulsed emissions. The three components are emitted across radio to $\gamma-$ray energies, however the amplitude of the emission varies across frequencies. The energy from the pulsar is dissipated through winds of ultrarelativistic cold plasma of electrons, positrons and ions \cite{Aharonian:2012zz} which then undergo \textit{Fermi acceleration} by crossing the shock fronts generated via interaction with the surrounding nebular medium.

\subsubsection{Dust filaments}
\begin{figure}[ht]
    \centering
    \includegraphics[width=0.8\textwidth]{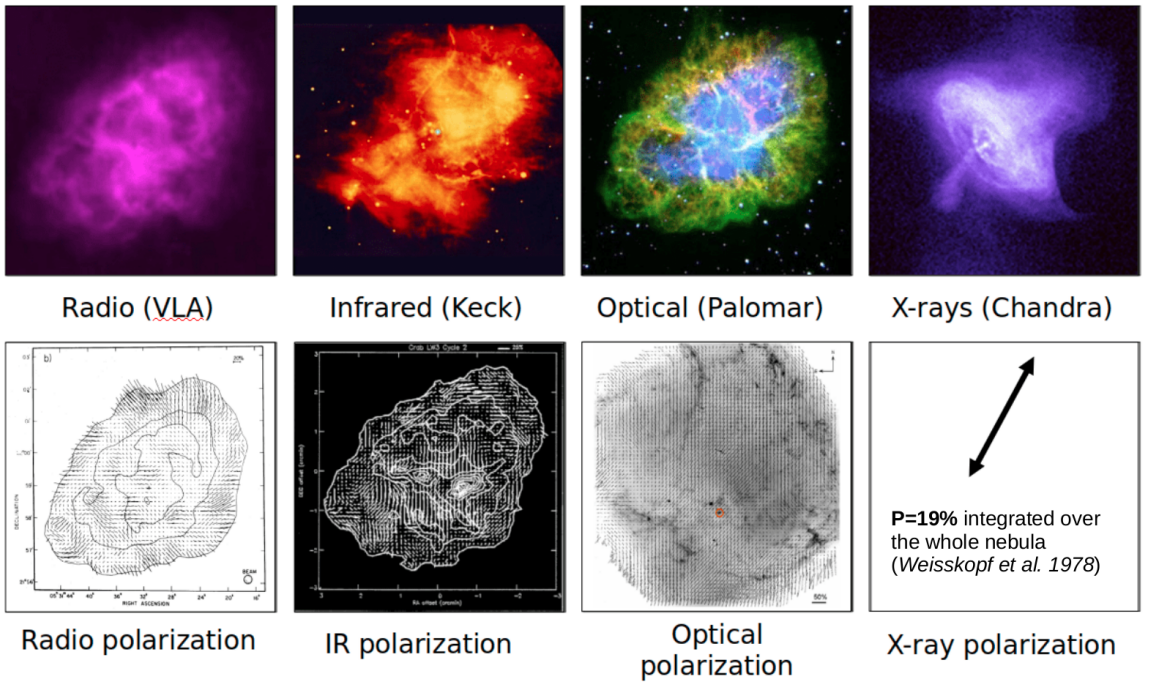}
    \caption{A multiwavelength view of the Crab Nebula. The synchrotron emission from particles ejected by the pulsar wind gyrating in the magnetic field are prominent in the radio and are further enhanced in the filamentary regions. The optical synchrotron emission (blue-green) is accompanied by emission lines from the filaments (red). The X-ray synchrotron emission originates from the jets and inner torus-like region surrounding the pulsar. Polarization maps from the radio to optical wavelengths reveal complex, spatially resolved fields. The images are not of the same scale as the size of the emitting regions decreases towards higher energies due to faster loss of energy by the non-thermal particles at very high energies. Image and edited caption from \cite{Ferrazzoli:2023bib}.}
    \label{fig:crab-multi}
\end{figure}

Another prominent feature of the Crab nebula is line emission in optical and thermal radiation in the infrared/sub-millimetre wavelength ranges. The line emission is prominent in filament-like structures \cite{Hester2008} that extend in the outer regions of the nebula. The line emission is the result of recombination of ions ionized by synchrotron radiation photoionizing the ejected material in the SNR \cite{Temim:2024gcz}.  Dust is prominent in regions of high gas density and emits thermal radiation in the sub-mm regime \cite{Gomez:2012}.
\paragraph{}
A multi-wavelength view of the Crab nebula is shown in Figure \ref{fig:crab-multi} which makes it clear that the different components of the nebula are not independent but complexly intertwined and interacting with each other. In this work, we study the polarization properties of the Crab nebula in the microwave (Chapter \ref{ch:tauA-pol-angle}) and X-ray (Chapter \ref{ch:x-ray}) regimes.

%% file: chapters/chp02.tex
\chapter{Multi-wavelength Probes of Cosmic rays}
\label{ch:observatories}
\begin{center}
    \begin{minipage}{0.75\textwidth}
        \textbf{Summary:} Non-thermal processes due to cosmic rays can be observed over a wide range of wavelengths: from radio to $\gamma-$rays. In this work, we focus on observations of the Inverse Compton and synchrotron emission due to cosmic ray electrons in the microwave (30$-$850 GHz range) and the X-rays ($2-8$ keV range), obtained by three observatories. The three observatories are described in this chapter.\\[3mm]
    \end{minipage}
\end{center}
Multi-wavelength observations of an astrophysical object can serve as a window into understanding the different components that comprise it. For example, with the Crab nebula, while its dust component is bright in the infrared regime, the pulsed emission from the pulsar and the interaction of the pulsar winds with the surrounding supernova remnant medium is bright in the X-ray regime. In this work, we are interested in the 
\begin{itemize}
    \item[-] study of the population of non-thermal electrons that scatter CMB photons to give rise to the ntSZ effect.
    \item[-] the study of polarized synchrotron emission in the Crab nebula across microwave and X-ray regimes in the context of extracting information of astrophysical significance and further application of this knowledge to calibrate instruments of CMB experiments (past and upcoming).
\end{itemize}
Each of the following sections in this chapter describe the observatories that are referred to in the subsequent chapters.
\section{\plk satellite}
\label{sec:ch4-planck}
\begin{figure}[ht]
    \centering
    \includegraphics[width=0.55\textwidth]{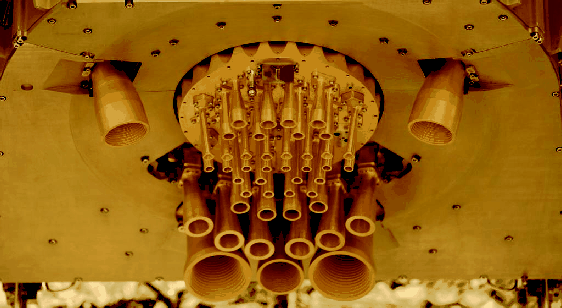}
    \caption{An image of the \textit{Planck} focal plane which shows the HFI horns in the circular structure in the centre surrounded by the LFI horns. Image adapted from \cite{Tauber:2010xyw}.}
    \label{ch2-planck-focalplane}
\end{figure}

\plk Surveyor \cite{Tauber:2010opu} was the European Space Agency's (ESA) third generation space telescope commissioned to observe the CMB. Initially proposed as two separate instruments named the Cosmic Background Radiation Anisotropy Satellite (COBRAS) and the Satellite for Measurement of Background Anisotropies (SAMBA), these projects were competing for a medium-sized mission slot in ESA’s Horizon 2000 program. The two collaborations were recommended to merge their efforts into a single telescope proposal. The combined COBRAS/SAMBA mission \cite{Tauber1994} was selected in 1996 as the third and final medium-sized mission of the Horizon 2000 program and was subsequently renamed \textit{Planck}. \plk was launched on May 14, 2009, together with the \textit{Herschel} Space Telescope aboard an Ariane 5 ECA launch vehicle, destined for the second Earth-Sun Lagrange Point (L$_2$). 
\paragraph{}
The \plk satellite operated from 2009 to 2013, mapping the CMB with two instruments onboard: the Low-Frequency Instrument (LFI, \cite{Mandolesi:2010yw}) and the High-Frequency Instrument (HFI,  \cite{Lamarre:2010meg}). The LFI consisted of actively-cooled pseudo-correlation radiometers that were maintained at a temperature of 20 K and was sensitive across three frequency bands: 30, 44, and 70 GHz. The HFI consisted of 50 bolometers maintained at a temperature of $0.1\,$K and was sensitive across six frequency bands: 100, 143, 217, 353, 545, and 857 GHz. All of the modules except for the 545 and 857 GHz modules were sensitive to linear polarization. \textit{Planck}'s off-axis Gregorian optics featured an unblocked aperture of 1.5 m \cite{Tauber:2010xyw} and both instruments shared the focal plane which can be seen in Figure \ref{ch2-planck-focalplane}.
\paragraph{}
The satellite completed five full mappings of the CMB before the HFI had to be decommissioned due to the depletion of Helium-3 and 4, which was essential for keeping the instrument cryogenics at temperatures of $0.1\,$K. The LFI continued to operate and performed two further all-sky surveys.
\paragraph{}
\textit{Planck} was mainly commissioned to measure the angular power spectrum of the primary temperature and polarization anisotropies of the CMB with higher sensitivity and spatial resolution than the previous missions: COBE and WMAP. However, it has served to provide a range of observations of a range of astrophysical objects from SNRs to galaxy clusters aiding science objectives of astrophysical and cosmological significance. 
\paragraph{}
We make use of \plk data for two objectives:
\begin{itemize}
    \item[-] Measurement of the ntSZ effect and a subsequent estimation of the magnetic field strength within galaxy clusters with radio halos in Chapter \ref{ch:ntSZ} which has been submitted as a paper.
    \item[-] Measurement of the position angle of polarization of the Crab nebula to perform a relative calibration of \plk detectors in Chapter \ref{ch:tauA-pol-angle}.
\end{itemize}

\section{Fred Young Submillimeter Telescope}
\label{sec:ch2-FYST}
The Fred Young Submillimeter Telescope (FYST, pronounced ``feest") \cite{CCAT-Prime2021} is a $6\,$m aperture telescope currently being constructed on the Cerro Chajnantor at an altitude of $\sim5600\,$m above sea level and $40\,$m below the summit. The site offers the best conditions for ground-based observations in the millimeter to mid-infrared wavelengths (terahertz range) \cite{Bustos2014}. The telescope consists of a high-throughput crossed-Dragone optical design \cite{Niemack:16} that is being implemented in a series of telescopes deployed as part of the ``Stage-IV" CMB project. The new optics design allows for high efficiency illumination of $> 10^5$ detectors along with an increase in the diffraction-limited field-of-view (FOV) \cite{Parshley2018}. FYST, for example, will have a focal plane of diameter $1.2\,$m with an FOV ranging from $7.8^\circ$ in diameter at 100 GHz ($3\,$mm) to $2^\circ$ at $860\,$GHz ($0.35\,$mm).
\paragraph{}

\begin{figure}[h!]
    \centering
    \begin{subfigure}[b]{0.5\textwidth}
       \includegraphics[width=\textwidth]{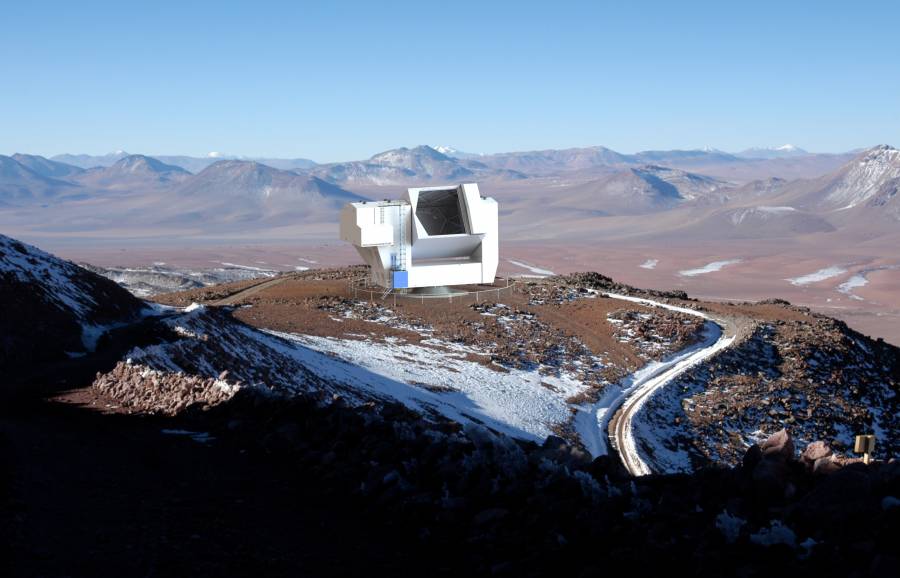}
       \caption{Artist’s impression of the assembled telescope close to the summit of Cerro Chajnantor at an elevation of $\sim 5600$ m above sea level and about $\sim 600$ m above the ALMA site. Image courtesy of Vertex Antennentechnik GmbH.}
    \end{subfigure}
    \centering
    \begin{subfigure}[b]{0.8\textwidth}
        \includegraphics[width=\textwidth]{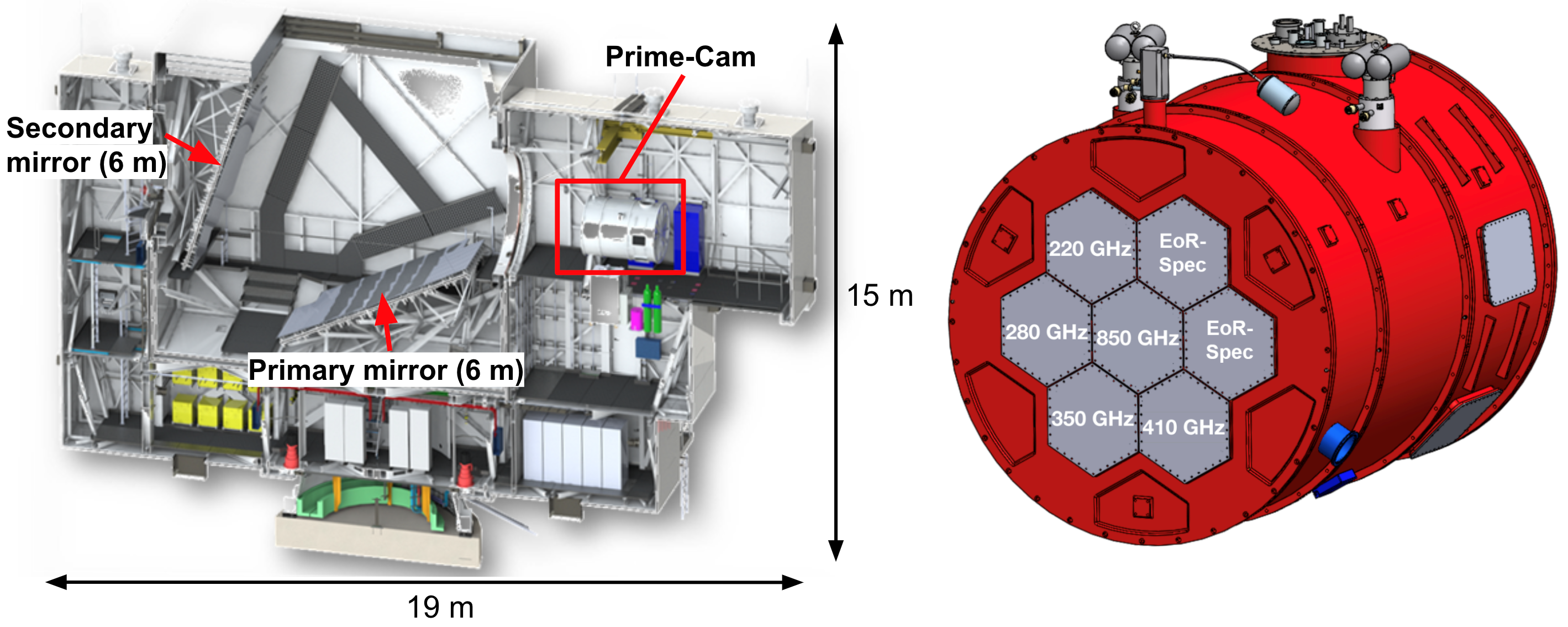}
        \caption{\textit{Left}: A cross-section of the FYST model reveals the $6\,$m primary and secondary mirrors, which focus light into the instrument space where Prime-Cam will be installed. \textit{Right}: A graphic depicting the seven module assembly that comprise the Prime-Cam cryostat. Image is from \cite{Vavagiakis:2022spj}.}
    \end{subfigure}
    \caption{A visualization of FYST and Prime-Cam cryostat.}
    \label{fig:fyst-pcam}
\end{figure}

FYST will host two instruments: Prime-Cam and CCAT Heterodyne Array Instrument (CHAI). Prime-Cam will house seven modules in a $1.8\,$m diameter cryostat \cite{Vavagiakis:2018gen}. It will host up to five broadband dual-polarization-sensitive modules for observations between 220 and 860 GHz with microwave kinetic inductance detector (MKID) arrays. The other two modules, referred to as the Epoch of Reionization Spectrometers (EoR-Spec), will host narrow-band detectors using a Fabry-Perot interferometer with the MKIDs operating between 190 and 450 GHz \cite{Huber:2022ztk}. CHAI is a dual colour (500 and 850 GHz) large format heterodyne array serves as a high resolution spectrometer. An artist's impression of FYST near the summit of Cerro-Chajnantor and a cross-section of the telescope along with a representation of the Prime-Cam cryostat with the frequency modules is shown in Figure \ref{fig:fyst-pcam}.
\paragraph{}
The CCAT collaboration, with the help of FYST employing novel optics, detectors equipped with the latest technology located on a site with good atmospheric transmission in the submillimeter observing wavelengths, would address the following seven science cases \cite{Stacey:2018yqe}:
\begin{enumerate}
    \item \textbf{Evolution of galaxy clusters:} FYST allows the observation of the thermal and kinetic SZ effects of several thousands of galaxy clusters with multi-frequency observations that complement the observations by Advanced Atacama Cosmology Telescope and Simons Observatory \cite{Mittal:2017hwf, Erler2018}. These observations can advance our understanding of dark matter and dark energy, tighten the constraints on $\sigma_8$ and the sum of neutrino masses, expose feedback processes in the ICM, and reveal the nature of cluster Far-Infrared emission. Forecasts for the constraining power of FYST on the non-thermal energy density and the magnetic field strength within the ICM are presented in Chapter \ref{ch:ntSZ}.
    \item \textbf{Modelling CMB foregrounds:} The signatures of primordial gravitational waves that are predicted by inflationary models are encoded in the CMB polarization signal, specifically in the B-modes. Current constraints are hindered by contamination from CMB foregrounds, specifically B-mode patterns introduced by Galactic dust. FYST can aid in better modelling of the polarized Galactic dust emission through the multi-frequency observations, resulting in improved constraints on the tensor-to-scalar ratio, $r$.
    \item \textbf{Epoch of re-ionization (EoR):} FYST will reveal the formation, evolution, and three dimensional large-scale clustering properties of the earliest star forming galaxies in the redshift range $6\leq z \leq 20$ by mapping the [CII]$\,158\,\mu m$ and [OIII]$\,88\,\mu m$ fine-structure lines over tens of square degrees using a Fabry-Perot interferometer (FPI) in conjunction with Prime-Cam. These spectral lines probe star formation in the early Universe which further traces the history of galaxy formation and the growth of large scale structure (LSS).
    \item  \textbf{Trace galaxy evolution from ``Cosmic Noon":} FYST will perform multi-frequency photometric measurements of dusty star forming galaxies (DSFGs) deeply up to the epoch of galaxy formation in the redshift range $1\leq z \leq 3$. The dust in DSFGs re-emits photons absorbed from emission from the stars in the far-infrared range which can be observed by FYST. This enables the reconstruction of the history of star formation and galaxy assembly by mapping the 3D distribution and the spectral energy distribution of dust of DSFGs from the EoR through to Cosmic Noon.
    \item \textbf{Galactic polarization:} Dust grains align with the local magnetic field resulting in linearly polarized dust emission in the millimeter/submillimeter range. The  degree of polarization and morphology encode information on the intrinsic grain properties and properties of the underlying magnetic field (such as strength and direction). FYST can observe this polarized dust emission from the different phases of the interstellar medium (ISM) from sub-pc to kpc scales. Further, star forming regions in the Milky Way and nearby galaxies will also be probed by observations of the [CI] and [CO] spectral lines by the CHAI instrument.
    \item \textbf{Rayleigh Scattering:} The frequency dependent signal produced by Rayleigh scattering of CMB photons by Neutral Hydrogen and Helium atoms during the epoch of recombination has so far eluded detection. A detection of the signal would enable improved constraints on cosmological parameters such as the effective number of relativistic species ($N_\mathrm{eff}$), sum of neutrino masses ($\sum m_\nu$) and aid in the search for primordial non-Gaussianities. With improved foreground mitigation techniques, complemented by data from other CMB experiments such as SO, ACT and \plk, and the multifrequency wide-field observations by FYST, a detection of this signal is now a possibility.
    \item \textbf{Time domain phenomena:} Shocks and energy deposition in the enveloping circumstellar medium and ISM by energetic transients such as supernovae, $\gamma-$ray bursts, X$-$ray binaries and tidal disruption events result in the variability of brightness in the submillimeter regime on timescales of seconds to many years. These non-thermal physical processes tend to be enshrouded by dust. A combination of broad, general survey approaches and specifically targeted campaigns with FYST will complement current established knowledge of the time domain Universe in the radio, ultraviolet, infrared and X$-$ray regimes.
\end{enumerate}
A forecast for all of the science cases with FYST are presented in further detail in \cite{CCAT-Prime2021}.
In order to achieve the goals that have been set for observations with FYST, one must understand how scanning and survey strategies, detector characteristics, and weather conditions affect the observations. This will further inform which are the optimal scanning strategies and map-making procedures to implement. A study of the same is performed with the help of TOD simulations which is the focus of Chapter \ref{ch:TOAST-TOD}.

\section{IXPE}
\label{sec:ch2-ixpe}
\begin{figure}[htbp]
    \centering
    \includegraphics[width=0.6\textwidth]{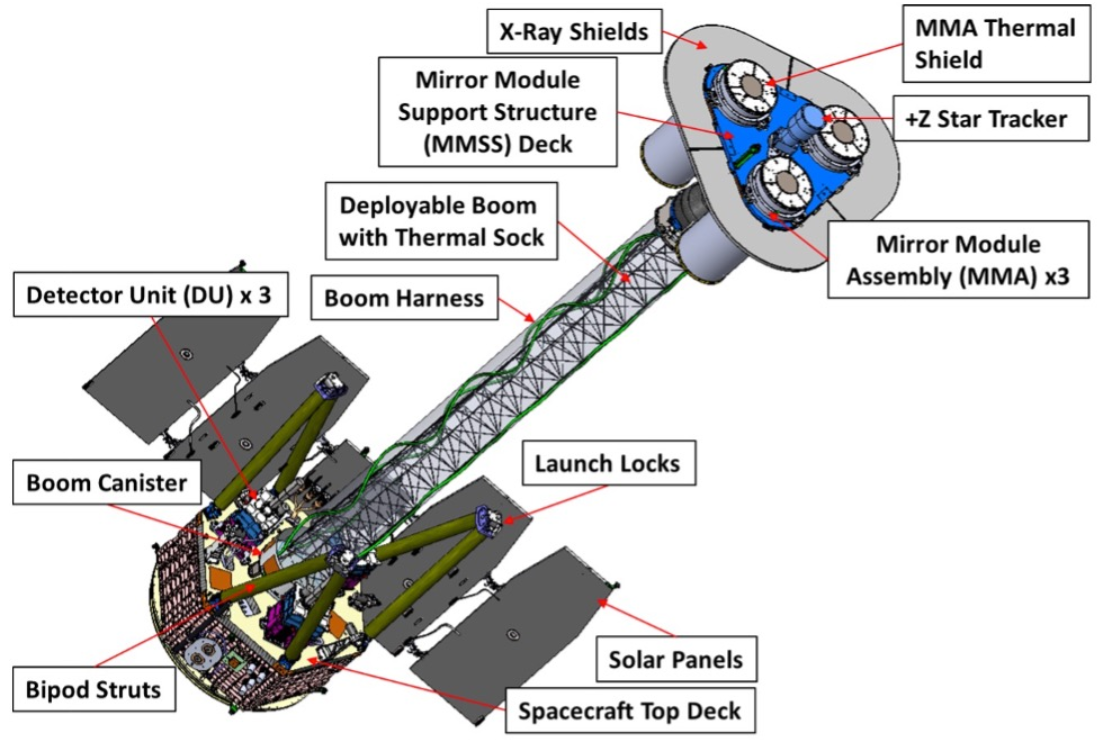}
    \caption{An illustration of the IXPE observatory with some of the payload components labelled. The observatory consists of three identical telescopes each with an MMA and a DU. The deployable boom aligns the MMA and its corresponding DU while maintaining the necessary focal length of each of the three telescopes. The X-ray shield prevents stray X-ray photons from entering the MMA while collimators attached to each DU ensure the on-axis photons are incident on the DUs. The thermal shields protect the MMA, which need to be maintained at $20^\circ\pm5^\circ$ C, from heat loss. Star trackers are used to collect pointing data.  Image adapted from \url{https://heasarc.gsfc.nasa.gov/docs/ixpe/about/}.}
    \label{fig:ixpe}
\end{figure}

The history of X$-$ray astronomy is interesting as the progression of spectroscopy was much faster than the progression of spectro-polarimetry. We shall review first a brief history of X$-$ray observatories and then understand the progression of polarimetry in the X$-$rays. 
\paragraph{}
The first hints of cosmic X-rays were captured with the observations of X-rays from the hot Corona of the Sun by Geiger counters placed on a captured German V-2 rocket by the team led by Herbert Friedman of the Naval Research Laboratory in 1949. The first X-ray detectors were based on Geiger counter technology that were aboard the Uhuru satellite \cite{Koch:1974zz} deployed in 1970 which observed more than 400 sources and helped classify the previously known \textit{X-ray stars} as Neutron stars or black holes in binary systems accreting matter. It also enabled the early detection of X-rays from galaxy clusters. The first X-ray telescope with mirrors was the \textit{Einstein} X-ray observatory \cite{Giacconi:1979kv} that was deployed in 1978 with four nested mirrors that remained operational until 1981. \textit{Einstein} was the first to image shock fronts in SNR and accurately located over 7000 X-ray sources. It also pioneered the study of dark matter in galaxies and clusters of galaxies.
\paragraph{}
During its operational lifetime between 1990 and 1999, the Roentgensatellite (ROSAT) \cite{Truemper:1982, Aschenbach:1988}, observed over 125,000 sources and investigated the atmospheres of stars, it enabled the creation of an extensive catalogue of galaxy clusters. With the Advanced Satellite for Cosmology and Astrophysics (ASCA) \cite{Tanaka:1994mt}, operational from 1993 to 2000, began the implementation of Charge Coupled Device (CCD) detectors, technology which continued to be implemented by XMM-Newton \cite{Lumb:2012wx} and \textit{Chandra} \cite{Weisskopf:2000tx} observatories (with improved sensitivity and precision). Nested mirrors\footnote{Nesting increases the area of an X-ray telescope.} are used on X-ray telescopes to focus X-ray photons onto the detectors (located at the focus). Each mirror is aligned at a small grazing angle with the incident stream of photons to ensure a collimated beam. \cite{Christensen:2022trl} provides a historical review of X-ray optics in Astrophysics.
\paragraph{}
Nine years after the discovery of X-rays by Roentgen, experiments on scattering of these newly discovered \textit{particles} by Charles Glover Barkla revealed that they followed the same principles of polarization of optical light, leading to the demonstration of X-rays as electromagnetic radiation. In the 1970s the exploration of X-ray polarimetry as a window into the study of astrophysical objects began. Ariel-5 and OSO-8, launched in 1974 and 1975 respectively, were the first X-ray satellites with polarimeters that were launched to measure polarized emission in X-rays. They exploited the phenomenon of Bragg diffraction \cite{Bragg:1913} wherein crystals\footnote{LiF and pyrolytic graphite in the case of Ariel-5 and OSO-8, respectively.} were mounted at a specific inclination angle with respect to incident photons to tune for the mounting angle and diffraction energies. The spinning satellite would be pointed slightly off-set from the source. This manoeuvre allowed photons from the source to impinge on the crystal at a slightly different angle during each phase of the spin. This enabled high-resolution spectroscopy of expected Si, S and Fe lines. At an angle of $45^\circ$, the diffractometer was used as a polarimeter around an energy of 2.5 keV. \cite{Costa:2022yaq} provide a historical review of X-ray polarimetry in Astrophysics.
\paragraph{}
Due to the low significance of detections of polarized emission, and complicated manoeuvering of the diffractometers based on Bragg diffraction, polarimetry in X-rays was largely disfavoured in favour of spectroscopy and imaging. In the early 2000s, the development of gas pixel detector (GPD) \cite{Bellazzini:2006bg} technology renewed calls for polarimetry in the X-rays. A GPD is a photoelectric polarimeter. A mixture of gases suspended in the GPD absorb incident X-ray photons to photoelectrons that further ionize the gas leaving photoelectron tracks in the medium. In simple words, the absorption point is used for imaging while the direction and length of the photoelectron tracks indicate the polarization properties and energy of the incident photons, respectively (more details can be found in Appendix \ref{app:gpd}). The GPD has now for the first time enabled spatial-spectral-time resolved X-ray polarimetry.
\paragraph{}
The Imaging  X-ray Polarimetry Explorer (IXPE) observatory \cite{Weisskopf:2016}, launched on December 9, 2021, is the first ever imaging polarimeter commissioned. An illustration of the observatory is shown in Figure \ref{fig:ixpe}. It was launched to acquire polarimetric imaging and spectroscopy from tens of astrophysical sources belonging to different classes. It is sensitive in the 2-8 keV range and was commissioned to observe specific targets over its planned two-year mission.
\paragraph{}
IXPE hosts three identical telescopes with Wolter-I type mirror module assemblies (MMA) \cite{Wolter:1952,VanSpeybroeck:72}, each focusing onto a detector unit (DU) which comprises the GPD \cite{Bellazzini:2006bg, Baldini:2021} with Diemethyl Ether\footnote{Gas with a small diffusion coefficient is a requirement to minimize smearing of photoelectron information.} as the suspended gas. A GPD consists of a gas cell with a Beryllium window, a Gas Electron Multiplier (GEM) that amplifies the charge of the photoelectrons emitted from the suspended gas by incident X-ray photons and a pixellated charge collection plane \cite{Sauli:1997qp} that is connected to readout electronics.
\paragraph{}
Polarized emission can be induced by ordered magnetic fields and by geometrical asymmetries resulting from highly dynamic components like jets, pulsar wind nebulae and disks. Polarimetry thus aids the study of the morphology and dynamical states of compact objects such as Active Galactic Nuclei (AGN), pulsars, blazars and extended sources such as Supernova Remnants (SNR). In this work, we analyzed observations by IXPE of the Crab nebula to measure its polarization properties, which is the focus of Chapter \ref{ch:x-ray}.

%% file: chapters/chp03.tex
\chapter{Physical Processes}
\label{ch:physical-processes}
\begin{center}
    \begin{minipage}{0.75\textwidth}
        \textbf{Summary:} Energy loss by non-thermal populations of electrons in radio halos and supernova remnants such as the Crab Nebula occur through Inverse Compton scattering and Synchrotron emission. The effects of new physics on CMB polarization, such as the cosmic birefringence, introduces a rotation of the plane of linear polarization. This chapter introduces each of these physical processes which will serve as a theoretical basis for upcoming chapters.\\[3mm]
    \end{minipage}
\end{center}

We discussed radio halos and the Crab nebula in Chapter \ref{ch:astro-objects}. Although the thermal emission from these objects is the dominant component across a wide frequency range, the non-thermal emission is also observed across multiple frequencies and we consider the Inverse Compton (IC) and synchrotron radiation processes to be the most relevant for energy loss by the non-thermal electron populations. These phenomena are discussed in Sections \ref{subsec:ch3-nonthermal-emission-theory} and \ref{subsec:ch3-IC-theory}.
\section{Non-thermal processes: Synchrotron emission}
\label{subsec:ch3-nonthermal-emission-theory}
Historically, synchrotron emission is one of the most important non-thermal continuum radiation from astrophysical sources, first discovered in the radio frequencies \cite{Shklovsky1954}. It is emitted by charged particles gyrating in the presence of magnetic fields at relativistic velocities. If the magnetic field is projected along the $z-$direction, the charged particle's path is helical with circular motion in the $x-y$ plane and a non-zero velocity in the $z-$direction due to the magnetic Lorentz force. From an $observer$ frame of reference this appears as radiation from a narrow cone with an opening angle $\theta \approx 1/\gamma$, where $\gamma$ is the Lorentz factor. $\gamma=\sqrt{1-\frac{v^2}{c^2}}$ where $v$ is the velocity of the electron and $c$ is the speed of light. The power emitted by a particle of mass $m$ scales as $\sim m^{-2}$, which causes synchrotron radiation from electrons to be $3\times10^6$ stronger than from protons. We thus associate synchrotron radiation to arise from relativistic electrons or positrons, commonly referred to as leptonic cosmic rays. 
\paragraph{}
This section follows the prescriptions defined in \cite{Rybicki} and \cite{blumenthal1970}. We first describe the electric and magnetic fields of radiation due to a moving charge and estimate the power radiated by an electron per frequency per unit solid angle. We then estimate the same for a distribution of non-thermal (\textit{relativistic}) electrons and compute the received power due to synchrotron emission.
\subsubsection{Motion of a particle in a Magnetic field}
Let us consider the radiation emitted by an electron of mass $m_\mathrm{e}$ and charge $e$ moving in a magnetic field $\mathbf{B}$ with velocity $\mathbf{v}$. The equations of motion of this system according to classical electrodynamics are 
\begin{subequations}
    \begin{equation}
        \frac{d}{dt}(\gamma m_\mathrm{e} \mathbf{v}) = \frac{e}{c}\mathbf{v}\times\mathbf{B},
        \label{eq:synch-vB}
    \end{equation}
    \begin{equation}
        \frac{d}{dt}(\gamma m_\mathrm{e} c^2)=e\mathbf{v\cdot E} = 0.
        \label{eq:synch-vE}
    \end{equation}
\end{subequations}
These can also be derived by equating Newton's equation for the the four-force with the equation for the Lorentz four-force as
\begin{equation}
    m_\mathrm{e}a^\mu = \frac{e}{c}F^\mu_\nu\,U^\nu,
\end{equation}
where $F_\nu^\mu$ is the electromagnetic field tensor, $U^\nu$ is the four-velocity. Eq.\ \eqref{eq:synch-vE} shows that $\gamma$ and the norm of the velocity $\vert \mathbf{v}\vert$ are constant. From eq.\ \eqref{eq:synch-vB} it is apparent that the velocity parallel to the magnetic field orientation ($v_\parallel$)is constant and consequantly so is $v_\perp$. Hence, the electron's motion is helical with the acceleration perpendicular to both magnetic field and the velocity vector. The electron's gyration frequency is
\begin{equation}
    \nu_g = \frac{e\,B}{2\pi \gamma m_\mathrm{e}c}.
    \label{eq:synch-nu-g}
\end{equation}
\paragraph{}
For non-relativistic accelerated charges, the total emitted power is given by Larmor’s formula
\begin{equation}
    P = \frac{2}{3}\frac{e^2\,\mathbf{a}^2}{c^3},
\end{equation}
where $\mathbf{a}$ is the three vector acceleration. To estimate the emitted power for relativistic particles we first consider the system in the instantaneous rest frame of the particle and find a covariant expression in this frame which would be valid in any frame. This is found to be
\begin{align}
    P &= \frac{2e^2}{3c^3}\gamma^4(a_\perp^2+\gamma\,a_\parallel^2)\\
    & = \frac{2e^2}{3c^3}r_0^2\gamma^2B^2\vert\mathbf{v}\vert^2\mathrm{sin}^2\theta,
\end{align}
where $r_0=\frac{e^2}{m_e\,c^2}$ is the classical radius of the electron, $\theta$ is the pitch angle (the angle between $\mathbf{v}$ and $\mathbf{B}$). An integration over all pitch angles yields for the total emitted power:
\begin{equation}
    P_\mathrm{sync}(\gamma) = \frac{4}{3}\sigma_T\,c\,\beta^2\gamma^2U_B,
\end{equation}
where $U_B=\frac{B^2}{8\pi}$ is the energy density of the magnetic field and $\sigma_T$ is the cross-section of Thomson scattering.
\paragraph{}
From the point of view of an external, not co-moving, \textit{observer}, the radiation is emitted from the electron in a forward direction into a
narrow cone with half opening angle $\theta=1/\gamma$. Due to geometry, the observer sees the orbit of the electron as an ellipse in projection. The radiation emitted by a single oscillating charge is thus elliptically polarized. The expression for the energy emitted per unit solid angle per frequency range along the two polarization states is given by \cite{Rybicki}
\begin{subequations}
    \begin{equation}
         \frac{dW_\perp}{d\nu} = \frac{\sqrt{3}e^2\gamma}{2c}[F(x)+G(x)],\\
    \end{equation}
   \begin{equation}
       \frac{dW_\parallel}{d\nu} = \frac{\sqrt{3}e^2\gamma}{2c}[F(x)-G(x)],
   \end{equation}
 \label{eq:dW-polstates}
\end{subequations}
where
\begin{equation}
    F(x) = x\int_{x}^\infty d\xi \, K_{5/3}(\xi),\quad G(x) = x\,K_{2/3}(x),
    \label{eq:synch-fx-gx}
\end{equation}
and $K_n(x)$ is the Bessel function of order $n$ and $x=\frac{\nu}{\nu_c}$. The power per frequency is estimated by dividing eqs.\ \eqref{eq:dW-polstates} by the orbital period $T = 2\pi\gamma\,m_e\,c/e\nu_B$ to obtain
\begin{subequations}
    \begin{equation}
        P_\perp(\nu) = \frac{\sqrt{3}e^3B\mathrm{sin}\theta}{4\pi m\,c^2}[F(x) + G(x)]
        \label{eq:synch-P-perp}
    \end{equation}
    \begin{equation}
        P_\parallel(\nu) = \frac{\sqrt{3}e^3B\mathrm{sin}\theta}{4\pi m\,c^2}[F(x) - G(x)].
        \label{eq:synch-P-par}
    \end{equation}
\end{subequations}
Finally, the total emitted power per frequency is
\begin{align}
    P(\nu) &= P_\perp(\nu) +  P_\parallel(\nu) \nonumber \\
    \label{eq:synch-total-1}
\end{align}
\subsubsection{Electron moving perpendicular to Magnetic field}
We shall now consider the specific case of an electron moving perpendicular to the magnetic field (i.e. $\theta=\pi/2$). 
The total instantaneous power radiated by such an electron, from eqs.\ \eqref{eq:synch-fx-gx} and \ref{eq:synch-total-1} is,
\begin{equation}
    P(\nu) = \frac{\sqrt{3}e^3\,B}{m_e\,c^2}\frac{\nu}{\nu_c}\int_{\frac{\nu}{\nu_c}}^\infty d\xi \, K_{5/3}(\xi), \quad \nu_c = \frac{3e\,B\,\gamma^2}{4\pi m_ec},
    \label{pressure1}
\end{equation}
where
$B=\vert\mathbf{B}\vert$ is the magnetic field strength, $\nu_c$ is the critical frequency and $\mathrm{K_{5/3}(x)}$ is the modified Bessel function of the second kind of order 5/3.
\subsubsection{Electron moving with arbitrary pitch angle}
In order to estimate the observed synchrotron emission from an electron with an arbitrary pitch angle, we first calculate the emission in the coordinate system (K$^\prime$) that is moving with a relative velocity $\beta_\parallel$ with respect to \textit{observer} frame (K). $\beta_\parallel$ is the velocity component of the electron which is parallel to the magnetic field. The Lorentz factor connecting K and K$^\prime$ is $\gamma \approx \mathrm{sin}^{-1}\theta$. The following quantities are now defined in frame K$^\prime$ as,
\begin{equation*}
    \gamma^\prime = \gamma \mathrm{sin}\theta,\quad \nu^\prime=\nu \mathrm{sin}\theta, \quad \Omega ^\prime = \Omega/\mathrm{sin}\theta.
\end{equation*}
where $\Omega$ is the rotation frequency. It can be shown that the electron energy loss rate is Lorentz invariant. Thus, $P_\mathrm{emitted} = P^\prime$ which implies
\begin{equation}
    P_\mathrm{emitted}(\nu) = \Big(\frac{d\nu^\prime}{d\nu}\Big)P^\prime(\nu^\prime)=\mathrm{sin}\theta\,P^\prime(\nu^\prime).
\end{equation}
Substituting for $\gamma^\prime,\,\nu^\prime$ in eq.\ \eqref{pressure1},
\begin{equation}
    P_\mathrm{emitted}(\nu) = \frac{\sqrt{3}e^3\,B\,\mathrm{sin}\theta}{m_\mathrm{e}\,c^2}\,x\int_x^\infty d\xi \, K_{5/3}(\xi),
    \label{pressure2}
\end{equation}
where $x=\frac{\nu}{\nu_c}$ and the critical frequency in eq.\ \eqref{pressure1} is redefined as
\begin{equation}
    \nu_c = \frac{3e\,B\,\gamma^2}{4\pi m_\mathrm{e}c}\,\mathrm{sin}\theta.
    \label{eq:ch4-critfreq}
\end{equation}
Finally, $P_\mathrm{received} = P_\mathrm{emitted}\,\mathrm{sin}^{-2}\theta$. The factor of $\mathrm{sin}^{-2}\theta$ however becomes irrelevant when the electrons are restricted to the volume of a medium (as is the case with all astrophysical sources). We shall now compute the synchrotron emission by a distribution of electrons.
\subsection{Total synchrotron emission from a distribution of electrons}
\begin{figure}
    \centering
    \begin{subfigure}{0.49\textwidth}
           \includegraphics[width=\textwidth]{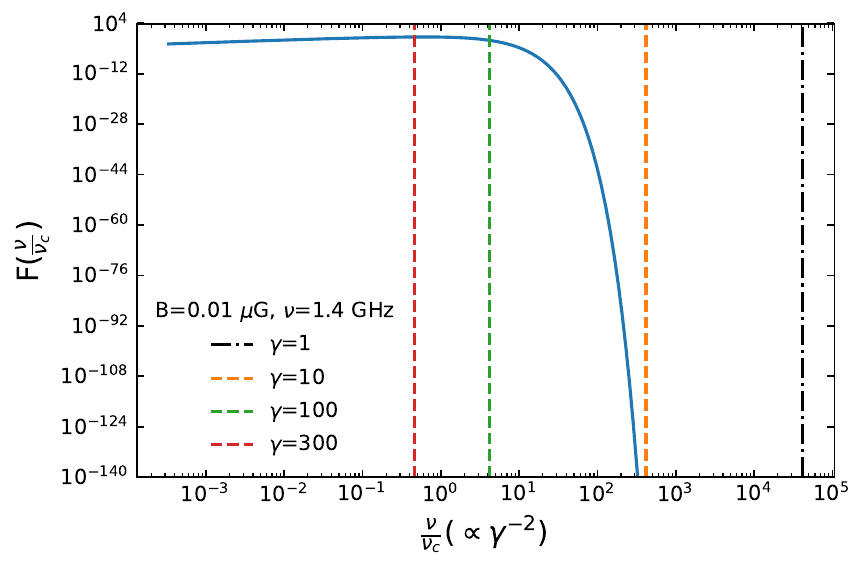}
    \end{subfigure}
    \begin{subfigure}{0.49\textwidth}
           \includegraphics[width=\textwidth]{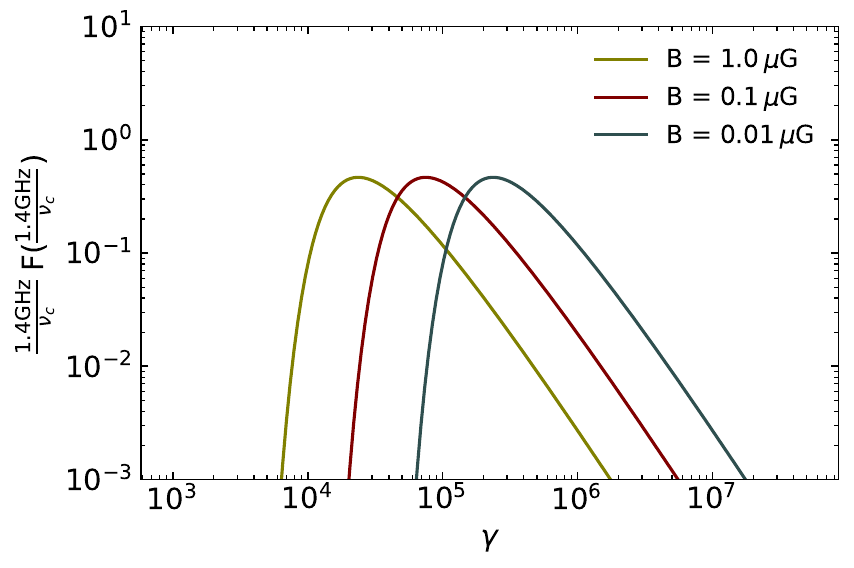}
    \end{subfigure}
    \caption{\textit{Left}: $F(x)$ is plotted as a function of $x$. The vertical dashed lines correspond to $x$($\gamma$). For the given $B$, the synchrotron emitting electrons at an observing frequency of 1.4\,GHz are $\gamma>10$ . \textit{Right}: The different curves correspond to a variation in the magnetic field strength. The curve shifts to the left for increasing B.}
    \label{fig:ch2-synch-fxb}
\end{figure}

Let $N_{obs}(\gamma,\theta)\,d\gamma\,d\Omega_\theta$ be the total number of observed electrons within a volume element with energy within $d\gamma$ and pitch angle within $d\Omega_\theta$. The received synchrotron spectrum is
\begin{equation}
    \frac{dW}{d\nu\,dt} = \int\int P_\mathrm{received}\,N_\mathrm{obs}(\gamma,\theta)\,d\gamma\,d\Omega_\theta.
        \label{dW1}
\end{equation}
The observed distribution of synchrotron emitting electrons is related to the true distribution of synchrotron emitting electrons as
\begin{equation}
    N_\mathrm{obs}(\gamma,\theta) = N(\gamma,\theta)\,\mathrm{sin}^2\theta.
\end{equation}
Rewriting eq.\ \eqref{dW1}, we have
\begin{equation}
    \frac{dW}{d\nu\,dt} = \int\int P_\mathrm{emitted}\,N(\gamma,\theta)\,d\gamma\,d\Omega_\theta.
    \label{dW2}
\end{equation}
Consider a power-law distribution of electrons written as\footnote{If the distribution is assumed to be locally isotropic and independent of pitch angle, it reduces to $N(\gamma)=k\gamma^{-\alpha}$}
\begin{equation}
    N(\gamma,\theta) = \frac{k}{4\pi}\,\gamma^{-\alpha}\,\frac{\mathrm{sin}\theta}{2}, \quad \gamma_1<\gamma<\gamma_2.
    \label{electronspec1}
\end{equation}
The total synchrotron emission spectrum per unit volume due to this power-law distribution of electrons is computed by substituting for $P_\mathrm{emitted}(\nu)$ from eq.\ \eqref{pressure2}, and $N(\gamma,\theta)$ in eq.\ \eqref{electronspec1} to obtain,
\begin{equation}
    \frac{dW}{d\nu\,dt} = \frac{\sqrt{3}k\,e^3B}{8\pi m_\mathrm{e}c^2}\int_0^\pi \int_{\gamma_1}^{\gamma_2} \mathrm{sin}^2\theta\,\gamma^{-\alpha}\,x\int_x^\infty d\xi\,K_{5/3}(\xi)\,d\gamma d\Omega_\theta.
    \label{eq:dW3}
\end{equation}
The $\gamma$ is related to the pitch angle through the critical frequency. Rewriting $\gamma$ in terms of $x$ starting from eq.\ \eqref{eq:ch4-critfreq},
\begin{equation*}
    \gamma = \sqrt{\frac{4\pi m_e\,c}{3e\,B}} \nu^{-\frac{1}{2}}(\mathrm{sin}\theta)^{-\frac{1}{2}}\,x^{-\frac{1}{2}}
\end{equation*}
\begin{equation}
    \Longrightarrow \gamma^{-\alpha}d\gamma = \Bigg(\frac{4\pi m_e\,c\,\nu}{3e\,B}\Bigg)^{-\frac{\alpha-1}{2}}\,x^{\frac{\alpha-3}{2}}(\mathrm{sin}\theta)^{\frac{\alpha-1}{2}}\,dx.
    \label{gdg}
\end{equation}
After the variable change, eq.\ \eqref{gdg}
is incorporated into eq.\ \eqref{eq:dW3} to obtain\footnote{The negative sign arises due to flipping of the limits on the integral with respect to $x$ i.e. by making $x$($\gamma_2$) and $x$($\gamma_1$) the lower and upper limits respectively.}
\begin{equation}
    \frac{dW}{d\nu\,dt} = -\frac{\sqrt{3}k\,e^3}{8\pi m_\mathrm{e}c^2}\,B^{\frac{\alpha+1}{2}}\Bigg(\frac{4\pi m_\mathrm{e}c\,\nu}{3e}\Bigg) ^{-\frac{\alpha-1}{2}}\int_0^\pi \int_{x(\gamma_2)}^{x(\gamma_1)}(\mathrm{sin}\theta)^{\frac{\alpha+3}{2}} x^{\frac{\alpha-1}{2}}\int_x^\infty d\xi\,K_{5/3}(\xi)\,dx d\Omega_\theta
    \label{eq:dW3-2}
\end{equation}
Making the assumption that the extreme parts of the electron energy spectrum do not contribute to the emission i.e. $\gamma_1$ and $\gamma_2$ are such that $\nu_c(\gamma_1)\ll \nu$ and $\nu_c(\gamma_2)\gg \nu$. This assumption implies that $0<\gamma< \infty$.
\paragraph{} Solving the integrals in eq.\ \eqref{dW1}, one obtains for the total synchrotron emission per unit volume,
\begin{equation}
    \frac{dW}{d\nu dt} = \frac{\sqrt{3}\sqrt{\pi}k\,e^3}{m_e\,c^2}B^\frac{\alpha+1}{2}\Bigg(\frac{3e}{2\pi m_ec}\Bigg)^\frac{\alpha-1}{2}\nu^{-\frac{\alpha-1}{2}}\,a(\alpha),
    \label{dW4}
\end{equation}
where
\begin{equation}
  a(\alpha) = \frac{\Gamma(\frac{\alpha}{4}+\frac{19}{2})\Gamma(\frac{\alpha}{4}-\frac{1}{12})\Gamma(\frac{\alpha+5}{4})}{(\alpha+1)\Gamma(\frac{\alpha+7}{4})}.
\end{equation}
\paragraph{}
To understand which part of the electron spectrum is contributing to the synchrotron emission, one can take a look at the integration with respect to the variable $x$. Let us consider $F(x)=x\int_x^\infty d\xi K_{5/3}(\xi)$ with $x=\frac{\nu}{\nu_c}$. The function is computed for $\nu = 1.4\,$GHz, $B=0.01\,\mu$G and pitch angle $\theta=\pi/2$, and plotted with respect to $x$ in the \textit{left} panel of Figure \ref{fig:ch2-synch-fxb}. It can be seen that $F(x)$ peaks around $x=1$ ($\nu \approx \nu_c$) and flattens towards lower $x$ (higher $\gamma$). The contribution to synchrotron emission occurs for $\gamma>10$ in this configuration which further informs our choice of the scattering electrons' distribution in momenta, namely $\gamma_1$ and $\gamma_2$ in eqs.\ \eqref{eq:dW3} and \ref{eq:dW3-2}. The variation of $xF(x)$ with respect to the magnetic field is shown in the \textit{right} panel of Figure \ref{fig:ch2-synch-fxb}. $xF(x)$ for a fixed electron momentum distributions falls significantly with lower $B$. Thus, as we consider populations of scattering electrons at lower energies, one would observe lower synchrotron flux at a given observing frequency with decreasing magnetic field strength.
\subsection{Polarization of synchrotron emission}
\label{subsec:pol-synch}
In the previous section we have estimated the total synchrotron power emitted by a population of non-thermal electrons in a spherically symmetric volume of medium. Due to the orientation of the underlying magnetic field, the emitted synchrotron radiation is linearly polarized. The state of the linear polarization over a period of time is dependent on the medium across which the photons traverse as disordered magnetic fields can also cause a \textit{de}-polarization of the linearly polarized synchrotron emission. We shall compute the linear polarized synchrotron radiation emitted by a population of non-thermal electrons with a power-law distribution in momenta. This is in the context of describing the observed polarized flux density from Crab nebula\footnote{We do not consider the short time-scales of energy loss due to synchrotron emission from turbulent regions such as knots in this section which occur due to the interaction of pulsar winds with the surrounding supernova remnant medium. We instead only consider the non-thermal electrons present in the bulk of the SNR with relatively longer time-scales} which are relevant to Chapters \ref{ch:tauA-pol-angle} and \ref{ch:x-ray}.
\paragraph{}
As described in eq.\ \eqref{eq:synch-fx-gx}, the polarized emission from one gyrating electron is given by
\begin{align}
    P(\nu)^\mathrm{pol}_\mathrm{emitted} &= \frac{\sqrt{3}e^3\,B}{m_e\,c^2}xG(x)\nonumber\\
    &= \frac{\sqrt{3}e^3\,B}{m_e\,c^2}\frac{\nu}{\nu_c}K_{2/3}(\frac{\nu}{\nu_c})
    \label{eq:synch-pol-1}
\end{align}
where $K_{2/3}(x)$ is the Bessel function of order $2/3$ and the polarized emission is perpendicular to the projected orientation of the magnetic field in the sky frame along the line-of-sight. For a power-law spectrum of electrons, one needs to consider the scenario of an ordered magnetic field as a disordered or randomly oriented magnetic field can result in zero polarized emission due to de-polarization across each of the randomly oriented magnetic field over the course of the electron's trajectory. 
Finally, the degree of polarization is estimated from eqs.\ \eqref{eq:synch-P-perp} and \eqref{eq:synch-P-par} as
\begin{equation}
    \Pi(\nu) = \frac{P_\perp(\nu)-P_\parallel(\nu)}{P_\perp(\nu)+P_\parallel(\nu)}=\frac{G(x)}{F(x)}
\end{equation}
and for a power-law distribution of electrons,
\begin{equation}
    \Pi = \frac{p+1}{p+\frac{7}{3}}.
\end{equation}
\section{Non-thermal processes: Inverse Compton}
\label{subsec:ch3-IC-theory}
Another process that should take place in the ICM is IC scattering. Non-thermal electrons inverse Compton scatter off CMB photons, thereby transferring some of their momentum to the photons. As a result, the CMB photons can be up-scattered to the up to the X-ray regime. In this section we will present the spectrum resulting from CMB photons scattering off a population of electrons.
The IC effect in the Thomson limit is discussed in detail in Chapter \ref{ch:ntSZ}. Here we shall consider the Klein-Nishina limit. This limit is an important electron energy loss mechanism for electrons with energies of the order $\langle\epsilon\rangle E >>m^2c^4$, where $\langle\epsilon\rangle$ is the characteristic photon energy field \cite{blumenthal1970}. The IC kernel function describing the resultant photon spectrum from scattering off of non-thermal electrons in the KN regime is
\begin{align}
    F_{IC} (\epsilon,\gamma, \epsilon_1) =& \frac{3\sigma_Tc}{4\gamma^2}\frac{n(\epsilon)}{\epsilon} \nonumber \\
    &\times\Bigg[2q\,\mathrm{ln}q+(1+2q)(1-q)+\frac{1}{2}\frac{(\Gamma q)^2}{1+\Gamma q}(1-q)\Bigg]
\end{align}
where $\epsilon_1$ is the energy of the scattered photons, $n(\epsilon)$ is the number of photons per unit volume per unit initial photon energy $\epsilon$ and,
\begin{equation}
    \Gamma = \frac{4\epsilon\gamma}{m_\mathrm{e}c^2}, \quad q=\frac{\epsilon_1}{\Gamma(\gamma m_\mathrm{e}c^2-\epsilon_1)}.
\end{equation}
For a power-law distribution of scattering electrons, the power per unit energy per unit volume is then
\begin{equation}
    P_{IC}(\epsilon_1) = \int_0^\infty\int_{\gamma_\mathrm{min}}^{\gamma_\mathrm{max}} N(\gamma)F_{IC}(\epsilon,\gamma, \epsilon_1)\,d\gamma d\epsilon.
\end{equation}

\section{Polarization of the CMB}
\begin{figure}
    \centering
    \includegraphics[width=0.3\textwidth]{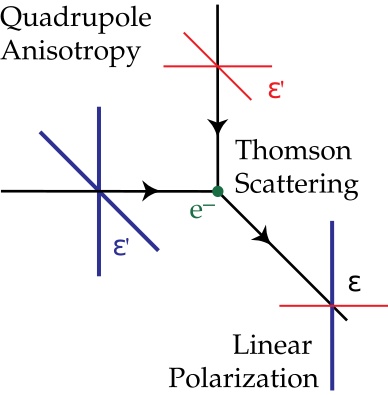}
    \caption{An illustration of the local quadrupole anisotropy of the intensity of a photon field incident on a free electron which undergoes Thomson scattering resulting in outgoing linearly polarized light. The \textit{blue} (\textit{red}) colour represents radiation of higher (lower) intensity. Image adapted from \cite{ODell:2001amw}.}
    \label{fig:cmb-thomson-scattering}
\end{figure}

Under the paradigm of the gravitational instabilities in the early Universe in which small anisotropies in the density field form the large scale structure observed today, the same anisotropies are theorized to introduce polarization of the anisotropies in the CMB. The Thomson scattering of CMB photons by free electrons can cause linearly polarized light. Polarization of outgoing radiation after Thomson scattering only occurs when the intensity of the incident radiation to a free electron locally has a non-vanishing quadrupole moment. Figure \ref{fig:cmb-thomson-scattering} shows an illustration of the Thomson scattering of an incident photon field with a local quadrupole moment by a free electron resulting in linearly polarized light.
\paragraph{}
The polarization of the CMB is governed by Thomson scattering and the presence of a quadrupole moment of the intensity of the photon radiation. High electron density prior to recombination meant the mean free path of photons was too short to result in a quadrupole anisotropy, as any anisotropies in the photon-baryon fluid would get \textit{erased} by subsequent interactions. After recombination, the electron density is too low for Thomson scattering. Thus, a small window of time during the epoch of last scattering, when the Universe was still ionized, was opportune for quadrupole anisotropies to develop in the photon flux and undergo Thomson scattering to introduce a polarized component of the CMB. Quadrupole anisotropies in the CMB photon field can arise due to velocity gradients in the baryon-photon fluid at the epoch of last scattering \cite{Zaldarriaga:2003bb} and due to gravitational waves (transverse-traceless perturbations to the metric) \cite{Hu:1997hv}. The polarization fraction is $\sim 10\%$ at scales of tens of arcminutes and is directly related to the time-scale over which last scattering occurred \cite{Hu:1997hv}. Late re-ionization is also known to introduce polarization of the CMB at large scales \cite{Zaldarriaga:1996ke, Kaplinghat:2002vt}.

\paragraph{}
The electric field vector of a monochromatic electromagnetic wave propagating along $z-$axis towards larger $z$ is \cite{Chandrashekhar1960}
\begin{subequations}
    \begin{align}
        E_x &= A_x\,\mathrm{cos}(\omega t),\\
        E_y &= A_y\,\mathrm{cos}(\omega t - \varphi),
    \end{align}
\end{subequations}
where the phase lag between the $E_x$ and $E_y$ components is indicated by $\varphi$  that determines the polarization state of the wave and $E_x$ ($E_y$) is proportional to the instantaneous electric field along the $x-$axis ($y-$axis). Linearly polarized light results from oscillations of the $E_x$ and $E_y$ in phase or in anti-phase i.e. when $\varphi=0$ or $\pi$. When $\varphi=\pm \pi/2$, the light is circularly polarized, and any other values of $\varphi$ result in elliptically polarized light. The Stokes parameters are also useful to describe the electromagnetic wave and are written as \cite{Rybicki}
\begin{subequations}
    \begin{align}
    I &= A_x^2 + A_y^2,\\
    Q &= A_x^2 - A_y^2,\\
    U &= 2\,A_xA_y\mathrm{cos}\varphi,\\
    V &= 2\,A_xA_y\mathrm{sin}\varphi.
\end{align}
\label{eq:ch4-stokes-IQUV}
\end{subequations}
Stokes $I$ is the intensity (or flux) of the wave, $Q$ and $U$ describe the linear polarization properties, while $V$ describes the circular polarization properties of the wave. The polarization introduced via Thomson scattering of the CMB photons is linearly polarized and hence, we do not consider the Stokes $V$ parameter while describing the polarization properties of the CMB.
\subsection{\pdfmath{E}- and \pdfmath{B}-mode  polarization}
The Stokes $Q$ and $U$ (defined in eq.\ \eqref{eq:ch4-stokes-IQUV}) parameters are variant under coordinate transformations, which do not make for ideal variables to describe CMB polarization. A clockwise rotation of the basis vectors of an angle $\psi$ transforms the $Q(\hat{n})$ and $U(\hat{n})$ as 
\begin{equation}\label{eqn: QiUtrans}
    Q(\hat{n})\pm iU(\hat{n}) \to \left[ Q(\hat{n})\pm iU(\hat{n})\right]e^{\pm2i\psi}\,.
\end{equation}
The combinations $Q(\hat{n})\pm iU(\hat{n})$ are also variant under transformation. One can express them in terms of the $_{\pm 2}Y_{\ell m}$ spin-weighted spherical harmonics as
\begin{equation}
    Q(\hat{n})\pm iU(\hat{n})=\sum_{\ell=2}^\infty\sum_{m=-\ell}^{\ell} {}_{\pm2}a_{\ell m}\,{}_{\pm2}Y_{\ell m}(\hat{n})\,,
    \label{eq:QiU-Ylm}
\end{equation}
with spherical harmonic coefficients
\begin{equation}\label{eqn:defAlm2}
    _{\pm 2}a_{\ell m}=\int \operatorname{d}^2\!\hat{n}\, \left[Q(\hat{n})\pm iU(\hat{n})\right] {}_{\pm{2}}Y_{\ell m}^*(\hat{n})\,.
\end{equation}
The summation in eq.\ \eqref{eq:QiU-Ylm} starts from $\ell=2$ as polarization results from quadrupole moments of the temperature (or intensity) of the CMB. The fields \pdfmath{E}$(\hat{n})$ and \pdfmath{B}$(\hat{n})$ are then the linear combinations of the spherical harmonic coefficients \cite{Zaldarriaga:1996xe, Kamionkowski:1996ks}:
\begin{subequations}
\begin{align}
      a_{\ell m}^E &\equiv -\Bigg[\frac{{_{+2}a_{\ell m}} + {_{-2}a_{\ell m}}}{2}\Bigg]\,,\\
      a_{\ell m}^B &\equiv i\Bigg[\frac{{_{+2}a_{\ell m}} - {_{-2}a_{\ell m}}}{2}\Bigg]\,,
\end{align}
\label{eqn:spinalm}
\end{subequations}
which are now invariant under coordinate transformations. These are appropriately used to describe CMB polarization. $E-$modes describe parity-even fields while $B-$modes are a parity-odd pseudoscalar field as they transform according to $a_{\ell m}^E\rightarrow a_{\ell m}^E (-1)^\ell$ and $a_{\ell m}^B \rightarrow a_{\ell m}^B(-1)^{\ell+1}$, respectively.
\paragraph{}
The statistics of the anisotropies in the CMB are described by the angular power spectra which are estimated as
\begin{equation}
    C_\ell^{XY} = \frac{1}{(2\ell +1)}\sum_{m=-\ell}^\ell a_{lm}^X\,a_{lm}^{Y^*}, 
    \label{eq:ch4-cell-xy}
\end{equation}
where $X$ and $Y$ can be $T$, $E$ or $B$. $TT$, $EE$, $BB$ and $TE$ correlations are parity-even while $TB$ and $EB$ are parity-odd due to the way that the $a_{lm}^E$ and $a_{lm}^B$ transform.

\subsection{Information encoded in the polarization of the CMB}
The polarization of the CMB encodes information about various physical processes that occurred in the early Universe and in the subsequent epochs. Some of the physical phenomena that are currently open questions in Cosmology are
\paragraph{$B-$modes from inflationary models}
The initial conditions for scalar, vector and tensor perturbations \cite{Grishchuk:1974ny, Starobinsky:1979ty} of the density are expected to be sourced by inflation. The scalar perturbations comprise the total density perturbations while the vector perturbations are damped upon re-entry into horizon. Tensor perturbations are predicted by several inflationary models to arise from primordial gravitational waves (PGW) \cite{Grishchuk:1974ny, Starobinsky:1979ty}. Many inflationary models predict the power spectra of scalar and tensor perturbations to be described by power-laws. The scalar perturbations source both temperature and $E-$mode anisotropies in the CMB while $B-$modes are sourced only from the PGW \cite{Seljak:1996gy, Kamionkowski:1996zd}. The relative amplitude of the ratio of the tensor-to-scalar modes, $r$, is model-dependent and can be inferred from primordial $B-$mode angular power spectra \cite{Kamionkowski:2015yta, Komatsu:2022nvu}. \cite{Tristram:2021tvh} find $r<0.032$ (at 95\% CL) from BICEP and \plk data.
\paragraph{$EB$ correlation due to cosmic birefringence}
Recent works have discovered a non-null $EB$ correlation produced by an isotropic cosmic birefringence angle, $\beta \approx 0.3^\circ$, in \plk and \textit{WMAP} data \cite{Minami:2019ruj, Minami:2020fin, Minami:2020odp, Diego-Palazuelos:2022cnh, Eskilt:2022wav, Eskilt:2022cff} that hint at signatures of parity-violating physics. Such a birefringence signal of cosmological origin can be attributed to a parity-violating coupling of a pseudo-scalar field to the CMB (an electromagnetic field). If the pseudo-scalar field is time-dependent, it would introduce a rotation of the plane of linear polarization of the CMB \cite{Carroll:1989vb, Lue:1998mq, Komatsu:2022nvu}. $\beta$ denotes the angle by which the rotation has occured from last scattering to today. A detection of $\beta$ would help constrain models of dark matter and dark energy \cite{Marsh:2015xka, Ferreira:2020fam}.
\paragraph{}
With upcoming CMB experiments, the statistical significance of the constraints on $r$ and $\beta$ are expected to improve. However, the $EB$ correlation of Galactic dust \cite{Clark:2021kze}, mixing of $E-$ and $B-$modes due to systematic effects and CMB lensing \cite{Zaldarriaga:1998ar}, and miscalibration of the orientation of polarization sensitivity of CMB detectors remain some of the limiting factors in this endeavour. 
\subsection{Miscalibration of polarization angle}
Experiments that aim to measure the CMB polarization to probe new physics, need to achieve calibration of instruments to great accuracy. Here we outline the effects of miscalibration of the orientation of polarization of CMB detectors on \pdfmath{E}- and \pdfmath{B}-mode  polarization.
\subsubsection{\pdfmath{E}- and \pdfmath{B}-mode mixing due to miscalibration}
\label{sec:ch2-cmbpol-ebmixing}
A miscalibration of the instrument, caused by perhaps a global rotation of the focal plane of the instrument or a relative rotation of the detectors with respect to the focal plane by a miscalibration angle $\alpha$ would introduce a clockwise rotation of the plane of linear polarization of the CMB.
Under a clockwise rotation of the plane of polarization by a miscalibration angle $\alpha$, Stokes $Q$ and $U$ transform according to
\begin{equation}
    Q^\mathrm{obs}(\hat{n})\pm i U^\mathrm{obs}(\hat{n}) = \big[Q(\hat{n})\pm iU(\hat{n})\big]\mathrm{e}^{\pm2i\alpha},
\end{equation}
where we have considered $\psi=\alpha$ from eq.\ \eqref{eqn: QiUtrans}, and $Q^\mathrm{obs}$ 
 and $U^\mathrm{obs}$ are the measured Stokes parameters. Further, the spherical harmonic coefficients transform as
 \begin{equation}
     _{\pm2}a_{\ell m}^\mathrm{obs} =\, _{\pm2}a_{\ell m}e^{\pm2i\alpha}.
 \end{equation}
 The observed \pdfmath{E}- and \pdfmath{B}-mode coefficients can now be written in terms of the intrinsic coefficients defined in eq.\ \eqref{eqn:spinalm} as,
\begin{subequations}
    \begin{equation}
    a_{lm}^{E,\mathrm{obs}} = a_{lm}^E\mathrm{cos}(2\alpha)-a_{lm}^B\mathrm{sin}(2\alpha),
    \end{equation}
    \begin{equation}
      a_{lm}^{B,\mathrm{obs}} = a_{lm}^B\mathrm{cos}(2\alpha)+a_{lm}^E\mathrm{sin}(2\alpha).
    \end{equation}
\end{subequations}
As described in eq.\ \eqref{eq:ch4-cell-xy}, the observed angular power spectra are written in terms of the intrinsic power spectra as
\begin{subequations}
    \begin{equation}
    C_\ell^{EE,\mathrm{obs}} = C_\ell^{EE}\mathrm{cos}^2(2\alpha)+C_\ell^{BB}\mathrm{sin}^2(2\alpha)-C_\ell^{EB}\mathrm{sin}(4\alpha),
    \end{equation}
    \begin{equation}
        C_\ell^{BB,\mathrm{obs}} = C_\ell^{EE}\mathrm{sin}^2(2\alpha)+C_\ell^{BB}\mathrm{cos}^2(2\alpha)+C_\ell^{EB}\mathrm{sin}(4\alpha),
    \end{equation}
    \begin{equation}
        C_\ell^{EB,\mathrm{obs}} = \frac{1}{2}\Big(C_\ell^{EE}-C_\ell^{BB}\Big)\,\mathrm{sin}(4\alpha)+C_\ell^{EB}\mathrm{cos}(4\alpha).
        \label{eq:Cell-EB-obs}
    \end{equation}
\end{subequations}
It is evident from eq.\ \eqref{eq:Cell-EB-obs} that systematic effects such as a global rotation of the focal plane of the instrument and/or a miscalibration of the relative orientation of polarization-sensitive detectors can result in non-zero $C_\ell^{EB,\mathrm{obs}}$. In such an instance, the self-calibration method of CMB experiments that involves setting $C_\ell^{EB,\mathrm{obs}}=0$ can introduce a bias on the observed angular power spectra due to miscalibration. The effect of $\alpha$ on $C_\ell^{EB,\mathrm{obs}}$ is degenerate with the effect introduced by $\beta$ as it introduces a similar rotation of the plane of polarization of the CMB. Thus, it is difficult to disentangle the rotation introduced by cosmic birefringence and that by instrumental systematics. These effects highlight the need for independent calibration methods.
\paragraph{}
One way of measuring $\alpha$ is by calibrating the CMB detectors with the measured position angle of polarization of an astrophysical source, such as the Crab nebula. With this intention, we measure the position angle of polarization of the Crab nebula in the microwave regime with \plk data and in the X$-$rays with IXPE data in Chapters \ref{ch:tauA-pol-angle} and \ref{ch:x-ray}, respectively.

%% file: chapters/chp04.tex
\chapter{Constraining the average magnetic field in galaxy clusters with current and upcoming CMB surveys}
\label{ch:ntSZ}
\section{Introduction}
\label{sec:intro}

Galaxy clusters (GCs) are the largest gravitationally bound aggregation of matter with masses up to $~10^{15}\,M_\odot$, formed at the nodes of filaments in the cosmic web. They are formed through mergers of smaller clusters and groups of galaxies, and through accretion \cite{Kravtsov2012} from the intergalactic medium surrounding filaments of galaxies (e.g., \cite{Peebles1970, Press1973, Voit2004}). The intracluster medium (ICM) is the primary reservoir of baryons within these nodes, accounting for roughly 15\% of the total cluster mass \cite{White1993, Vikhlinin2006}, and exhibiting a complex interplay between hot ionized plasma, turbulence, and an underlying extended magnetic field \cite{Sarazin1986}.
\paragraph{}

These astrophysical processes determine the observable properties of GCs and multi wavelength observations are necessary to understand the roles of the different ICM components. The dominant (and cosmologically relevant) ICM component is the bulk plasma following Maxwell-Boltzmann distributions within a range of temperatures, made visible by the thermal bremssstrahlung emission in the X-rays, or from the thermal Sunyaev-Zeldovich (tSZ) effect \cite{Sunyaev:1970bma, Sunyaev:1972eq} distortion in the cosmic microwave background (CMB). To a large extent, this dominant thermal component remains unaffected by the presence of the non-thermal particles and magnetic fields, although in specific, localized regions such as radio lobes, the latter can dominate the plasma dynamics \cite{Prokhorov2010, Battaglia2011, Eckert2018, Bykov2019}. However, as the sensitivity of our measurements improve, the modeling uncertainties arising from these non-thermal components will play an increasingly important role. Further, their understanding can offer new insights to probe the origin and dynamics of the large-scale structure of the Universe.
\paragraph{}

The main evidence that non-thermal electrons and magnetic fields exist in the inter-galactic space in GCs comes from different types of observations of diffuse synchrotron emission at radio wavelengths \cite{Brunetti2014, vanweeren2019}. Typically, the observed morphology of diffuse synchrotron emission can be classified into (i) cluster radio relics which are of irregular shape and trace merger shocks, (ii) radio halos which are centrally located and generally much more extended than the relics, and (iii) revived active galactic nucleus (AGN) fossil plasma sources which trace AGN plasma re-energised by various physical processes in the ICM. In this work we focus on the radio halos (RHs), as these are the only truly cluster-wide non-thermal emission whose morphologies have been shown to follow closely that of the ICM (e.g., \cite{Govoni2001, Rajpurohit2018, Botteon2020}). 
While the origin of RHs remains unclear, a general consensus has arisen behind a turbulent re-acceleration model, in which populations of seed electrons are locally re-accelerated due to turbulent states of the ICM, following the case of GC mergers (e.g., \cite{Pasini2022, Ruszkowski2023}). Despite its observational success over competing theories, the turbulent re-acceleration model suffers from uncertainty about the source and the energy distribution of the seed electrons that need to be fixed \textit{posteriori} from observational data. A direct measurement or constraints on the cluster-wide non-thermal electron spectral energy distribution (SED) is therefore a much-valued quantity.
\paragraph{}

Magnetic field strength in the diffuse ICM is measurable from the observations of the Faraday Rotation Measure (FRM) (which is inferred from observations of polarized synchrotron emission at multiple wavelengths) of the embedded or background radio sources with intrinsic polarization \cite{Heiles1976, Verschuur1979}. The synchrotron emission alone cannot be used to directly estimate the magnetic field, as it depends on the product of the non-thermal particle density and some power of the magnetic field strength. The FRM data remains sparse due to a lack of suitably positioned background sources at cosmological distances, and is also sensitive to the local environment of the polarized sources, susceptible to biases arising from the location of polarized sources, and foregrounds \cite{Johnson2020, Osinga2022}. 
In this regard, measurement of the inverse-Compton (IC) emission in combination with the synchrotron emission has been considered as the most promising way to constrain cluster-wide magnetic fields, as the former depends only on the SED of the non-thermal electrons (when the incoming radiation source is known), and helps to break the degeneracy with the magnetic field strength in the synchrotron data. The predominant case of incoming radiation is the CMB, which when scattered by the $\sim$GeV energy non-thermal electrons, results in the excess IC emission that extends to X-ray and gamma-ray regimes \cite{Sarazin1999, Wik2012, Ackermann2015, Xi2017}.

\paragraph{}

Measurement of the excess IC emission in the X-rays has been a decades-long endeavour, with mixed success \cite{Rephaeli2006, Million2008, Ota2013, Mernier2023}. The main difficulty lies in the limited sensitivity of the X-ray instruments in the hard X-ray energies, which is absolutely critical for distinguishing the IC emission component from the multi-temperature and multi-keV plasma's thermal emission \cite{Wik2012, Wik2014, Cova2019, RojasBolivar2020}. In this regard, the measurement of the same IC effect in the millimeter/submillimeter domain is now poised to make a decisive contribution, in light of the unprecedented depth of many recent and upcoming CMB sky surveys. The relevant physical phenomenon is the non-thermal Sunyaev-Zeldovich (ntSZ) effect, which, in contrast to the dominant tSZ effect, concerns the scattering of CMB photons from the non-thermal electron populations \cite{Ensslin2000, Mroczkowski:2018nrv}. There is a long history of ntSZ research in the context of GCs and radio lobes of AGN \cite{Ensslin2000, Colafrancesco2003, Colafrancesco2008, Colafrancesco2012, Acharya2020, Marchegiani2021}, although no direct detection has been made of the global ntSZ signal in GCs, apart from one measurement localized to known X-ray cavities in the ICM \cite{Abdulla2019}.
\paragraph{}

Our goal in this paper is to show that the current and upcoming CMB data are very close to making a measurement of the global IC excess,  and we place meaningful constraints on the magnetic fields in GCs from these data. Specifically, we study whether the \textit{Planck} satellite’s all-sky survey data and newer catalogs of radio halo clusters (i) can provide any constraints on the SED of the non-thermal electrons from modelling the ntSZ signal, and (ii) can potentially place constraints on the magnetic field strength by combining these constraints with the existing synchrotron flux measurements. From there, we explore the constraining power of upcoming CMB experiments such as Simons Observatory (SO) \cite{SimonsObservatory2018} and Fred Young Submillimeter Telescope (FYST) \cite{CCAT-Prime2021} on the ntSZ effect and further on the magnetic field strengths.
\paragraph{}

The rest of this paper is organized as follows. In Section \ref{sec:theory}, we formulate the modelling of the ntSZ effect and synchrotron emission in GCs, and discuss the assumptions we have made in this work. In Section \ref{sec:data}, we discuss the data and simulated microwave sky maps from which current and future constraints on the ntSZ signal, the non-thermal electron density, and magnetic field strength are measured, respectively. We discuss the methods implemented in the extraction of the SZ spectrum from \textit{Planck} data, and the fitting procedure in Section \ref{sec:fit}. We present the results in Section \ref{sec:results} which are then discussed in Section \ref{sec:summary}. We assume a flat $\Lambda$CDM cosmological model with the parameter values $\Omega_m = 0.308$ and $H_0 = 67.8\,\mathrm{km}\,\mathrm{s}^{-1}\mathrm{Mpc}^{-1}$ \cite{Planck2015cosmo} throughout this paper.  


\section{Theoretical basis}
\label{sec:theory}
This section describes the theoretical framework of our analysis. As outlined in the introduction, our method of finding the signature of the ntSZ signal or putting upper limits on non-thermal electron density (which translates to lower limits on the magnetic field strength) is based on three assumptions:
\begin{enumerate}
\item non-thermal electron pressure follows the same radial distribution as the thermal electron pressure,
\item non-thermal electrons follow one single power-law momentum distribution throughout the cluster volume, and 
\item the magnetic field energy density follows the number density of non-thermal electrons.
\end{enumerate}

These assumptions can be considered too simplistic to capture the complexity of the ICM, but they simplify the data analysis and allow us to place meaningful first constraints. Furthermore, there is some degree of theoretical and observational support for at least the first and third assumptions. Below we describe our motivation behind adopting these three criteria.

The first assumption on spatial distribution of non-thermal electrons allows us to create a 2D matched filter (Section \ref{sec:fit}) to optimally extract the cluster ntSZ signal, along with the thermal SZ (tSZ) signal, from the maps. Combined with the second assumption, this also enables us to obtain the density profile of non-thermal electrons by assuming that they have the same pseudo-temperature (Section \ref{sec:pseudoT}) throughout the emitting volume. Evidence that the non-thermal pressure density closely follows that of thermal electrons have been shown in several simulations of CR transport (\cite{Pinzke:2010st, Zandanel:2013wja}). We specifically refer to the results from \cite{Pinzke:2010st} which show that the ratio $X_\mathrm{CR} = P_\mathrm{CR}/P_\mathrm{th}$ stays approximately constant, within a narrow range, for a Coma-like disturbed cluster out to a large fraction of the virial radius. Since our sample of RH clusters are generally all disturbed systems, it will be reasonable to assume that the non-thermal pressure profile thus closely follows that of the thermal pressure.
\paragraph{}
The second assumption is merely a tool for simplifying the calculations, although it can easily be relaxed for more complicated models. By assuming a uniform, global power-law distribution we ignore the effects of electron ageing, reacceleration etc., however, we do compare results for four different power-law distributions to assess the impact of this simplistic assumption.
\paragraph{}
Lastly, the third assumption of a universal magnetic field radial profile in RH clusters is not critical for our analysis, but it enables a more realistic calculation of the synchrotron power and comparison with radio data, as opposed to assuming a constant $B$ value. This energy equipartition argument leads to magnetic field strength scaling roughly to the square-root of the thermal electron density, $B(r) \propto n_\mathrm{e,th}(r)^{0.5}$. Observational evidence for such a scaling have been found by \cite{Murgia2004,Bonafede2010}
and discussed in the context of MHD simulations by \cite{Vazza2018}.

\subsection{Characteristics of the ntSZ effect}
\label{sec:ntsz}
The distortion in specific intensity due to the ntSZ effect can be written as
\begin{equation}
    \delta i(x) = (j(x) - i(x))\: I_0 \: \tau_{\mathrm{e,nth}}, \quad \tau_{\mathrm{e,nth}} = \sigma_T\int n_{\mathrm{e,nth}} dl,
    \label{eq:di1}
\end{equation}
where $x=\frac{h\nu}{k_\mathrm{B}T_{\mathrm{CMB}}}$, $I_0 = 2\frac{(k_\mathrm{B}T_{\mathrm{CMB}})^3}{(hc)^2}$ is the specific intensity of the CMB, $i(x) =\frac{x^3}{e^x-1}$ is the Planck spectrum attributed to the CMB spectrum, $\tau_{\mathrm{e,nth}}$ is the optical depth due to non-thermal electrons, and $j(x)$ is the flux scattered from other frequencies to frequency x. For a given isotropic electron momenta distribution $f_\mathrm{e}(p)$ (where p is the normalized electron momentum, $p=\frac{p_{\mathrm{phys}}}{m_\mathrm{e}c}$ and $p_{\mathrm{phys}}=\beta_\mathrm{e}\gamma_\mathrm{e}$) with normalization $\int_0^\infty f_\mathrm{e}(p)p^2\,dp = 1$, the ntSZ effect can be described as \cite{Ensslin2000}
\begin{equation}
  \delta i(x)  = \Bigg[\Bigg(\int_{p_1}^{p_2}\int_{-s_\mathrm{m}(p_1)}^{s_\mathrm{m}(p_2)}f_\mathrm{e}(p) K(e^\mathrm{s};p)\,e^\mathrm{s}\, \frac{(x/e^\mathrm{s})^3}{(e^{x/e^\mathrm{s}}-1)}\, ds\, dp \Bigg) - \frac{x^3}{e^x-1}\Bigg]I_0\tau_{\mathrm{e,nth}},
  \label{eq:di2}
\end{equation}
where $s_\mathrm{m} (p) = \mathrm{ln}\Big[\frac{1+\beta_\mathrm{e}}{1-\beta_\mathrm{e}}\Big]$ is the maximum logarithmic shift in energy with $\beta_\mathrm{e}=\frac{p}{\sqrt{1+p^2}}$ and the photon scattering kernel \cite{Ensslin2000}
\begin{equation}
\begin{split}
    K(e^\mathrm{s};p) = &-\frac{3(1-e^\mathrm{s})}{32p^6e^\mathrm{s}}\big[1+(10+8p^2+4p^4)e^\mathrm{s}+e^{2s}\big]\\
    &+\frac{3(1+e^\mathrm{s})}{8p^5}\Bigg[\frac{3+3p^2+p^4}{\sqrt{1+p^2}}-\frac{3+2p^2}{2p}(2\mathrm{arcsinh}(p)-\arrowvert s\arrowvert)\Bigg].
\end{split}
    \label{eq:ksp}
\end{equation}
The amplitude and shape of the spectrum of the ntSZ effect is dependent on the number density and the momentum distribution of the scattering non-thermal electrons. In this work, we consider power-law and broken power-law models for the scattering non-thermal electrons with different minimum and maximum momenta.

\subsubsection{Power-law distribution} 
The simplest and most commonly used distribution of non-thermal electron momenta would be a negative power-law, with fixed minimum ($p_1$) and maximum ($p_2$) momenta, and power-law index ($\alpha$). Imposing the normalization of $\int_0^\infty f_\mathrm{e}(p)p^2\,dp = 1$, this power-law is written as
\begin{equation}
    f_\mathrm{e}(p;\alpha,p_1,p_2) = A(p_1, p_2, \alpha)p^{-\alpha}, \quad \mathrm{where} \quad A(p_1, p_2, \alpha)=\frac{(\alpha -1)}{(p_1^{1-\alpha} - p_2^{1-\alpha})}.
    \label{eq:electrondist1}
\end{equation}
With the assumption that the same scattering electrons cause synchrotron radiation, $\alpha$ is related to the spectral index of synchrotron emission, $\alpha_{\mathrm{synch}}$, as $\alpha = 2\alpha_{\mathrm{synch}}+1$ \cite{Rybicki}.

This simple case can be improved by considering a broken power-law to mimic radiative energy losses at the low-energy end of the spectrum. 
For modelling the distribution of non-thermal electron momenta with a broken power-law, we fix the minimum ($p_1$), break ($p_{\mathrm{br}}$) and maximum ($p_2$) momenta, and take $\alpha_1$ and $\alpha_2$ as the indices of the flat and power-law parts of the model. This broken power-law can then be written as \cite{Colafrancesco2003}
\begin{equation}
f_\mathrm{e}(p;p_1,p_2,p_{\mathrm{br}},\alpha_1,\alpha_2)=
   C(p_1,p_2,p_{\mathrm{br}},\alpha_1,\alpha_2)
    \begin{cases}
      p^{-\alpha_1} & p_1 < p < p_{\mathrm{br}} \\
      p_{\mathrm{br}}^{-\alpha_1+\alpha_2}p^{-\alpha_2} & p_{\mathrm{br}} < p < p_2\\
      \end{cases}.
      \label{eq:electrondist2}
\end{equation}
As with the power-law model, we consider $\alpha_2 = 2\alpha_\mathrm{synch}+1$, and the normalization factor $C(p_1,p_2,p_{\mathrm{br}},\alpha_1,\alpha_2)$ arises due to the condition that $\int_0^\infty f_\mathrm{e}(p)p^2\,dp = 1$. We choose a small, non-zero power-law index for the ``flat'' part of the broken power-law model for ease of numerical integration.
\paragraph{Adopted model parameters:} 
We consider four different cases of electron momentum distribution in this paper: Two single power-law distributions with $p_1=30$ and 300, respectively (cases S1 and S2); and two broken power-law distributions with $p_\mathrm{br}$=300 and 1000, respectively (cases B1 and B2). We fix $\alpha_\mathrm{synch}=1.3$, meaning the indices of the power-laws are fixed to $\alpha = \alpha_2 = 3.6$ in Eqs. (\ref{eq:electrondist1}) and (\ref{eq:electrondist2}). Together with a dominant thermal component with ICM temperature 8 keV, whose momentum is characterized by a Maxwell-J\"{u}ttner distribution (Appendix \ref{app:rSZ}), these model parameters are used in turn to fit the match-filtered peak signal. These model parameters are summarized in the Table \ref{tab:models} below.

\begin{figure}[ht]
    \centering
    \includegraphics[width=1.0\textwidth]{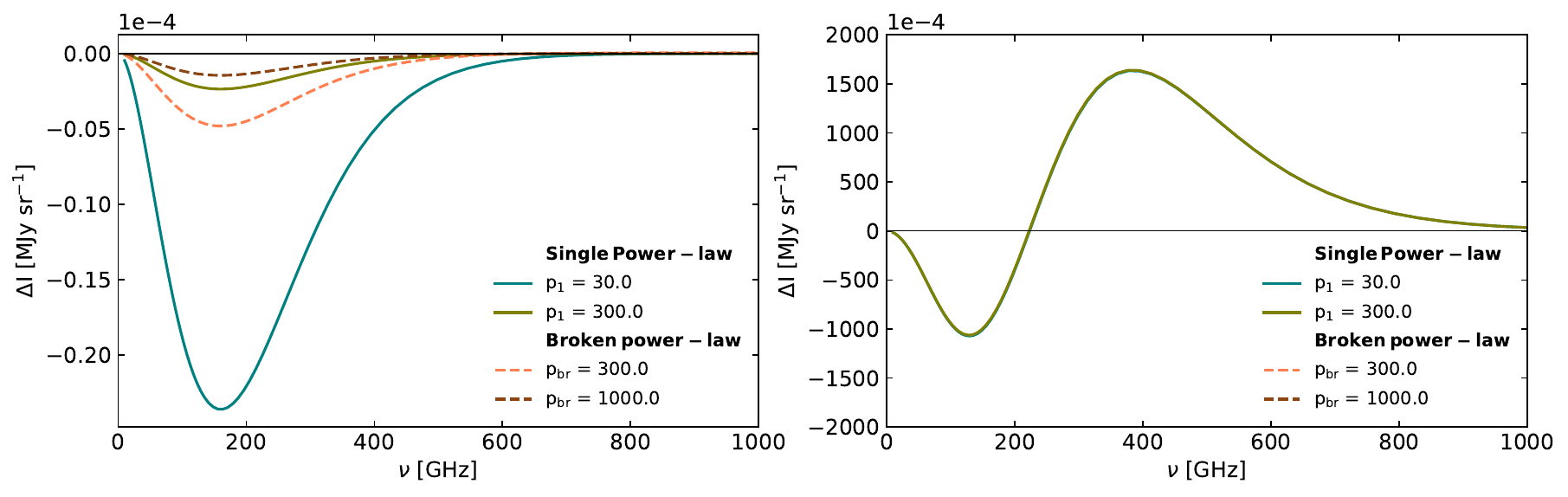} 
    \caption{\textit{Left}: Distortion in the CMB specific intensity introduced by the ntSZ effect due to \textit{single} power-law and \textit{broken} power-law non-thermal electron models with $y_\mathrm{nth}=10^{-6}$. Amongst the models considered, the amplitude of the distortion is largest for the single power-law with $\mathrm{p_{1}}=30$. \textit{Right}: The total SZ spectrum that consists of the tSZ$_\mathrm{rel}$ with $k_\mathrm{B}T_\mathrm{e}=8.0\,$keV and $y_{th}=10^{-4}$, kSZ effect with $v_\mathrm{pec}$ derived from a Gaussian distribution with $\sigma=100\, \mathrm{km\,s}^{-1}$, and ntSZ effect estimated for each of the non-thermal electron models with $y_\mathrm{nth}=10^{-6}$. The distinction from the dominant tSZ effect is not visible in this linear-scale plot.}
    \label{fig:ntSZ-spectra}
\end{figure}

\begin{table}[ht]
        \centering
    \begin{tabular}{ccc}
    \hline
     Components & Model & Parameters \\
    \hline
    &  & \\
    $\mathrm{tSZ_{rel}}$ ($k_\mathrm{B} T_\mathrm{e}$ = 8 keV)  &   \hspace{-1em}\rdelim\{{2}{*}
    S1 & $p_1 = 30$,  $p_2 = 10^5$, 
    $\alpha=3.6$ \\
    
    $+$ {\it single} power-law &   S2 & $p_1 = 300$,  $p_2 = 10^5$, $\alpha=3.6$ \\ 
    
    & & \\ 
    
    $\mathrm{tSZ_{rel}}$ ($k_\mathrm{B} T_\mathrm{e}$ = 8 keV)  &   \hspace{-1em}\rdelim\{{2}{*}
    B1 & $p_1 = 1$,  $p_\mathrm{br} = 300$, $p_2 = 10^5$,  $\alpha_1=0.05$, $\alpha_2 = 3.6$ \\
    
    $+$ {\it broken} power-law &   B2 & $p_1 = 1$,  $p_\mathrm{br} = 1000$, $p_2 = 10^5$, 
    $\alpha_1=0.05$, $\alpha_2 = 3.6$  \\    
    & & \\ 
    \hline    
       \end{tabular}
            \caption{Adopted parameters for the single and broken power-law models, along with the fixed-temperature thermal component, that are used in the spectral fitting.}
    \label{tab:models}
\end{table}
\paragraph{} 
The distortion in the CMB specific intensity introduced by the ntSZ effect, with the assumption of non-thermal electron models described in Table \ref{tab:models}, is shown in Figure \ref{fig:ntSZ-spectra}. The distortion due to the S1 model is the largest as, under our definition of the normalization of the electron momenta distributions, more non-thermal electrons are available to scatter the CMB photons. We also notice that the spectra are shallower for higher $p_1$ or $p_\mathrm{br}$ and the frequency at which the distortion is zero is shifted to higher frequencies. In the \textit{right} panel of Figure \ref{fig:ntSZ-spectra}, we also show the total SZ effect (tSZ$_\mathrm{rel}$, kSZ (described in Appendix \ref{app:ksz}) and ntSZ) wherein we see the characteristic shape of the spectrum of the SZ effect with a decrement in the specific intensity of the CMB at frequencies $<217$ GHz and an increment at frequencies $> 217$ GHz. The tSZ$_\mathrm{rel}$ is the dominant effect and thus, it is difficult to disentangle the distortions due to the other SZ effects.

\subsubsection{The zero-crossing frequency}
\begin{figure}[h]
    \centering
    \includegraphics[width=\textwidth]{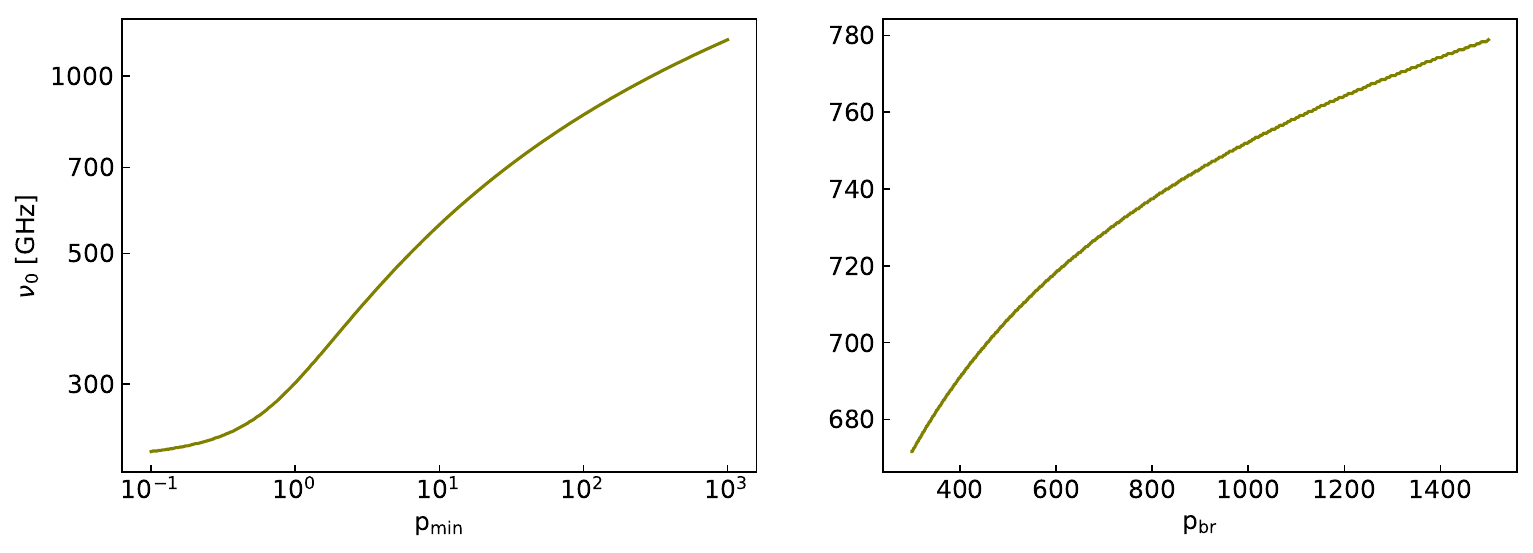}
    \caption{Frequency of zero-distortion in  specific intensity as a function of p\textsubscript{1} (left) and p\textsubscript{br} (right) which are parameters used to define the power-law and broken power-law models to compute the ntSZ effect [Eqs.  (\ref{eq:electrondist1}$-$\ref{eq:electrondist2})].}
    \label{fig:nuzero}
\end{figure}

Observations at submillimeter frequencies (roughly, above 300 GHz) are important for finding the spectral signature of the ntSZ signal. A characteristic feature of any inverse-Compton spectral distortion is the frequency at which there is no net distortion. For the ntSZ effect, this zero-crossing frequency is sensitive to the lower momentum cut-off of the electron momentum distribution, essentially the energy density of the non-thermal electrons (as shown in Figure \ref{fig:nuzero}). By measuring the zero-crossing frequency from observed spectra, one can distinguish between the energy densities of the thermal and non-thermal electron populations in the ICM. With prior information on the temperature of the population of thermal electrons, constraints on the momentum distribution of non-thermal electrons can be obtained.

\subsubsection{Pseudo-temperatures and non-thermal pressure}
\label{sec:pseudoT}
Analogous to the Comptonization parameter associated with the thermal SZ effect, we can express $\tau_{\mathrm{e,nth}}$ in terms of $y_{\mathrm{nth}}$ as \cite{Ensslin2000}
\begin{equation}
    \tau_{\mathrm{e,nth}} = \frac{m_\mathrm{e}c^2}{\langle k_\mathrm{B}\tilde{T}_\mathrm{e}\rangle}\:y_{\mathrm{nth}},
    \label{taunth}
\end{equation}
where
\begin{equation}
  y_{\mathrm{nth}}  = \frac{\sigma_T}{m_\mathrm{e}c^2} \int n_{\mathrm{e,cr}} k_\mathrm{B}\tilde{T}_\mathrm{e} dl,
  \label{ynth}
\end{equation}
is the integral of the non-thermal electron pressure along line-of-sight and
\begin{equation}
    k_\mathrm{B}\tilde{T}_\mathrm{e} = \frac{P_{\mathrm{e,nth}}}{n_{\mathrm{e,cr}}} = \int_0^\infty f_\mathrm{e}(p)\frac{1}{3}p\: v(p)\: m_\mathrm{e}c\:dp.
    \label{pseudo}
\end{equation}
Here, $k_\mathrm{B}\tilde{T}_\mathrm{e}$ is a pseudo-temperature attributed to the non-thermal electrons and $n_{\mathrm{e,cr}}$ is the normalization of the number density of non-thermal electrons. An analytical expression for Eq.\ \eqref{pseudo} which is given by \cite{Ensslin2000},
\begin{equation}
    k_\mathrm{B}\tilde{T}_\mathrm{e} = \frac{m_\mathrm{e}c^2(\alpha-1)}{6[p^{1-\alpha}]_{p_{2}}^{p_{1}}}\Bigg[\mathrm{B}_{\frac{1}{1+p^2}}\Big(\frac{\alpha-2}{2},\frac{3-\alpha}{2}\Big)\Bigg]_{p_{2}}^{p_{1}},
    \label{pseudo2}
\end{equation}
where  $\mathrm{B_x}(a, b)$ is the incomplete beta function. Rewriting Eq. (\ref{eq:di2}) in terms of $y_\mathrm{{nth}}$, we obtain
\begin{equation}
    \delta i(x) = \Bigg[\Big(\int_{p_1}^{p_2}\int_{-s_\mathrm{m}(p_1)}^{s_\mathrm{m}(p_2)}f_\mathrm{e}(p)\:K(e^\mathrm{s};p)\,e^\mathrm{s}\,ds\: dp \Big) \,-\, \frac{x^3}{e^x-1}\Bigg]\, I_0\frac{m_\mathrm{e}c^2}{\langle k_\mathrm{B}\tilde{T}_\mathrm{e}\rangle}y_\mathrm{{nth}}.
    \label{eq:di3}
\end{equation}

Since the pseudo-temperature is fixed by the choice of the power-law momentum distribution, the non-thermal electron distribution is "isothermal" in our analysis. Correspondingly, the density profile follows that of the assumed GNFW model (Section \ref{sec:gnfw}) of ICM pressure, which is then converted into a synchrotron emissivity profile using a magnetic field-strength model.

\subsection{Synchrotron emission}
\label{sec:sync}
The energy lost by an electron with an arbitrary pitch angle ($\theta$) in the presence of a magnetic field with strength $B$ is \cite{Rybicki}
\begin{equation}
P_{\mathrm{emitted}}(\nu) = \frac{\sqrt{3}e^3\,B\,\mathrm{sin}\theta}{m_\mathrm{e}\,c^2}\,x\int_x^\infty d\xi \, K_{5/3}(\xi),
    \label{eq:synch1}
\end{equation}
where
\begin{equation}
   x = \frac{\nu}{\nu_c}, \quad  \nu_c = \frac{3e\,B\,\gamma^2}{4\pi m_\mathrm{e}c}\,\mathrm{sin}\theta,
   \label{eq:nu-c}
\end{equation}
and $K_{5/3}(\xi)$ is the modified Bessel function of second kind of order 5/3. Consider a power-law distribution of electrons written as\footnote{If the distribution is assumed to be locally isotropic and independent of pitch angle, it reduces to $N(\gamma)=k\gamma^{-\alpha}$.}
\begin{equation}
    N(\gamma,\theta) = \frac{k}{4\pi}\,\gamma^{-\alpha}\,\frac{\mathrm{sin}\theta}{2}, \quad \gamma_1<\gamma<\gamma_2.
    \label{electronspec}
\end{equation}
The total synchrotron emission per unit volume for such a distribution of electron momenta is then given by
\begin{equation}
\begin{split}
    \frac{dW}{d\nu\,dt} &= \int\int P_{\mathrm{emitted}}(\nu)\,N(\gamma,\theta)\,d\gamma\,d\Omega_\theta \\
    &= \frac{\sqrt{3}k\,e^3B}{8\pi m_\mathrm{e}c^2}\int_0^\pi \int_{\gamma_1}^{\gamma_2} \mathrm{sin}^2\theta\,\gamma^{-\alpha}\,x\int_x^\infty K_{5/3}(\xi)\,d\xi\,d\gamma d\Omega_\theta.\\
\end{split}
    \label{eq:dW1}
\end{equation}
Upon comparison with the Eq. (\ref{eq:electrondist1}), $k = n_\mathrm{e,cr}\,A(\alpha,\gamma_1,\gamma_2)$. Further, assuming a radial profile for the magnetic field strength, Eq. (\ref{eq:dW1}) is re-written as
\begin{equation}
    \frac{dW(r)}{d\nu\,dt} = \frac{\sqrt{3}\,n_\mathrm{e,cr}\,A(\alpha,\gamma_1,\gamma_2)\,e^3B(r)}{8\pi m_\mathrm{e}c^2}\int_0^\pi \int_{\gamma_1}^{\gamma_2} \mathrm{sin}^2\theta\,\gamma^{-\alpha}\,x(r)\int_{x(r)}^\infty K_{5/3}(\xi)\,d\xi\,d\gamma d\Omega_\theta.
    \label{eq:dW2}
\end{equation}

We use a cluster sample (Table \ref{tab:catalogue}) where the synchrotron fluxes are scaled to a fixed observing frequency of 1.4 GHz. To get to the rest-frame emissivity following the Eq. (\ref{eq:dW2}) above, we use the cluster redshifts to convert to emission-frame frequencies. This is then integrated out to a fixed radius of $R_{500}$ to match the reported luminosity values. 


\subsection{Radial profiles of electrons and the magnetic field}

Finally, we describe the radial profiles used for matched-filtering the cluster SZ signal and model the synchrotron emissivity profiles. We assume the same pressure profiles for thermal and non-thermal electrons. Under the additional assumption of isothermal electrons (pseudo-temperature in the non-thermal case, Section \ref{sec:pseudoT}), the pressure profile also gives the density profile. The magnetic field strength is then related to this electron density profile by assuming that their energy densities will have the same radial dependence (see  \cite{Ensslin1998}).

\subsubsection{Pressure profile}
\label{sec:gnfw}
The spatial profile of the SZ effect is determined by the radial profiles of the Compton-y parameters, $y_\mathrm{th}$ (r) and $y_\mathrm{nth}$(r) [see Eq. (\ref{eq:di3})]. In order to model the radial profiles of $y_\mathrm{th}$(r) and $y_\mathrm{nth}$(r), we use the generalised Navarro-Frenk-White (GNFW) profile of the thermal electrons \cite{Nagai2007, Arnaud2010} with a fixed choice of the shape parameters. The only determining factors for the cluster pressure profile are then its mass and redshift.

The GNFW profile is used for modelling the distribution of thermal pressure within the ICM and is expressed as
\begin{equation}
    \frac{P(r)}{P_{500}} = \frac{P_0}{(c_{500}\frac{r}{R_{500}})^\gamma[1+(c_{500}\frac{r}{R_{500}})^\alpha]^{(\beta-\gamma)/\alpha}},
    \label{eq:pr}
\end{equation}
where 
\begin{equation}
R_{500} = \Big(\frac{3\,M_{500}}{4\pi\:500\rho_{\mathrm{crit}}}\Big)^{1/3}.
\end{equation}
Here, $\rho_{\mathrm{crit}}$ is the critical density at cluster redshift $z$, $c_{500}$ is the gas-concentration parameter, $P_0$ is the amplitude of pressure, and $\gamma$, $\alpha$, and $\beta$ describe the inner, intermediate and outer slopes of the profile. The slope parameter $\alpha$ should not be confused with the power-law index of the electron energy distribution [Eq. (\ref{eq:electrondist1})]. The parameters ($c_{500}$, $\gamma$, $\alpha$, $\beta$) are referred to as shape parameters. We adopt 
\begin{equation}
    P_{500} = 1.65\times10^{-3}\:h(z)^{8/3}\times \Bigg[\frac{M_{500}}{3\times10^{14}h_{70}^{-1} \: M_\odot}\Bigg]^{2/3 +\alpha_\mathrm{p}+\alpha_\mathrm{p}'(r)}h_{70}^2 \: \mathrm{keV}\: \mathrm{cm}^{-3}
    \label{eq:p500},
\end{equation}
presented in \cite{Arnaud2010} with their best-fit parameters of $P_0=8.403\,h_{70}^{-3/2}$, $\alpha_\mathrm{p}=0.120$, $\alpha=1.051$, $\beta=5.4905$, $\gamma=0.3081$, $c_{500} = 1.177$ and $\alpha_\mathrm{p}'(r)=0$. In Eq. (\ref{eq:p500}), $h(z)=\frac{H(z)}{H_0}$ is the reduced Hubble parameter at cluster redshift $z$, and $h_{70} = \frac{H_0}{70\,\mathrm{km}\,\mathrm{s}^{-1}\mathrm{Mpc}^{-1}}$.
\paragraph{}
The GNFW pressure profile is then integrated along the line-of-sight (los) to compute the radial profile of the Compton-y parameter, 
\begin{equation}
    y(r) = \frac{\sigma_T}{m_\mathrm{e}c^2}\int_{\mathrm{los}}P_\mathrm{e}(r)\: dl,
    \label{eq:yr}
\end{equation}
where $P_\mathrm{e}(r)$ is described by Eqs. (\ref{eq:pr}) and (\ref{eq:p500}). This projection is done numerically with the assumption of spherical symmetry, and the resulting $y$-profile (for each individual cluster) is taken as the template for optimally extracting the cluster tSZ$+$ntSZ signal via matched filtering.

\subsubsection{Magnetic field}
In order to compute the radial profile of the synchrotron emission (Eq. (\ref{eq:dW2})) for a given radio halo, we need radial profiles of the non-thermal electrons and the magnetic field in the ICM. With the relation $P_\mathrm{e}(r) = n_\mathrm{e}(r)\langle k_\mathrm{B}T_\mathrm{e}\rangle$, we assume the deprojected GNFW profile for $n_\mathrm{e}(r)$, and the corresponding distribution of the magnetic field is
\begin{equation}
    B(r) = B_0\Bigg(\frac{n_\mathrm{e}(r)}{n_{\mathrm{e,0}}}\Bigg)^{0.5},
    \label{eq:Br}
\end{equation}
where $B_0$ and $n_{\mathrm{e,0}}$ are the central magnetic field strength and electron number density, respectively. As discussed in Section \ref{sec:theory}, this radial dependence follows from an energy equipartition argument wherein the magnetic field energy density and the relativistic electron density have the same radial scaling \cite{Ensslin1998}. Observational evidence of this power-law dependence has been demonstrated by \cite{Murgia2004, Bonafede2010}. While the exact value of the power-law index is not critical for our analysis, a profile where the magnetic field strength scales down with radius is necessary to compute a realistic estimate of the synchrotron power.

\begin{figure}[ht]
    \centering
    \includegraphics[width=0.8\textwidth]{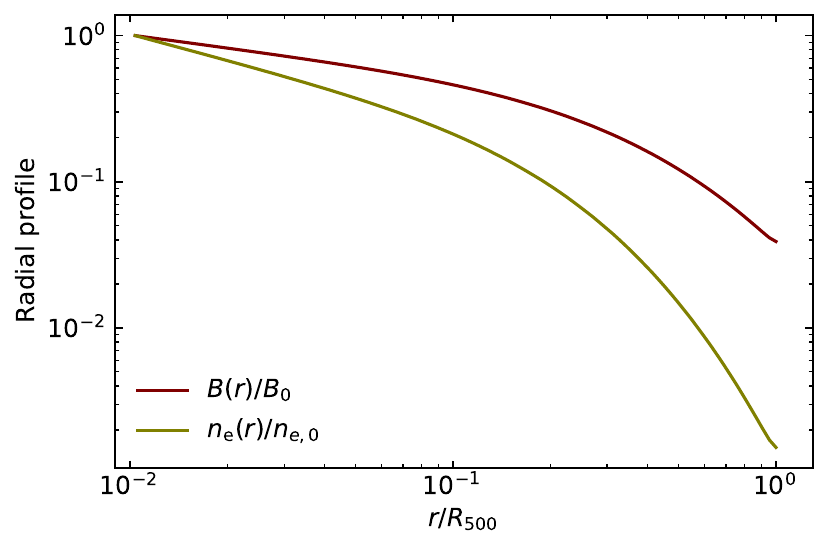}
    \caption{Radial profiles of the electron number density (which traces the GNFW pressure profile) and the magnetic field strength as a function of $r/R_{500}$. Both profiles are normalized to unity to highlight the diverging rate of radial fall-off.}
    \label{fig:neBprofile}
\end{figure}

The radial profiles of the $n_\mathrm{e}(r)$ and $B(r)$ are plotted in Figure \ref{fig:neBprofile}. We find that the magnetic field profile is significantly flatter than the electron density profile (the latter having identical shape as the thermal pressure profile under the assumption of isothermality). This translates into different factors of improvement on the electron number density and magnetic field constraints, when a future experiment with improved sensitivities is considered. We fit for the central magnetic field strength via Eqs.\ \eqref{eq:Br} and \eqref{eq:dW2}, and present the estimated central and volume-averaged magnetic field strength for each of the non-thermal electron models considered in Section \ref{sec:results}.

\section{Data and simulations}
\label{sec:data}
Observations in the mm/sub-mm wavelength range are necessary to exploit the difference in spectral shapes of the tSZ and ntSZ signals. The zero-crossing frequency, which is dependent on the non-thermal electron momenta distribution, also lies in this range. CMB surveys offer data in exactly this regime. 
\subsection{Radio halo cluster sample}
Our sample of GCs hosting radio halos are compiled from \cite{vanweeren2019}. 62 such radio halos are selected and their coordinates, $M_{500}$ and redshift estimates are obtained from the second \textit{Planck} catalogue of SZ sources \cite{Planck2015}. The synchrotron radiation flux measurements at 1.4 GHz for a sub-sample of 32 GCs \cite{Yuan2015, Gennaro2021} and the associated spectral index of the power-law describing the synchrotron emission are obtained from literature. These characteristics of our sample of GCs are tabulated in Table \ref{tab:catalogue}. A mean synchrotron power of $1.54\times10^{31}\,\mathrm{erg\,s}^{-1}\mathrm{Hz}^{-1}$ is assumed to obtain constraints on the magnetic field strength in GCs.

\begin{longtable}{cccc}
\hline \multicolumn{1}{c}{\textbf{Cluster}} & \multicolumn{1}{c}{\textbf{$z$}} & \multicolumn{1}{c}{\textbf{$M_{500}$}} &
\multicolumn{1}{c}{\textbf{log$_{10}$($P_{1.4\,\mathrm{GHz}}$})} \\ 
& & ($\times 10^{14} \mathrm{M}_\odot$) & \\
\hline 
\endfirsthead
\multicolumn{4}{c}
{{\bfseries \tablename\ \thetable{} -- continued from previous page}} \\
\hline \multicolumn{1}{c}{\textbf{Cluster}} & \multicolumn{1}{c}{\textbf{$z$}} & \multicolumn{1}{c}{\textbf{$M_{500}$}} &
\multicolumn{1}{c}{\textbf{log$_{10}$($P_{1.4\,\mathrm{GHz}}$})} \\ 
& & ($\times 10^{14} \mathrm{M}_\odot$) & \\
\hline 
\endhead
\hline \multicolumn{4}{r}{{Continued on next page}} \\ \hline
\endfoot
\endlastfoot
        Coma&0.023&7.165297&-0.19$\pm$0.04\\
        A3562&0.049&2.443&-0.95$\pm$0.05\\
        A754&0.054&6.853962&-0.24$\pm$0.03\\
        A2319&0.056&8.735104&0.24$\pm$0.02\\
        A2256&0.058&6.210739&-0.08$\pm$0.01\\
        A399&0.072&5.239323&-0.7$\pm$0.06\\
        A401&0.074&6.745817& \\
        A2255&0.081&5.382814&-0.06$\pm$0.02\\
        A2142&0.089&8.771307&-(0.72$\pm$1.22)\\
        A2811&0.108&3.647853&  \\
        A2069&0.115&5.30745&  \\
        A1132&0.137&5.865067&-(0.79$\pm$1.09)\\
        A3888&0.151&7.194754&0.28$\pm$0.69\\
        A545&0.154&5.394049&0.15$\pm$0.02\\
        A3411-3412&0.162&6.592571&-(0.57$\pm$1.0)\\
        A2218&0.171&6.585151&-0.41$\pm$0.01\\
        A2254&0.178&5.587061& \\
        A665&0.182&8.859059&0.58$\pm$0.02\\
        A1689&0.183&8.768981&-0.06$\pm$0.15\\
        A1451&0.199&7.162284&-(0.19$\pm$1.15)\\
        A2163&0.203&16.116468&1.24$\pm$0.01\\
        A520&0.203&7.80038&0.26$\pm$0.02\\
        A209&0.206&8.464249&0.24$\pm$0.02\\
        A773&0.217&6.847479&0.22$\pm$0.05\\
        RXCJ1514.9-1523&0.223&8.860777&0.14$\pm$0.1\\
        A2261&0.224&7.77852&-(0.17$\pm$1.15)\\
        A2219&0.228&11.691892&1.06$\pm$0.02\\
        A141&0.23&5.672555&  \\
        A746&0.232&5.335297&0.43$\pm$0.11\\
        RXCJ1314.4-2515&0.247&6.716546&-(0.17$\pm$0.62)\\
        A521&0.248&7.255627&0.07$\pm$0.04\\
        A1550&0.254&5.877626&  \\
        PSZ1G171.96-40.64&0.27&10.710258&0.58$\pm$0.05\\
        A1758&0.28&8.217337&0.72$\pm$0.11\\
        A697&0.282&10.998416&0.08$\pm$0.04\\
        RXCJ1501.3+4220&0.292&5.869359& \\ 
        Bullet&0.296&13.100348&1.16$\pm$0.02\\
        A2744&0.308&9.835684&1.21$\pm$0.02\\
        A1300&0.308&8.971329&0.58$\pm$0.16\\
        RXCJ2003.5-2323&0.317&8.991968&1.03$\pm$0.03\\
        A1995&0.318&4.924279&0.1$\pm$0.08\\
        A1351&0.322&6.867679&1.01$\pm$0.06\\
        PSZ1G094.00+27.41&0.332&6.776592&0.58$\pm$0.02\\
        PSZ1G108.18–11.53&0.335&7.738726&  \\
        MACSJ0949.8+1708&0.383&8.23875& \\
        MACSJ0553.4–3342&0.407&8.772141 & \\
        MACSJ0417.5–1154&0.443&12.250381 & \\
        MACSJ2243.3–0935&0.447&9.992374 & \\
        MACSJ1149.5+2223&0.544&10.417826 & \\
        MACSJ0717.5+3745&0.546&11.487184& \\
        ACT-CLJ0102–4915&0.87&10.75359 & \\
        AS1121&0.358&7.193831&  \\
        ZwCl0634+4750803&0.174&6.652367&-(0.51$\pm$1.69)\\
        PLCKG004.5-19.5&0.54&10.356931&  \\
        CL0016+16&0.5456&9.793704&0.76$\pm$0.07\\
        PLCKESZG285–23.70&0.39&8.392523 & \\
        RXCJ0256.5+0006&0.36&5.0 & \\
        MACSJ1752.0+4440&0.366&4.3298&1.1$\pm$0.03\\
        A800&0.2472&3.1464&  \\
        CL1446+26&0.37&2.70015& \\ 
        CIZAJ2242.8+5301&0.192&4.0116&1.16$\pm$0.05\\
        MACSJ0416.1–2403&0.396&4.105336&  \\
\hline
\caption{Cluster identifiers, redshift ($z$), mass ($M_{500}$) and synchrotron power at 1.4 GHz (in $10^{24}\,$W/Hz) that are used in this work.}
\label{tab:catalogue}
\end{longtable}

\subsection{{\it Planck} all-sky maps}
The \textit{Planck} satellite observed the sky for four years with two instruments. The Low-frequency Instrument (LFI) was sensitive in the 30 -- 70 GHz range \cite{Planck2018LFI} and the High-frequency Instrument (HFI) was sensitive in the 100 -- 857 GHz range \cite{Planck2018HFI}. We used the 2018 release of the {\it Planck} all-sky multi-frequency maps \cite{Planck2018overview}. 70 GHz maps from the LFI and maps from all six bands from the HFI are used. The maps are available in \texttt{HEALPix}\footnote{\url{http://healpix.sourceforge.net}} format \cite{Gorski2005} with N\textsubscript{side} = 2048 for the HFI channels and N\textsubscript{side} = 1024 for the LFI channels. We chose to work in units of surface brightness (MJy$\,\mathrm{sr}^{-1}$) and this required the 70 -- 353 GHz maps, which are originally available in units of K\textsubscript{CMB}, to be converted to surface brightness maps using the Unit Conversion - Colour Correction (UC-CC)\footnote{\url{https://wiki.cosmos.esa.int/planckpla2015/index.php/UC\_CC\_Tables}} tables. The resolution of the maps and the respective UC values are tabulated in Table \ref{tab:ucres}.

\begin{table}[ht]
    \centering
    \begin{tabular}{ccc}
    \hline
         Frequency band & FWHM & UC\\
         (GHz) & (arcmin) & ($\mathrm{MJy\,sr}^{-1}\,\mathrm{K_{CMB}}^{-1}$)\\
         \hline
         70 & 13.31& 129.187\\
         100 & 9.68 & 244.096\\
         143 & 7.30 & 371.733\\
         217 & 5.02 & 483.687\\
         353 & 4.94 & 287.452\\
         545 & 4.83 & 58.036\\
         857 & 4.64 & 2.2681 \\
         \hline
    \end{tabular}
    \caption{Resolution of \textit{Planck}'s 70 GHz LFI band and all HFI frequency bands. The unit conversion coefficients are used to convert the 70 -- 353 GHz maps from units of $\mathrm{K_{CMB}}$ to $\mathrm{MJy\,sr}^{-1}$.}
    \label{tab:ucres}
\end{table}

\subsection{Simulated microwave sky maps}
\label{subsec:simsky}
In order to quantify the constraining power of upcoming CMB experiments in obtaining upper limits on the non-thermal electron number density, an estimate of the noise covariance matrix is required. We first simulate the microwave sky maps at the observing frequencies and convolve them with a Gaussian beam (Table \ref{tab:sensitivities}).
\paragraph{}
The simulated maps comprise of the following components:
\begin{itemize}
    \item Galactic foregrounds (dust, synchrotron, anomalous microwave emission, and free-free emission) which are simulated using the Python Sky Model (PySM) software \cite{PySM}.
    \item Cosmic Infrared Background (CIB), CMB, radio point sources, tSZ, and kinetic SZ (kSZ; \cite{Sunyaev1980}) components which are simulated using the Websky extragalactc CMB simulations (\cite{Websky}).
\end{itemize}
Finally, white noise with variance given by the sensitivites  of the instruments \cite{Choi2019, SimonsObservatory2018} (listed in Table \ref{tab:sensitivities}) are added to the maps to represent the detector noise and any residual atmospheric noise. Since we consider small areas of the sky, the dominant noise component is the white noise and we choose to ignore the $1/f$ noise component. \texttt{ccatp\_sky\_model}\footnote{\url{https://github.com/MaudeCharmetant/CCATp_sky_model}} \texttt{Python} package, which incorporates the PySM and Websky simulations, is used to simulate the microwave sky maps in this work.
\begin{table}[ht]
        \centering
    \begin{tabular}{ccc}
    \hline
    Frequency band & FWHM & Sensitivity\\
     (GHz) & (arcmin) & ($\mu$K-arcmin)\\
    \hline
    27 & 7.4 & 71 \\
    39 & 5.1 & 36 \\
    93 & 2.2 & 8 \\
    145 & 1.4 & 10 \\
    225 & 1.1 & $\left[ 22^{-2} + 15^{-2} \right]^{-1/2}$ \\
    280 & 1.1 & $\left[ 54^{-2} + 27^{-2} \right]^{-1/2}$ \\
    350 & 1.1 & 105 \\
    405 & 1.1 & 372 \\
    860 & 1.1 & $5.75\times 10^5$ \\
    \hline
        \end{tabular}
            \caption{Sensitivities of FYST, SO and the combined sensitivities of SO+FYST configuration. For the common frequencies between these two experiments (225 and 280 GHz) the noise values are added in quadrature (following inver-variance weighting), where the first number inside the parenthesis corresponding to the SO, and the second value corresponding to FYST.
            }
    \label{tab:sensitivities}
\end{table}
\section{Methods}
\label{sec:fit}

We first extract $10^\circ \times 10^\circ$ cluster fields from the all-sky multi-frequency maps centred around the coordinates of each of the GC in our sample using the \texttt{gnomview()} function of the \texttt{healpy}\footnote{\url{https://healpy.readthedocs.io/en/latest/index.html}} \cite{healpy} module. To improve the Signal-to-Noise Ratio (SNR) of the SZ effect and minimize the amplitude of contaminants in the cluster fields, we employ the following methods:
\begin{enumerate}
    \item Matched-filtering (MF): This method is used to minimize the noise and other astrophysical contaminants to optimally extract the cluster signal, assuming a fixed spatial template.
    
    \item Stacking: Stacking cluster fields ensures amplification of the tSZ and ntSZ signals by averaging the uncorrelated noise, and minimizes the kSZ signal from individual clusters.
\end{enumerate}
In the following subsections we discuss these methods in detail.

\subsection{Matched-filtering method}
Matched filters are designed to optimally extract a signal in the presence of Gaussian noise and have been shown to be effective for extracting cluster SZ signals (e.g. \cite{Haehnelt1996, Melin2006, Erler2018, Zubeldia2021}). Presence of small non-Gaussian noise components will not cause a bias but the solution may not be optimal \cite{Melin2006}. 
In practice, setting up a matched filter is extremely easy as it only requires that we know the spatial template of the emitting sources. In the flat-sky approximation, this filter function (a vectorized map) can be written as the following in the Fourier space \cite{Schaefer2006, Erler2019}:
\begin{equation}
\boldmath{\Psi} = \left[ \mathbf{\tau}^\mathrm{T} \mathbf{C}^{-1} \mathbf{\tau}\right]^{-1} \mathbf{\tau}\mathbf{C}^{-1},
\end{equation}
where $\boldsymbol{\tau}$ is the Fourier transform of the 2D $y$-profile model, and $\mathbf{C}$ is the azimuthally-averaged noise power spectrum of the unfiltered map. We use the publicly available \texttt{PYTHON} implemention of MF, called \texttt{PyMF}\footnote{\url{https://github.com/j-erler/pymf}} \cite{Erler2019}, to filter the $10^\circ \times 10^\circ$ maps. The noise power spectra are computed from the same fields after masking the GCs, or, in the case of forecasts, from random empty fields.

\subsection{Stacking}
\begin{figure}[ht]
    \centering
    \includegraphics[width=1.0\textwidth]{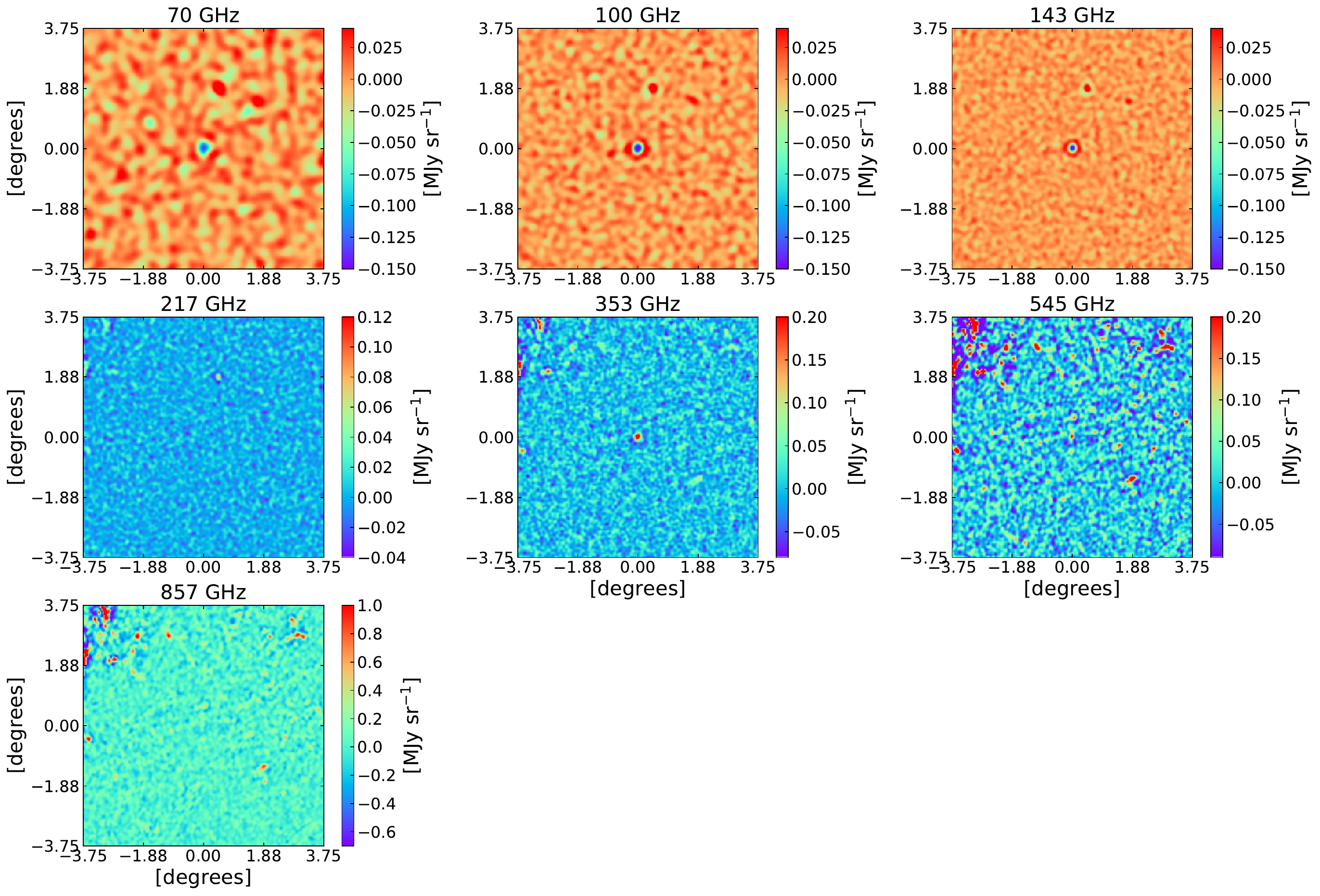}
    \caption{$7.5^\circ \times 7.5^\circ$ stacked matched-filtered maps of the 62 galaxy clusters in the sample for each of the 70 GHz \textit{Planck} LFI and HFI channels. The SZ effect signal is clearly seen in the centre. There is still some residual dust emission visible in the HFI maps. The color scale is intentionally set differently for each map in order to enhance any features in the map.}
    \label{fig:stackedmf}
\end{figure}

Matched-filtered cluster fields are stacked to obtain an \textit{average} filtered map at each observed frequency. Since the noise properties are practically Gaussian after filtering, this leads to a suppression of noise by roughly a factor of $1/\sqrt{62}$, where 62 is the number of clusters in our radio halo sample. Stacking also has the additional advantage of suppressing the kSZ signal by the same factor, which acts as a random source of noise at the cluster location. 
The stacked matched-filtered maps are shown in Figure \ref{fig:stackedmf}.

\begin{figure}[ht]
    \centering
    \includegraphics[width=0.8\textwidth]{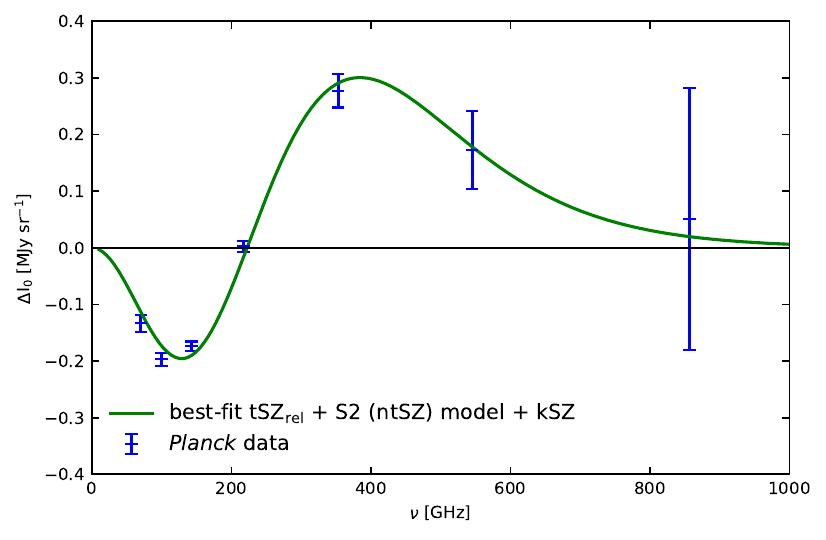}
    \caption{Spectrum of the SZ effect extracted from the stacked matched-filtered maps of the \plk data. The values correspond to the central pixel value in the stacked matched-filtered maps. The error bars correspond to the extent of the variance of foregrounds in the maps.}
    \label{fig:extractedspec}
\end{figure}

The amplitude of the SZ signal at each frequency channel is in the central pixel of the stacked matched-filtered map, and this value is extracted to obtain a spectrum of the SZ effect. The extracted spectrum is displayed in Figure \ref{fig:extractedspec}. It shows the characteristic shape of the SZ effect with a decrement in specific intensity at frequencies $\nu <$ 217 GHz and an increment at higher frequencies. The error bars correspond to the variance of astrophysical emission in the stacked matched-filtered maps. The uncertainties for the high-frequency channels are larger as the mean contribution from dust emission is still prominent in the maps and the HFI at $\nu = 353,\, 545\,$ and 857 GHz, in general, are relatively noisy.

\subsection{Spectral fitting}
The extracted amplitude of the SZ signal can be decomposed into the distortions due to $\mathrm{tSZ_{rel}}$, kSZ and ntSZ effects as\footnote{The subscript 0 refers to the fact that the amplitudes correspond to the central pixel in the maps.}
\begin{equation}
\begin{split}
    \Delta I_{\mathrm{0,\nu}} &= \delta i_{\mathrm{0,\nu}}^{\mathrm{th}} + \delta i_{\mathrm{0,\nu}}^{\mathrm{nth}} + \delta i_{\mathrm{0,\nu}}^{\mathrm{kSZ}}\\
    &=\Bigg[\Bigg(\int_{p_1}^{p_2}\int_{-s_{\mathrm{m}}(p_1)}^{s_{\mathrm{m}}(p_2)} f_\mathrm{e,th}(p;\Theta) K(e^\mathrm{s};p)\,e^\mathrm{s}\: \frac{(x/e^\mathrm{s})^3}{(e^{x/e^\mathrm{s}}-1)}\: ds\: dp\Bigg) - \frac{x^3}{e^x-1}\Bigg] I_0\frac{m_\mathrm{e} c^2}{k_\mathrm{B}T_\mathrm{e}}\,y_0^\mathrm{th} \\
    &+ \Bigg[\Bigg(\int_{p_1}^{p_2}\int_{-s_\mathrm{m}(p_1)}^{s_\mathrm{m}(p_2)}f_\mathrm{e}(p)\:K(e^\mathrm{s};p)\,e^\mathrm{s}\: \frac{(x/e^\mathrm{s})^3}{(e^{x/e^\mathrm{s}}-1)}\,ds\: dp \Bigg) \,-\, \frac{x^3}{e^x-1}\Bigg]\, I_0\frac{m_\mathrm{e}c^2}{\langle k_\mathrm{B}\tilde{T}_\mathrm{e}\rangle}\,y_0^\mathrm{{nth}} \\
    & -I_0\: \frac{x^4e^x}{(e^x-1)^2}\, y_0^\mathrm{kSZ},  
    \label{eq:excomp1}
\end{split}
\end{equation}

where $\Delta I_\mathrm{0,\nu}$ is the amplitude of the SZ effect signal from stacked matched-filtered map of frequency $\nu$, $\delta i_{\mathrm{0,\nu}}^\mathrm{th}$, $\delta i_{\mathrm{0,\nu}}^\mathrm{kSZ}$ and $\delta i_{\mathrm{0,\nu}}^\mathrm{nth}$ are distortions due to the tSZ$_\mathrm{rel}$, kSZ and ntSZ effects, respectively; $f_\mathrm{e,th}(p;\Theta)$ is the Maxwell-J\"{u}ttner distribution used to describe the thermal distribution of electrons in terms of the normalized thermal energy parameter, $\Theta = \frac{k_\mathrm{B}T_\mathrm{e}}{m_\mathrm{e}c^2}$, and $y_0^\mathrm{kSZ}$ is analogous to $y_0^\mathrm{th}$.  Appendix \ref{sec:modelsz} can be referred to for more information on how we compute the $\mathrm{tSZ_{rel}}$ and kSZ spectra. We thus fit the extracted SZ spectrum with a three-component model consisting of the tSZ$_\mathrm{rel}$, kSZ and ntSZ signals, using the MCMC sampling method.  While fitting this three-component model to \textit{Planck} data, we use bandpass corrected spectra (a description of which can be found in Appendix \ref{sec:bandpass}). 

The shape of the $\mathrm{tSZ_{rel}}$ spectrum is fixed by using a single $T_\mathrm{e}$ to represent our stack of clusters and fit only for the amplitude, $y_\mathrm{0,th}$. The spectral distortion for $\mathrm{tSZ_{rel}}$ is described in Appendix \ref{app:rSZ}. Since we work with the stacked signal for spectrum fitting, we adopt a single, median temperature from all the clusters for $T_\mathrm{e}$, where individual cluster temperatures are obtained from a mass-temperature relation as given in \cite{Reichert2011}. The median temperature (energy) is approximately 8 keV and is used for computing the relativistic corrections to the tSZ signal. 

The $y_\mathrm{{kSZ}}$ is marginalized over by drawing $v_\mathrm{{pec}}$ from a Gaussian distribution of zero mean and a standard deviation corresponding to the expected line-of-sight velocity after stacking. This marginalization amplitude is $100/\sqrt{62}$ \,kms$^{-1}$ in each step of the chain. We also fix the shape of the ntSZ spectrum by assuming specific values for $p_1$ and $p_2$ and fit for the amplitude of the ntSZ effect, $y_{\mathrm{0,nth}}$. Finally, the posterior probability distributions of $y_{\mathrm{th}}$ and $y_\mathrm{{nth}}$ are estimated for four different models of the non-thermal electron momentum distribution. To fit the stacked spectrum we need a frequency-to-frequency noise covariance. In the case of \textit{Planck} data, the covariance matrix is computed empirically from the stacked matched-filtered maps as
\begin{equation}
    C_{ij} \equiv \frac{1}{N_\mathrm{pix}-1}\sum_{p=1}^{N_\mathrm{pix}} (I_{i}(p)-\overline{I}_{i})(I_{j}(p)-\overline{I}_{j}),
    \label{eq:covmatrix}
\end{equation}
where $N_\mathrm{pix} = 1600$ denotes the number of pixels of pixel size $1.5'$ in a $10^\circ \times10^\circ$ field and $\mathbf{I}(p)$ denotes the value of pixel $p$ in intensity map \textbf{I}. The covariance matrix is estimated by masking the cluster region in the stacked matched-filtered maps. In the case of SO$+$FYST forecasts, the frequency covariance is computed in a similar way from randomly located empty fields.

\begin{figure}
    \centering
    \includegraphics[width=1.0\textwidth]{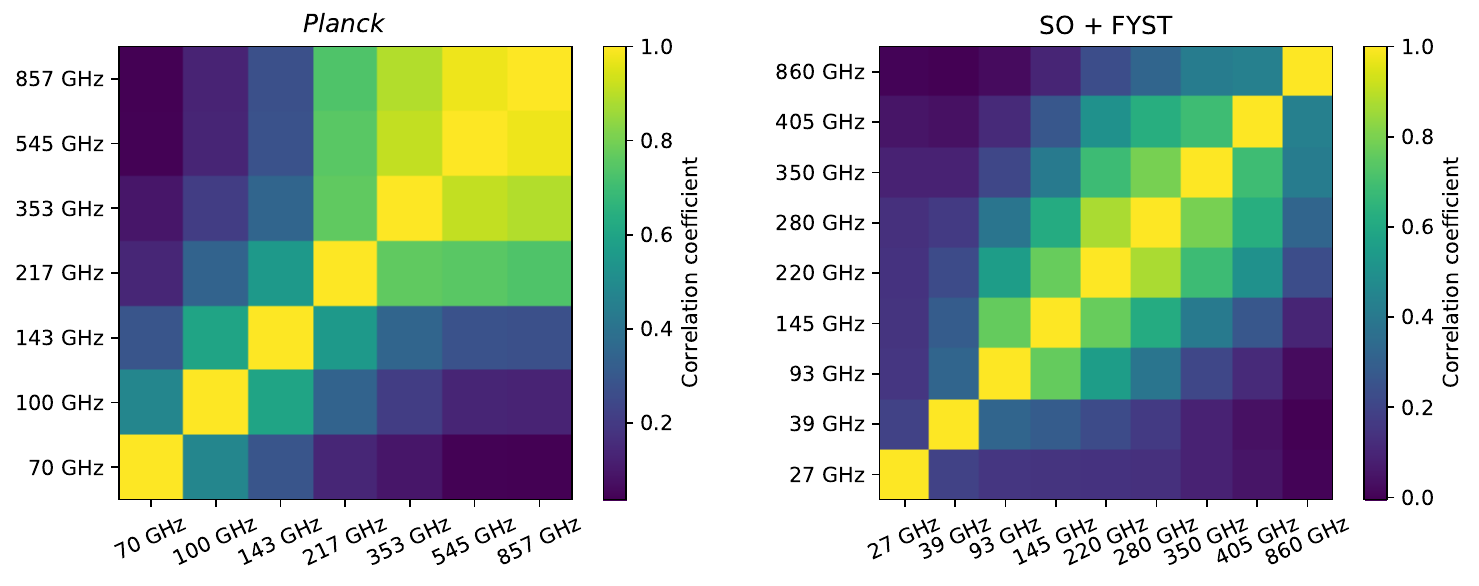}
    \caption{The spectral correlation between the different frequency maps from \textit{Planck} data (\textit{left}) and simulated frequency maps with SO+FYST sensitivities (\textit{right}).}
    \label{fig:correlationcoeff}
\end{figure}

\paragraph{}
For our forecasts, we follow a similar procedure of computing the noise covariance matrices using stacked matched-filtered maps extracted from simulated maps described in Section \ref{subsec:simsky}. Specifically, 100 $10^\circ \times10^\circ$ fields centered around coordinates sampled from a uniform distribution are extracted from a full-sky simulated map, and a mean of these 100 fields is computed to represent foregrounds in one simulated cluster field. We then compute 62 such cluster fields, perform matched-filtering and stack the matched-filtered fields to get one simulated stacked matched-filtered cluster field. This procedure is performed at each observing frequency. These simulated stacked fields are then used to compute the noise covariance matrix described in Eq. (\ref{eq:covmatrix}). Figure \ref{fig:correlationcoeff} shows the correlation matrices computed from \textit{Planck} data and the simulated maps with the SO+FYST configuration.

\section{Results}
\label{sec:results}

Due to the presence of the dominant tSZ$_\mathrm{rel}$ effect and the constraining power of the sensitivities of \textit{Planck}, we are able to obtain upper limits on the $y_\mathrm{nth}$ and $n_\mathrm{e,nth}$, and further, lower limits on the magnetic field strength.  

\subsection{Current constraints from the {\it Planck} data}
The constraints on the amplitude of the tSZ$_\mathrm{rel}$ and ntSZ effects are shown in Table \ref{tab:results1}. With the assumption of isothermality and a median $k_\mathrm{B}T_\mathrm{e} = 8.0\,$keV for the stack of GCs, the $y_0^{\mathrm{th}}$ is well constrained by \textit{Planck} data. However, for $y_0^\mathrm{nth}$, we are only able to obtain upper limits with the data. Models of electron momenta that assume higher energies result in higher upper limits for $y_\mathrm{nth}$. The average electron number density remains consistent for all models. A large variation in constraints (for a fixed synchrotron flux density at 1.4 GHz) on the central and volume-averaged magnetic field strengths is observed.

\begin{table}[ht]
        \centering
    \begin{tabular}{ccccccc}
    \hline
    Model & Obs & $y_0^\mathrm{th}$& $y_0^\mathrm{nth}$& $\Bar{n}_\mathrm{e,nth}$ & $\Bar{B}$ & B$_0$\\
    &  & ($\times 10^{-4}$) & ($\times 10^{-4}$) & ($\times 10^{-6}$ cm$^{-3}$)&($\mu$G)&($\mu$G)\\
    \hline
         & & & & & &\\
         S1& \textit{Planck}& $1.83^{+0.09}_{-0.10}$ & $<4.81$ & $<2.06$& $>0.24$ & $>1.05$\\
         & & & & & &\\
        S2& \textit{Planck} & $1.82^{+0.09}_{-0.10}$ & $<48.61$ & $<2.08$ & $>0.02$ & $>0.08$\\
        \hline
        & & & & & &\\
         B1& \textit{Planck} & $1.82^{+0.09}_{-0.10}$ & $<23.83$ & $<2.09$& $>0.03$ & $>0.15$\\
        & & & & & & \\
        B2& \textit{Planck} & $1.82^{+0.09}_{-0.10}$ & $<77.70$ & $<2.05$& $>0.01$ & $>0.03$\\
        & & & & & &\\
        \hline
        \end{tabular}
            \caption{$y_0^\mathrm{th}$ and upper limits on $y_0^\mathrm{nth}$ obtained from the \textit{Planck} data. The number density and magnetic field strength are volume-averaged quantities within the $r_{500}$. For a central magnetic field strength of 1$\mu$G, and $\alpha=3.6$ as the index of the power-laws describing the electron momentum distribution, the volume-averaged value is $0.23\,\mu$G.}
    \label{tab:results1}
\end{table}
\subsection{Upcoming constraints from SO and FYST}
\begin{figure}[ht]
    \centering
    \includegraphics[width=0.8\textwidth]{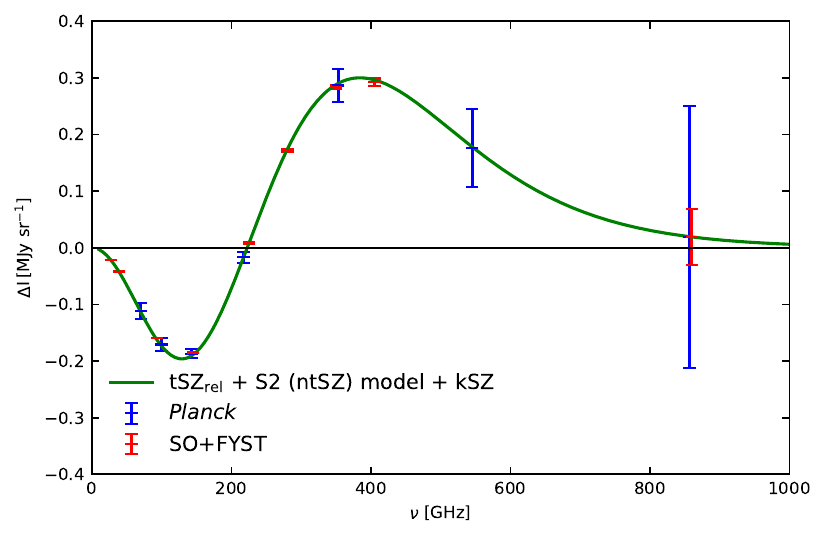}
    \caption{A comparison of noise variance expected for SO+FYST for a stack of 62 fields with \textit{Planck}. Observations of more radio halos could further improve the constraints estimated in this study.}
    \label{fig:noise_forecast}
\end{figure}

Assuming the sensitivities of SO+FYST configuration tabulated in Table \ref{tab:sensitivities}, we check for the constraining capabilities of upcoming CMB experiments on the amplitude of ntSZ effect. Further, the lower limits on magnetic field strength are computed and tabulated in Table \ref{tab:results2}. The expected variance in the measurement of the SZ spectrum for a stack of 62 cluster fields is plotted in Figure \ref{fig:noise_forecast} compared to the variance from \textit{ Planck} data. The error bars are significantly smaller at all frequencies due to the combined sensitivities of SO and FYST.

\begin{table}[ht]
        \centering
    \begin{tabular}{cccccc}
    \hline
    Model & Obs & $y_0^\mathrm{nth}$& $\Bar{n}_\mathrm{e,nth}$ & $\Bar{B}$ & B$_0$\\
    &  & ($\times 10^{-4}$) & ($\times 10^{-6}$ cm$^{-3}$)&($\mu$G)&($\mu$G)\\
    \hline
         & & & &\\
         S1& SO+FYST& $<1.10$ & $<0.47$ & $>0.46$ & $>2.0$\\
         & & & & \\
        S2& SO+FYST & $<11.37$ & $<0.49$ & $>0.04$ & $>0.16$\\
        \hline
        & & & &\\
         B1& SO+FYST & $<5.07$ & $<0.47$ & $>0.07$ & $>0.28$\\
        & & & &\\
        B2& SO+FYST & $<17.29$ & $<0.48$ & $>0.02$ & $>0.07$\\
        & & & &\\
        \hline
        \end{tabular}
            \caption{Same as Table \ref{tab:results1} but for SO+FYST.}
    \label{tab:results2}
\end{table}

\begin{figure}[ht]
    \centering
    \includegraphics[width=1.0\textwidth]{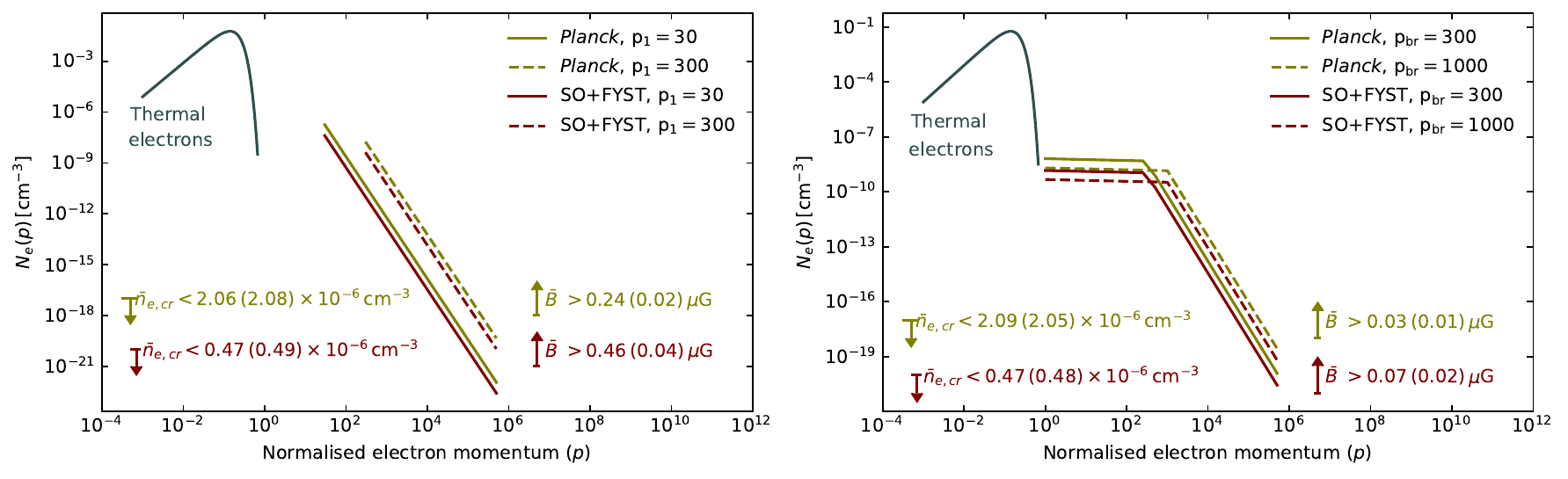}
    \caption{Thermal and non-thermal populations of scattering electrons in the ICM. $k_\mathrm{B}T_\mathrm{e} = 8.0$ keV is assumed for the distribution of thermal electrons. Left: Power-law distributions of electrons are assumed for the non-thermal electrons with constraints estimated from \textit{Planck} data and SO+FYST configuration. Right: Broken power-law is assumed for the non-thermal electrons with constraints from \textit{Planck} and SO+FYST configuration. Full and broken lines correspond to different minimum (broken) momenta for power-law (broken) distributions.}
    \label{fig:fullnedist}
\end{figure}
\begin{figure}[ht]
    \centering
    \includegraphics[width=1.\linewidth]{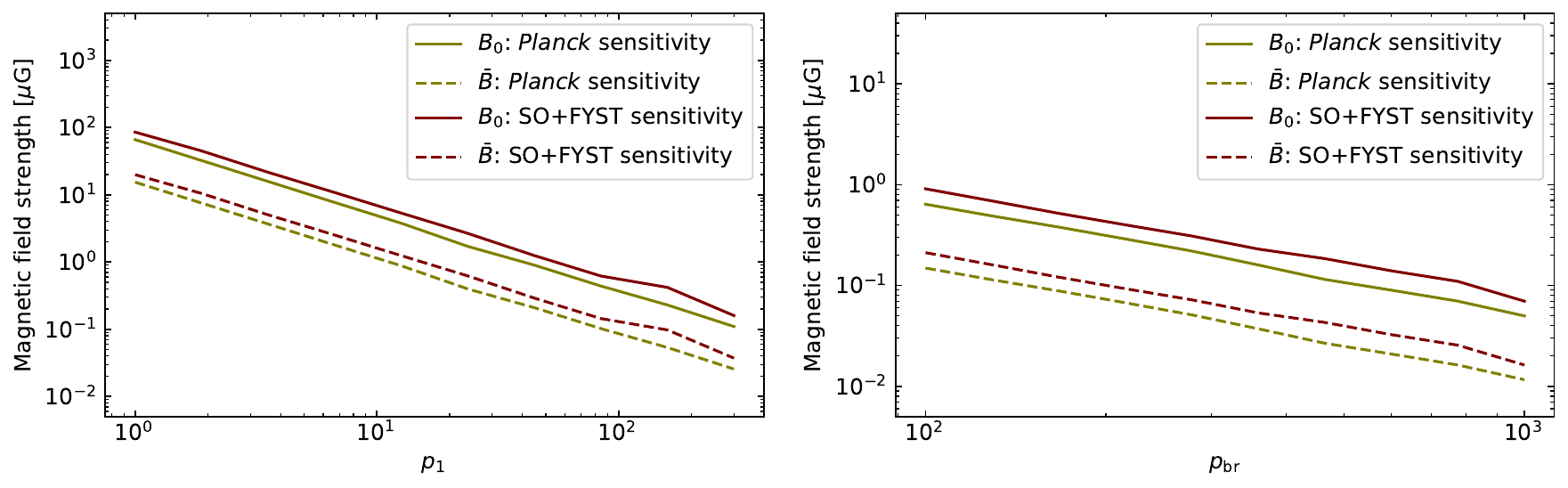}
    \caption{Lower limits on the central magnetic field strength ($B_0$) and volume-averaged field strength ($\bar{B}$) for different values of $p_1$ (\textit{left}) and $p_\mathrm{br}$ (\textit{right}) for \plk (in \textit{olive}) and SO+FYST (in \textit{maroon}) sensitivities.}
    \label{fig:p1-pbr-B}
\end{figure}

\paragraph{}
We also estimate the central and volume-averaged magnetic field strength assuming a mean synchrotron power with \plk and SO+FYST sensitivities as a function of $p_1$ for single power-law and $p_\mathrm{br}$ for the broken power-law models describing the non-thermal electron momentum distributions [Eqs. \eqref{eq:electrondist1} and \eqref{eq:electrondist2}]. The estimated lower limits on $B_0$ and $\bar{B}$ are plotted in Figure \ref{fig:p1-pbr-B}. The dependence of the synchrotron emission on the electron momentum is determined by the synchrotron kernel $x\,\int_x^\infty K_{5/3} (\xi)\,d\xi$ in Eqs. \eqref{eq:synch1}, \eqref{eq:dW1} and \eqref{eq:dW2}. Depending on the magnetic field strength (and thus the critical frequency, Eq.\ \eqref{eq:nu-c}), the synchrotron kernel probes different regions of the non-thermal electron momentum distribution. For synchrotron emission at 1.4 GHz, it is only the higher-momentum tail of the distribution that contributes to the emission. When we consider lower $p_1$ (or $p_\mathrm{br}$) and a fixed synchotron power, higher magnetic field strengths will be estimated as there are lower number of high-momentum non-thermal electrons emitting synchrotron radiation at 1.4 GHz and this is evident in Figure \ref{fig:p1-pbr-B} wherein $B_0$ increases with decreasing $p_1$ and $p_\mathrm{br}$.

\section{Discussion and conclusions}
\label{sec:summary}

The amplitude of the $\mathrm{tSZ_{rel}}$ effect is well constrained by \textit{Planck} under the assumption of an isothermal ICM, and simultaneously, for the first time, upper limits on the ntSZ amplitude and non-thermal electron number density for different models of the non-thermal electron momenta are obtained. The resulting number densities of the different populations of electrons in the ICM, as derived from the constraints, are plotted in Figure \ref{fig:fullnedist}. While $y_\mathrm{nth}$ and $N_\mathrm{e}(p)$ are sensitive to the choice of models of $f_\mathrm{e}(p)$, there is no significant variation in the volume-averaged number density of the non-thermal electrons. This can be attributed to fixing $\int_{p_1}^{p_2}f_\mathrm{e}(p)\,p^2\,dp=1$ and the resulting normalization of $f_\mathrm{e}(p)$. The corresponding lower limits of the magnetic field strength are estimated from the known synchrotron power, and found to be well within the limits of measurements through Faraday rotation. To estimate the diffuse magnetic field strength, we use the equipartition argument to relate the magnetic field strength with the non-thermal particle density, and assume a radially dependent $B(r)$ to estimate a volume-averaged field strength within a spherical volume of radius $0.5\,$Mpc. The limit on the magnetic field strength increases with a decrease in the electron momentum considered, since the electrons with lower energy require larger magnetic fields to produce the same synchrotron flux density at 1.4 GHz, as seen from the formulation in Section \ref{sec:sync}.

\paragraph{}

The resulting magnetic field profile (Figure \ref{fig:neBprofile}) is shallower than the commonly assumed beta-model profile in the literature, staying within the same order-of-magnitude inside the volume bounded by the $r_{500}$. This has followed from our choice of the GNFW profile to model the electron pressure and subsequently $y_\mathrm{nth}(r)$ [Eqs. (\eqref{eq:pr}, \eqref{eq:yr})], which is steeper than the beta model within the $r_{500}$. We have also assumed isothermality for the electrons (pseudo-temperature for the non-thermal electrons), i.e., the same power-law distribution throughout the cluster volume. These assumptions on the GNFW profile and isothermality do not reflect the dynamic environments within the ICM. Rather, we simply consider them as a reasonable set of assumptions considering the current state of knowledge, and the spatial resolution and sensitivities of the CMB experiments.
 
\paragraph{}

Our forecasts indicate that with upcoming experiments such as SO and FYST, improved constraints on $y_\mathrm{nth}$ and $B$ can be obtained. For the most simplistic single-slope power law models (such as S1 with $p_1 = 30$), the prospective constraints on $y_\mathrm{nth}$ with upcoming data would require a central $B_0$ value $>2\,\mu$G to reconcile with the observed synchrotron power (Table \ref{tab:results2}). This is at the limit of some of the recent measurements of the central magnetic field strength using FRM, e.g., \cite{Govoni2017} infer $B_0=1.5\pm 0.2 \;\mu$G in the GC Abell 194.
\paragraph{}
When we consider lower $p_1$ (or $p_\mathrm{br}$), it results in fewer electrons in the higher-momentum tail of the non-thermal electron distribution which emit synchrotron radiation at 1.4 GHz, thus requiring higher magnetic field strengths when assuming the same synchrotron power. Hence, lowering $p_1$ results in higher $B$ estimates which are in tension with the current data. The lower limits on $B_0$ and $\bar{B}$ estimated as a function of $p_1$ and $p_\mathrm{br}$, assuming \plk and SO+FYST sensitivities, and a fixed synchrotron power at 1.4 GHz, are plotted in Figure \ref{fig:p1-pbr-B}. The assumption of a single power-law with $p_1<30$ for the non-thermal electron momentum distribution results in lower limits of $B_0>3\,\mu$G assuming SO+FYST sensitivities. We can thus state that SO$+$FYST data will be able to rule out some of these simplistic models for the non-thermal particle distribution in GCs. This can prove to be extremely useful in discerning the acceleration mechanisms and physical extent of the non-thermal electron population in GCs within the next few years. 

\paragraph{}

It is worth highlighting that these future constraints, with upcoming CMB survey data, are obtained with the parameters of the same 62 galaxy clusters, in other words, assuming a RH sample of 62 clusters within a similar mass and radio power range. As new observations with LOFAR and other low-frequency surveys are rapidly improving the number of known RHs, both in the lower-mass regimes and at higher redshifts (e.g. \cite{Botteon2022} \cite{DiGennaro2021}), better statistical accuracy will be available with larger RH sample when leveraging the future CMB data for the ntSZ effect. However, more accurate forecasts utilizing a larger cluster sample size would require careful modelling of the scaling of the RH power with both cluster mass and redshift, which we have left out for a future study.

%% file: chapters/chp05.tex
\chapter{TOD Simulations with \texttt{TOAST}}
\label{ch:TOAST-TOD}
\begin{center}
    \begin{minipage}{0.75\textwidth}
        \textbf{Summary:} Detectors of sub-mm experiments sample the sky in time by employing certain scan strategies, and the raw observations are called time-ordered data (TOD). While the signal-to-noise ratio of each sample is low, information can be retained by reconstructing the sky with a combination of samples collected by many detectors over a period of time. The reconstruction of information is dependent on 
        \begin{itemize}[topsep=2mm, itemsep=-.5mm]
            \item Science objectives that dictate the field of observation
            \item Fields of observation dictate the telescope's scanning strategy
            \item Noise properties (eg. power, spatial/temporal correlations) given a scanning strategy
        \end{itemize}       
        All of the above factors inform the survey strategy that a telescope needs to employ and the map-making methods. TOD simulations aid in this study. In this chapter, we describe the software tools used for TOD simulations and the ingredients necessary for such a simulation in the context of FYST (discussed in Section \ref{sec:ch2-FYST}). We then explore the method of destriping for reconstruction of maps from TOD and the characteristics of residual correlated noise in the resulting maps based on the choice of baseline length adopted for destriping.\\[3mm]
    \end{minipage}
\end{center}

\section{Simulations of TOD}
\label{sec:Simulations-of-TOD}
The signal observed by a detector that is sensitive to an orientation of linear polarization can be approximated as
\begin{equation}
    s = I + \eta.(Q\,\mathrm{cos}2\psi+U\,\mathrm{sin}2\psi)+n,
    \label{eq:sky-sig}
\end{equation}
where $I$, $Q$ and $U$ are the Stokes parameters of the incident radiation, $\eta$ is the polarization efficiency of the detector, $n$ is the noise component and $\psi$ is the orientation in which the detector is sensitive to polarization. TOD consist of such samples of the signal observed by multiple detectors over a period of time. In order to simulate TOD, we approximate the sky as a pixelized map under an assumed pixelization scheme. We then estimate which detectors in the focal plane are sampling a pixel $p$ at a given time while considering the scanning strategy of the telescope. 
\paragraph{}
Assuming that the TOD are linearly dependent on an \textit{input} map $m_{p,\mathrm{in}}=[I_p,Q_p,U_p]^T$ of the sky, they can be described as \cite{Tegmark:1997vs}
\begin{equation}
    y_{p,t} = P_{p,t}\,m_{p,\mathrm{in}} + n_{p,t}
    \label{eq:ch-5-ypt}
\end{equation}
where $y_{p,t}$ is a vector of samples from sky pixel $p$ at time $t$, $n_{p,t}$ is the noise associated with each sample and $P_{p,t}$ is the pointing matrix that contains information on which detectors are sampling pixel $p$ at time $t$ that maps an input map (Section \ref{sec:ch5-skysignal}) into TOD.
\paragraph{Pointing Matrix} The mapping of the detector pointings to a sky pixel involves two operations: the projection of the detector's line-of-sight (LoS) with respect to the boresight and the projection of the boresight onto the sky pixel, the former is described by the detector quaternions; the latter is then described by three angles: $\theta_t$, $\phi_t$ and $\psi_t$. $\theta_t$ and $\phi_t$  describe the boresight direction in sky coordinates and $\psi_t$ is the angle on the plane orthogonal to the direction of observation between the telescope coordinates and sky coordinates  i.e. $\psi_t=\psi_0+\psi^\prime_t$ where $\psi_0$ is the intrinsic polarization orientation that the detector is sensitive to and $\psi^\prime_t$ is the relative orientation of the sky and the focal plane at time $t$. The $\psi_t$ described here is the same as in eq.\ \eqref{eq:sky-sig} but in sky coordinates. $P_{p,t}$ is a sparse matrix as not all detectors in the focal plane are sampling pixel $p$ at a time $t$.
\paragraph{} \texttt{Time Ordered Astrophysics Scalable Tools (TOAST)}\footnote{\url{https://toast-cmb.readthedocs.io/en/latest/}} is a software framework that offers tools for simulation and processing of telescope timestreams. \toast handles data that is organized into individual observations, each of which is independent from the others. An observation consists of co-sampled detectors over a specific time period, with the intrinsic detector noise assumed to be stationary during that period. Typically, other factors such as elevation and weather conditions (eg. precipitable water vapour (PWV) in the atmosphere, wind properties) remain constant within an observation.
\paragraph{} The following subsections are all the aspects that comprise a sub-mm telescope observation, and the ingredients necessary for a TOD simulation.
\subsection{Focal plane}
The focal plane of the telescope is simulated using the \texttt{sotodlib} module currently being developed by the Simons Observatory (SO) collaboration. A focal plane simulated by \texttt{sotodlib} is essentially an \texttt{Ordered Dictionary} containing the detector quaternions\footnote{A quaternion $q$ is defined as $q = (q_0, \mathbf{q}) = q_0 + q_1 i + q_2 j + q_3 k$ where $q_0$ is a scalar and the $\mathbf{q} = (q_1, q_2, q_3)$ vector component comprises orthogonal imaginary quantities with the $i$, $j$, $k$ imaginary axes satisfying $i^2 = j^2 = k^2 = -1$, $ij = -ji = k$, $jk = -kj = i$, and $ki = -ik = j$. A unitary quaternion $q$ provides a compact representation of a rotation around an arbitrary axis. A rotation of a vector $\mathbf{v}$ about an axis is calculated by $\mathbf{v'}=\mathbf{qvq}^{-1}$ while considering the vector as an imaginary quaternion, $v = (0,\mathbf{v})$.} with respect to the boresight and the variables describing the detector noise. 
\begin{figure}[ht]
    \centering
    \includegraphics[width=0.8\textwidth]{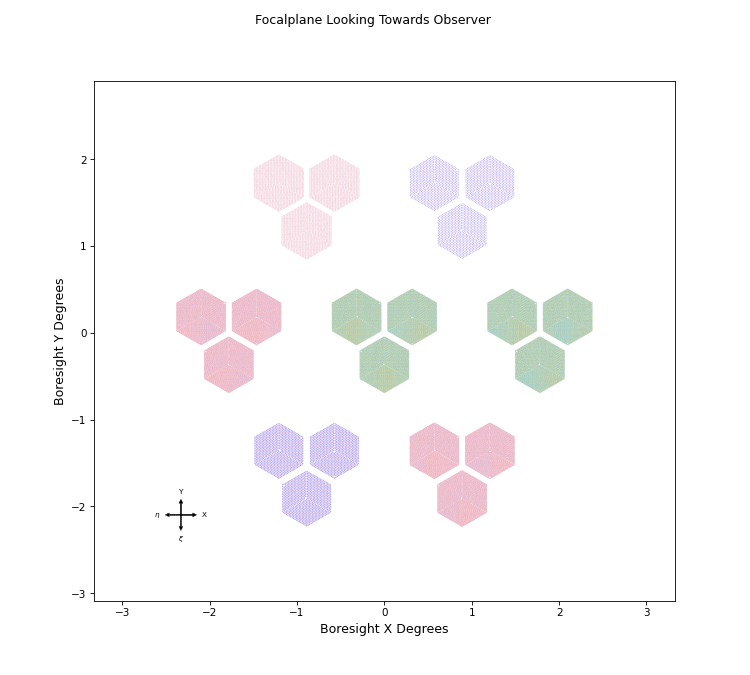}
    \caption{The FYST Prime-Cam focal plane as seen from the sky when the telescope is scanning at an elevation of $60^\circ$. Each hexagon represents one wafer of detector arrays and three wafers comprise a frequency module. \textit{Top row}: 220 (\textit{left}) and 280 GHz module (\textit{right}); \textit{Middle row}: EoR-Spec module (260 or 360 GHz), 850 and 350 GHz modules (\textit{left} to \textit{right}); \textit{Bottom row}: EoR-Spec module (260 or 360 GHz) (\textit{left}) and 410 GHz (\textit{right}) modules. The positioning of the EoR-Spec modules in the cryostat is tentative.}
    \label{fig:focplane}
\end{figure}

\texttt{sotodlib}\footnote{\url{https://github.com/ccatobs/sotodlib}} was modified to accommodate the FYST configuration of detector arrays. A projection of the focal plane in the XY plane centered at the boresight as seen from the sky is shown in Figure \ref{fig:focplane}. 

\begin{wrapfigure}[10]{r}{0.33\textwidth}
    \centering
    \begin{center}
        \includegraphics[width=0.26\textwidth]{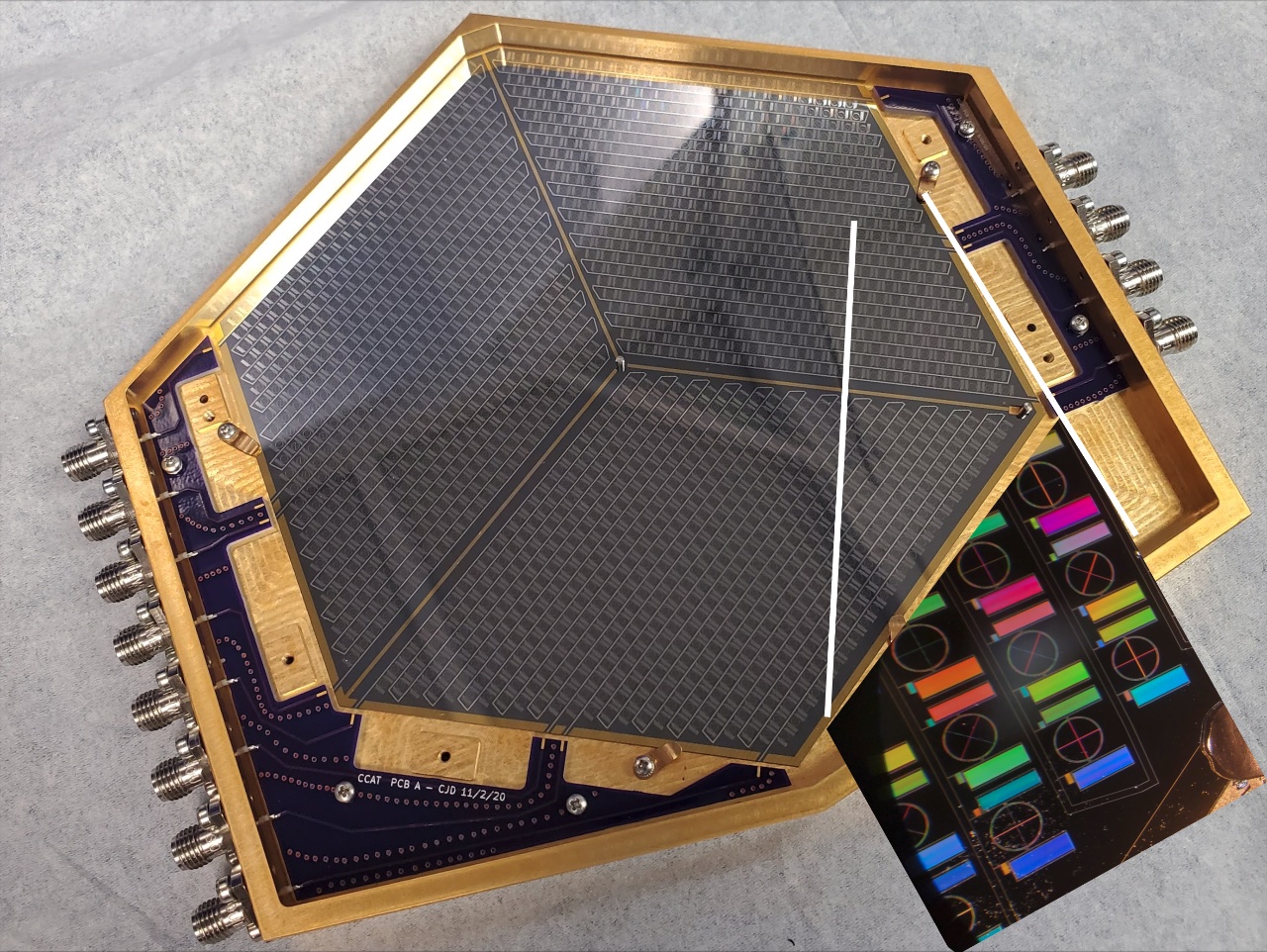}
    \end{center}
    \caption{280 GHz array assembled on gold-plated aluminum
base. Image from \cite{2022JLTP..209..849C}.}
    \label{fig:choi-280array}   
\end{wrapfigure}

Each frequency module consists of three detector array \textit{wafers} with each \textit{wafer} consisting three rhombi of detectors. The detectors in each rhombus are oriented at $45^\circ,\, 90^\circ,\,135^\circ$ and $180^\circ$ with respect to the center of the \textit{wafer}, yielding four unique orientations that the detectors are sensitive to per rhombus. The rhombi are aligned at an angle of $120^\circ$ with respect to each other to accommodate the readout electronics (Figure \ref{fig:choi-280array}).
While the orientations of polarization sensitivity of two rhombi now mirror each other, the detectors in the third rhombus are sensitive along $15^\circ,\,60^\circ,\,105^\circ$ and $150^\circ$. Thus, an alignment at an angle of $120^\circ$ of the rhombi with each other within the \textit{wafer} enables each \textit{wafer} to accommodate 8 unique orientations that the polarization sensitive detectors are sensitive to. This ensures that each sky pixel is sampled at least by four different orientations which is the minimum requirement to reconstruct the $I$, $Q$ and $U$ Stokes parameters for a sky pixel (as seen in Equation \ref{eq:sky-sig}). 

\subsection{Scanning and Survey strategies}
Given the wide range of scientific goals planned for FYST (presented in Section \ref{sec:ch2-FYST}), different scanning and survey strategies need to be employed. A survey strategy corresponds to the region of the sky to be observed/mapped and the period of observation. A scanning strategy refers to the way that a telescope needs to observe the region of interest in order to achieve smaller pixel variance.
One can place the following demands on a scanning strategy to achieve minimal noise residuals in the course of map-making:
\begin{itemize}
    \item Move the telescope at a speed where it samples the sky faster than the changes in atmosphere while accounting for the physical logistics of telescope movements.
    \item Maximal exposure of detectors to each sky pixel of interest such that each sky pixel is observed by multiple detectors to make corrections for instrumental variations, noise and gain fluctuations possible.
    \item The detectors should sample each sky pixel in different directions in order to make the data less susceptible to temporal variations of noise (such as atmosphere and gain fluctuations).
\end{itemize}
\subsubsection{Constant Elevation Scan strategy} 
\begin{figure}[ht]
    \centering
    \begin{subfigure}{0.49\textwidth}
        \includegraphics[width=\textwidth]{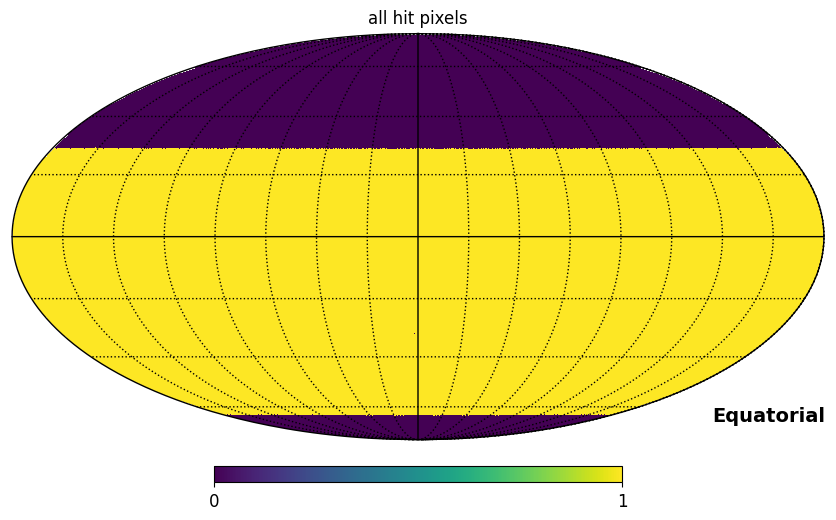}
    \end{subfigure}
    \begin{subfigure}{0.49\textwidth}
        \includegraphics[width=\textwidth]{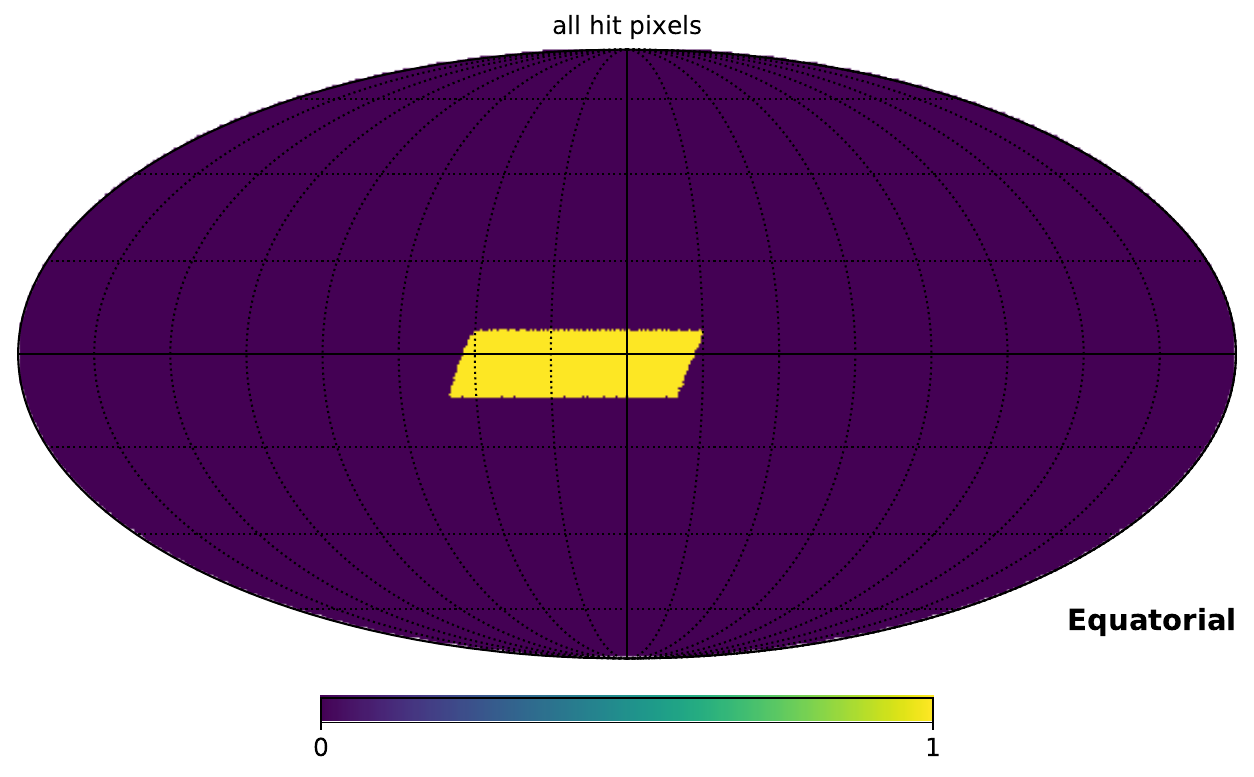}
    \end{subfigure}
    \caption{WFS (\textit{left}) and \textit{deep56} (\textit{right}) fields in Equatorial coordinates.}
    \label{fig:hitpixels}
\end{figure}

With the above considerations, the Constant Elevation Scan (CES) strategy is currently being considered for the wide-field survey (WFS) which would map $\sim$20,000 deg$^2$ of the sky over an estimated 4000 hours (a visualization of the WFS field is shown in the left panel of Figure \ref{fig:hitpixels}). CES corresponds to scanning of the field in the Azimuth back-and-forth at a constant elevation. This ensures that the telescope is observing through a constant air mass, the telescope cryogenics are stable, and the local environment and instrument offsets are sampled in a consistent manner. The scanning speed and acceleration of the telescope are dictated by instrumental effects, atmospheric changes, and the mechanical constraints of the telescope (such as vibrations of the optics). Further, in order to sample all the Fourier modes of the signal of interest (for example, with the CMB) and to minimize systematic effects in the map-making process, cross-linking technique is employed. Cross-linking involves observing at complementary central azimuthal angles to capture both rising and setting skies.
\subsubsection{Modulated High Cadence scanning strategy}
\begin{figure}[ht]
    \centering
    \includegraphics[width=0.8\textwidth]{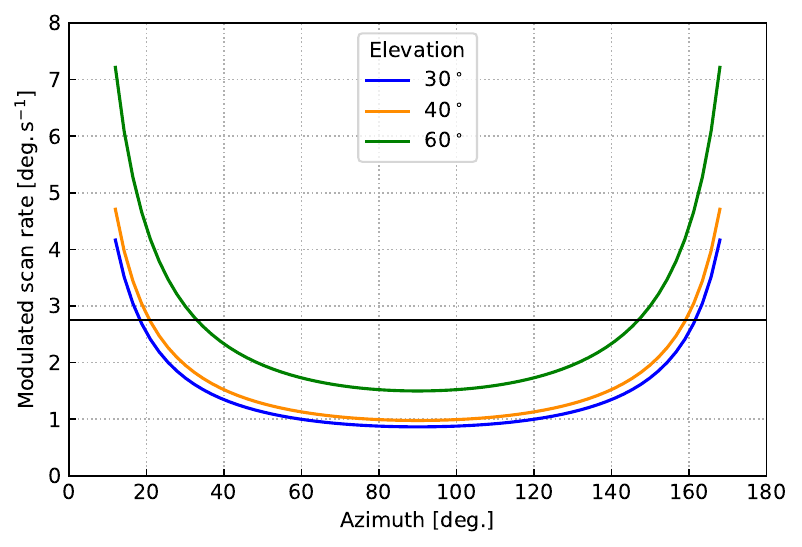}
    \caption{Modulated scan rate at fixed elevations of $30^\circ$, $40^\circ$ and $60^\circ$ plotted as a function of Azimuth. The Azimuth range has been constrained to [$12^\circ$, $178^\circ$] in this plot as the scan range diverges significantly beyond this range. The black horizontal line indicates the maximum telescope base scan rate of $2.75^\circ\,\mathrm{s}^{-1}$. For each constant elevation, a specific Azimuth range needs to be chosen to stay below the maximum base scan rate.}
    \label{fig:modscan-azimuth}
\end{figure}

The Modulated High Cadence (MHC) scanning strategy developed by \cite{Ebina:2022ojw} is also currently being considered for the WFS. This strategy is proposed to provide uniform observation and depth. It is essentially a CES with a slight modification: modulation of the scanning rate of the telescope. The scan rate is modified by
\begin{equation}
    \omega = \frac{\omega_0}{\mathrm{cos}\beta\,\mathrm{sin}\alpha},
\end{equation}
where $\omega$ is the azimuthal scan rate, $\omega_0$ is the base scan rate, $\beta$ is the elevation and $\alpha$ is the Azimuth. The modulated scan rate is thus determined by the Azimuth at which the telescope is planned to observe. For a base scan rate of $\omega_0=0.75^\circ\,\mathrm{s}^{-1}$, the modulated scan rate for different Azimuth is shown in Figure \ref{fig:modscan-azimuth}. Thus, keeping the maximum scan rate imposed, $\omega<2.75^\circ\mathrm{s}^{-1}$, the Azimuth range that the telescope is allowed to scan is limited to elevations of $30^\circ - 60^\circ$ for $\omega_0=0.75^\circ\,\mathrm{s}^{-1}$. This further informs the survey strategy employed by FYST.

\paragraph{}
In this work, we focus on the implementation of CES and MHC scanning strategies during rising skies.
\subsubsection{Survey Strategy}
The survey strategy is informed by the scanning strategy discussed previously. For the science case that involves measuring the CMB polarization signal, the WFS strategy will be implemented. For the purposes of TOD simulation, we consider the Atacama Cosmology Telescope's (ACT) \cite{Swetz:2010fy, ACTpol} \textit{deep56} \cite{DeBernardis:2016uxo} field to observe. The \textit{deep56} field covers $\sim 700\,\text{deg}^2$ while the WFS will cover $\sim 20,000\,\text{deg}^2$, including an overlap with the \textit{deep56} field, allowing for synergies with ACT. The survey fields are plotted in Equatorial coordinates in Figure \ref{fig:hitpixels}.

\subsection{Sky Signal}
\label{sec:ch5-skysignal}
\begin{figure}[ht]
    \centering
    \begin{subfigure}{1.\textwidth}
        \includegraphics[width=\textwidth]{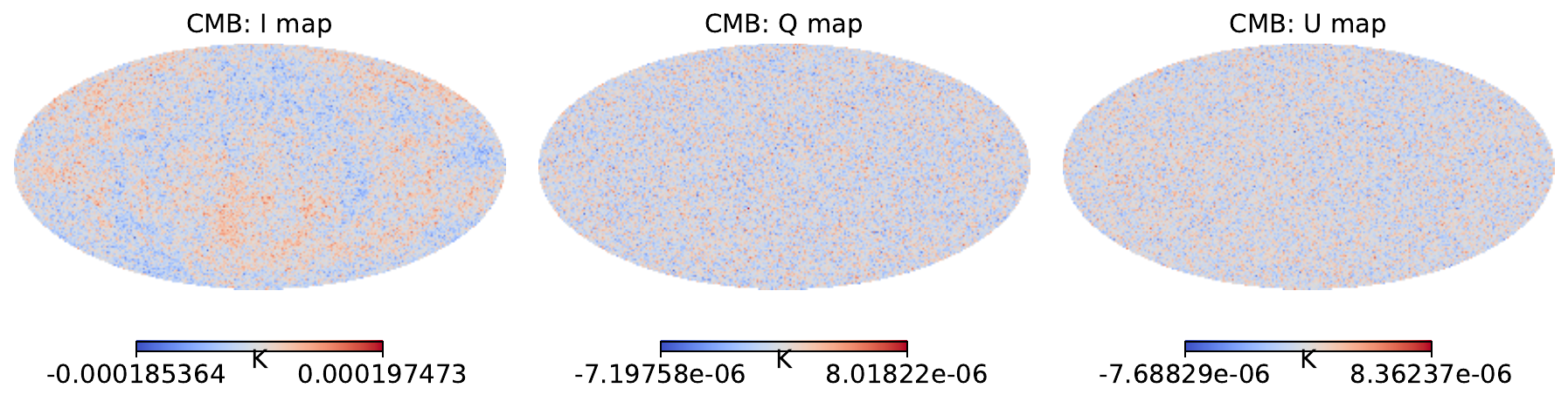}    
    \end{subfigure}
    \begin{subfigure}{1.\textwidth}
        \includegraphics[width=\textwidth]{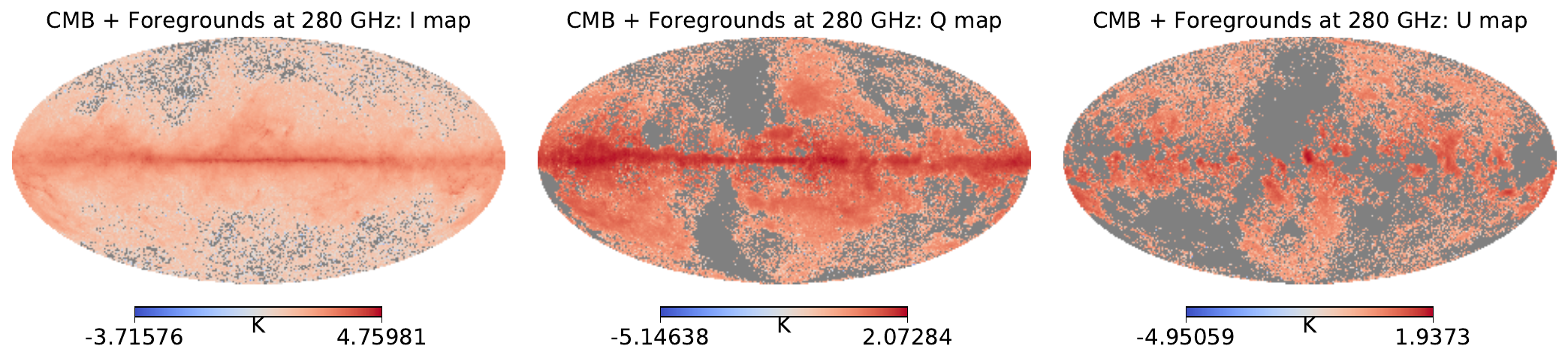}
    \end{subfigure}
    \caption{CMB-only (\textit{upper row}) and CMB+foregrounds (\textit{bottom row}) input maps convolved with $0.78'$ Gaussian symmetric beam at 280 GHz with \texttt{Nside}=1024.}
    \label{fig:ch-5-input-maps}
\end{figure}

We require a template of the astrophysical signal in the sky that will be observed in the form of Stokes $I$, $Q$, $U$ maps in \texttt{HEALPix} pixelization scheme \cite{Gorski2005}, hereon referred to as input maps (represented by $m_{p,\mathrm{in}}$ in eq.\ \eqref{eq:ch-5-ypt}). We consider two instances: CMB and CMB with foregrounds. The CMB-only maps are simulated by first generating theoretical $C_\ell$'s using the \texttt{Python} implementation of \texttt{CLASS} \cite{Blas2011} called \texttt{classy}\footnote{\url{https://pypi.org/project/classy/}}. The CMB+foregrounds map is generated using \texttt{PySM} \cite{PySM} with ``c1", ``d5", ``s1" and ``f1" models for CMB, dust, synchrotron and free-free emission. We consider \texttt{Nside}=1024 which corresponds to $\ell_\text{max}=2048$. To account for the beam of the detectors, the maps are convolved with a Gaussian symmetric beam of FWHM $0.78^\prime$ for the 280 GHz module. The maps are shown in Figure \ref{fig:ch-5-input-maps}.
\subsection{Detector noise}
\begin{figure}[ht]
    \centering
    \includegraphics[width=0.7\textwidth]{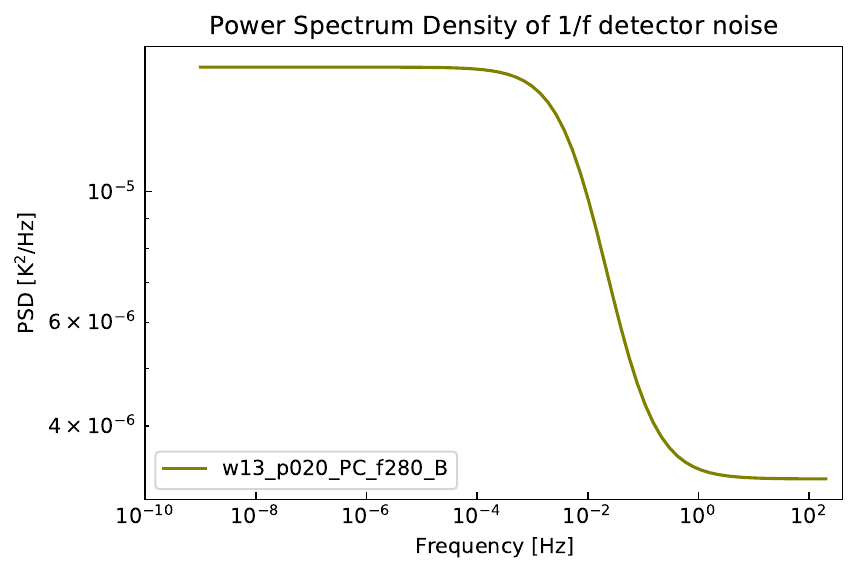}
    \caption{PSD of the detector noise (as described by eq.\ \eqref{eq:ch5-noise-psd}) assumed for the detectors as a function of frequency.}
    \label{fig:ch-5-detnoise-psd}
\end{figure}

We assume the noise in TOD due to the instrument is a linear combination of white noise and a low-frequency (temporally correlated) $1/f$ noise. The detector noise power spectrum density (PSD) is modelled as
\begin{equation}
    P(f) = \mathrm{NET}^2\Bigg(\frac{f^\alpha+f_\mathrm{knee}^\alpha}{f^\alpha+f_\mathrm{min}^\alpha} \Bigg).
    \label{eq:ch5-noise-psd}
\end{equation}
The Noise Equivalent Temperature (NET) is a measure of the sensitivity of the detector; $f$ is the temporal frequency; the knee frequency $f_\mathrm{knee}$ denotes the frequency at which the variance of the $1/f$ noise component is equivalent to the variance of the white noise component; a minimum frequency $f_\mathrm{min}$ is included such that $P(f)$ flattens for $f<f_\mathrm{min}$\footnote{Without the inclusion of $f_\mathrm{min}$, the PSD diverges for $f=0$. The inverse of the lifetime of the Universe determines the lowest observable frequency \cite{Milotti, Monelli:2024eyn} which is $\sim 10^{-17}$ Hz.}; $\alpha$ is typically positive in the definition of the PSD as described in eq.\ \eqref{eq:ch5-noise-psd}. In our simulations, we consider  $f_\mathrm{knee}=50\,$mHz, $f_\mathrm{min}=0.01\,$Hz, sampling frequency $f_\mathrm{s} = 400\,$Hz, NET = 1803.1 $\mu \mathrm{K}^2\,\sqrt{\mathrm{s}}$ and $\alpha=1.0$ for each detector \cite{CCAT-Prime2021}. The PSD of the detector noise is plotted in Figure \ref{fig:ch-5-detnoise-psd}. 
\subsection{Atmosphere}
The total background emission in intensity observed by a given instrument (in the submm regime) onboard a ground-based telescope is essentially the CMB, emission from foregrounds, emission from the telescope optics and the atmosphere:
\begin{equation}
    I(\nu) = B_\nu(T_\mathrm{CMB})+I_\mathrm{FG}(\nu)+\epsilon_\mathrm{tel}(\nu)\,B_\nu(T_\mathrm{tel}) + \big[1-\varepsilon(\nu)\big]\,B_\nu(T_\mathrm{atm}),
\end{equation}
where $B_\nu(T)$ describes the thermal emission of a blackbody of temperature $T$ with $T_\mathrm{CMB},\, T_\mathrm{tel}$, and $T_\mathrm{atm}$ being the temperature of the CMB, temperature of the telescope and temperature of the atmosphere, respectively; $I_\mathrm{FG}$ is the emission from foregrounds; $\epsilon_\mathrm{tel}$ is the emissivity of the telescope optics (mirrors or reflectors) that is approximated as $\epsilon_\mathrm{tel}\sim \nu^\gamma$. Further,
\begin{equation}
    \varepsilon(\nu) = \mathrm{e}^{-m(90^\circ - \theta)\,\tau_0},
\end{equation}
is the atmospheric transmission written in terms of the airmass $m$ at elevation $\theta$ and $\tau_0$ is the optical depth at zenith. Apart from contributing to the background emission, the physical fluctuations in the atmosphere contribute to fluctuating emission in time and direction of observation. Modelling the atmospheric fluctuations is complex and is dependent on the time of observation (temperature, pressure, etc.) and the observation site (air column density, water vapour, etc.). Displacement of the atmosphere by winds can introduce non-stationary effects. In order to simulate the atmospheric TOD, \toast implements the following method:
\begin{itemize}
    \item A cuboidal volume of the atmosphere is assumed and the detectors' LoS are estimated after displacing them by a certain factor due to assumed wind parameters i.e. direction and amplitude of the wind. 
    \item The volume is compressed for computational purposes by only retaining the volume elements corresponding to the beam size and a length scale of turbulence relevant to the fluctuations under consideration.
    \item An element-by-element covariance matrix is constructed following the prescription proposed in \cite{POLARBEAR:2015mbo} which is then projected onto the compressed volume of the atmosphere.
    \item The detector TOD are simulated by integrating along the LoS through this volume.
\end{itemize}
The atmosphere introduces a $1/f$ low-frequency noise component correlated across time. The modelling of this atmospheric component of the noise is essential to ground-based telescopes. \cite{POLARBEAR:2015mbo} find a polarization fraction of $<1\%$ for the atmosphere from POLARBEAR Year-one observations \cite{POLARBEAR:2014hgp}. While the linearly polarized component of atmospheric emission can be minimal, instrumental effects such as beam mismatch can cause leakage from intensity to polarization \cite{Shimon:2007au}. 

\section{Map-Making methods}
\label{sec:ch05-mapmaking}
While one can extract information from TOD through a likelihood analysis, it is computationally intensive (especially for science cases with large surveys such as the CMB). Reduction of the TOD into maps assuming a pixelization scheme is thus recommended.
\paragraph{}
We first define the mapmaking problem and then describe the various methods employed to solve it. While the reconstruction of the sky signal (or any signal of interest) can be performed in pixel space or harmonic space, we only consider the pixel space approach in this work. Assuming a linear relation between the TOD sampled from a specific sky pixel (under a given pixelization scheme) by detectors and the map corresponding to said pixel that we are interested in reconstructing, the TOD, 
$\mathbf{d}$, sampled at time $t$ within sky pixel $p$ can be expressed as (eq.\ \eqref{eq:ch-5-ypt})
\begin{equation}
    \mathbf{d} = \textbf{P\,m} + \mathbf{n},
    \label{eq:tod}
\end{equation}
where $\mathbf{n}$ is the noise vector associated with the TOD vector \textbf{d}, $\textbf{m}$ corresponds to the Stokes parameters ($T$/$I$, $Q$, $U$) maps that we are interested in reconstructing, and $\textbf{P}$ is the response matrix that contains information on the detector pointings. $P_{p,t}$ essentially maps the detector's quaternion pointing to a pixelized representation of the sky.
\paragraph{}

\paragraph{}

Map-making is described as a \textit{problem} due to the sheer dimensions of the matrices involved in eq. \eqref{eq:tod} and the behaviour of the correlated noise for a given scanning strategy. $\mathbf{P}$ has dimensions ($N_\mathrm{obs}$, 3$N_\mathrm{pix}$) where $N_\mathrm{obs}=n_\mathrm{det}n_\mathrm{samples}$ is the size of the $\mathbf{d}$ vector (the number of samples $n_\mathrm{samples}$ collected by $n_\mathrm{det}$ number of detectors over the course of observation) and $N_\mathrm{pix}$ is the number of pixels that comprise the reconstructed pixelized sky. A generalized representation of the map-making solution in the linear case can simply be expressed as \cite{Tegmark:1996qs}
\begin{equation}
    \widehat{\mathbf{m}} = M\cdot\mathbf{d},
    \label{eq:ch-5-general-mapmaker}
\end{equation}
where the definition of the matrix $M$ depends on the mapmaking method and $\widehat{\mathbf{m}}$ is the reconstructed Stokes $I$, $Q$, $U$ maps. In the upcoming subsections, we shall look at a few mapmaking methods and the observed noise properties in the context of FYST configuration.
\subsection{Bin-averaging}
\label{subsec:binning}
\begin{figure}[htbp]
    \centering
    \includegraphics[width=0.8\textwidth]{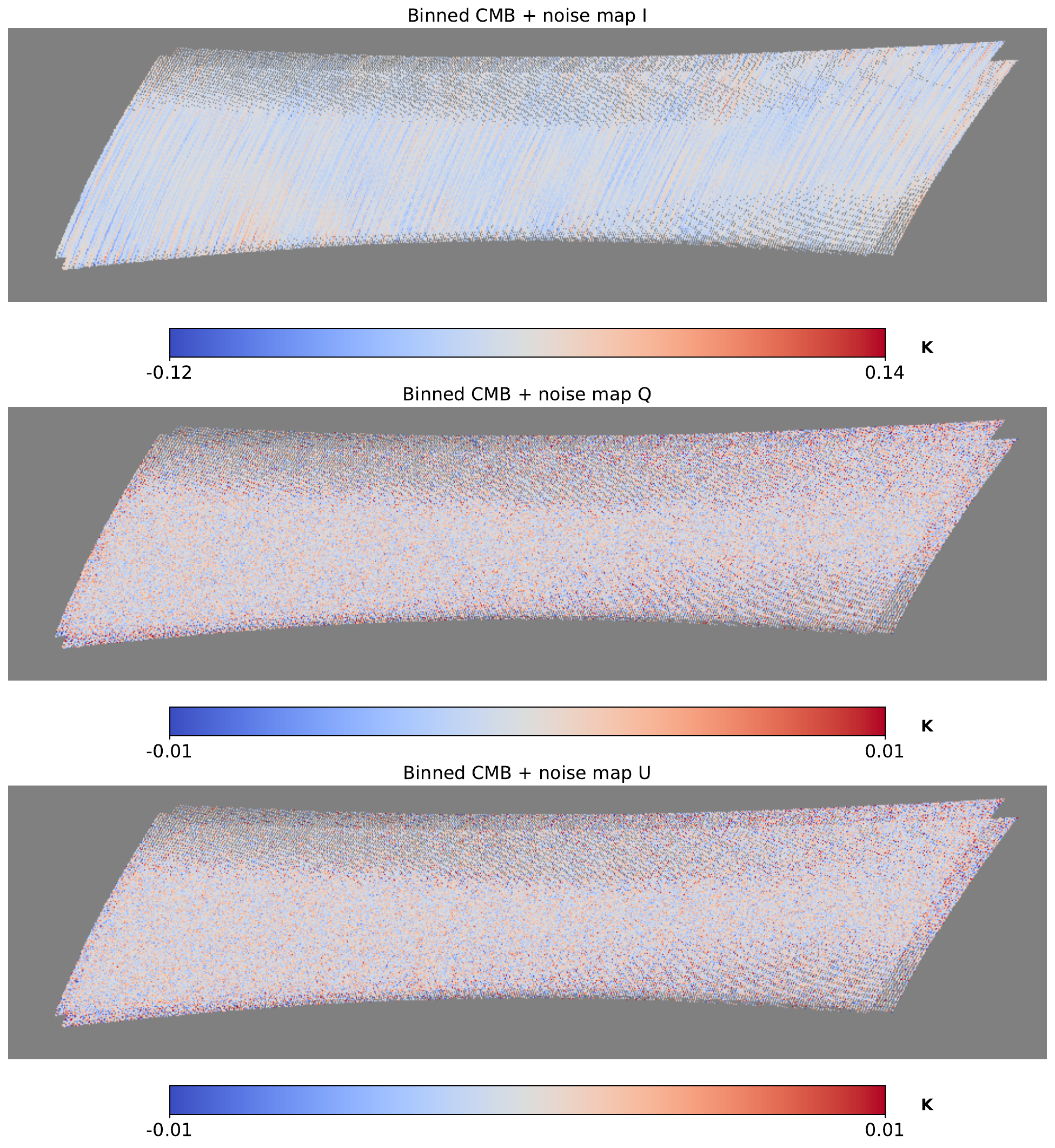}
    \caption{Binned maps reconstructed from simulated TOD comprising the CMB, detector and atmospheric noise at \texttt{Nside}=1024. Each \textit{row} corresponds to the Stokes $I$ (\textit{top}), $Q$ (\textit{middle}) and $U$ (\textit{bottom}) maps. The correlated noise causes the striping observed in the maps. The small gaps in the maps are recovered when the fields are observed for longer periods of time or by decreasing the \texttt{Nside}.}
    \label{fig:binnedmap_stripe}
\end{figure}

Bin-averaging (or binning) is one of the least computationally intensive map-making methods that can be implemented to reconstruct $\widehat{\mathbf{m}}$ from eq.\ \eqref{eq:ch-5-general-mapmaker}. Bin-averaging requires $M=(\widehat{P}^\mathrm{T} \widehat{P})^{-1} \widehat{P}^\mathrm{T}$ such that
\begin{equation}
    \widehat{\mathbf{m}} = (\widehat{P}^\mathrm{T} \widehat{P})^{-1} \widehat{P}^\mathrm{T} \cdot \mathbf{d},
\end{equation}
and $\widehat{P}$ is the response matrix assumed by the map-maker. When $\widehat{P} = P$, one is able to recover the sky maps, up to a noise term:
\begin{align}
    \widehat{\mathbf{m}} &= M \cdot \mathbf{d} = (P^\mathrm{T} P)^{-1} P^\mathrm{T} \cdot \left(P\cdot \mathbf{m} + \mathbf{n}\right) \nonumber\\
    &= \mathbf{m} + (P^\mathrm{T} P)^{-1} P^\mathrm{T} \cdot \mathbf{n}\,.
\end{align}
The binned $I$, $Q$ and $U$ maps generated from simulated TOD are shown in Figure \ref{fig:binnedmap_stripe}. One of the most glaring features observed here are stripes along the maps. This is due to $1/f$ low-frequency noise (instrument noise and atmosphere) which is correlated in time. Binning is thus not an ideal map-maker for experiments that involve instruments which generate such correlated noise and ground-based experiments which observe through the atmosphere.
\subsection{Destriping}
\label{subsec:destriping}
\begin{figure}[htbp]
    \centering
    \includegraphics[width=0.8\textwidth]{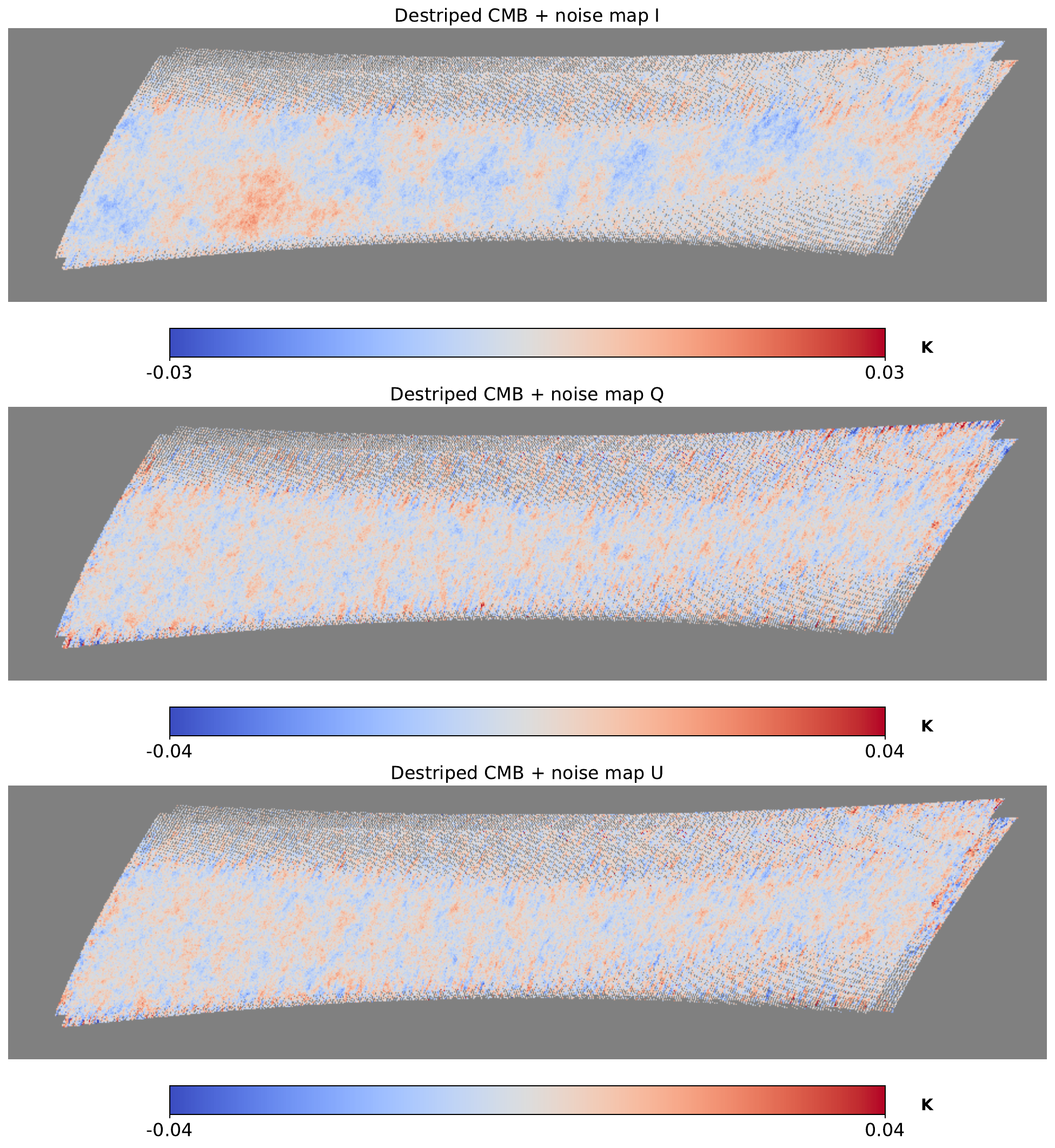}
    \caption{Destriped maps reconstructed from simulated TOD comprising the CMB, detector and atmospheric noise at \texttt{Nside}=1024. Each \textit{row} corresponds to the Stokes $I$ (\textit{top}), $Q$ (\textit{middle}) and $U$ (\textit{bottom}) maps.}
    \label{fig:destripedmap_cmbfg}
\end{figure}

\begin{figure}[htbp]
    \centering
    \includegraphics[width=0.8\textwidth]{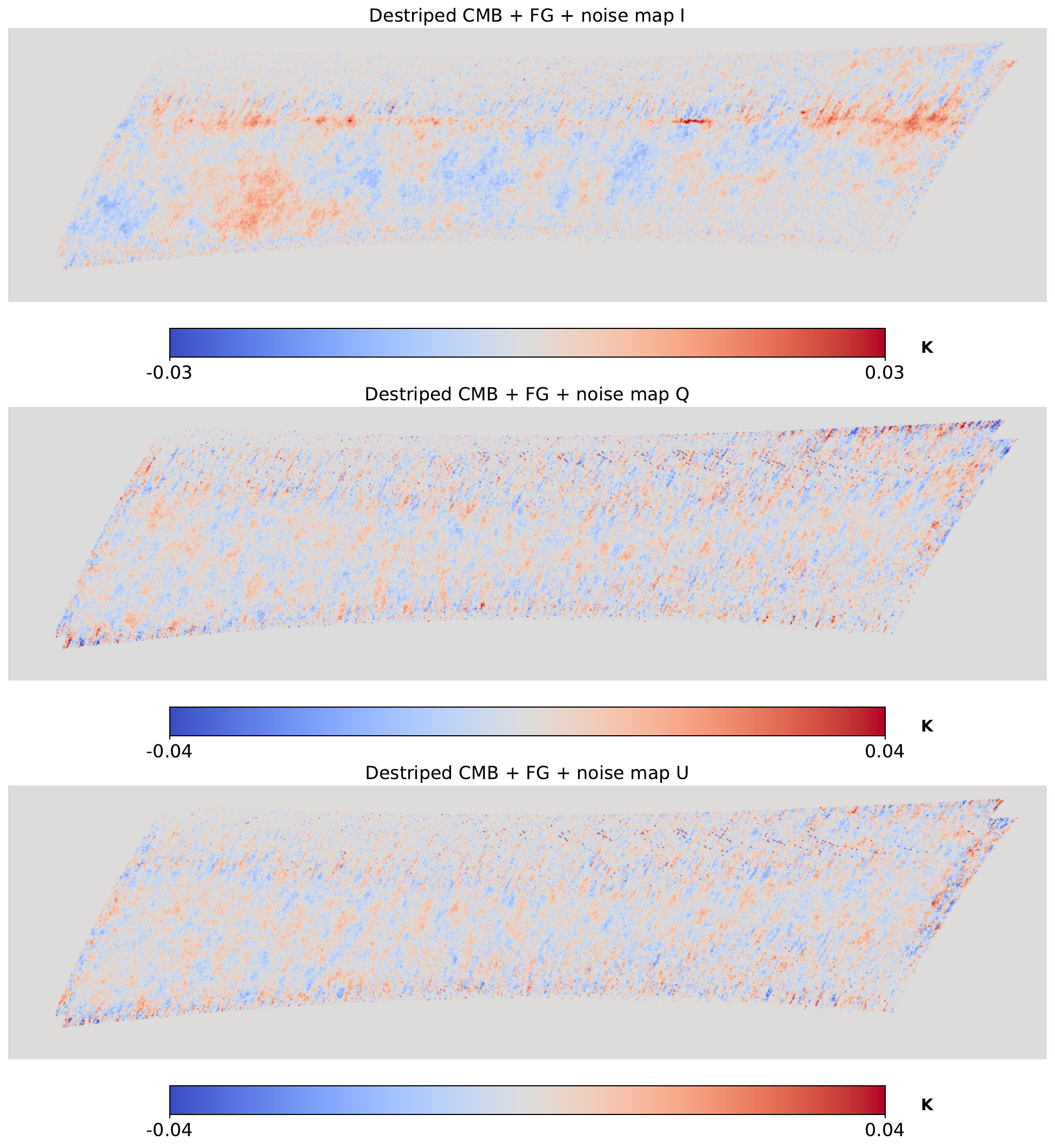}
    \caption{Destriped maps reconstructed from simulated TOD comprising the CMB, foreground emission, detector and atmospheric noise at \texttt{Nside}=1024. Each \textit{row} corresponds to the Stokes $T$ (\textit{top}), $Q$ (\textit{middle}) and $U$ (\textit{bottom}) maps. The galactic foreground emission is prominent in the temperature map.}
    \label{fig:destripedmap_cmbfg2}
\end{figure}

One of the ways to tackle the low-frequency $1/f$ noise is to implement the destriping map-making method. We divide the TOD into $n_b$ segments of equal length $n_{\text{base}}$; thus, $n_t = n_b \cdot n_{\text{base}}$. For each segment, we define an offset, called the baseline. The baselines model the $1/f$ noise, which will be subtracted from the TOD. The remaining noise is approximated as white noise.
Thus, the noise term in eq.\ \eqref{eq:tod} can be written as a combination of the correlated $1/f$ component modelled by a sequence of uniform baselines and a white noise component as \cite{Kurki-Suonio:2009clj, Keihanen:2009tj}
\begin{equation}
    \mathbf{n} = \mathbf{F\,a} + \mathbf{w}.
\end{equation}
Vector $\mathbf{a}$ contains the unknown amplitudes of the baselines, and matrix $\mathbf{F}$ maps them into TOD. The matrix $\mathbf{F}$ of size ($N_\mathrm{obs}$, $n_b$) is slightly modified from $\mathbf{P}$ as we now consider baselines instead of every sample i.e. it is a sparse matrix with zeroes and ones indicating which samples belong to which baseline. Assuming that the white noise component and the $1/f$ component are independent, the total noise covariance in the time domain can be expressed as
\begin{equation}
    \mathbf{C_t = \langle nn}^\mathrm{T} \rangle = \mathbf{F C_a F}^\mathrm{T} + \mathbf{C_w},
\end{equation}
where $\mathbf{C_w} = \langle \mathbf{ww}^\mathrm{T}\rangle$\footnote{The $\langle\rangle$ operation corresponds to the expectation value} is the covariance of the white noise component which is diagonal (with elements $\sigma_t^2$), and $\mathbf{C_a} = \langle \mathbf{aa}^\mathrm{T}\rangle$ is the covariance matrix of the baseline amplitudes $\mathbf{a}$.
\paragraph{}
Framing the problem in terms of the maximization of the likelihood, we need to minimize the $\chi^2$ which is written as
\begin{equation}
    \chi^2 = (\mathbf{d-Fa-Pm})\mathbf{C_w}^{-1}(\mathbf{d-Fa-Pm})+\mathbf{a^TC_a^{-1}a}.
    \label{eq:ch-5-chi2-destripe}
\end{equation}
The $\chi^2$ defined above in eq.\ \eqref{eq:ch-5-chi2-destripe} needs to be minimized with respect to both $\mathbf{a}$ and $\mathbf{m}$.
Minimizing with respect to \textbf{m} yields
\begin{equation}
    \mathbf{m} = (\mathbf{P^TC_n^{-1}P})^{-1}\mathbf{P^TC_n}^{-1}(\mathbf{d-Fa}).
    \label{eq:destripe-m1}
\end{equation}
Eq.\ \eqref{eq:destripe-m1} is substituted into eq.\ \eqref{eq:ch-5-chi2-destripe} to enable minimization with respect to \textbf{a} to obtain the baseline amplitudes. The solution for \textbf{a} is given by
\begin{equation}
    (\mathbf{F^TC_n^{-1}ZF}+\mathbf{C_a^{-1}})\mathbf{a} = \mathbf{F^TC_n^{-1}Zd},
    \label{eq:destripe-a1}
\end{equation}
where
\begin{equation}
    \mathbf{Z} = \mathbf{I}-\mathbf{P}(\mathbf{P^TC_n^{-1}P})^{-1}\mathbf{P^TC_n^{-1}}.
    \label{eq:destripe-a2}
\end{equation}
Eqs.\ \eqref{eq:destripe-m1} and \eqref{eq:destripe-a1} can be considered the destriping solutions. The minimization of the two equations are performed iteratively until the minima are estimated. Destriped maps consisting of CMB and noise consisting atmosphere and detector noise are shown in Figure \ref{fig:destripedmap_cmbfg}. In comparison with Figure \ref{fig:binnedmap_stripe}, one can notice how destriping removed the stripe-like features that appear in maps due to the $1/f$ noise. In Figure \ref{fig:destripedmap_cmbfg2}, the same configuration with the inclusion of foregrounds in the input is shown. The Galactic foreground emission is prominent in the Stokes $T$ map.
\subsubsection{Baseline Lengths}
\begin{figure}[ht]
    \centering
    \begin{subfigure}[b]{0.49\textwidth}
    \includegraphics[width=\textwidth]
    {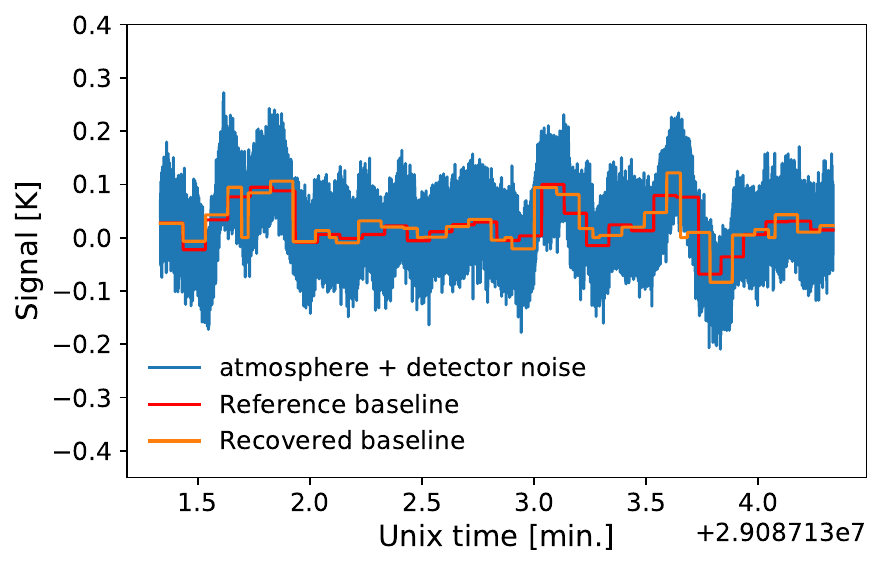}
    \end{subfigure}
    \centering
    \begin{subfigure}[b]{0.49\textwidth}
    \includegraphics[width=\textwidth]
    {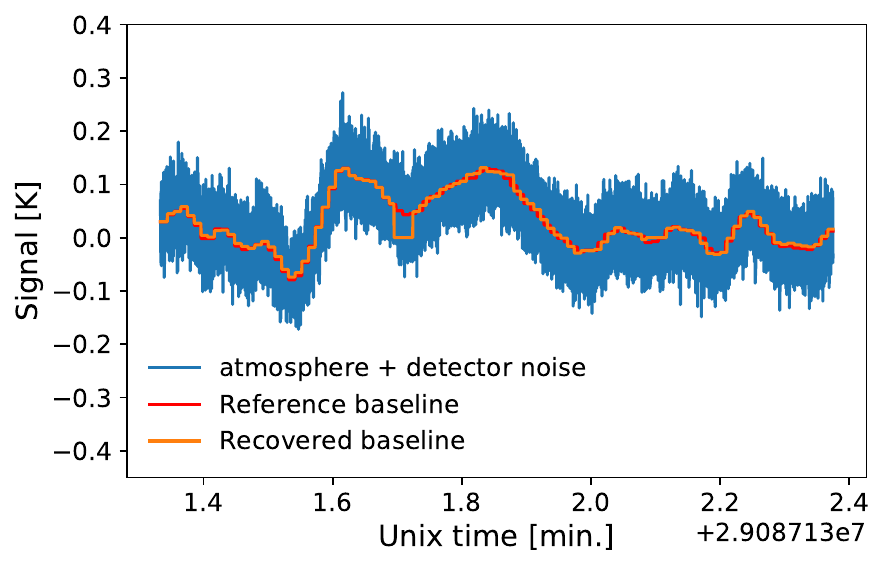}
    \end{subfigure}
    \caption{Recovered baselines (in \textit{orange}) from destriping of the timestreams from a detector when considering CMB, detector and atmospheric noise as input signal with baseline lengths $6\,$s (\textit{left}) and $0.6\,$s (\textit{right}). Also shown is the reference baseline (in \textit{red}) which is the noise timestream averaged over the baseline length considered. For the configuration of parameters considered, the shorter baseline appears to capture the $1/f$ correlated noise better.}
    \label{fig:recon-destripe-baselines}
\end{figure}

\begin{figure}[ht]
    \centering
    \includegraphics[width=0.5\textwidth]{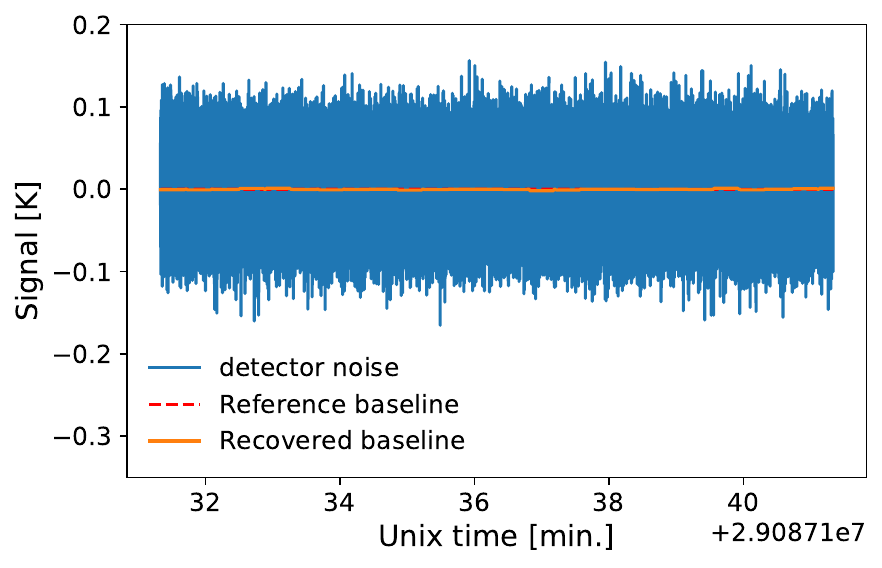}
    \caption{Recovered baselines (in \textit{orange}) from destriping of the timestreams from a detector when considering CMB and detector noise as input signal with baseline length of 300s. Also shown is the reference baseline (in \textit{red}) which is the noise timestream averaged over the baseline length considered. The detector noise timestreams are in \textit{blue}.}
    \label{fig:recon-destripe-detnoisebaselines}
\end{figure}

The $1/f$ component of noise that is temporally correlated is modelled over a time period called a baseline or baseline length. A minimization of the linear system described by Equation \ref{eq:tod} yields offsets to each sample. These offsets are then subtracted from the TOD to result in destriped TOD. The destriped TOD are then binned at the respective pixels to produce destriped maps. It is crucial to choose the right baseline length, which scales as $1/f_\mathrm{knee}$, over which one can assume the noise components to be correlated in order to model the $1/f$ copmonent of noise accurately and destripe the TOD resulting in minimal residuals in the reconstructed maps.
\paragraph{}
In Figure \ref{fig:recon-destripe-baselines}, a snippet of the recovered baselines for a baseline length of $6\,$seconds (\textit{left} panel) and $0.625\,$s (\textit{right} panel) from destriping for the case of input signal consisting of CMB, detector and atmospheric noise is shown. Along with the recovered baselines, the reference baselines, which are estimated by averaging the noise timestreams over the baseline length considered, are also plotted. In Figure \ref{fig:recon-destripe-detnoisebaselines}, the recovered baselines from destriping TOD consisting CMB and detector noise with a baseline length of $300\,$s is shown. The detector noise can be considered as stationary and thus, longer baselines can be used to approximate the detector $1/f$ noise. In the case of the atmospheric noise, since it is temporally variable, shorter baseline lengths ($<0.5\,$s which correspond to $f_\mathrm{knee}\sim 2\,$Hz) are able to recover the $1/f$ component better. The $f_\mathrm{knee}$ of the PSD describing atmospheric noise is dependent on the PWV and the size of the atmospheric structures that the detectors are exposed to. The $f_\mathrm{knee}$ increases with increasing PWV and \cite{Dunner:2012vp} find it to be in the range $1-5\,$Hz from observations at 148 GHz with the Atacama Cosmology Telescope (ACT) in Chile.

\section{Angular power spectra}
\begin{figure}[htbp]
    \centering
    \includegraphics[width=0.9\textwidth]{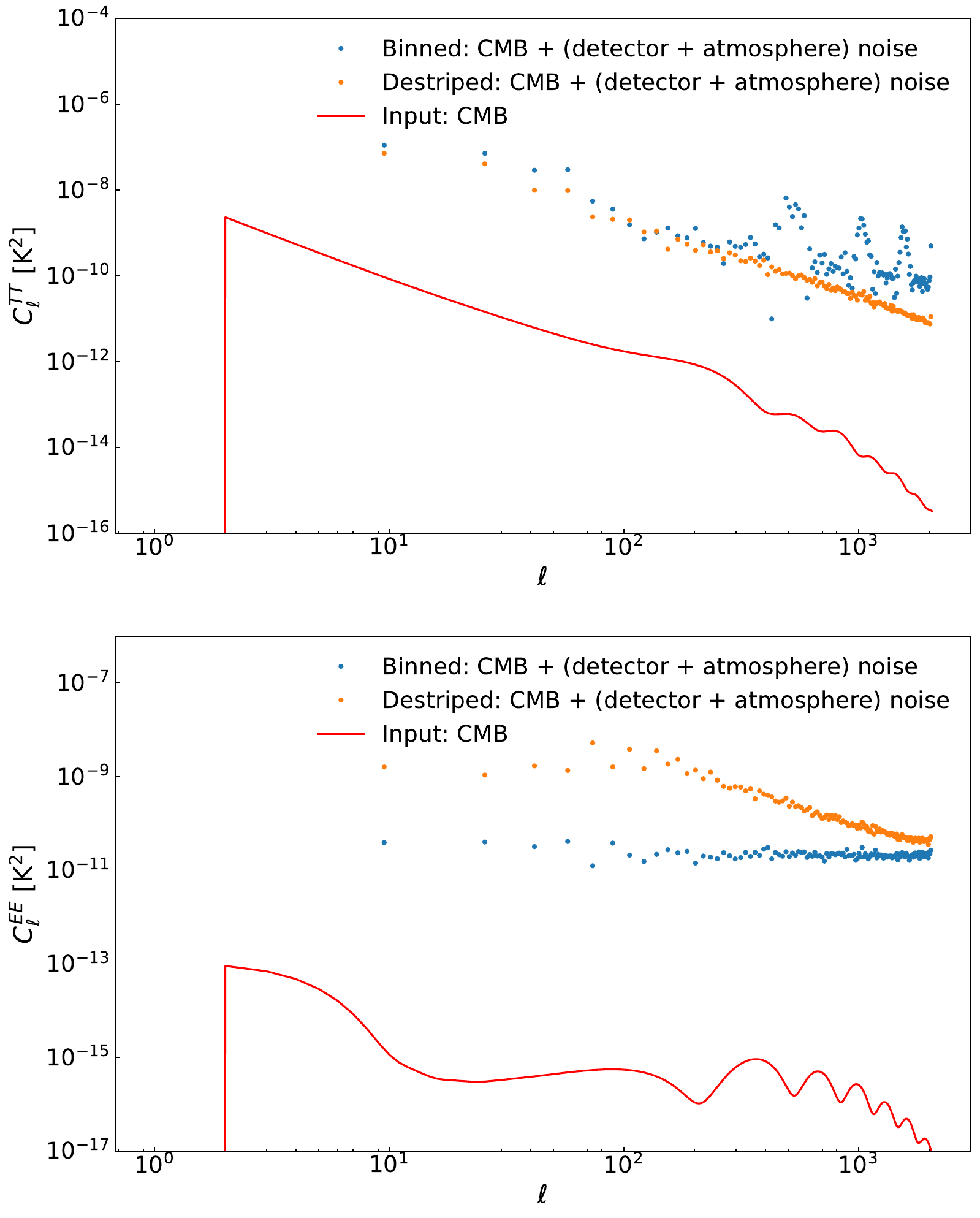}
    \caption{$C_\ell^{TT}$ (\textit{top}) and $C_\ell^{EE}$ (\textit{bottom}) estimated from simulated destriped (\textit{blue dots}) and binned (\textit{orange dots}) maps containing detector noise, atmosphere and CMB in comparison with input CMB $C_\ell^{TT}$ and $C_\ell^{EE}$ (\textit{red curves}). A baseline length of $0.1\,$s was considered in destriping the simulated TOD. A reduction in power at all scales is observed in the $C_\ell^{TT}$ from destriped map in comparison with those from binned map. The $1/f$ correlated noise is however still present in the destriped map. $C_\ell^{EE}$ from destriped map displays a $1/f$ feature which is absent in the binned map which is dominated by white noise indicating a bias introduced by destriping in the Stokes $Q$ and $U$ maps.}
    \label{fig:ch5-cmbdetnoise-angspectra}
\end{figure}

\begin{figure}[htbp]
    \centering
    \includegraphics[width=0.9\textwidth]{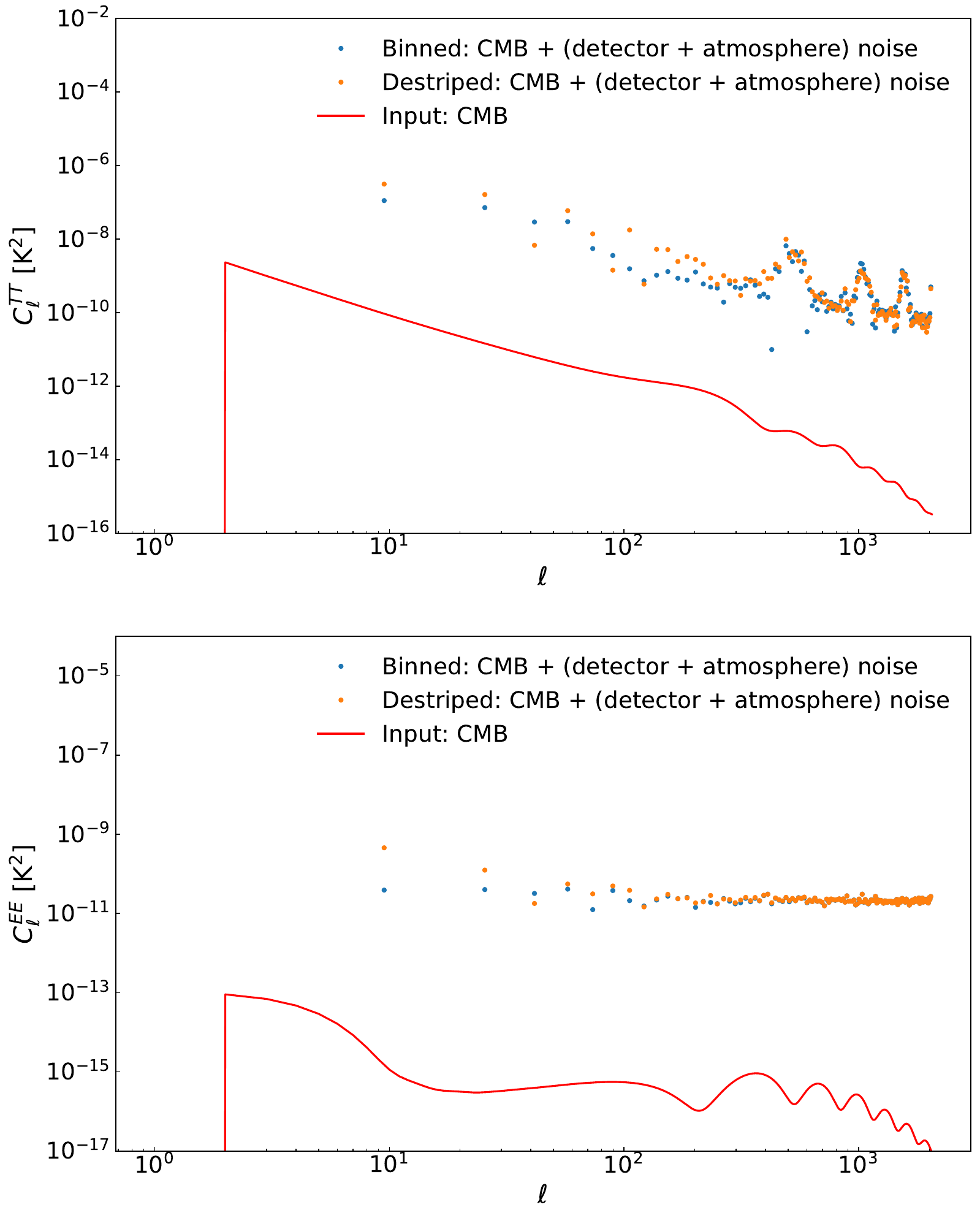}
    \caption{$C_\ell^{TT}$ (\textit{top}) and $C_\ell^{EE}$ (\textit{bottom}) estimated from simulated destriped (\textit{blue dots}) and binned (\textit{orange dots}) maps containing detector noise, atmosphere and CMB in comparison with input CMB $C_\ell^{TT}$ and $C_\ell^{EE}$ (\textit{red curves}). A baseline length of $300\,$s was considered in destriping the simulated TOD. The bias due to correlated $1/f$ noise is very prominent in the $C_\ell^{TT}$ from destriped maps as well as with those estimated from the binned maps. $C_\ell^{EE}$ from destriped map displays a $1/f$ feature at low$-\ell$ which is then dominated by white noise at intermediate and high$-\ell$. Long baseline length might be favourable in minimizing the $1/f$ residual noise in the $E-$mode polarization power spectra.}
    \label{fig:ch5-cmbdetnoise-angspectra-longbs}
\end{figure}

In a real CMB experiment, once maps (or $a_{lm}$, in the case of reduction of the TOD in Harmonic space) are generated, the next step is to perform foreground cleaning to mitigate contamination from foreground emission and to get an estimate of the CMB. Further, various summary statistics are implemented for cosmological inference from the estimate of the CMB. Several foreground cleaning methods such as template fitting, component separation and parametric methods (a review of the methods can be found in \cite{Ichiki:2014qga}) are available which exploit the differences between the frequency dependence of the SED of various polarized components of foreground emission (synchrotron and dust emission being the dominant components in the microwave regime) and the CMB. Foreground-cleaned maps, $a_{lm}$ coefficients or $C_\ell^{XY}$ can then be analyzed to constrain cosmological parameters \cite{Gerbino:2019okg} or extract other information depending on the scientific goals.
\paragraph{}
In this work, we estimate the temperature ($C_\ell^{TT}$) and $E-$mode polarization ($C_\ell^{EE}$) angular power spectra , from  binned and destriped maps that contain detector and atmospheric noise, and the CMB to compare the properties of the residual $1/f$ noise. We are working with a \textit{partial} sky field\footnote{This is often the case with ground-based telescopes that are limited by ground obscuration. Certain survey strategies and map-based analyses avoid the galactic plane to minimize contamination by foreground emission.} which leads to undefined $a_{lm}$ for certain $\ell$ and $m$ that causes coupling of modes. Pseudo-$C_\ell$ estimators \cite{Peebles:1973, Wandelt:2000av} are one of the methods introduced to overcome the issue of incomplete sky coverage to recover $a_{lm}$ from \textit{partial} sky fields with minimal loss of information. The Pseudo-$C_\ell$ algorithm implemented in the \texttt{NaMaster}\footnote{\url{https://github.com/LSSTDESC/NaMaster?tab=readme-ov-file}} module \cite{Alonso:2018jzx} is used to compute the $C_\ell^{XY}$ from simulated maps in this work.
\paragraph{}
We compare the $C_\ell^{TT}$ and $C_\ell^{EE}$ estimated from destriped and binned maps containing detector noise, atmosphere and CMB with the input CMB angular power spectra in Figure \ref{fig:ch5-cmbdetnoise-angspectra}. The destriped map is created from a baseline length of $0.1\,$s. There is an overall increase in power in the $C_\ell^{TT}$ due to the detector and atmospheric noise. Destriping has resulted in a small improvement in the reduction of the $1/f$ noise at large scales. However, destriping appears to have introduced correlated noise in the Stokes $Q$ and $U$ maps, resulting in the shape of the $C_\ell^{EE}$ that we observe in Figure \ref{fig:ch5-cmbdetnoise-angspectra} whereas the $C_\ell^{EE}$ from binned map indicates that the $Q$ and $U$ maps are dominated by white noise. This is also evident in the Stokes $Q$ and $U$ destriped maps in Figure \ref{fig:destripedmap_cmbfg} which show features that are absent in the binned maps in Figure \ref{fig:binnedmap_stripe}. In Figure \ref{fig:ch5-cmbdetnoise-angspectra-longbs}, we show the angular power spectra estimated from destriped maps reconstructed from a baseline length of $300\,$s. Here we observe that the $1/f$ correlated noise is not reduced in the temperature maps, however this long baseline length appears favourable for the reduction of $1/f$ noise in the polarization maps. In order to understand this differing trend in the temperature and polarization maps and subsequently in the angular power spectra, we shall look into the dependence of the residual correlated noise in the destriped maps on the choice of baseline lengths.

\subsubsection{Effect of Baseline lengths}
\label{subsec:angspectra-baseline}
\begin{figure}[ht]
    \centering
    \begin{subfigure}[b]{0.49\textwidth}
        \includegraphics[width=\textwidth]{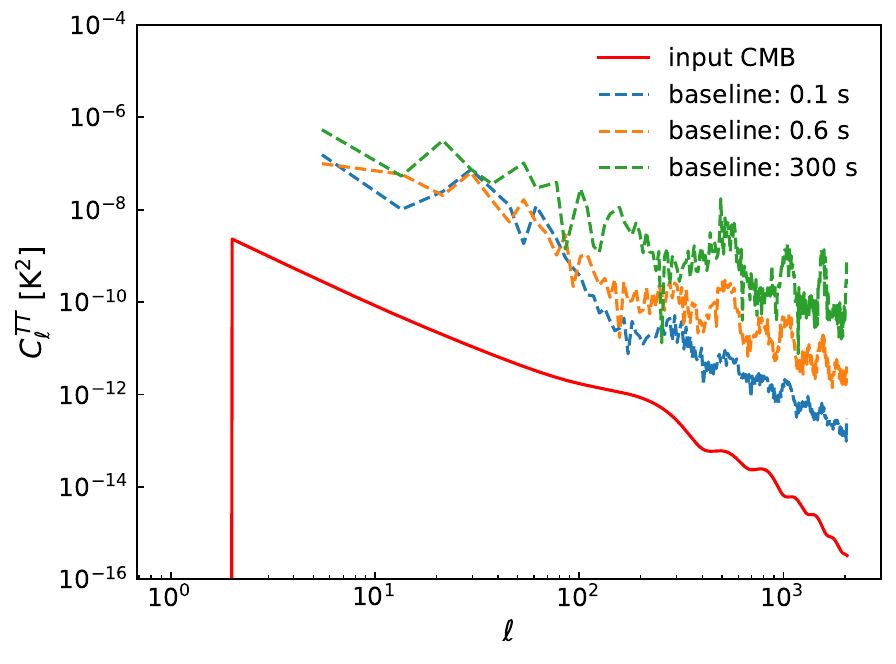}
    \end{subfigure}
    \centering
    \begin{subfigure}[b]{0.49\textwidth}
        \includegraphics[width=\textwidth]{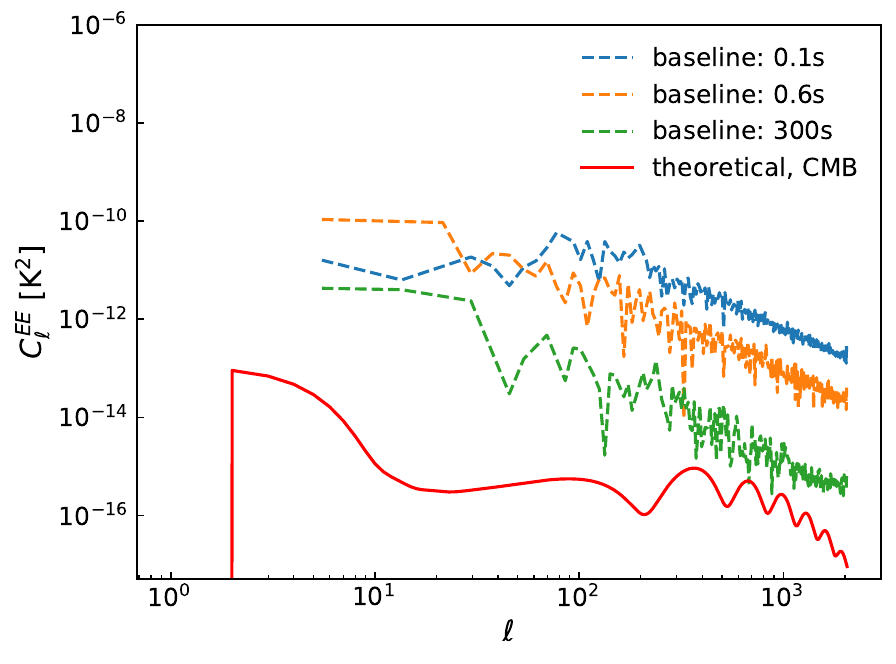}
    \end{subfigure}
    \caption{$C_\ell^{TT}$ and $C_\ell^{EE}$ of the residual correlated noise estimated from CRN maps reconstructed from atmospheric and detector noise timestreams in comparison with the input CMB (\textit{red curve}). The different colours correspond to correlated residual noise from destriping at different baseline lengths. The residual $1/f$ noise in temperature maps and the noise bias in $C_\ell^{TT}$ decreases with decreasing baseline lengths. However, the residual $1/f$ noise increases with decreasing baseline lengths in the polarization maps resulting in larger noise bias in the $C_\ell^{EE}$.}
    \label{fig:ch5-ang-powspectra-detatmosnoise}
\end{figure}

\begin{figure}[ht]
    \centering
    \begin{subfigure}[b]{0.49\textwidth}
        \includegraphics[width=\textwidth]{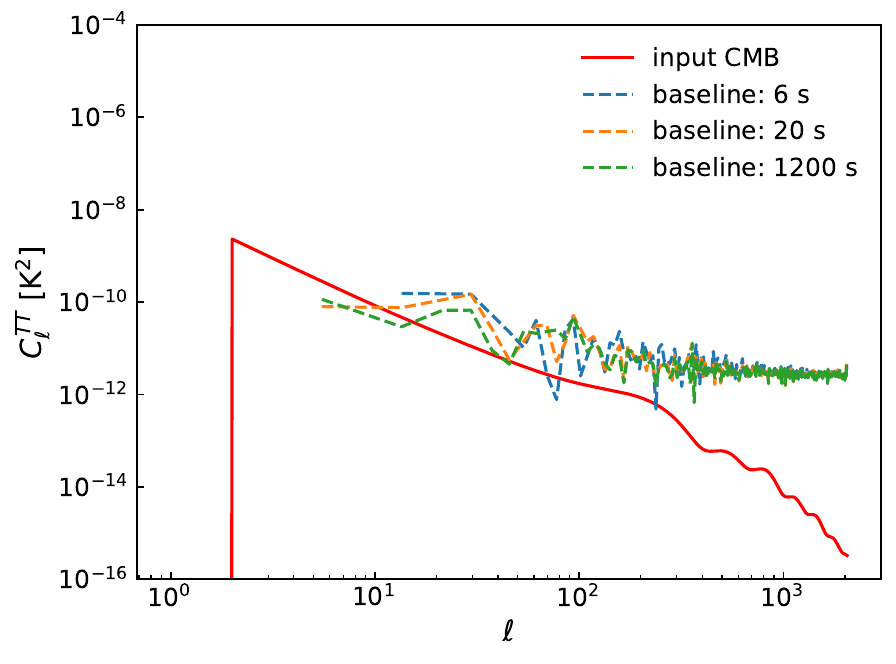}
    \end{subfigure}
    \centering
    \begin{subfigure}[b]{0.49\textwidth}
        \includegraphics[width=\textwidth]{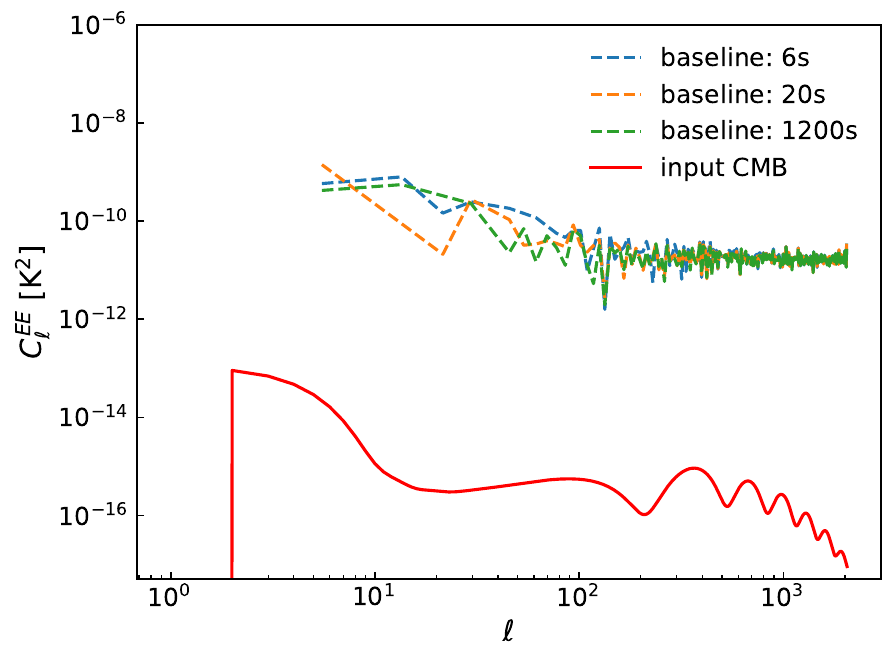}
    \end{subfigure}
    \caption{$C_\ell^{TT}$ and $C_\ell^{EE}$ of the residual correlated noise estimated from CRN maps reconstructed from detector noise timestreams in comparison with the input CMB (\textit{red curve}). The different colours correspond to correlated residual noise from destriping at different baseline lengths. White noise is a large contributing factor to noise bias in this instance.}
    \label{fig:ch5-ang-powspectra-detnoise}
\end{figure}

\begin{figure}[ht]
    \centering
    \begin{subfigure}[b]{0.49\textwidth}
        \includegraphics[width=\textwidth]{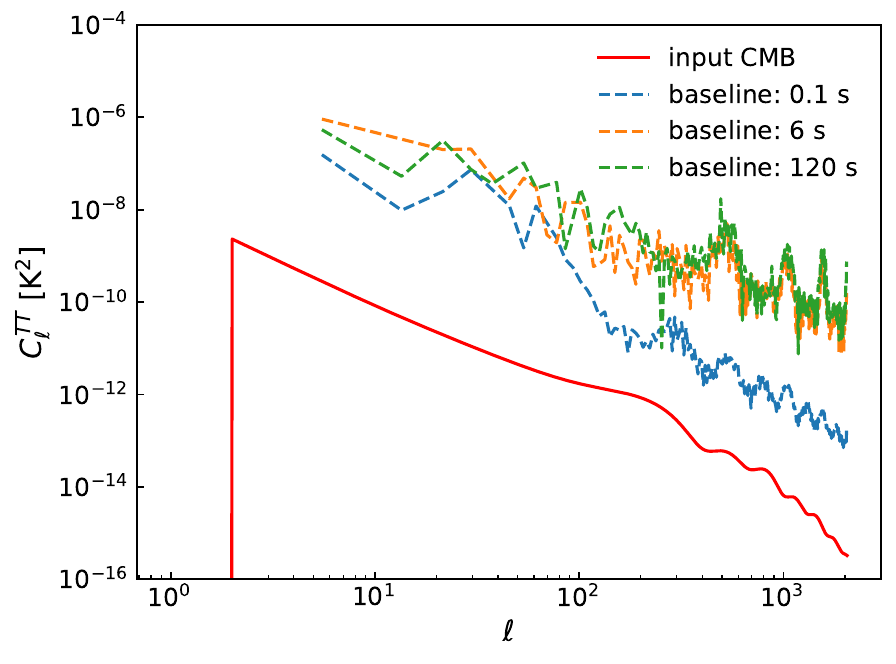}
    \end{subfigure}
    \centering
    \begin{subfigure}[b]{0.49\textwidth}
        \includegraphics[width=\textwidth]{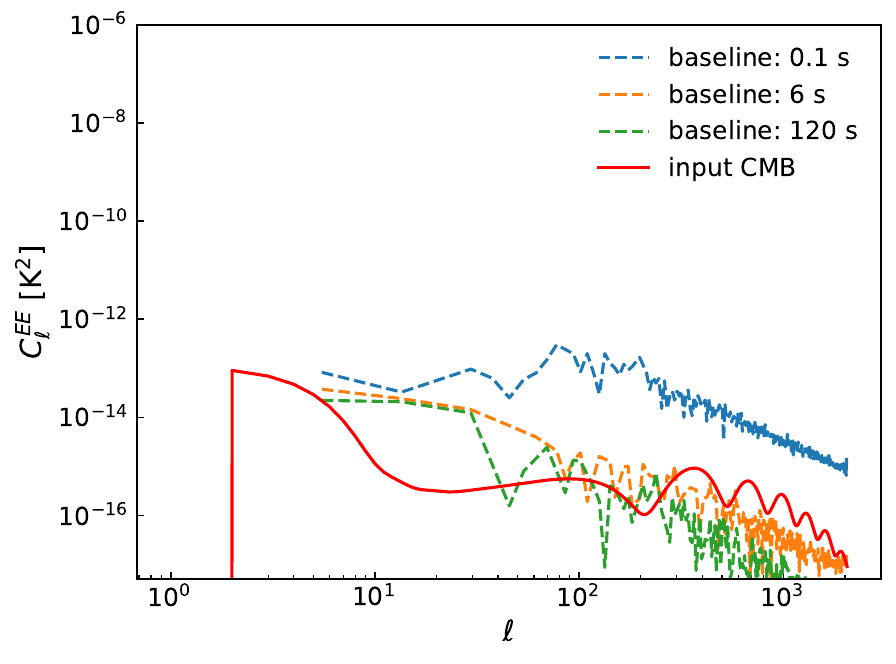}
    \end{subfigure}
    \caption{Same as Figure \ref{fig:ch5-ang-powspectra-detatmosnoise} but with lower NET used to describe the PSD of instrument noise used to simulate noise TOD.}
    \label{fig:ch5-ang-powspectra-detnoise-reduced}
\end{figure}

The residual correlated noise in destriped maps is dependent on the choice of baseline length. To understand this dependence, we estimate the $C_\ell^{TT}$ and $C_\ell^{EE}$ from \textit{correlated residual noise} (CRN) maps. CRN maps are computed by subtracting detector white noise binned maps from destriped $1/f$ noise (atmospheric and detector) maps. Figure \ref{fig:ch5-ang-powspectra-detatmosnoise} shows the $C_\ell^{TT}$ (\textit{top} panel) and $C_\ell^{EE}$ (\textit{bottom} panel) estimated from the CRN maps in comparison with the angular power spectrum of the input CMB. We find overall that the shorter baselines considered result in lower noise bias in $C_\ell^{TT}$. However, $C_\ell^{EE}$ appear to have higher noise bias with decreasing baseline length. This can be due to the shorter baselines capturing the $1/f$ component due to atmospheric noise well leading to lower residual correlated noise in the temperature maps. But since atmospheric emission is not the dominant noise component in polarization, destriping with baseline lengths that are shorter than the baseline length appropriate for $1/f$ due to detector noise will introduce a noise bias in the polarization maps and further in $C_\ell^{EE}$. We test this reasoning by considering the following two scenarios:
\begin{description}
    \item[Detector noise only TOD:] In this scenario we ignore atmospheric noise and only consider detector noise while simulating TOD. They are then destriped considering baseline lengths of $0.1\,$s, $0.6\,$s and $300\,$s. The CRN map for each baseline is estimated and the angular power spectra are computed, which are plotted in Figure \ref{fig:ch5-ang-powspectra-detnoise}. The noise bias in $C_\ell^{TT}$ at low$-\ell$ is reduced for increasing baselines (the opposite trend from when atmospheric noise was included) with white noise becoming dominant at high$-\ell$. The noise bias in $C_\ell^{EE}$ shows a similar trend, however a baseline length of $20\,$s appears to reduce variance due to correlated  noise well at low$-\ell$.
    \item[CMB + detector noise with lower NET + atmosphere TOD:] In this scenario we consider an NET (necessary to define the PSD of the detector noise, eq.\ \eqref{eq:ch5-noise-psd}) of $13\,\mu\mathrm{K\,\sqrt{s}}$ \cite{CCAT-Prime2021} to describe the detector TOD which is expected when the focal plane array is populated with the planned number of detectors. TOD consisting, detector (with improved sensitivity) and atmospheric noise are simulated and destriped assuming baseline lengths of $0.1\,$s, $6\,$s and $120\,$s. CRN maps are reconstructed and angular power spectra are estimated, which are shown in Figure \ref{fig:ch5-ang-powspectra-detnoise-reduced}. The noise bias in $C_\ell^{TT}$ is not different from the case with higher NET in Figure \ref{fig:ch5-ang-powspectra-detatmosnoise} and appears to decrease with shorter baselines. This final test indicates that atmospheric noise is the dominant noise component in temperature (or intensity) and requires very short baseline lengths to capture the $1/f$ correlation for destriping. Due to the lower NET considered,the noise bias in the $C_\ell^{EE}$ is significantly reduced in comparison with the case of instrument noise with higher NET in Figure \ref{fig:ch5-ang-powspectra-detatmosnoise}. The noise bias is the least for the longest baseline considered for destriping. This further indicates that the instrument noise and bias introduced by destriping at very short baselines that capture atmospheric noise would be the dominant component contributing to the noise bias in the $C_\ell^{EE}$.
\end{description}


\section{Discussion}
We have built a TOD simulation pipeline for the Prime-Cam module that will operate on FYST using the \texttt{TOAST} software framework. In a real CMB experiment, the raw TOD are first calibrated to estimate the measured signal in units of thermodynamic temperature, beam of the instruments and characterisitics of the detectors such as gain and polarization efficiency. Our simulation ignores these pre-processing aspects of the experiment by assuming that the measured signal is in units of temperature and that the beam of the detectors follow a Gaussian symmetric beam with FWHM of $0.78^\prime$ at 280 GHz (by providing input maps in K or K$_\mathrm{CMB}$ units convolved with the assumed beam). 
\paragraph{}
We have used the binning and destriping map making methods to reconstruct Stokes $I/T$, $Q$ and $U$ maps from the TOD. The effects of $1/f$ noise due to detector and atmospheric noise, and the choice of baseline length for destriping on the angular power spectra, specifically $C_\ell^{TT}$ and $C_\ell^{EE}$ are studied. Further, the characteristics of the noise bias in the angular power spectra due to correlated residual noise in destriped maps are studied for different baseline lengths considered for destriping.
\paragraph{}
We find the atmospheric noise to be the dominant noise component in temperature maps and the instrument noise in $Q$ and $U$ maps. Destriping with baseline lengths that capture the $1/f$ component of atmospheric noise appear to introduce noise bias in $C_\ell^{EE}$. Long baselines that capture the detector noise do not lead to a reduction in the noise bias in $C_\ell^{TT}$. These characteristics indicate that when using the destriping method to reconstruct maps, in the presence of atmospheric noise, one might need to consider different baseline lengths for reconstructing Stokes $I/T$ and Stokes $Q$, $U$ (polarization) maps. The characteristics of the residual correlated noise at low$-\ell$ indicate that a filtering of the TOD in the Fourier domain before destriping might be necessary for efficient removal of the correlated $1/f$ noise. Finally, the high sensitivity of FYST and the scan strategies being considered appear favourable for capturing polarized emission in the sky at high accuracy.
\paragraph{}
 Such TOD simulations can further be used to estimate the computing resources required for data acquisition and map-making for first-light and subsequent observations with FYST. First-light data can be used to study atmospheric noise properties. The characteristics of $1/f$ noise observed in the simulations can inform the data acquisition and analyses techniques that will be employed with FYST.

%% file: chapters/chp06.tex
\chapter{New determination of polarization angle of the Crab nebula from the \plk data release 4}
\label{ch:tauA-pol-angle}
\begin{center}
    \begin{minipage}{0.75\textwidth}
        \textbf{Summary:} Polarization of the CMB is sensitive to new physics. It is imperative to achieve absolute calibration of the polarization-sensitive detectors of CMB experiments to enable accurate measurement of the polarization of CMB and discern the effects of parity-violating physics, such as cosmic birefringence, on the polarization of CMB. In this work, we explore the possibility of achieving relative and absolute polarization angle calibration of \plk detectors with observations of a bright astrophysical source: the Crab Nebula. We also attempt an absolute calibration of the measured position angle of polarization of the Crab nebula with constraints on \plk miscalibration angles measured by a calibration against Galactic foreground emission by \cite{Eskilt:2022cff}. We make use of timestreams from \plk Public Release 4 for the same.\\[3mm]
    \end{minipage}
\end{center}

\section{Introduction}
\label{sec:npipe-intro}
Polarization of the cosmic microwave background (CMB) is sensitive to new physics that violates parity symmetry under inversion of spatial coordinates \cite{Lue:1998mq,Komatsu:2022nvu}. One such effect, cosmic birefringence, is the rotation of the plane of linear polarization of photons due to a new parity-violating interaction in the electromagnetic sector \cite{Carroll:1989vb,Carroll:1991zs,Harari:1992ea}. 

To probe cosmic birefringence, it is essential to know the absolute position angle of linear polarization measured by the detectors with respect to that in the sky. A tantalizing hint of the rotation of the plane of linear polarization of the CMB at the level of $0.3^\circ$ has been reported using the \plk (discussed in Section \ref{sec:ch4-planck}) and \textit{WMAP} data \cite{Minami:2020odp,Diego-Palazuelos:2022dsq,Eskilt:2022wav,Eskilt:2022cff} with polarized Galactic foreground emission as an absolute calibrator for the position angle of linear polarization \cite{Minami:2019ruj,Minami:2020xfg,Minami:2020fin}. Although the known instrumental systematics of the \plk and \textit{WMAP} data have been shown to have negligible effects on cosmic birefringence \cite{Diego-Palazuelos:2022cnh,Cosmoglobe:2023pgf}, unknown systematics cannot be excluded. To make progress, independent measurements from independent experiments using independent calibration methods that do not rely on Galactic foreground emission and that are as accurate as $0.1^\circ$ or better are needed \cite{Cornelison:2022zrc,Murata:2023heo,Murphy:2024fna,COSMOCal:2024tbx}. 

The observed CMB polarization fields can be decomposed into parity eigenstates, called the $E$- and $B$-mode polarization fields \cite{Kamionkowski:1996ks,Zaldarriaga:1996xe}. The cross correlation of the $E$ and $B$ modes, $C_\ell^{EB}$, and that of the temperature and $B$ modes, $C_\ell^{TB}$, are sensitive to violation of the parity symmetry \cite{Lue:1998mq}. Self-calibration of CMB experiments by setting the $C_\ell^{EB}$ and $C_\ell^{TB}$ to zero and derotating the angular power spectra might be susceptible to potential biases due to: (i) non-zero $C_\ell^{EB}$ and $C_\ell^{TB}$ introduced by foregrounds and systematics (ii) mis-interpretation of non-zero $C_\ell^{EB}$ and $C_\ell^{TB}$ due to new physics. Ground-based CMB experiments make use of man-made calibrators such as polarizing dielectric sheets and polarization-selecting wire grids but these are limited by near-field response, uncertain stability over long time periods and difficulty of implementation \cite{BICEP1:2013rur,Cornelison:2022zrc}. An alternative  independent calibration method is the use of astrophysical sources \cite{Aumont2009, Kaufman:2016mcp}. The Crab nebula (also known as Taurus A or Crab nebula), a supernova remnant that emits synchrotron radiation, is the brightest polarized source in the sky \cite{Hester2008}. If the position angle of linear polarization of the Crab nebula is known precisely, it can be used to calibrate the instrument and measure the polarization of the target, such as the CMB. However, to achieve this goal, the instrument for measuring the polarization of the Crab nebula must first be calibrated well, which has not yet been achieved \cite{Aumont2018}.
\paragraph{}
The position angle at 90 GHz, as observed using the XPOL instrument with the IRAM 30m telescope \cite{Thum2008}, was found to be $149.0^\circ \pm 1.4^\circ$ in Equatorial coordinates using the angle definition of the International Astronomical Union (IAU)
within the area exhibiting the highest polarization flux \cite{Aumont2009}. Observations with the \textit{NIKA} instrument \cite{Monfardini:2010mv, Adam:2017gba} in the same region yielded a position angle of $140^\circ \pm 1.0^\circ$
\cite{Ritacco2018} at 150 GHz and $-87.195^\circ \pm 0.806^\circ$
in Galactic coordinates at 260 GHz ($150.43^\circ$ in Equatorial coordinates) \cite{Ritacco2021}. The POLARBEAR experiment measured $150.75^\circ \pm 0.16^\circ$
at 150 GHz \cite{POLARBEAR:2024vel}. Furthermore, the Korean VLBI Network (KVN) measured position angles of $154.2^\circ \pm 0.3^\circ$, 
$151^\circ \pm 0.2^\circ$,
$150^\circ \pm 1.0^\circ$,
and $151.3^\circ\pm 1.1^\circ$
at 22, 43, 86, and 94 GHz, respectively \cite{MKAM2022}. Such high-resolution observations, where the beam size is much smaller than the angular size of Crab nebula, are susceptible to spatial variations of the polarized emission.
\paragraph{}
In this chapter, we investigate the feasibility of achieving relative calibration of the orientation of polarization of \plk detectors with the measured position angle of polarization of Crab Nebula from \plk PR4 data (also referred to as \npp data) in the microwave regime. We also attempt an absolute calibration of the measured position angle of polarization of the Crab nebula with the estimates of \plk miscalibration angles by \cite{Eskilt:2022cff}.
\paragraph{}
This chapter is structured as follows. We describe the data products used in this work in Section \ref{sec:npipe}, followed by a description of methods employed to measure the position angle and the corresponding uncertainties in Section \ref{sec:npipe-methods}. The results and a discussion of our findings are presented in Sections \ref{sec:npipe-results} and \ref{sec:npipe-conclusion}, respectively. The position angle of polarization of Crab nebula are presented in IAU convention in galactic coordinates in this work, unless stated otherwise.

\section{\npp data products}
\label{sec:npipe}
\begin{figure}[ht]
    \centering
        \includegraphics[width=0.8\textwidth]{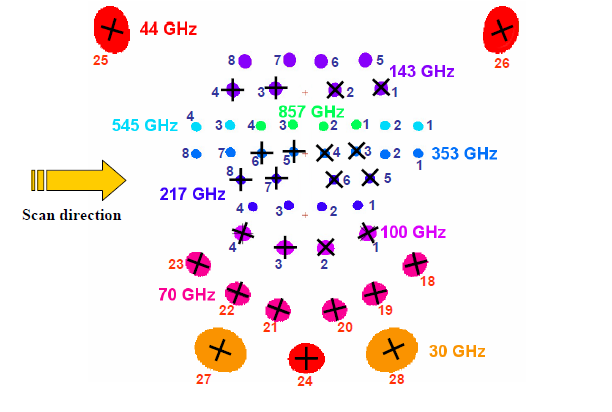}
    \caption{A representation of the \plk focal plane as seen by an observer at infinity where cross represent the orientation of polarization sensitivity and the size of the coloured spots indicate the relative resolution. Image adapted from \cite{Rocha:2022gku}.}
    \label{fig:ch-4-planckfocalplane}
\end{figure}

\plk data were published over four releases from 2013 to 2020. We first work with the public release 4 (PR4) \cite{Planck:2020olo} and compare our results with the data from PR2 \cite{Planck:2015mrs} and PR3 \cite{Planck:2018nkj}. PR4 comprises the \npp processed data which is designed to process both HFI and LFI raw data in one framework. It includes simulations, destriped TOD (more information on the destriping method can be found in Section \ref{subsec:destriping}) and bandpass-corrected all-sky maps which are used in this work.
\paragraph{}
\plk PR4 differs from the previous releases as the framework involved one framework to process both LFI and HFI raw data into calibrated, bandpass-corrected, noise-subtracted frequency
maps. The data release\footnote{All of the data products mentioned are available at National Energy Research Scientific Computing Center (NERSC) under \texttt{/global/cfs/cdirs/cmb/data/planck2020}} comprised calibrated timelines, reprocessed destriped TOD, high-fidelity Monte Carlo simulations and all-sky frequency maps. It also includes updated instrument model with improved corrections to polarization angles and efficiencies of the HFI detectors.
\paragraph{}
The \npp reprocessing involved the inclusion of data acquired during repointing of the satellite between stable science scans. Inclusion of these 4-minute periods lead to a 9\% increase in the integration time and improvement of sky sampling from the addition of detector samples between regular scanning paths which were separated by two arcminutes. Degree-scale noise in the maps are suppressed by \npp by modelling the low-frequency (temporally correlated) $1/f$ noise fluctuations with offsets of 167 ms (the characteristics of $1/f$ noise at the TOD and map levels are discussed in Chapter \ref{ch:TOAST-TOD}). Corrections to detector pointings lead to improved suppression of high frequency noise in the repointing manoeuvre data. An improved method of deconvolution of the time response of bolometers also leads to a reduction of the leakage of signal from flagged to science data that previously contributed to several percent of the small-scale noise variance in PR3 maps. \npp processing also corrected for systematic effects such as gain fluctuations, far-sidelobe pickup, and bandpass mismatch from both continuum emission and CO.
\paragraph{}
Another feature of \npp data products is the availability of a data split where systematics between the split are expected to be uncorrelated. The horns in the focal plane (shown in Figure \ref{fig:ch-4-planckfocalplane}) are split into two independent sets, A and B, and reprocessing is performed independently on each set. Due to the scanning \plk strategy at least two polarized horns are necessary to solve for an all-sky polarized map. In the case of 30 and 40 GHz, the data split is across time instead of the focal plane due to a lack of redundant polarized horns. Thus, with such a time-wise split, the instrument noise and gain fluctuations will remain uncorrelated while the same may not be true for other systematics as the initial beam and bandpass mismatch are shared between the two subsets. Table \ref{tab:ch4-detectors-AB-split} contains the details of which sets of horns comprise a data split.
\begin{table}[ht]
    \centering
    \begin{tabular}{ccc}
     Frequency (GHz)   &  Set A & Set B\\
     \hline
     \hline
       30 & Years 1 and 3 & Years 2, 4 and start of 5\\
       44 & Years 1 and 3 & Years 2, 4 and start of 5\\
       70 & Horns 18, 20, and 23 & Horns 19, 21, and 22\\
       100 & Horns 1 and 4 & Horns 2 and 3\\
       143 & Horns 1, 3, 5, and 7 & Horns 2, 4 and 6\\
       217 & Horns 1, 3, 5, and 7 & Horns 2, 4, 6 and 8\\
       353 & Horns 1, 3, 5, and 7 & Horns 2, 4, 6 and 8\\
       545 &  Horn 1 & Horns 2 and 4\\
       857 & Horns 1 and 3 & Horns 2 and 4\\
       \hline
    \end{tabular}
    \caption{Subsets of detectors that comprise splits A and B. The horn numbers mentioned in the table correspond to the horn numbers in the illustration of the \plk focal plane in Figure \ref{fig:ch-4-planckfocalplane}.}
    \label{tab:ch4-detectors-AB-split}
\end{table}
\subsubsection{\npp Simulations}
\label{subsubsec:ch4-npipe-sims}
The \npp simulations are Monte Carlo simulated maps that consist of CMB, Galactic foregrounds, noise and systematic effects. Simulations of the A/B detector splits are also available.
\paragraph{}
\npp simulations make use of the PR3 simulations for CMB realizations. The \texttt{Commander} model is evaluated for each of the target frequencies from an earlier version of the \npp data for simulations of the Galactic foregrounds. The thermal dust emission is described by a one-component modified blackbody while the SED of synchrotron emission is described by a power-law. In instances where the components of the foreground model are measured at higher angular resolution than the target frequency (e.g., dust at 30 GHz), those components are convolved with an azimuthally-symmetric beam calculated specifically for the \npp data set \cite{Hivon:2016qyw}. Divergences in the deconvolution process are prevented by maintaining the Gaussian beam with a full-width-at-half-maximum (FWHM) of 5 arcmin present in \texttt{Commander}’s dust component at 217 and 353 GHz. Static zodiacal emission is accounted for by the addition of the same nuisance templates that \texttt{Commander} is marginalized over. Noise maps account for instrumental effects with the incorporation of beam systematics, gain calibration and bandpass mismatches, analogue-to-digital conversion non-linearities, and transfer-function corrections. Noise maps also account for the non-linear response of the instrument and the \npp processing pipeline, satisfying the necessity of generating the simulations to capture the non-linear couplings between signal and noise that are introduced by the \npp processing.

\subsubsection{\npp Time Ordered Data}
Each detector produces one data point for an observation which is the polarization-modulated flux density that can be expressed as \cite{Rocha:2022gku}
\begin{equation}
    s_{d,p} = I_{d,p}+\eta_d\Big[Q\mathrm{cos}(2(\psi+\psi_{d,p}))+U\mathrm{sin}(2(\psi+\psi_{d,p}))\Big]
\end{equation}
where $s_{d,p}$ is the polarization-modulated flux density measured by detector $d$ from pixel $p$, $I,\,Q,\,U$ are the Stokes parameters constituting the source flux density, $\eta$ is the polarization efficiency of the detector\footnote{An ideal detector sensitive to polarized emission would have $\eta_d$=1.}, $\psi$ is the angle between the source coordinate system and the focal plane reference frame and $\psi_d$ is the orientation in detector coordinates along which the detector is sensitive to linearly polarized emission. A discussion on TOD can be found in Section \ref{sec:Simulations-of-TOD}.
\paragraph{} For $n_\mathrm{obs}$ number of observations and $n_\mathrm{det}$ number of detectors, one can construct a vector of these polarization-modulated flux densities expressed in terms of the response matrix $\textbf{P}$ is (Chapter \ref{ch:TOAST-TOD})
\begin{equation}
    \mathbf{d} = \mathbf{Pm}+\mathbf{n}
    \label{eq:ch4-tod-def},
\end{equation}
where each row of $\textbf{P}$ is constructed from pointing weights as $\big[1,\, \eta\, Q\,\mathrm{cos}(2(\psi+\psi_d)),\, \eta\,U\,\mathrm{cos}(2(\psi+\psi_d))\big]$ per detector, the map $\textbf{m}$ is a vector containing the three Stokes parameters per detector as $m=\Big[I,\,Q,\,U\Big]^\mathrm{T}$, and the noise vector $\textbf{n}$ describes the noise component associated with the measurement. Destriping is performed to reduce the low-frequency (temporally correlated) $1/f$ noise $\textbf{n}$. An extensive discussion of TOD and destriping map-making method is presented in Section \ref{sec:ch05-mapmaking}. Destriped time ordered data (TOD) from a region of four deg$^2$ centred at the coordinates $l=184.557^\circ,\, b=-05.784^\circ$ (in Galactic coordinate system) of Crab Nebula are analysed in this work. The destriped \npp processed TOD files contain the following information:
\begin{itemize}
    \item[-] Detector pointings projected onto the sky coordinates.
    \item[-] Stokes $I$, $Q$, $U$ weights along with the polarization efficiency that enables the construction of the response matrix $\textbf{P}$ defined in eq.\ \eqref{eq:ch4-tod-def}.
    \item[-] $f_{d}$ and the corresponding time associated with the sample.
    \item[-] Ring number for each $f_d$ that corresponds to a full scan circle traced by \plk.
    \item[-] Orientation of the detector along which it is sensitive to linearly polarized emission.
\end{itemize}

\section{Methods}
\label{sec:npipe-methods}
\begin{figure}[htbp]
    \centering
    \includegraphics[width=0.85\textwidth]{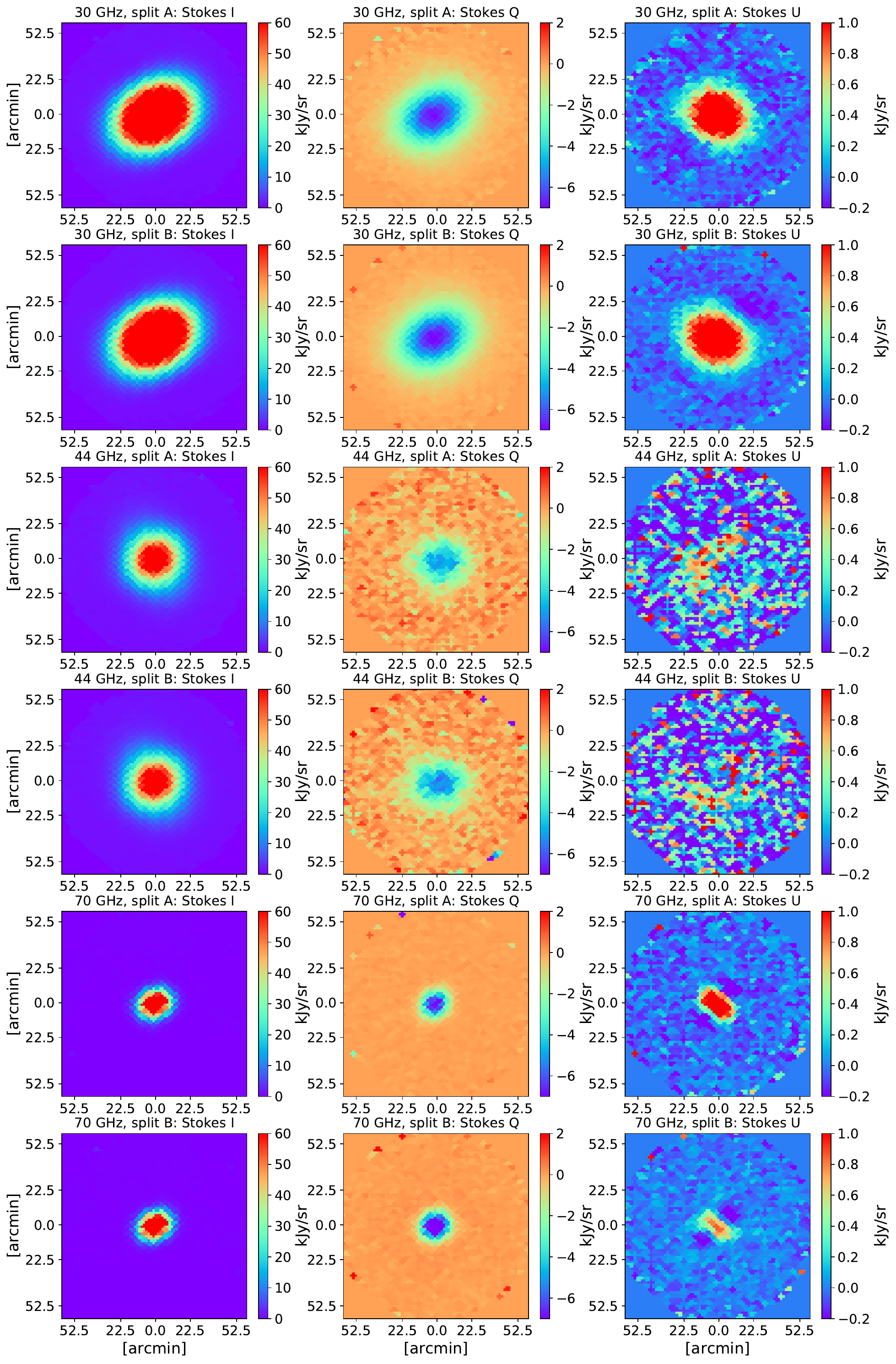}
    \caption{Stokes $I$, $Q$, $U$ A/B split maps (gnomonic projections of the \texttt{Healpix} maps centred at the coordinates $l=184.557^\circ,\, b=-05.784^\circ$) of Crab nebula at each of the polarization sensitive \plk frequency band of the LFI in units of kJy$\,$sr$^{-1}$. Every two \textit{rows} correspond to a frequency band of the LFI.}
    \label{fig:lfi-AB-allstokes}
\end{figure}

\begin{figure}[htbp]
    \centering
    \includegraphics[width=1.\textwidth]{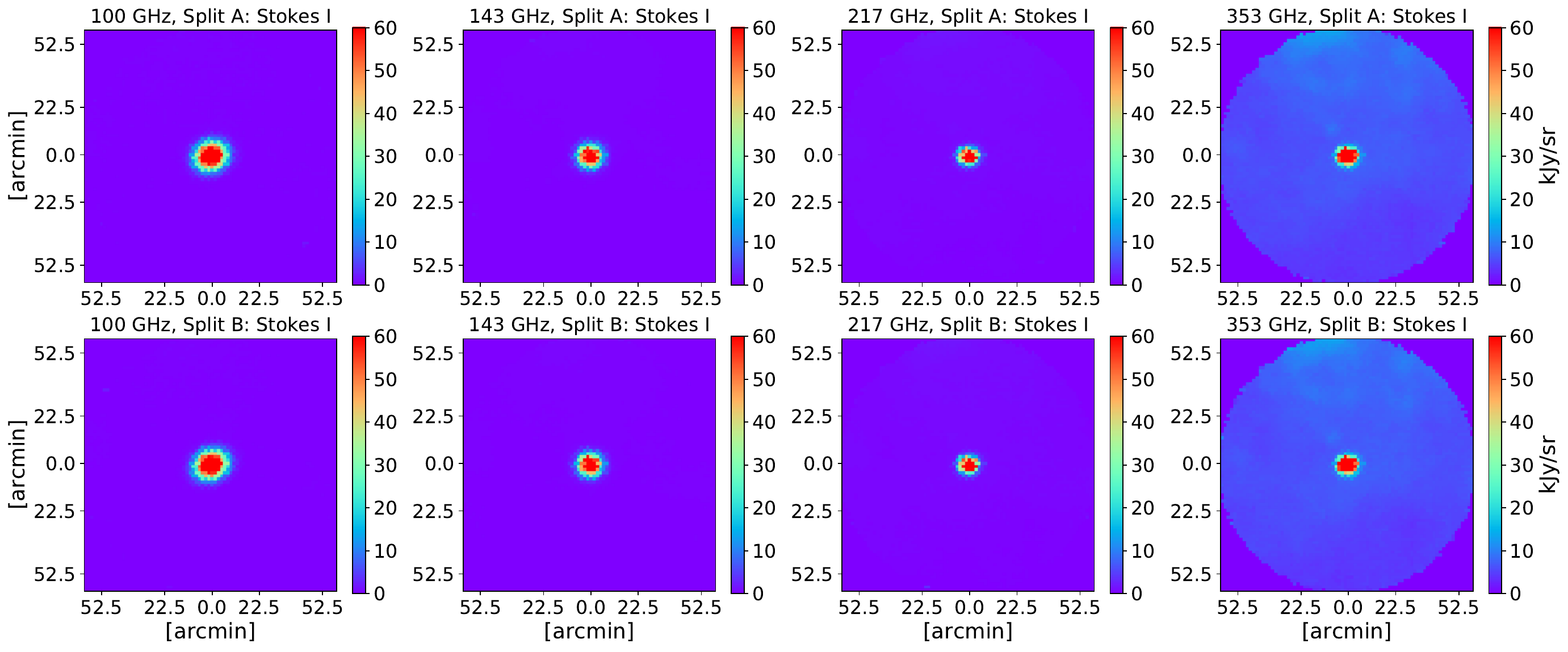}
    \caption{Stokes $I$ A/B split maps of Crab nebula at each of the polarization sensitive \plk frequency band of the HFI in units of kJy$\,$sr$^{-1}$. Each \textit{column} corresponds to a HFI frequency band and each \textit{row} corresponds to a split.}
    \label{fig:hfi-AB-stokesI}
\end{figure}

\begin{figure}[htbp]
    \centering
    \includegraphics[width=1.\textwidth]{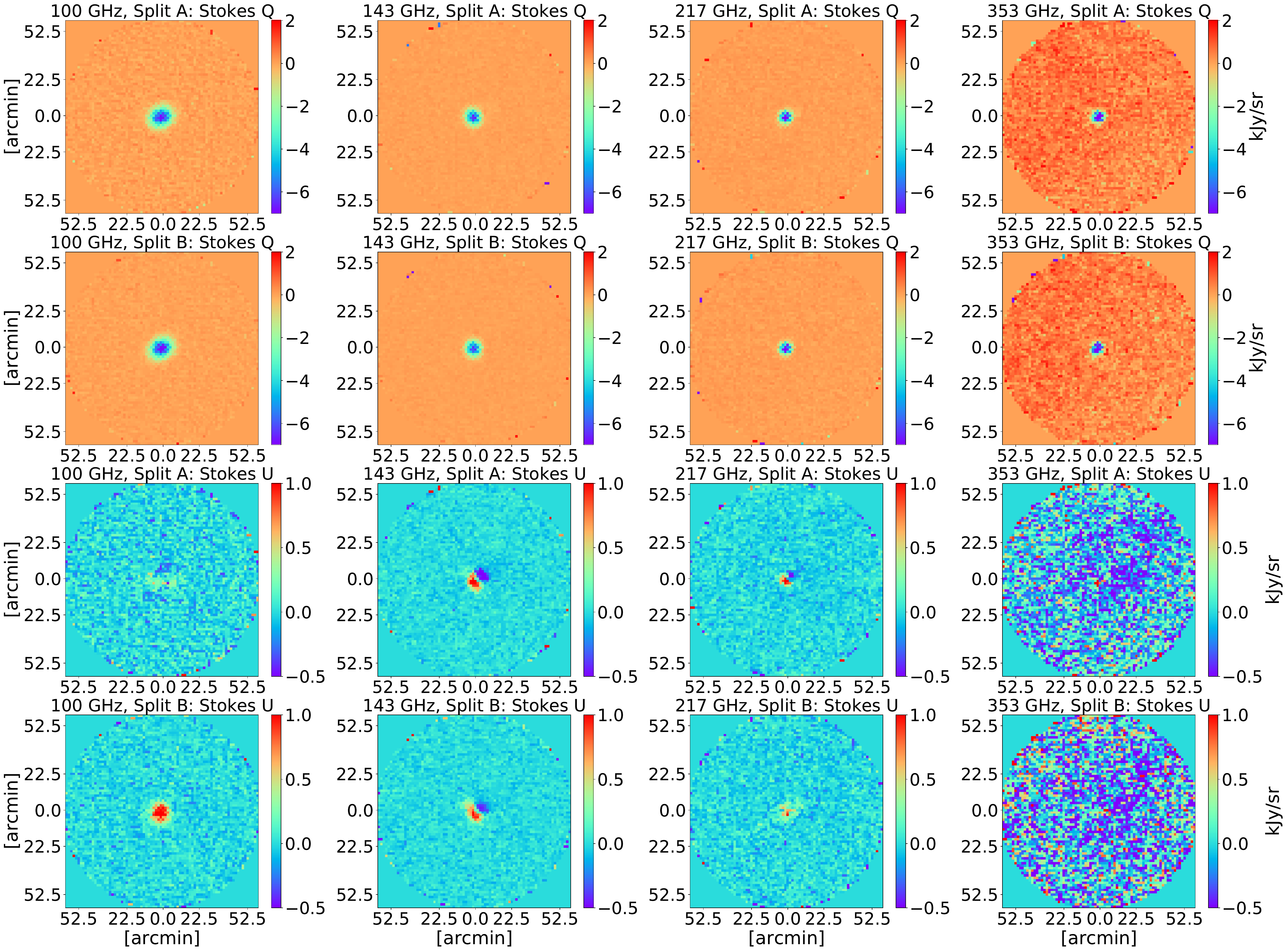}
    \caption{Same as Figure \ref{fig:hfi-AB-stokesI} but for Stokes $Q$ and $U$ A/B split maps. Intensity to polarization leakage due to (i) bandpass mismatch is apparent in Stokes $U$ split B map at 100 GHz. (ii) beam asymmetries is apparent in the Stokes $Q$ and $U$ maps at 143 and 217 GHz}
    \label{fig:hfi-AB-stokesQU}
\end{figure}

The measurement of the position angle of polarization of the Crab nebula can be made at the TOD level \cite{POLARBEAR:2014hgp} or at the map level. We choose to reduce the TOD into the Stokes $I$, $Q$ and $U$ maps from which the position angle of polarization of the Crab nebula is measured. Since the \npp TOD are already pre-processed and destriped, they can be bin-averaged to reconstruct the Stokes $I$, $Q$, $U$ maps in \texttt{HEALPix} pixelization scheme \cite{Gorski2005} at \texttt{Nside}=1024 for observations with the LFI  and \texttt{Nside}=2048 for observations with the HFI. The binned map is estimated as (Section \ref{subsec:binning})
\begin{equation}
    \widehat{\mathbf{m}} = (\mathbf{P^TP})\mathbf{P}^{-1}\cdot\mathbf{d},
    \label{eq: ch-4-bin}
\end{equation}
where $\mathbf{P}$ is the response matrix and $\mathbf{d}$ is the \npp pre-processed destriped TOD.
\paragraph{}
The destriped TOD are binned by grouping the individual detectors according to the A/B splits (Table \ref{tab:ch4-detectors-AB-split}) to produce A/B split maps (hereon referred to as A/B maps). Gnomonic projections of the A/B maps from LFI and HFI centered at the coordinates of Crab nebula are shown in Figures \ref{fig:lfi-AB-allstokes}, \ref{fig:hfi-AB-stokesI} and \ref{fig:hfi-AB-stokesQU}. Total maps are also produced by making maps from destriped TOD from all the detectors combined (shown in Figures \ref{fig:stokesIQUmaps-LFI} and \ref{fig:stokesIQUmaps-HFI}).


\paragraph{}
We measure the flux densities from the reconstructed $I$, $Q$, $U$ maps to estimate the position angle of polarization of Crab nebula. The destriped TOD (and thus our maps) in the $30-353\,$GHz range are available in units of the CMB temperature (K\textsubscript{CMB}) which need to converted to units of MJy$\,$sr$^{-1}$. The band-averaged unit conversion coefficients extracted from the \plk UC CC Tables\footnote{\url{https://wiki.cosmos.esa.int/planckpla2015/index.php/UC_CC_Tables}} are applied to the Stokes $I$, $Q$, $U$ maps to convert them into units of MJy$\,$sr$^{-1}$. The coefficients used are tabulated in Table \ref{tab:apchars} and a further discussion on the coefficients can be found in Section \ref{subsec:colour-correction}.

\subsection{Aperture Photometry}
\label{subsec:photometry}
\begin{figure}[htbp]
    \centering
    \includegraphics[width=1.\textwidth]{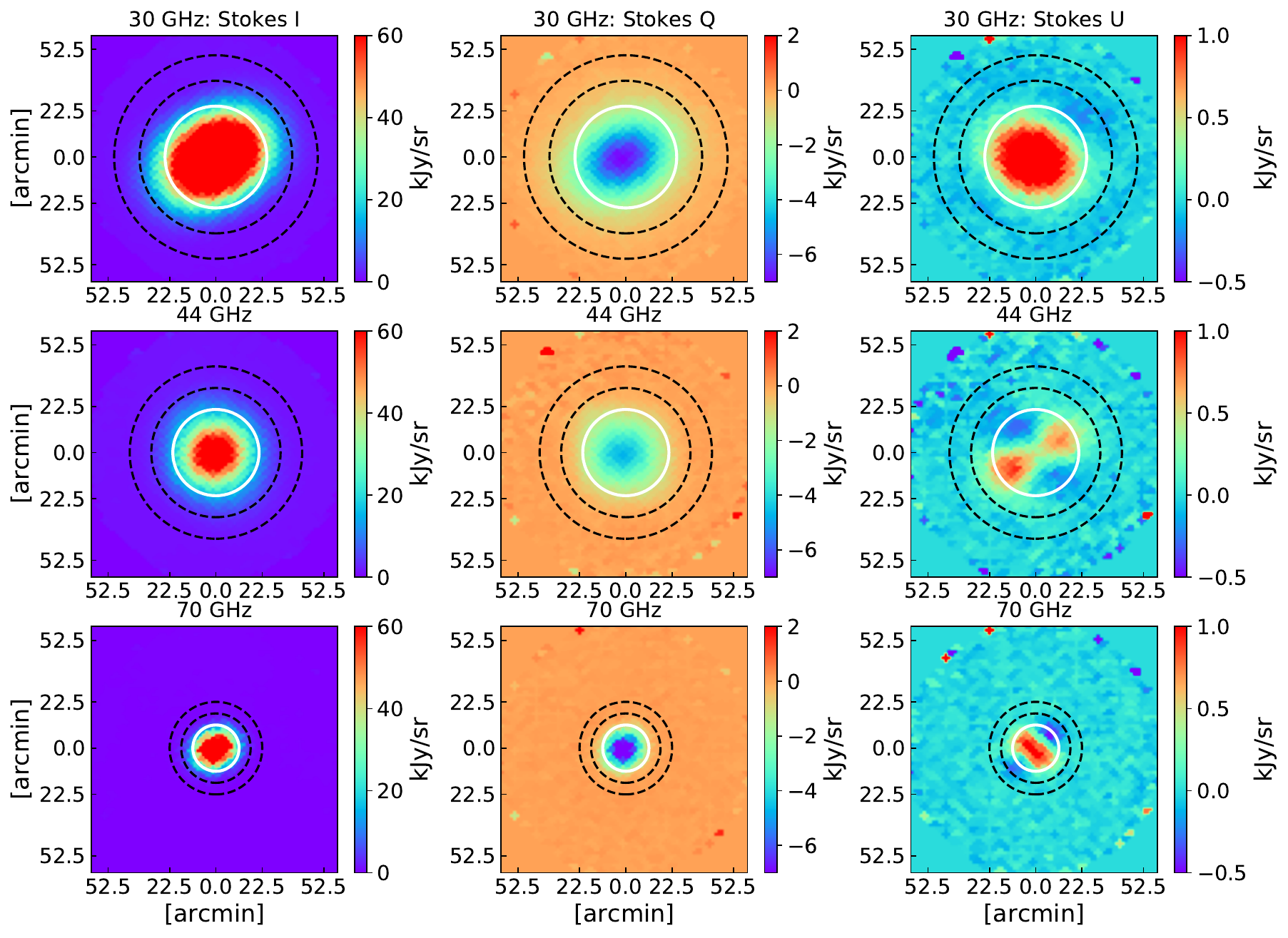}
    \caption{Stokes $I$, $Q$, $U$ total maps (gnomonic projections of the \texttt{Healpix} maps centred at the coordinates $l=184.557^\circ,\, b=-05.784^\circ$) of Crab nebula at each of the polarization sensitive \plk frequency band of the LFI in units of kJy$\,$sr$^{-1}$. Each \textit{row} corresponds to a frequency band of the LFI. The white solid circle marks the circular aperture over which the integrated flux is computed. The black-dashed annulus corresponds to the region from which the background flux is measured to apply background correction.}
    \label{fig:stokesIQUmaps-LFI}
\end{figure}

\begin{figure}[htbp]
    \centering
    \includegraphics[width=1.\textwidth]{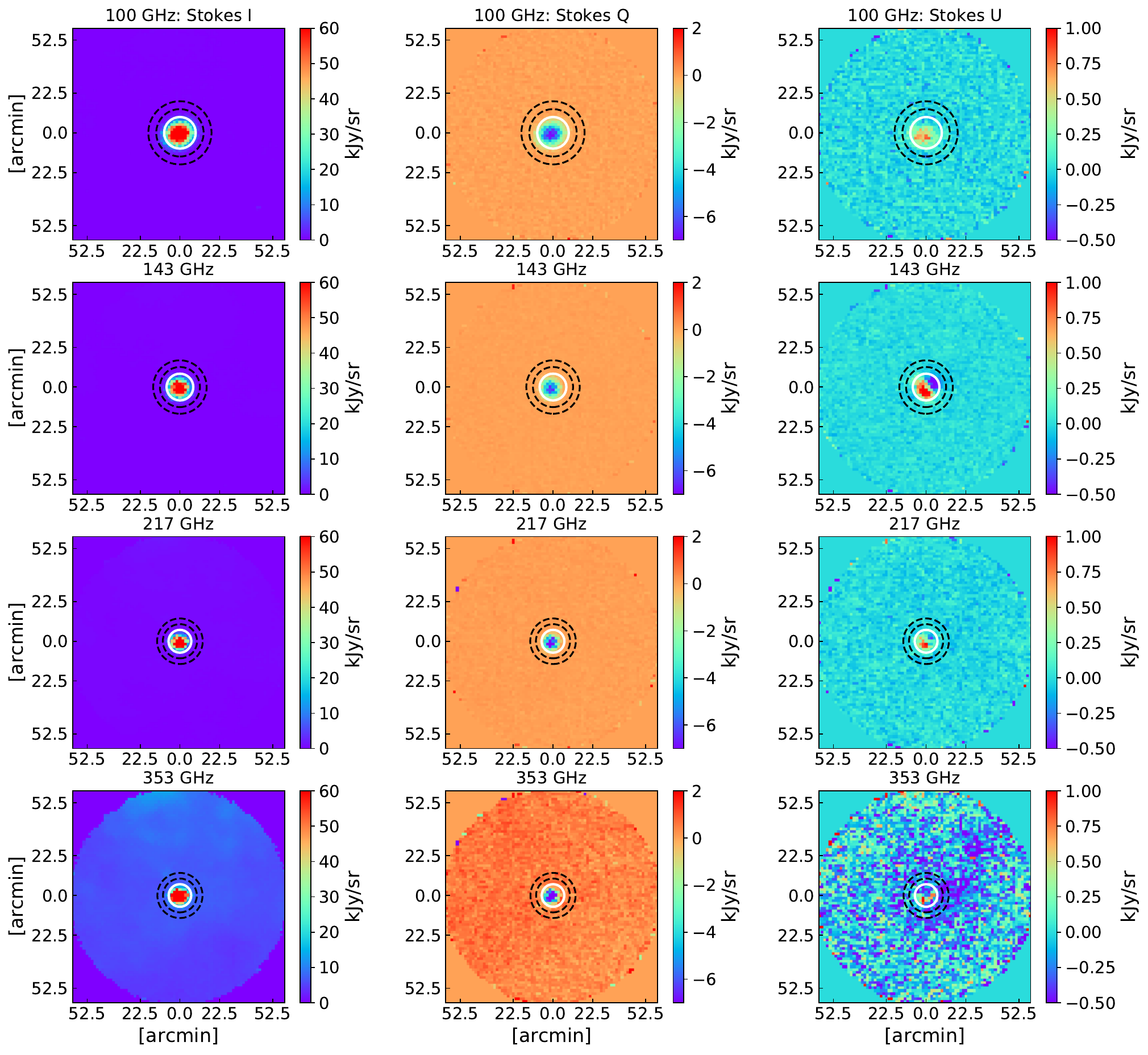}
    \caption{Same as Figure \ref{fig:stokesIQUmaps-LFI} but for each of the polarization sensitive \plk frequency band of the HFI.}
    \label{fig:stokesIQUmaps-HFI}
\end{figure}

We perform circular aperture photometry to estimate the flux densities at each frequency band. The center of the aperture is chosen to be at the coordinates $l=184.557^\circ,\, b=-05.784^\circ$. The aperture size, $\theta_{\mathrm{ap}}$, is chosen to be
\begin{equation}
    \theta_{\mathrm{ap}} = 1.5\times\sqrt{\theta_{\mathrm{Crab}}^2 + \theta_{\mathrm{FWHM}}^2},
\end{equation}
where $\theta_{\mathrm{Crab}} = 7'$ is the intrinsic angular size of Crab nebula \citep{Green2009} and $\theta_{\mathrm{FWHM}}$ is the FWHM at the respective frequency band. The $\theta_{\mathrm{FWHM}}$ and the corresponding $\theta_{\mathrm{ap}}$ considered for each frequency band are tabulated in Table \ref{tab:apchars}. The flux density at a given frequency band within this aperture is then
\begin{equation}
    F_\nu = \sum_{\theta_{\mathrm{ap}}} X_\nu \times \frac{\theta_\mathrm{pix}^2}{3600}\Big(\frac{\pi}{180^\circ}\Big)^2\,f_\mathrm{A},
    \label{eq:apflux}
\end{equation}
where $F_\nu$ is measured in units of Jy, $X_\nu=\{I, Q, U\}$ at the corresponding frequency band, $\theta_\mathrm{pix} = 1.5'$ is the pixel size and $f_\mathrm{A}=1.4$ is the aperture correction that needs to be applied to account for any loss of flux that is outside of the aperture \cite{Planck:2014loa}. We then apply a background correction by estimating the median background flux within an annulus with inner radius of  $1.5\,\theta_{\mathrm{ap}}$ and outer radius of $2\,\theta_{\mathrm{ap}}$ scaled to the on-source aperture as
\begin{equation}
    F^\mathrm{bg}_\nu = X^\mathrm{bg}_\nu \times \frac{n_\mathrm{pix}\theta_\mathrm{pix}^2}{3600}\Big(\frac{\pi}{180^\circ}\Big)^2\,f_\mathrm{A},
    \label{eq:annflux}
\end{equation}
where $X^\mathrm{bg}_\nu$ is the median $\{I,Q,U\}$ in units of Jy within the background annulus and $n_\mathrm{pix}$ is the number of pixels within the on-source aperture. The uncertainties on the measured flux densities are then estimated as
\begin{equation}
    \sigma_{X_\nu} = \frac{\sigma_{X_\nu,\mathrm{bg}}\times n_\mathrm{pix}}{\sqrt{n^\mathrm{bg}_\mathrm{pix}}}\,f_\mathrm{A},
\end{equation}
where $\sigma_{X_\nu,\mathrm{bg}}=\{\sigma_\mathrm{I,bg}, \sigma_\mathrm{Q,bg}, \sigma_\mathrm{U,bg}\}$ is the standard deviation within the background annulus in units of Jy at a given frequency band and $n^\mathrm{bg}_\mathrm{pix}$ is the number of pixels within the background annulus. Figures \ref{fig:stokesIQUmaps-LFI} and \ref{fig:stokesIQUmaps-HFI} can be referred to for a visualization of the variation of the apertures and background annuli for each frequency band.
\begin{table}[ht]
    \centering
    \begin{tabular}{cccc}
    \hline
    Frequency & Unit Conversion & $\theta_\mathrm{FWHM}$ & $\theta_\mathrm{ap}$\\
    (GHz) & (MJy sr$^{-1}$K$_\mathrm{CMB}^{-1}$) &  & \\
    \hline
       30 & 23.5099 & $32.3'$ & $49.57'$\\
       44 & 55.7349 & $27.1'$ & $41.98'$\\
       70 & 129.1869 & $13.3'$ & $22.54'$\\
       100 & 244.0960 & $9.66'$ & $17.89'$\\
       143 & 371.7327 & $7.27'$ & $15.14'$\\
       217 & 483.6874 & $5.01'$ & $12.91'$\\
       353 & 287.4517 & $4.86'$ & $12.78'$\\
    \hline
    \end{tabular}
    \caption{Frequency bands and their corresponding unit conversion, effective FWHM of the beam and diameter of the circular aperture considered for aperture photometry. The unit conversion factors in this table are the band average corrections estimated by calibrating against the CMB dipole (more information is presented in \ref{subsec:colour-correction}).}
    \label{tab:apchars}
\end{table}
\subsection{Position angle of Polarization}
The position angle of polarization of Crab nebula is estimated from the Stokes $Q$ and $U$ parameters as\footnote{The negative sign against $\widehat{U}$ is to obtain angles in IAU convention as opposed to the COSMO convention that \npp data ascribe to.}
\begin{equation}
    \psi = \frac{1}{2}\mathrm{tan}^{-1}\Bigg(\frac{-\widehat{U}}{\widehat{Q}}\Bigg),
    \label{eq:ch4-polangle}
\end{equation}
where $\widehat{Q}$ and $\widehat{U}$ are the background subtracted aperture integrated fluxes measured from the Stokes $Q$, $U$ maps using eq.\ \eqref{eq:apflux}. $\psi_\nu^\mathrm{split}$ corresponds to the position angle of polarization measured from $Q$ and $U$ split maps as
\begin{equation}
    \psi_\nu^\mathrm{split} = \frac{1}{2}\mathrm{tan}^{-1}\Bigg(\frac{-\widehat{U}_\nu^\mathrm{split}}{\widehat{Q}_\nu^\mathrm{split}}\Bigg)
    \label{eq:ch4-split-nu}
\end{equation}
where $\widehat{Q}_\nu^\mathrm{split}$ and $\widehat{U}_\nu^\mathrm{split}$ are the background-subtracted Stokes $Q$ and $U$ flux estimated at each frequency band $\nu$ from the A/B maps.
\paragraph{}
Once we have an estimate of $\psi$, we compute the uncertainties associated with it from \npp simulations (described in Section \ref{subsubsec:ch4-npipe-sims}) which, essentially, would capture any systematics and other errors introduced by the \npp pre-pocessing. While the simulated maps for LFI contain emission from Crab nebula, the simulations for HFI are missing Crab nebula. We thus extract the region required for aperture photometry from the A/B maps and inject them to the \npp simulated A/B maps. Once added, the usual method of background subtraction, and position angle estimation are performed (eqs.\ \eqref{eq:apflux} and \eqref{eq:annflux}). The position angle is calculated for each of the 100 (Crab nebula injected) simulations. The uncertainty associated with $\psi$ is then the standard deviation of the estimated position angles from the simulations.
\paragraph{} 
Further, for an uncertainty $\Delta\psi$ on the measured polarization angle $\psi$, we estimate the inverse noise-weighted mean of $\psi_\nu^\mathrm{split}$ for each \plk frequency as
\begin{equation}
    \Bar{\psi}_\nu = \frac{\sum_{n=A,B}\psi_\nu^\mathrm{split}/(\Delta\psi_\nu^\mathrm{split})^2}{{\sum_{n=A,B} 1/(\Delta\psi_\nu^\mathrm{split})^2}}\pm\sqrt{\frac{1}{\sum_{n=A,B}(\Delta\psi_\nu^\mathrm{split})^2}}.
    \label{eq:ch4-mean-psisplit}
\end{equation}
\section{Results}
\label{sec:npipe-results}
\begin{figure}[ht]
    \centering
    \includegraphics[width=1.\textwidth]{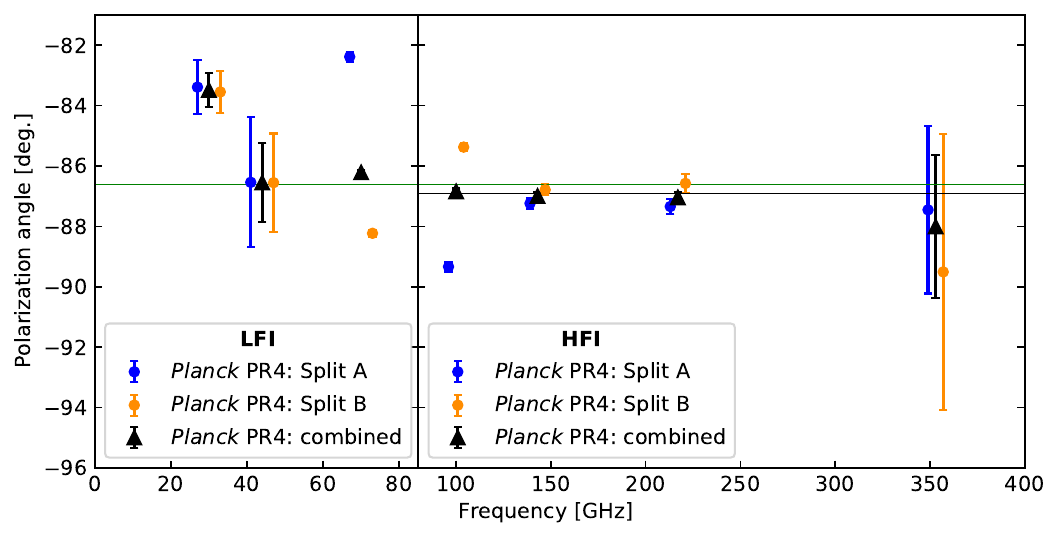}
    \caption{Position angles of polarization of Crab nebula at \plk frequencies measured from A/B maps generated from destriped \npp TOD. Black (green) horizontal line corresponds to inverse noise-weighted mean from HFI (LFI+HFI). The angles appear stable across the HFI.}
    \label{fig:polangleall}
\end{figure}

The position angles of polarization of Crab nebula at each frequency and A/B split are presented in Table \ref{tab:results-angles}. The uncertainty associated with the measured $\psi_\nu^\mathrm{split}$ increases with frequency as the pixel-pixel variance is larger and the contamination from foregrounds (especially dust) is prominent at frequencies $>217\,$GHz. The measured inverse noise-weighted mean $\Bar{\psi}_\nu$ at each LFI and HFI frequency along with their associated uncertainties measured from the \npp simulated maps are plotted in Figure \ref{fig:polangleall}. The measurements appear stable across the HFI frequencies within the uncertainties, however the same is not true for the LFI. It is apparent that the measurement at 30 GHz is inconsistent with the measurements at other frequencies. The angles between A/B splits measured at 70 and 100 GHz, are inconsistent between each other within the estimated uncertainties. We shall consider the LFI and HFI separately from hereon and explore these inconsistencies further.
\begin{table}[htbp]
    \centering
    \begin{tabular}{|c|c|c|c|c|c|c|}
    \hline
    & Split & $\psi_\nu^\mathrm{split}$ & $\Bar{\psi}_\nu$ & $\alpha^\mathrm{IAU}$ & $\Tilde{\psi}^\mathrm{split}_\nu$ & $\Bar{\Tilde{\psi}}_\nu$\\
      &  & (deg) & (deg) & (deg)& (deg)& (deg)\\
    \hline
    & & & & &  & \\
     &  30$-$A  & $-83.39\pm 0.88$ &  &  &  & \\
     &   30$-$B  & $-83.55\pm 0.71$ & $-83.49\pm 0.55$& &  & \\
     LFI &  44$-$A  & $-86.54 \pm 2.15$ & & $-$ & $-$ & $-$\\
     &  44$-$B  & $-86.55\pm 1.63$ & $-86.55\pm 1.30$& &  & \\
     &  70$-$A  & $-82.38 \pm 0.16$ & & &  & \\
     &  70$-$B  & $-88.23\pm 0.12$ & $-86.21\pm 0.09$& &  & \\
       & & & & & & \\
       \hline
       \hline
       & & & & & & \\
      & 100$-$A  & $-89.33 \pm 0.16$ & & $0.28\pm 0.13$& $-89.33\pm0.21$ & \\
      & 100$-$B  & $-85.37\pm 0.12$ & $-86.84\pm 0.10$& $0.41\pm 0.13$& $-85.37\pm0.19$ & $-87.14\pm 0.14$\\
       & & & & & & \\
       \cline{2-7}
       & & & & & & \\
     &  143$-$A  & $-87.24 \pm 0.19$ & & $-0.05\pm 0.11$& $-87.23\pm0.24$ & \\
      HFI & 143$-$B  & $-86.79\pm 0.18$ & $-87.0\pm 0.13$& $-0.18\pm 0.11$ & $-86.79\pm 0.26$ & $-87.03\pm 0.18$\\
       & & & & & & \\
       \cline{2-7}
       & & & & & & \\
      & 217$-$A  & $-87.34 \pm 0.25$ & & $0.04\pm 0.11$& $-87.34\pm 0.32$ & \\
      & 217$-$B  & $-86.58\pm 0.31$ & $-87.04\pm 0.20$& $0.06\pm 0.11$& $-86.58\pm0.37$ & $-87.02\pm 0.24$\\
       & & & & & & \\
       \cline{2-7}
       & & & & & & \\
      & 353$-$A  & $-87.45 \pm 2.77$ & & $0.19\pm 0.10$& $-87.45\pm2.75$ & \\
      & 353$-$B  & $-89.51\pm 4.56$ & $-88.01\pm 2.37$& $0.22\pm 0.11$ & $-89.51\pm 4.52$ & $-88.01\pm 2.35$\\
       & & & & & & \\
    \hline
    \hline
        & & & & & & \\
    HFI & & Inverse noise-& $-86.92\pm 0.06$ & $-$ & $-$ &$-87.08 \pm 0.10$ \\
    & & weighted mean &$(150.71\pm 0.06)$ & &  & $(150.55\pm 0.10)$\\
    \hline
    \end{tabular}
    \caption{Polarization angle per split at each frequency channel, mean polarisation angle per frequency channel, miscalibration angle per split per frequency channel in degrees estimated in galactic coordinates and in IAU convention. The mean values are inverse-weighted averages estimated with statistical uncertainties. $\alpha^\mathrm{IAU}$ are miscalibration angles from \cite{Eskilt:2022cff} in IAU convention. $\Tilde{\psi}_\nu$ ($\Bar{\Tilde{\psi}}_\nu$) are polarization angles (inverse noise-weighted mean) measured after applying a \textit{de-rotation} of the Stokes $Q$ and $U$ maps by $\alpha$ at each split. In the bottom-most row the inverse noise-weighted mean in Galactic (Equatorial) coordinates from the HFI measurements is shown.}
    \label{tab:results-angles}
\end{table}
\subsubsection{Bandpass mismatch}
\label{sec:res-bandpassmismatch}
\begin{figure}[htbp]
    \centering
    \includegraphics[width=0.8\textwidth]{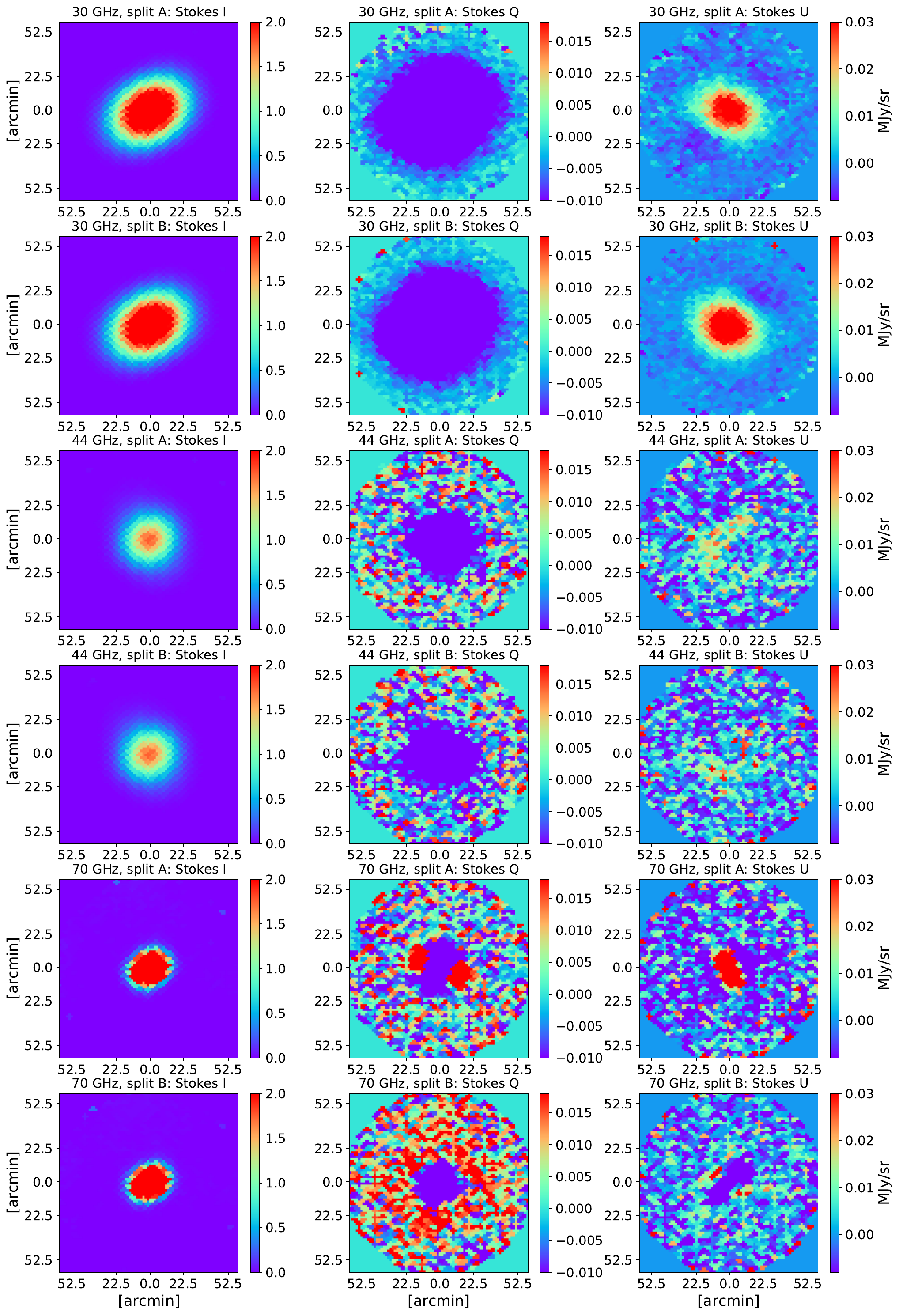}
    \caption{Stokes $I$, $Q$, $U$ A/B maps reconstructed from background-subtracted colour corrected TOD observed by the LFI. Every two \textit{rows} correspond to a frequency band of the LFI. Intensity to polarization leakage due to beam asymmetry is present in the Stokes $Q$ and $U$ Split A maps at 70 GHz, and in the Stokes $U$ A/B maps at 30 GHz.}
    \label{fig:lfi-AB-allstokes-ccbgcorr}
\end{figure}

\begin{figure}[ht]
    \centering
    \includegraphics[width=1.\textwidth]{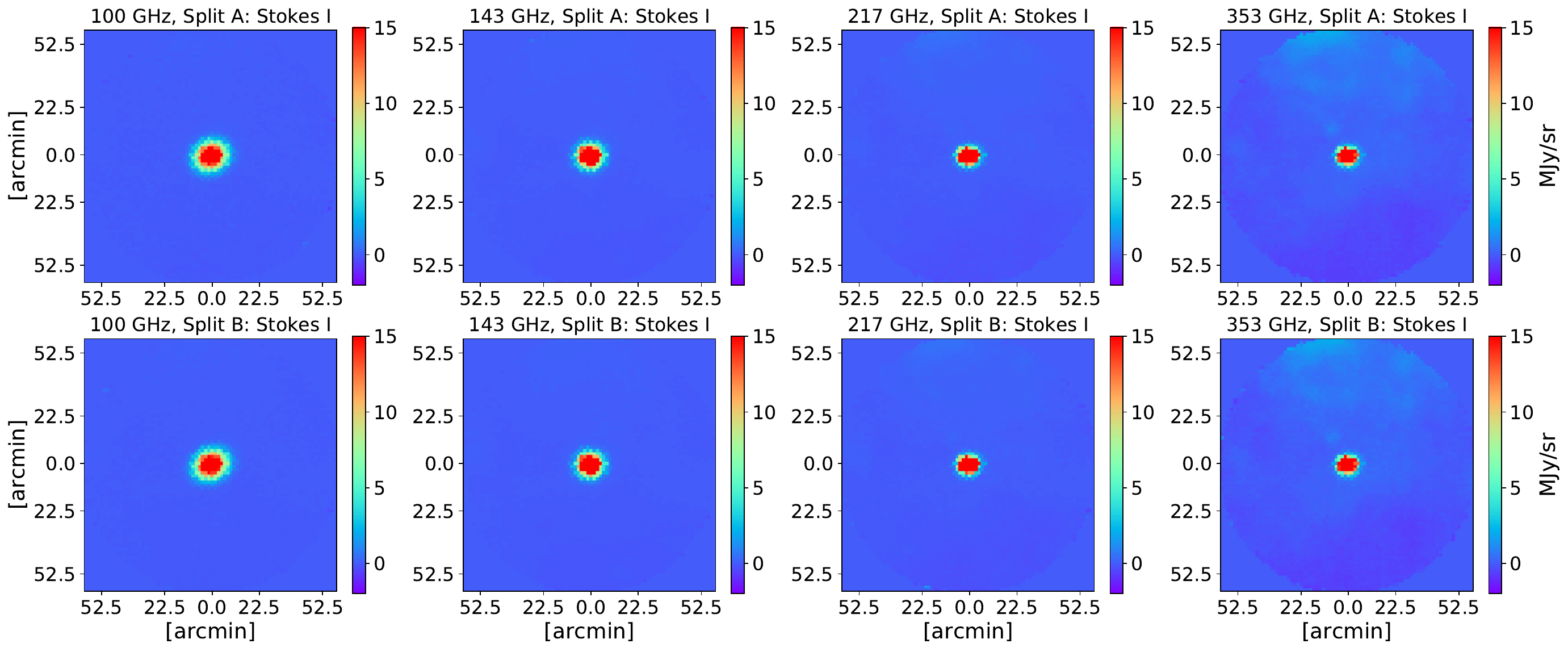}
    \caption{Stokes $I$ A/B maps reconstructed from background subtracted colour corrected TOD of the HFI. Each \textit{column} corresponds to a frequency band of the HFI.}
    \label{fig:hfi-AB-I-ccbgcorr}
\end{figure}

\begin{figure}[ht]
    \centering
    \includegraphics[width=1.\textwidth]{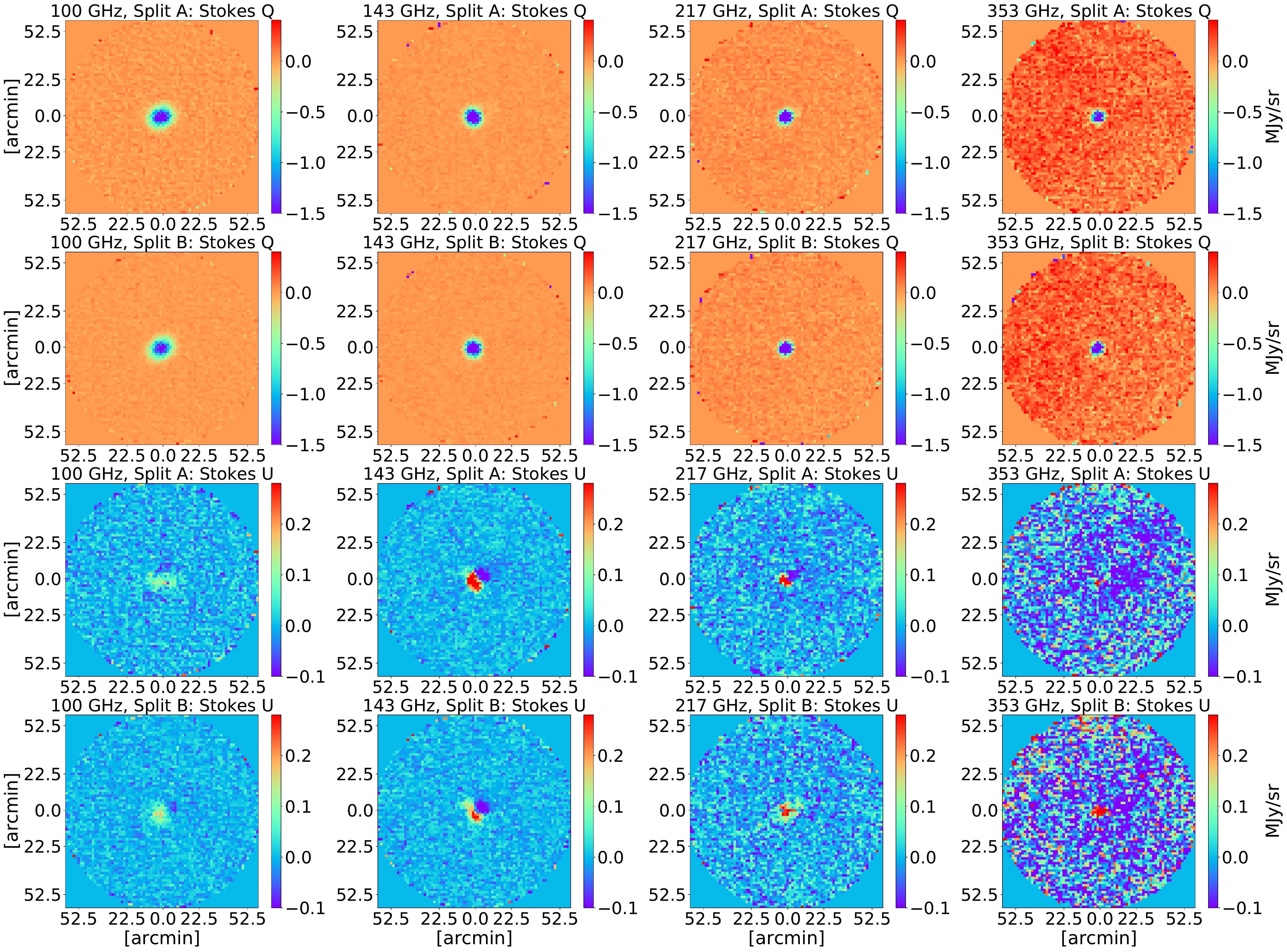}
    \caption{Same as Figure \ref{fig:hfi-AB-I-ccbgcorr} but Stokes $Q$ (\textit{top} two \textit{rows}) and Stokes $U$ (\textit{bottom} two \textit{rows}) A/B maps from the HFI. Leakage from intensity to polarization due to beam asymmetry is prominent in the 143 and 217 GHz A/B Stokes $U$ maps in the dipole feature that is present in the center of the field.}
    \label{fig:hfi-AB-allstokes-ccbgcorr}
\end{figure}

One of the systematics that is apparent in Figures \ref{fig:lfi-AB-allstokes} and \ref{fig:hfi-AB-stokesQU} is the leakage from intensity to polarization due to bandpass mismatch that could explain the discrepancy in the measured polarization angles from 70 and 100 GHz A/B maps. The gain of the instruments are calibrated against the CMB dipole. Since the bandpasses of the two arms of the instruments are not identical, unpolarized foreground emission which can have a spectrum distinct from the CMB, will still appear with varying amplitudes in the two arms, leading to polarization leakage \cite{Leahy:2010ihb, Planck:2015qep}. For example, the leakage from intensity to polarization is evident in the Stokes $U$ map of the 100 GHz split B map shown in the \textit{lowest} panel of Figure \ref{fig:hfi-AB-stokesQU}. This effect is referred to as bandpass mismatch and is corrected for by the application of colour correction. The colour correction, described in Section \ref{subsec:colour-correction}, for the variable response of the detectors within a frequency band to the SED of Crab nebula is applied at the TOD level after background subtraction. This is described by re-writing eq.\ \eqref{eq: ch-4-bin} as
\begin{equation}
    \widehat{\mathbf{m_c}} = (\mathbf{P^TP})\mathbf{P}^{-1}\cdot\big[(\mathbf{d}-F_{\nu,bg})\mathbf{C})\big],
    \label{eq:cc-bgsub-tod}
\end{equation}
where $\widehat{\mathbf{m_c}}$ is now the map reconstructed by bin-averaging the background-subtracted colour-corrected TOD, $F_{\nu,bg}$ is an estimate of the background contribution from the background annuli and $\textbf{C}$ is a vector of colour (and unit) conversions for each detector. $\textbf{C}$ is estimated separately for HFI and LFI from the reduced instrument model (RIMO) provided with the PR4 by assuming the SED of Crab nebula follows a power-law $A\nu^{-\alpha}$. The method employed to estimate the unit conversion and colour correction for each detector is described in Appendix \ref{subsec:colour-correction}. The A/B split maps reconstructed from the method described by eq.\ \eqref{eq:cc-bgsub-tod} are shown in Figures \ref{fig:lfi-AB-allstokes-ccbgcorr}, \ref{fig:hfi-AB-I-ccbgcorr} and \ref{fig:hfi-AB-allstokes-ccbgcorr}.
\subsection{LFI}
In order to understand the inconsistency of the measurement of $\Bar{\psi}_\nu$ at 30 GHz with other frequencies and the inconsistency of $\psi_\nu^\mathrm{split}$ between A/B maps at 70 GHz, we consider
\begin{itemize}
    \item[-] the application of colour correction and background subtraction at TOD level.
    \item[-] the estimation of the position angle of polarization using aperture photometry on \texttt{Cosmoglobe} data.
\end{itemize}
\subsubsection{LFI: Bandpass mismatch}
The flux measured at the LFI frequencies are fit with a power-law SED (assumed for the emission from the Crab nebula) of the form $A\nu^{-\alpha}$ to obtain the best-fit $A=1001.01\,$Jy and $\alpha=-0.29$. These estimates are used to calculate the colour corrections for the LFI (Section \ref{subsec:colour-correction}). As described in Section \ref{sec:res-bandpassmismatch}, background subtraction and colour correction are then applied at the TOD level from which maps are reconstructed using the bin-averaging mapmaking method (eq.\ \eqref{eq:cc-bgsub-tod}). The angles estimated from maps reconstructed from background subtracted colour corrected TOD are plotted in Figure \ref{fig:pa-all-aftercc}. After colour correction, the measurements at 44 and 70 GHz appear inconsistent by $2.8^\circ$ and $-2.8^\circ$ with the measurements from HFI. The A/B split measurements at 70 GHz are inconsistent with each other by $6.8^\circ$ while at 44 GHz the A/B map measurement is discrepant with the measurment from total map by $2.3^\circ$. These inconsistencies can be explained by the presence of intensity to polarization leakage due to beam asymmetries that is apparent, for example, in the Stokes $Q$ and $U$ Split A maps at 70 GHz shown in Figure \ref{fig:lfi-AB-allstokes-ccbgcorr}.

\subsubsection{Position angle of polarization with \texttt{Cosmoglobe} data}
\begin{figure}[htbp]
    \centering
    \includegraphics[width=0.8\textwidth]{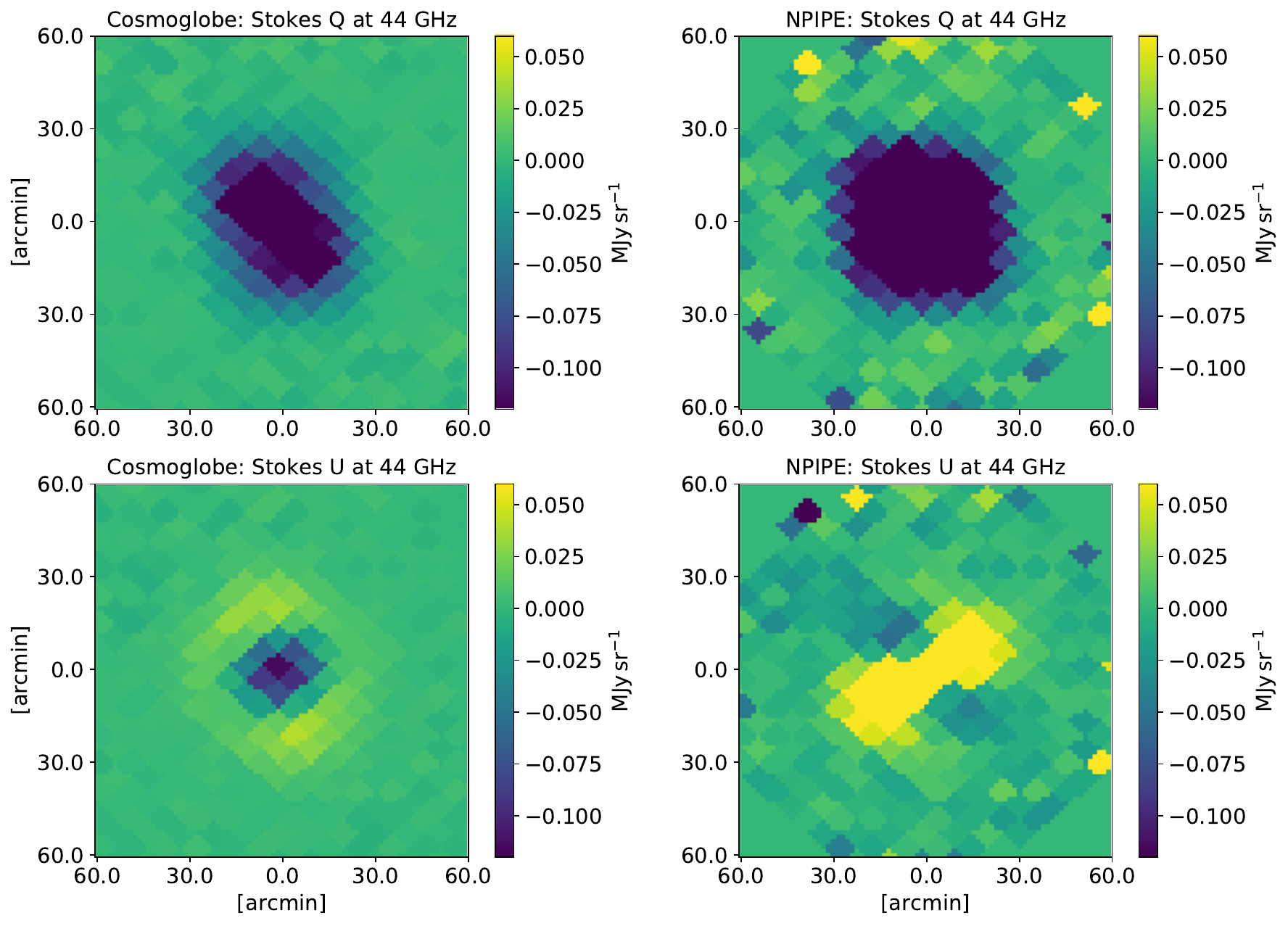}
    \caption{Stokes $Q$ (\textit{top} row) and $U$ (\textit{bottom} row) maps of Crab nebula at $44\,$GHz from \texttt{Cosmoglobe} DR1 (\textit{left}) and \npp (\textit{right}) data.}
    \label{fig:cosmo-npipe-44map}
\end{figure}

\begin{figure}[ht]
    \centering
    \includegraphics[width=0.7\textwidth]{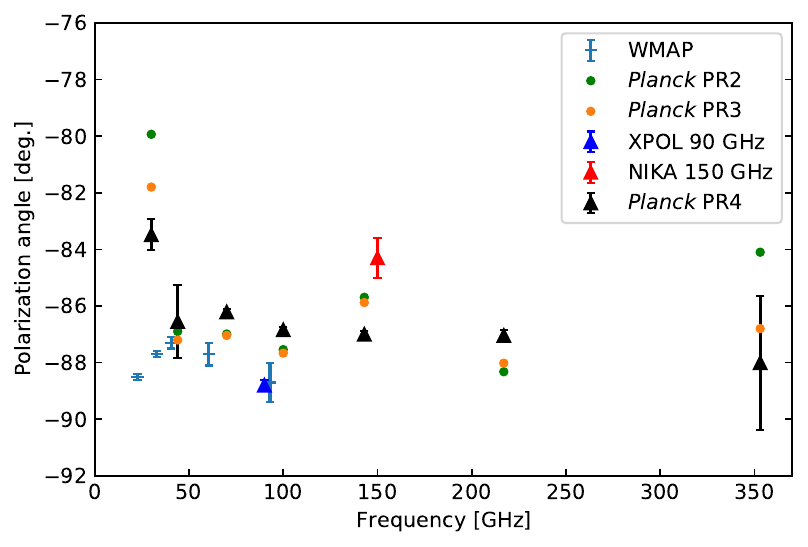}
    \caption{Position angle of polarization measurements from various experiments in Galactic coordinates in comparison with those measured in this work (labelled \npp). Apart from the measurements from this work, estimates from XPOL \cite{Aumont2009}, NIKA \cite{Ritacco2018} and \textit{WMAP} \cite{Weiland:2010ij} are also included. \npp measurements appear stable across the HFI.}
    \label{fig:polangle}
\end{figure}

We performed the same analysis of measuring the background subtracted Stokes $\widehat{Q}_\nu$ and $\widehat{U}_\nu$ parameters from aperture photometry as described in Section \ref{subsec:photometry} and estimating the position angle of polarization using eq.\ \eqref{eq:ch4-polangle} with \texttt{Cosmoglobe} Data Release 1 (CG1) \cite{Cosmoglobe:2023pgf} maps\footnote{All CG1 data products are publicly available at \url{https://cosmoglobe.uio.no}.}. The \texttt{Cosmoglobe} initiative aims to describe the radio, microwave and sub-millimeter sky by combining all available experiments. CG1 involved an end-to-end Bayesian analysis of the full \textit{WMAP} and LFI TOD processed within the \texttt{Commander} framework \cite{Eriksen:2004ss, Eriksen:2007mx}. The approach involved combining complementary sets of data and  breaking degeneracies between calibration, map-making and component separation systematics through Gibbs sampling of the global posterior distribution.
\paragraph{}
The angles measured using CG1 maps are tabulated in Table \ref{tab:cosmo-prs-pa} in comparison with measurements from \plk data. While the measurement at 30 GHz is now consistent with measurements at other frequencies, the measurement at 44 GHz is discrepant by $\sim -8^\circ$. Upon inspection of the maps at 44 GHz, Figure \ref{fig:cosmo-npipe-44map}, it is apparent that the Stokes $U$ in CG1 is positive rather than negative as with the \npp data revealing a systematic in CG1 maps that could occur for compact sources possibly attributed to systematics such as beam leakage and bandpass mismatch \cite{Svalheim:2022mks}.
\begin{table}[htbp]
    \centering
    \begin{tabular}{cccc}
       Dataset & 30 GHz  & 44 GHz & 70 GHz \\
        & (deg) & (deg) & (deg) \\
       \hline
        \textit{Planck~}PR2 & -79.93 & -86.90 & -86.94 \\
        \textit{Planck~}PR3 & -81.80 & -87.20 & -87.04 \\
        \npp & -83.53 & -87.03 & -86.99 \\
        \texttt{Cosmoglobe} DR1 & -87.04 & -95.18 & -87.27 \\
        \hline
    \end{tabular}
    \caption{Position angle of polarization of Crab nebula in Galactic coordinates in IAU convention measured from LFI data of different datasets. For a comparison with \texttt{Cosmoglobe} DR1, the \npp measurements at 30 and 44 GHz have been made using maps downgraded to \texttt{Nside}$=512$.}
    \label{tab:cosmo-prs-pa}
\end{table}
\paragraph{}
While the measured position angle of polarization at 30 GHz became less susceptible to systematics (in terms of its consistency with measurements at HFI frequencies) from data with each PR of \plk (Figure \ref{fig:polangle}), the estimates for the 44 and 70 GHz A/B splits are now discrepant. It is apparent that intensity to polarization leakage due to beam asymmetries is one of the possible reasons for this trend as seen in Figure \ref{fig:lfi-AB-allstokes-ccbgcorr} of the 70 GHz Stokes $Q$ and $U$ Split A maps. A more cautious treatment of the systematics of the LFI is thus necessary. 
\subsection{HFI}
In order to understand the discrepancy in the angles between the A/B splits at 100 GHz, we consider 
\begin{itemize}
    \item[-] the application of colour correction and background subtraction at TOD level.
    \item[-] a novel method of absolute calibration of the measured $\psi_\nu^\mathrm{split}$ with estimates of miscalibration angles from \cite{Eskilt:2022cff} (in Section \ref{sec:ch7-abscalib}).
\end{itemize}
\subsubsection{HFI: Bandpass mismatch}
\begin{figure}
    \centering
    \includegraphics[width=0.8\textwidth]{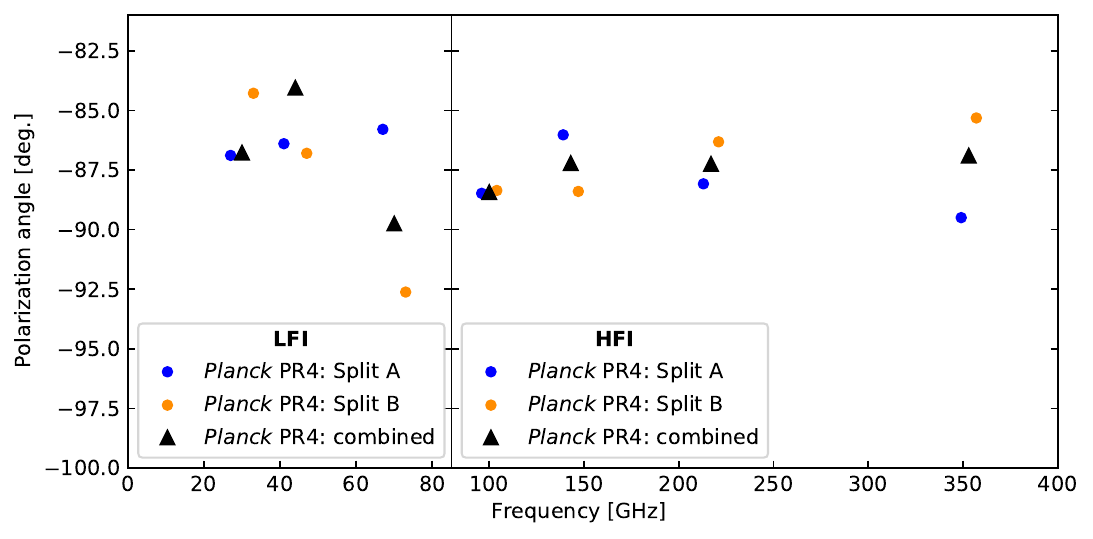}
    \caption{Position angle of polarization measured from maps (A/B split and total maps) reconstructed from background subtracted colour corrected TOD observed by LFI and HFI. The \textit{blue points} (\textit{orange points}) are angles measured from Split A (Split B) maps and the \textit{black triangles} correspond to measurements from total maps.}
    \label{fig:pa-all-aftercc}
\end{figure}

\begin{figure}[ht]
    \centering
    \begin{subfigure}{0.49\textwidth}
       \includegraphics[width=\textwidth]{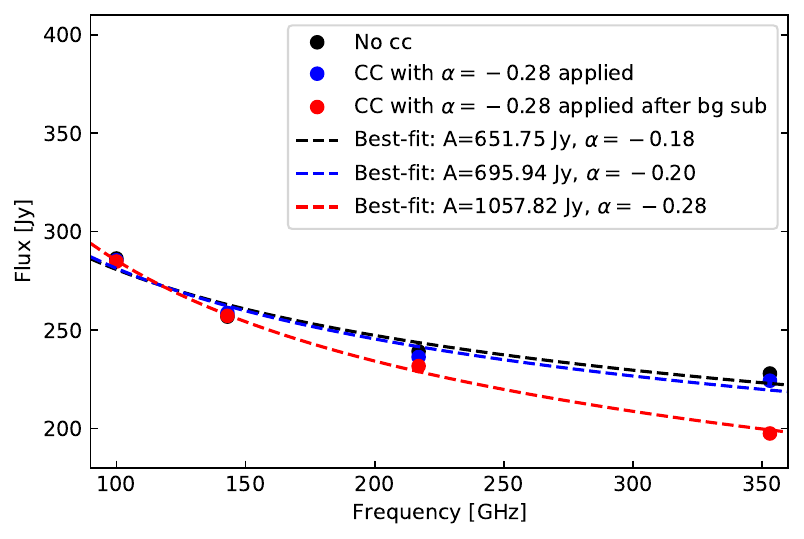} 
    \end{subfigure}
    \begin{subfigure}{0.49\textwidth}
       \includegraphics[width=\textwidth]{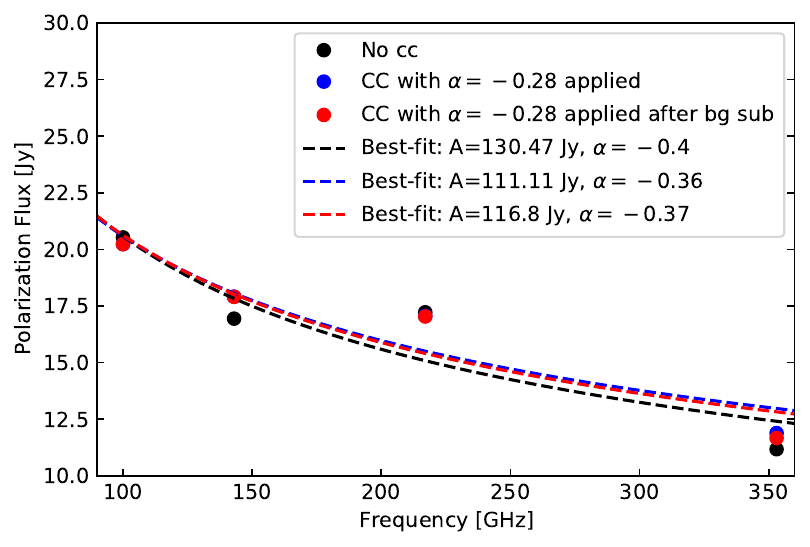} 
    \end{subfigure}
    \caption{Estimates of the flux (\textit{left}) and Polarization flux (\textit{right}) measured from aperture photometry of HFI maps. \textit{Black} points correspond to flux measured from maps without colour correction. \textit{Blue} points correspond to flux measured from colour correction applied at TOD level and background subtraction at map-level. The \textit{red} points correspond to maps reconstructed from background-subtracted and colour corrected TOD (as described by eq.\ \eqref{eq:cc-bgsub-tod}). The dashed curves represent the best-fit SED.}
    \label{fig:ch4-flux-withfit}
\end{figure}

 We measure the flux densities across HFI frequencies and fit a SED described by a power-law of the form $A\nu^{-\alpha}$. We find the best-fit $A=1057.8\,$Jy and $\alpha=-0.28$ for the flux SED which is used to estimate the colour corrections. \cite{Ritacco2018} find $A=1010.2\pm 3.8\,$Jy with $\alpha=-0.323$\footnote{We do verify that there is not a significant difference in the estimated colour corrections by assuming $\alpha=-0.28$ or $\alpha=-0.32$.} for the flux SED and $A_{pol}=78.98 \pm 7.82$ Jy with $\alpha=-0.347$ for the polarization flux SED from \plk PR3 data. The flux and polarization flux as a function of frequency measured from HFI data before and after colour correction are plotted in Figure \ref{fig:ch4-flux-withfit}. Application of colour correction resulted in an increase in the normalization and a steepening of the flux SED. The effect of the correction is more pronounced in the 217 and 353 GHz channels which could be attributed to Galactic dust emission that becomes dominant at these frequencies.
\paragraph{}
The angles computed from maps (A/B split and total maps) reconstructed from background-subtracted colour-corrected TOD are plotted in Figure \ref{fig:pa-all-aftercc}. In comparison with Figure \ref{fig:polangleall}, the A/B split measurements at 100 GHz are now consistent with each other to $0.12^\circ$. The measurements from the A/B maps at 143 and 217 GHz are now different from the estimates from the total maps (shown in Figure \ref{fig:polangleall}) which could be due to the leakage from intensity to polarization from beam asymmetries as Crab nebula is an extended source. For example, in the Stokes $U$ A/B maps of the 217 and 143 GHz maps, the presence of leakage from polarization to intensity due to beam asymmetry is evident (shown in Figure \ref{fig:hfi-AB-allstokes-ccbgcorr}).

\section{Position angle of Crab nebula and miscalibration angle}
\label{sec:ch7-abscalib}
There are instances of a global rotation of the focal plane or a rotation of the orientation of polarization sensitivity of a detector relative to the focal plane which can result in a miscalibration of the instrument. This miscalibration angle is degenerate with the cosmic birefringence angle and in our case, introduces a rotation of the measured Stokes $Q$ and $U$ parameters. $E-$ and $B-$ mode mixing due to miscalibration and its effect on the observed angular power spectra of the CMB are presented in Section \ref{sec:ch2-cmbpol-ebmixing}. The measured position angle of polarization of Crab nebula is thus\footnote{We do not distinguish between a global and relative miscalibration angle in this case.}
\begin{equation}
    \psi = \psi_\mathrm{Crab}+\alpha,
\end{equation}
where $\psi_\mathrm{Crab}$ is the intrinsic position angle of polarization of Crab nebula and $\alpha$ is the miscalibration angle due to any errors in calibration of the orientation of the polarization sensitive detectors of \plk. We attempt an absolute calibration of the measured position angle of polarization of the Crab nebula by assuming $\alpha$ estimated by \cite{Eskilt:2022cff}. \cite{Eskilt:2022cff} exploit the fact that for certain models of time-dependent pseudo-scalar fields, while CMB photons undergo a rotation of the plane of polarization due to the phenomenon of cosmic birefringence by an angle $\beta$, the galactic foreground emission is not susceptible to this phenomenon. While the plane of polarization of CMB photons is rotated by $\beta+\alpha$, the plane of polarization of polarized galactic foreground emission undergoes rotation only by $\alpha$. \cite{Eskilt:2022cff} calibrate $\alpha$ against galactic foreground emission and obtain constraints on $\beta$. We assume $\psi_\mathrm{Crab}=\Bar{\tilde{\psi}}_\mathrm{HFI}$, where $\Bar{\tilde{\psi}}_\mathrm{HFI}$ is the inverse noise-weighted mean of the polarization angles measured across all frequency channels and splits of HFI, and perform an absolute calibration of $\psi_\nu^\mathrm{split}$.
\paragraph{}
The absolute calibration is performed by first \textit{de-rotating} the Stokes $Q_\nu$ and $U_\nu$ maps as
\begin{align}
    \widetilde{Q}_\nu &= Q_\nu\,\mathrm{cos}(2\alpha_\nu)+U_\nu\,\mathrm{sin}(2\alpha_\nu),\\
    \widetilde{U}_\nu &= -Q_\nu\,\mathrm{sin}(2\alpha_\nu)+U_\nu\,\mathrm{cos}(2\alpha_\nu),
\end{align}
where $\widetilde{Q}_\nu$ and $\widetilde{U}_\nu$ are now the \textit{de-rotated} Stokes $Q_\nu$ and $U_\nu$ maps by a miscalibration angle $\alpha_\nu$. Then the position angle of polarization of Crab nebula from the \textit{de-rotated} split maps ($\Tilde{\psi}_\nu^\mathrm{split}$) are estimated. A similar \textit{de-rotation} is applied to the \npp simulations injected with Crab nebula to measure the uncertainties on ($\Tilde{\psi}_\nu^\mathrm{split}$) and finally estimate the inverse noise-weighted mean ($\Bar{\Tilde{\psi}}_\nu$). The measured absolutely calibrated position angles of polarization of Crab nebula at each frequency and split of the HFI, $\Tilde{\psi}_\nu^\mathrm{split}$, and the corresponding inverse noise-weighted mean at each HFI frequency are presented in Table \ref{tab:results-angles}. Accounting for the fact that the $\alpha_\nu$ are correlated across frequency, $\Bar{\Tilde{\psi}}_\nu$ and its associated uncertainty, $\sigma_{\bar{\tilde{\psi}}_\nu}$, are re-computed from the covariance matrix $C_{ij}$ as
\begin{equation}
    \Bar{\tilde{\psi}}_\nu = \sum_i\omega_i\tilde{\psi}_\nu^\mathrm{split},
    \label{eq:derot-weight-mean}
\end{equation}
where
\begin{equation}
    \omega_i = \frac{\sum_k(C^{-1})_{ik}}{\sum_j\sum_k(C^{-1})_{jk}}\quad\mathrm{and}\quad \sigma_{\bar{\tilde{\psi}}_\nu}^2 = \frac{1}{\sum_j\sum_k(C^{-1})_{jk}}.
    \label{eq:derot-weight-uncertainty}
\end{equation}
$C_{ij}$ is constructed from the $\alpha_\nu$ posteriors computed by \cite{Eskilt:2022cff}. The weighted mean position angle of polarization of the Crab nebula from absolute calibration of the HFI against constraints on $\beta$ is found to be $-87.54^\circ\pm0.09^\circ$ ($150.09^\circ\pm 0.09^\circ$) in Galactic (Equatorial) coordinates, estimated using eqs.\ \eqref{eq:derot-weight-mean} and \eqref{eq:derot-weight-uncertainty}.
\paragraph{}
When we compare the estimates for $\Tilde{\psi}^\mathrm{split}_\nu$ with $\psi^\mathrm{split}_\nu$ in Table \ref{tab:results-angles}, it is evident that a miscalibration of the \plk detectors by $\alpha_\nu$ from \cite{Eskilt:2022cff} does not account for the discrepancy we observe between the A/B split angles at 100 GHz.

\section{Conclusion}
\label{sec:npipe-conclusion}
In this Chapter, we measured the polarization properties of the Crab nebula from the PR4 release of \plk TOD in the microwave regime. This measurement was to aid in the relative calibration of the orientation of polarization sensitivity of detectors of \plk and an absolute calibration of other CMB experiments using a bright (in the microwave) source such as the Crab nebula. We measure the position angle of polarization of Crab nebula from A/B split maps for LFI and HFI of \plk mission and find an inverse noise-weighted mean of $-86.92^\circ\pm 0.06^\circ$ ($150.71^\circ\pm0.06^\circ$) in Galactic (Equatorial) coordinates. We however find this estimate to be biased by systematic effects and thus, a relative calibration of the \plk detectors and an absolute calibration of detectors of other CMB experiments with this measurement is not feasible.
\paragraph{}
An inspection of the discrepancy of measured angles at 30 GHz revealed that beam leakage and bandpass mismatch could play a role in the inconsistency. We find the same to be the case for the HFI where we observe an inconsistency between the angles measured in the A/B split maps at 100 GHz. Upon the application of colour correction and background subtraction at the TOD-level, the discrepancy at 100 GHz disappears but a slight difference in the $\psi_\nu^\mathrm{split}$ at 143 and 217 GHz of the HFI is now observed (Figure \ref{fig:pa-all-aftercc}). Intensity to polarization leakage, potentially due to beam asymmetries, is observed in the Stokes $Q$ and $U$ A/B split maps (Figures \ref{fig:lfi-AB-allstokes-ccbgcorr} and \ref{fig:hfi-AB-stokesQU}). A more detailed study of the systematics (such as beam asymmetries that cause leakage from intensity to polarization) is thus necessary before considering the measured position angles of polarization for absolute calibration of CMB experiments. This would involve the application of a model for the beam for each detector at the TOD level. 
\paragraph{}
We further attempt a novel method of absolute calibration of the position angle of polarization of the Crab nebula with constraints on the cosmic birefringence angle, $\beta$, and the associated miscalibration angles from \cite{Eskilt:2022cff} to obtain $\psi_\mathrm{Crab}=-87.54^\circ\pm0.09^\circ$ ($150.09^\circ\pm 0.09^\circ$) in Galactic (Equatorial) coordinates in IAU convention. This estimate, however, is biased by systematic effects such as beam asymmetries and bandpass mismatch. Potential future measurements (or constraints) of $\beta$ estimated at a higher significance using methods as in \cite{Minami:2020fin, Diego-Palazuelos:2022dsq, Eskilt:2022cff} can be used to absolutely calibrate the measured position angle of polarization of the Crab nebula observed by upcoming CMB experiments.
\paragraph{}
Independent methods of calibration of orientation of polarization sensitivity of detectors of CMB experiments are necessary for accurate measurement of the polarization of CMB to detect or constrain signatures of new physics at higher significance. For an absolute calibration of the instruments of CMB experiments with an astrophysical source such as Crab nebula, dedicated observations from ground-based telescopes are required where systematic effects can be modelled accurately for a robust measurement of the position angle of linear polarization of Crab nebula. 

%% file: chapters/chp07.tex
\chapter{Observations of the Crab Nebula with IXPE}
\label{ch:x-ray}

\begin{center}
    \begin{minipage}{0.75\textwidth}
        \textbf{Summary:} The Crab nebula is bright in X$-$rays. Its polarized emission, primarily consisting of polarized synchrotron emission, was observed by the IXPE. In this chapter, we discuss the analysis of IXPE data of the nebula and the reconstruction of the Stokes $I$, $Q$, and $U$ parameters. Polarization properties of the nebula are estimated. We finally compare the measured position angle of polarization and polarization degree with the results obtained previously from \plk data in the microwave regime.\\[3mm]
    \end{minipage}
\end{center}
The emission of most high energy astrophysical objects peaks in the X-ray band ($2-6\,$keV range). X-ray polarimetry is a powerful and unique tool in the study of the magnetic field, geometry and emission mechanisms within these objects \cite{Kallman:2004, Soffitta:2013hla}. The polarized emission observed in X-rays is due to synchrotron emission from non-thermal electrons gyrating in the underlying magnetic field, the description of which can be found in Section \ref{subsec:pol-synch}. The emission observed from the Crab nebula can be assumed to have two components: a temporally-variable emission coming from shock accelerated regions of the pulsar wind nebula and a stable emission arising from the SNR component of the Crab nebula. While the temporally-variable emission is of particular interest for the study of the compact object and its influence on the underlying magnetic field strength through an interplay of electrically charged jets with the surrounding medium, we are interested in measuring the emission attributed to the SNR component. We need polarization measurements from components of the Crab nebula that are stable across time for absolute calibration of orientation of polarization-sensitive detectors of CMB experiments. 
\paragraph{}
The position angle of polarization measured in X-rays is susceptible to lower systematics than with the observations from microwaves due to the characteristics of the instruments and calibration accuracies. With the assumption of stability of the polarization angle across different frequencies, we would like to explore the possibility of using the position angle of polarization measured in the X-rays of the Crab nebula for absolute calibration of \plk detectors and possibly set a precedent for upcoming CMB experiments.
\paragraph{}
X-ray polarimetry of the Crab nebula in the past has been performed by two missions: OSO-8 mission in the 1970s and the PolarLight instrument onboard the CubeSat mission in 2020. The Bragg polarimeter onboard OSO-8 measured polarized emission from the Crab nebula in the 2.6 and 5.2 keV bands in 1976 and 1977. Following the IAU convention and in Equatorial coordinates, the position angle of polarization was found to be $156.36^\circ\pm 1.44^\circ$ at 2.6 keV and $152.59^\circ\pm 4.04^\circ$ at 5.2 keV \cite{Weisskopf:1978}, measured from emission from the SNR component of the Crab nebula.
\paragraph{}
In this work, we make use of the publicly available data from the IXPE observatory (described in Section \ref{sec:ch2-ixpe}) to measure the polarization properties, namely the position angle of polarization, of the Crab nebula in the $2-8\,$keV range. If the measured position angle of polarization in the X-rays is at a lower uncertainty than the angles measured in the microwave regime (Chapter \ref{ch:tauA-pol-angle}), an absolute calibration of the orientation of the polarization sensitivity of the \plk detectors can be performed by measuring the \plk miscalibration angles at lower uncertainty. This would further demonstrate the possibility of using multiwavelength polarimetric measurements of bright astrophysical sources such as the Crab nebula for calibration of CMB experiments.
\paragraph{}
This chapter is structured as follows. We first present the GPD and the definition of Stokes parameters in Section \ref{sec:x-ray-Stokes}. We shall then discuss the IXPE observations in Section \ref{sec:x-ray-obs} followed by methods of analysis in Section \ref{sec:x-ray-methods}. The results and a discussion of the same are presented in Sections \ref{sec:x-ray-results} and \ref{sec:x-ray-discussion}.
\section{Stokes parameters}
\label{sec:x-ray-Stokes}
The extraction of the Stokes parameters from observations with the GPD is quite unique and will be the focus of this section. We shall first understand the process of a detection within the GPD and the statistics of the reconstruction of the Stokes parameters.
\subsection{Modulation factor}
\begin{figure}[ht]
    \centering
    \includegraphics[width=0.7\textwidth]{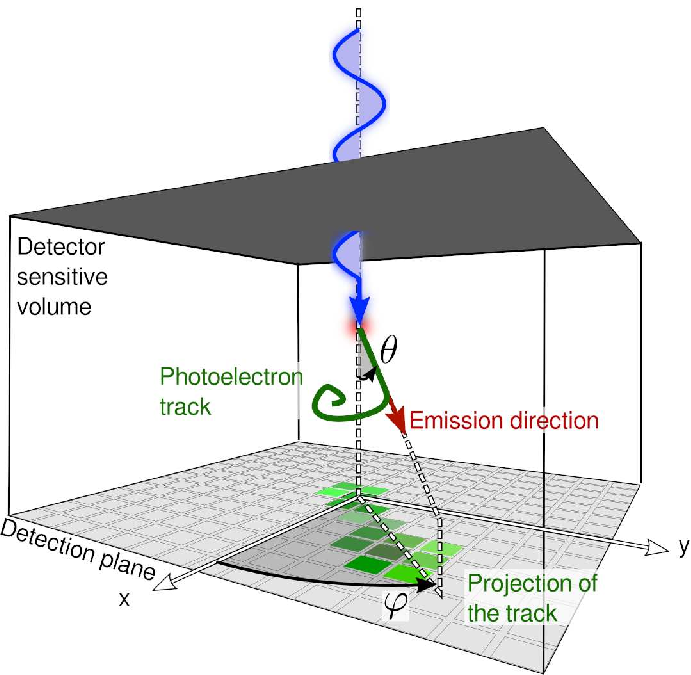}
    \caption{A graphic depicting the photon track within a GPD extracted from \cite{Baldini:2021}. The angle $\varphi$ formed by the direction of emission of the photoelectron is dependent on the direction of the incident photon electric field.}
    \label{fig:gpd-photontrack}
\end{figure}

\begin{figure}[ht]
    \centering
    \includegraphics[width=0.8\textwidth]{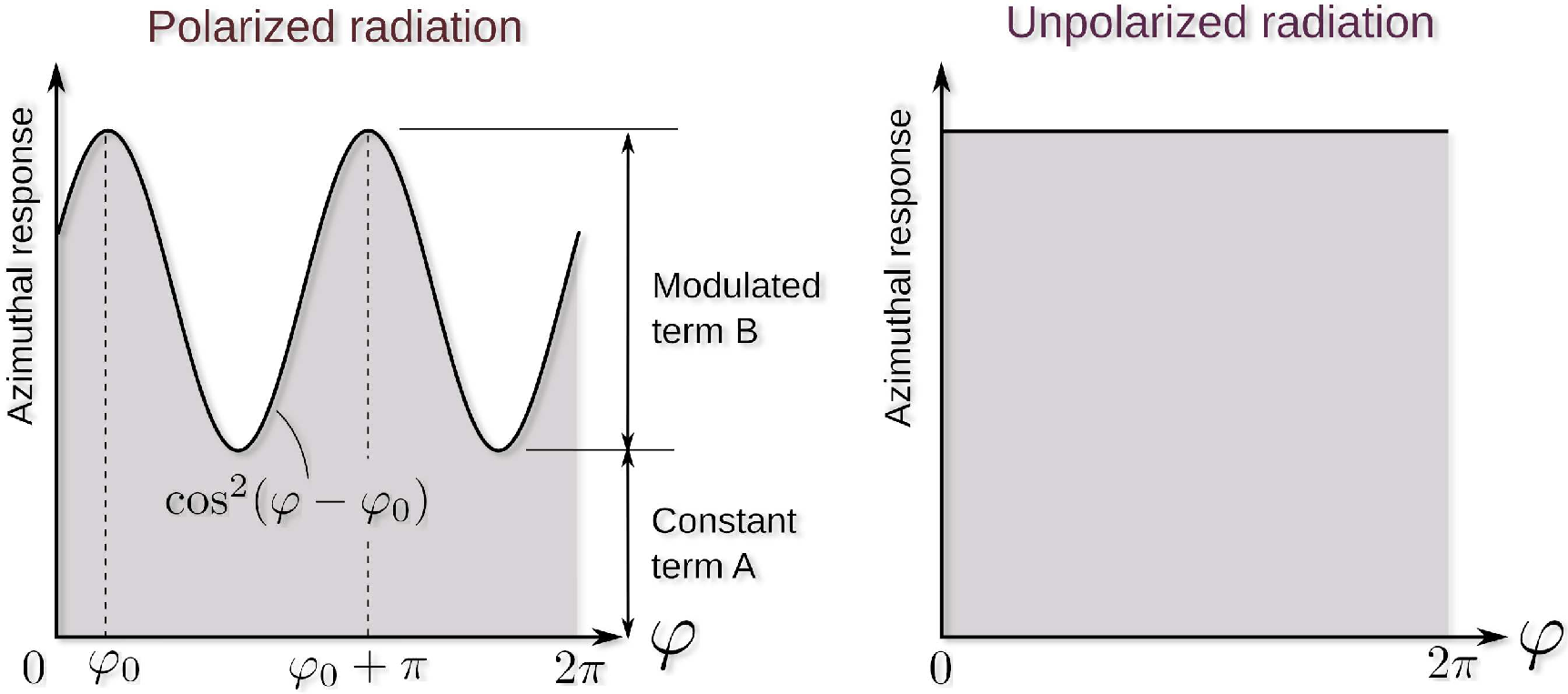}
    \caption{An illustration of the cosine-square modulation of the differential cross-section of emitted photoelectrons in the GPD in response to incident polarized emission (\textit{left}) and the same response due to incident unpolarized emission (\textit{right}). Image is from \cite{Muleri2022}.}
    \label{fig:modulation-illustr.}
\end{figure}

When a photoelectron within the GPD is emitted, the direction of emission or the \textit{measured} azimuthal angle ($\varphi$) when the direction is projected onto the detector plane (the illustration in Figure \ref{fig:gpd-photontrack} shows the geometry of this event) is dependent on the electric field vector of the incident photons. If one plots a histogram of these azimuthal angles over many recorded events for incident linearly polarized light, a $\mathrm{cos}^2$ modulation is observed which is a function of the detector response, polarization state and energy of the incident photons. Such a modulation is not observed for unpolarized light. This is illustrated in Figure \ref{fig:modulation-illustr.}.
\paragraph{}
The differential cross-section of the emission of photoelectrons follows a cosine-square modulation (Figure \ref{fig:modulation-illustr.}). The amplitude of the modulation is proportional to the polarization degree and is maximum in the direction of the polarization angle. The differential cross-section of the photoelectrons emitted by incident linearly polarized photons in a GPD is anisotropic and can be described as \cite{heitler1984quantum, Costa:2001mc}
\begin{equation}
    \frac{d\sigma_\mathrm{ph}}{d\Omega} = r_0^2\frac{Z^5}{137^4}\Bigg(\frac{mc^2}{h\nu}\Bigg)^{7/2}\frac{4\sqrt{2}\mathrm{cos}^2\varphi\,\mathrm{sin}^2\theta}{(1-\beta \mathrm{cos}\theta)^4},
    \label{eq:photoelectron-diff-cross-sec}
\end{equation}
where $\theta$ is the angle between the direction of incident photon and the emitted photoelectron, $\varphi$ is the azimuthal angle of the emitted photoelectron in detector plane which depends on the electron field vector of the incident photon, $r_0$ is the classical electron radius, $Z$ is the atomic number of the gas that the incident photons interact with, $\beta$ is the photoelectron velocity in units of $c$ and $\varphi$ is the azimuthal angle on the plane orthogonal to the incident direction. A visualization of the geometry of this interaction is shown in Figure \ref{fig:gpd-photontrack}. 
Thus, the distribution of the azimuthal angles contains most of the information about the polarization properties of the incident photons and this is quantified by the modulation factor $\mu_\mathrm{E}$ which accounts for the detector response and the energy of the incident photons.
\subsection{Reconstruction of Stokes parameters: weighted analysis}
\begin{figure}[htbp]
    \centering
    \includegraphics[width=0.6\textwidth]{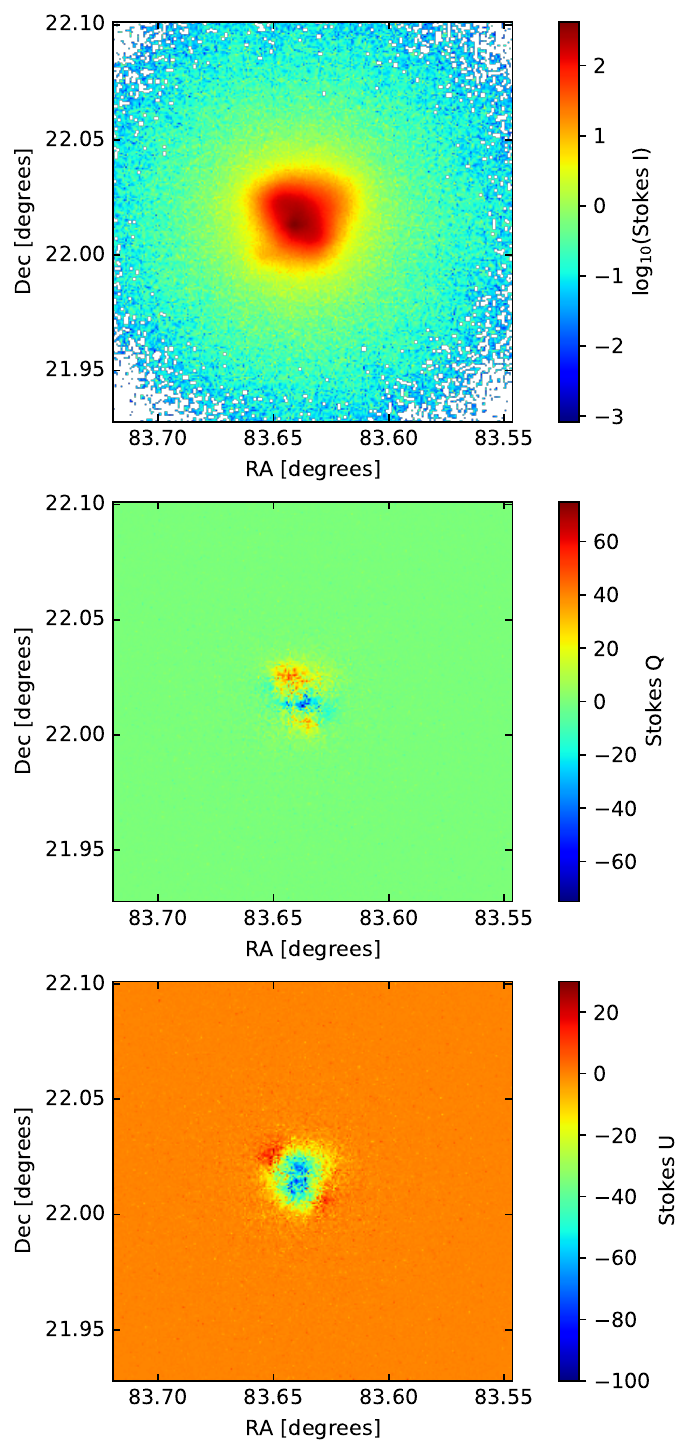}
    \caption{Binned Stokes $I$, $Q$, $U$ maps of the Crab nebula in the $2.0-8.0\,$keV range as observed by IXPE.}
    \label{fig:xray-binned-IQU}
\end{figure}

An observation from the GPD consists of a 2-D \textit{image} of the photoelectron tracks (\textit{right} panel of Figure \ref{fig:app-gpd-graphic}) from which information of the event is extracted. A photoelectron reconstruction algorithm \cite{Bellazzini:2003a, Bellazzini:2003b, Fabio-Muleri:2014, Baldini:2021} was developed in-house for the IXPE GPD which estimates the initial direction of emission of the photoelectron, the position of the photon absorption, and track properties such as track length and total energy. A description of the reconstruction algorithm can be found in \cite{DiMarco:2022}.
\paragraph{}
The reconstructed Stokes parameters for each event with the reconstructed emission angle, $\varphi_k$ of the $k$-th photoelectron are\footnote{A factor of 2 is multiplied in the definition of $q$ and $u$ parameters in eq.\ref{eq:iqu} to arrive at the conventional definition of the polarization degree and angles later.}
\begin{equation}
    \begin{split}
        i_{k} &= 1,\\
        q_{k} &= 2\,\mathrm{cos}(2\varphi_{k}),\\
        u_{k} &= 2\,\mathrm{sin}(2\varphi_{k}).\\
    \end{split}
    \label{eq:iqu}
\end{equation}
for an azimuthal angle distribution of
\begin{equation}
    f(\varphi) = \frac{1}{2\pi}\Big[1+\Pi\,\mu\,\mathrm{cos}(2(\varphi-\varphi_0))\Big]
    \label{eq:xray-fpsi-mod}
\end{equation}
where $\Pi$ is the polarization degree, $\varphi_0$ is the position angle of polarization and $\mu$ is the modulation factor which is the response of the instrument to 100\% polarized radiation.
\paragraph{}
However, \cite{DiMarco:2022} find that the sensitivity of the reconstructed Stokes parameters can be improved by assigning weights to each event. The weights are estimated from a moment analysis of the photoelectron tracks. When specific events whose photoelectrons tracks are better reconstructed are weighted more, the Minimum Detectable Polarization (MDP, described in Section \ref{app:mdp}) decreases (detector response and thus, the sensitivity increases) for the course of an observation despite a decrease in the number of events. Thus, following the prescription of the weighted analysis presented in \cite{Kislat:2014sdf, DiMarco:2022}, which is outlined below, we perform a weighted analysis to extract the Stokes $I$, $Q$, and $U$ maps of the Crab nebula in the $2-8\,$ keV energy bin.
\paragraph{}
According to the weighted analysis prescription, eq.\ \eqref{eq:iqu} needs to be modified by introducing the weights $w_{k}$ as a multiplicative factor. In order to account for the detector response, sensitivity and the dependence of the modulation factor on energy, we also introduce the corrections applied. The reconstructed Stokes parameters are given by
\begin{equation}
    \begin{split}
        \Tilde{i}_{k} &= \frac{w_{k}}{A_{\mathrm{eff}(E_k)}},\\
        \Tilde{q}_\mathrm{k} &= \frac{w_{k}}{A_{\mathrm{eff}(E_k)}\,\mu_{E_k}}\,q_{k},\\
        \Tilde{u}_{k} &= \frac{w_{k}}{A_{\mathrm{eff}(E_k)}\,\mu_{E_k}}\,u_{k},\\
    \end{split}
\end{equation}
where $w_{k}$ are statistical weights obtained from moment analysis, $A_{\mathrm{eff}(E_k)}$ and $\mu_{E_k}$ are the effective area and modulation factor at energy ${E_k}$, respectively.
\paragraph{}
The overall measurements of the Stokes parameters over $N$ events assuming one energy bin are then
\begin{equation}
    \begin{split}
        I &= \varepsilon\sum_{k=1}^N \Tilde{i}_\mathrm{k},\\
        Q &= \varepsilon\sum_{k=1}^N \,\Tilde{q}_{k},\\
        U &= \varepsilon\sum_{k=1}^N \Tilde{u}_{k},
    \end{split}
    \label{eq:xray-IQU}
\end{equation}
with $\varepsilon=\frac{1}{T} \Big(\frac{\sum_{k=1}^{N} w_k}{\sum_{k=1}^{N} (w_k)^2} \Big)$ where $T$ is the on-source time. The effect of the application of weights is to increase the modulation factor with the disadvantage of a reduction of \textit{effective} number of events. This effect is quantified by introducing \cite{Muleri2022}
\begin{equation}
    N_\mathrm{eff} = \frac{(\sum_k w_k)^2}{\sum_k w_k^2} = \frac{I^2}{\sum_k w_k^2}
    \label{eq:xray-neff}
\end{equation}
where we have used the definition of $I$ from eq.\ \eqref{eq:xray-IQU}.
The expected values of $Q$ and $U$, under the ideal conditions of $A_{\mathrm{eff}(E_k)}=1,\,\mu_{E_k}=1$, are estimated by introducing eq.\ \eqref{eq:xray-fpsi-mod} into eq.\ \eqref{eq:xray-IQU} and integrating over $\varphi$ as 
\begin{equation}
    \begin{split}
        \langle Q\rangle &= \sum_{k=1}^N  2\varepsilon\,w_k\,\int_0^{2\pi}\mathrm{cos}(2\varphi)f(\varphi)d\varphi = I\,\Pi\,\mu\,\mathrm{cos}(2\varphi_0),\\
        \langle U\rangle &= \sum_{k=1}^N  2\varepsilon\,w_k\,\int_0^{2\pi}\mathrm{sin}(2\varphi)f(\phi)d\varphi = I\,\Pi\,\mu\,\mathrm{sin}(2\varphi_0),      
    \end{split}
    \label{eq:xray-QU-expvalues}
\end{equation}
from which we can deduce the polarization degree as
\begin{equation}
    \Pi = \frac{\sqrt{Q^2+U^2}}{\mu I} = \frac{\sqrt{Q_\mathrm{N}^2+U_\mathrm{N}^2}}{\mu },
    \label{eq:xray-poldegree}
\end{equation}
where we have introduced normalized (by $I$) Stokes parameters $Q_\mathrm{N}$ and $U_\mathrm{N}$.
\paragraph{}
In order to make Stokes $I$, $Q$, and $U$ maps, the reconstructed Stokes parameters described in eq.\ \eqref{eq:iqu} can be summed over each pixel as in eq.\ \eqref{eq:xray-IQU} while considering one energy bin from $2.0\,\mathrm{keV}\leq E\leq 8.0\,\mathrm{keV}$ and a pixel size. The choice of pixel size is dependent on the choice of statistics and involves a possibility of mixing contributions from different components of the nebula into a given pixel with the choice of large pixel size. Since we are interested in the emission from the SNR component of the nebula, a relatively large pixel size of $5.2^{\prime \prime}$ is chosen.
\paragraph{}
Further, the position angle of polarization ($\varphi_0$ in eqs.\ \eqref{eq:xray-fpsi-mod} and \eqref{eq:xray-QU-expvalues}) is estimated as
\begin{equation}
   \varphi_0 = \psi = \frac{1}{2}\mathrm{tan}^{-1}\Bigg(\frac{U}{Q}\Bigg).
    \label{eq:pa}
\end{equation}
The binned Stokes $I$, $Q$, $U$ maps from IXPE are shown in Figure \ref{fig:xray-binned-IQU}.

\section{Data}
\label{sec:x-ray-obs}
The first IXPE Crab nebula observation (ObsID 01001099) was conducted in two parts: the first from February 21-22, 2022, with a spacecraft roll angle of $158.0^\circ$ (East of North), and the second from March 7-8, 2022, with a roll angle of $158.3^\circ$. Each segment had an on-source time of approximately 43 ks and 49 ks, respectively. Since the offset between the optical axis and the spacecraft axes had not yet been measured and accounted for during the target pointing, the optical axis was displaced from the target by about $2.74^\prime$. The \texttt{ixpecalcarf} tool of \texttt{HEASoft V6.33.2} \cite{heasoft2014} was used to generate effective area and modulation response functions that account for this offset, using the latest on-axis response files in the HEASARC CALDB database (XRT version 20231201, GPD version 20240125).
\paragraph{}
The second IXPE Crab observation (ObsID 02001099) was also conducted in two parts: the first segment took place from February 22-23, 2023, with a roll angle of $158.0^\circ$, and the second segment occurred from April 1-3, 2023, with a roll angle of $158.9^\circ$. Each segment had an on-source time of $\sim 74$ ks.
\paragraph{}
The third IXPE Crab observation (ObsID 02006001) was conducted in one segment from October $9-10$, 2023 with a roll angle of $339.0^\circ$ and $\sim 60$ ks on-source time. In this work, we make use of observations from the third observation cycle. This choice of data is due to two factors: (i) It could not be established that the correction for displacement of the optical axis for the first observation was successful and requires further investigation; (ii) The application of the barrycenter correction to two segments within an observation was not successfully implemented for the first two sets of observations at the time of this study.
\paragraph{}
Three identical telescopes consisting of a mirror module assembly (MMA) and a detector unit (DU) collected data onboard the IXPE. Each DU is at the focus of a MMA. Each observation consists of data collected from the three DUs, which comprises the GPD (Appendix \ref{app:gpd}). Thus, the effective area and modulation factors are estimated from the response files for each DU separately.

\section{Methods}
\label{sec:x-ray-methods}
\begin{figure}[htbp]
    \centering
    \includegraphics[height=5.5cm, width=9.5cm]{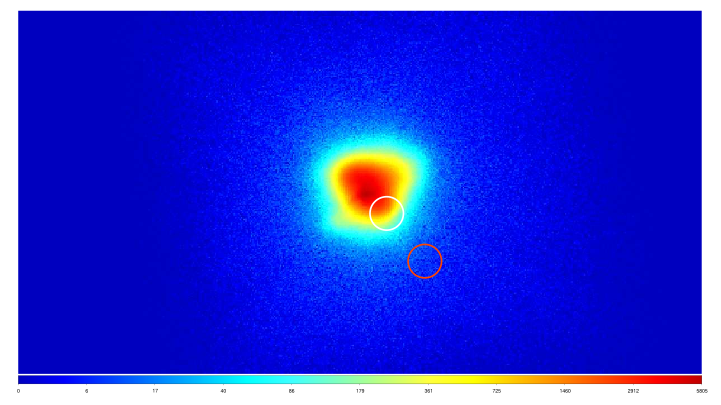}
    \caption{Photon events map (in log scale) of the Crab nebula in the $2-8\,$ keV range observed by DU3 with the \textit{source} (in \textit{white}) and \textit{background} (in \textit{fuchsia}) apertures chosen for polarimetric analysis marked. }
    \label{fig:xray-ds9-DU3-aps}
\end{figure}

We make use of the Level-2 data files that are publicly available from the IXPE archive hosted on the HEASARC website\footnote{\url{https://heasarc.gsfc.nasa.gov/docs/ixpe/archive/}}. The Level-2 event lists are calibrated and filtered data. They contain the pulse invariant event energy channel, time of photon arrival, X and Y coordinates of the photon incidence in the gnomonic projection of the detector coordinates in the plane tangential to the sky coordinates, and the weights from the moment analysis. 
\paragraph{}
We make use of the tools publicly available in \texttt{HEASoft V6.33.2} \cite{heasoft2014} for Spectro-polarimetric analysis of the data. We use the \texttt{ixpecalcarf} tool to first generate the ARF and MRF response files (Appendix \ref{app:xray-data-products}) for each dataset to account for the vignetting effects and apply aperture correction for the aperture of $0.5^\prime$ that we have chosen to analyse. The \texttt{barycorr} tool is then used to apply the barycentric correction with the JPL-DE430 solar ephemeris and source coordinates of RA$=83.633^\circ$ and DEC$=22.014^\circ$. Next, we make use of the \texttt{xpphase} tool available with the \texttt{ixpeobssim} \cite{Baldini:2022nve} software to compute the pulse profiles. This is useful to filter out events that are attributed to the main and secondary pulses as our focus is to analyse events from the off-pulse phase emmitted by the nebular region. 
\paragraph{}
We make use of the \texttt{ds9} tool to extract \textit{source} and \textit{background} regions that are necessary for spectral fitting. The \textit{source} region chosen for this work encompasses the region reported as having peak polarized flux measured at 90 GHz with the XPOL instrument on the IRAM telescope by \cite{Aumont2009}. A \textit{background} region is chosen in the vicinity of the \textit{source} aperture outside of the region with MDP$<0.4$ (Figure \ref{fig:app-mdp-99} can be referred to for a map of the MDP). Once we have extracted the regions of interest using \texttt{ds9}, the \texttt{xselect} tool is used to isolate the events from these regions and extract the \texttt{source} and \texttt{background} $I$, $Q$, $U$ spectra for all three DUs of one dataset. Figure \ref{fig:xray-ds9-DU3-aps} shows the regions that correspond to the \texttt{source} and \texttt{background} in the DU3 field of view. These are then fit first separately and then simultaneously (i.e. $I$, $Q$, $U$ spectra from each DU) with a \texttt{constant$\times$tbabs(polconst$\times$powerlaw)} model using the XSPEC \texttt{fit} tool. The \texttt{tbabs} factor accounts for the absorption of flux through the Hydrogen column density ($N_\mathrm{H}$), \texttt{powerlaw} models the energy dependence and with the \texttt{polconst} we assume constant polarization properties across the region of interest. Since we are considering only one energy bin, the \texttt{powerlaw} component would act as a normalization factor that scales the model to fit the data.

\section{Results}
\label{sec:x-ray-results}
The binned Stokes $I$, $Q$, $U$ maps of the Crab nebula measured from IXPE data is shown in Figure \ref{fig:xray-binned-IQU}. The Stokes $I$ map shows a similar shape of the Crab nebula as observed by the $Chandra$ mission with a jet-like feature extending in the South-West direction from the center of the nebula. A dipole feature is observed in the $Q$ and $U$ maps. 

\begin{figure}[htbp]
    \centering
    \includegraphics[width=0.56\textwidth]{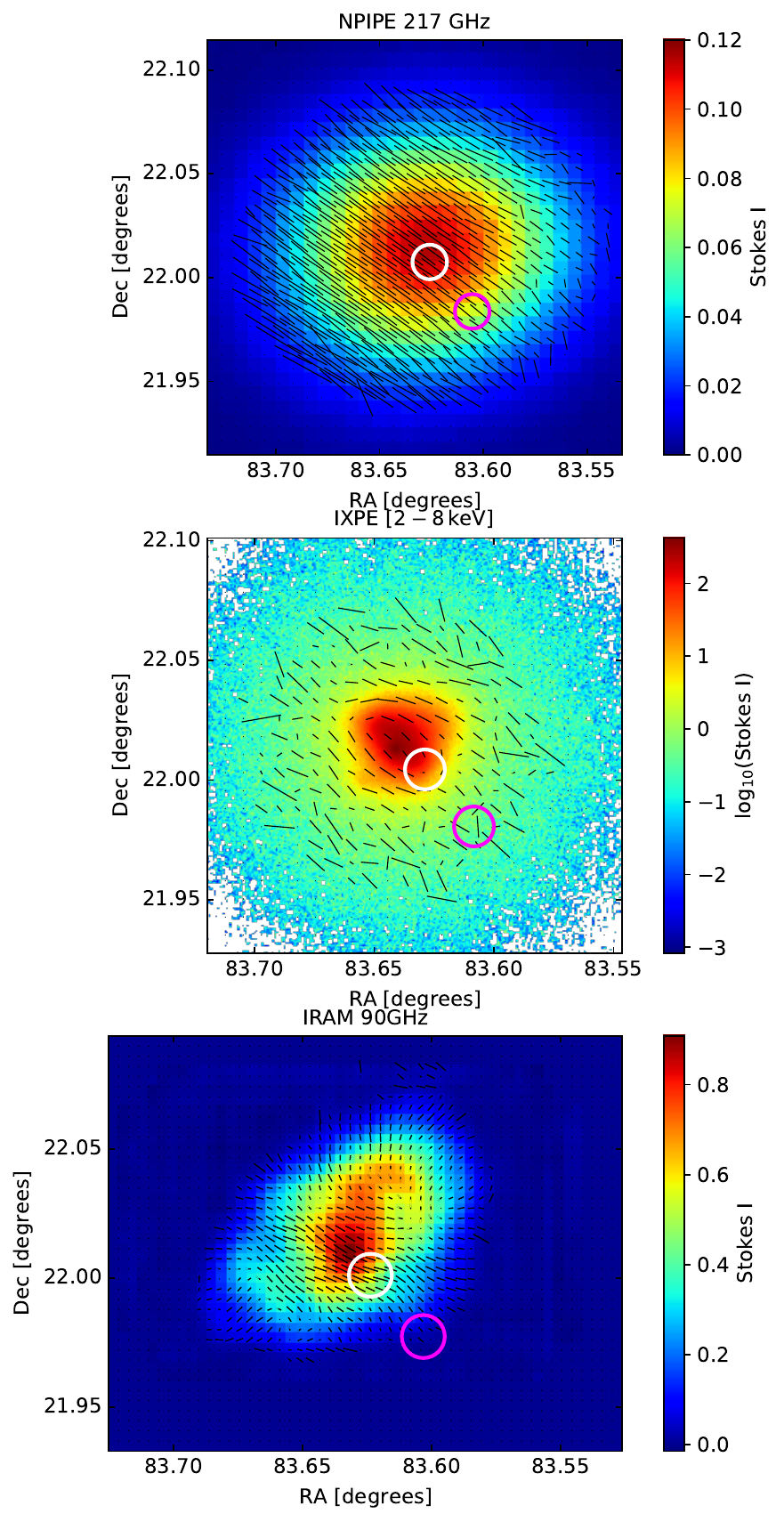}
    \caption{Polarization vectors overplotted on the Stokes $I$ maps computed from \npp (\textit{top}), IXPE (\textit{middle}) and IRAM (\textit{bottom}) data. The \textit{source} (in \textit{white}) and \textit{background} (in \textit{fuchsia}) apertures chosen for polarimetric analysis of the IXPE data are overplotted for comparison. While the polarization vectors estimated from \npp data appear uniform, certain features are apparent in the IXPE and IRAM data. The polarization vectors do appear to be in agreement when considering the region within low MDP in the X-rays.}
    \label{fig:crab-stokesi-polvec}
\end{figure}

\paragraph{}
The preliminary results from the fit are summarised in Table \ref{tab:xspec-results}. The errors correspond to $1\sigma$ estimated by the \texttt{fit} tool using the Fisher information matrix. While the measured photon index across the detector units remains stable, there is a large variation in the measured polarization properties which could be attributed to differences in photon statistics across the DUs and other systematics. The measured position angle of polarization within the \textit{source} aperture from the simultaneous fit of spectra from the three DUs is found to be $148.17^\circ \pm1.32^\circ$ where the uncertainty corresponds to $1\sigma$ estimated using the Fisher information matrix by the \texttt{fit} tool of XSPEC.
\begin{table}[h]
    \centering
    \begin{tabular}{ccccc}
         DU&Absorption density&Photon index&$\Pi$&$\psi$ \\
         &($\times 10^{22}\,\mathrm{cm}^{-2}$)& & \% & (deg.)\\
         \hline
         & & & & \\
         1 & $0.086\pm0.0521$ & $2.05\pm0.02$ & $12.3\pm0.93$ & $147.46\pm2.16$\\
         2 & $0.238\pm0.0547$ & $2.07\pm0.03$ & $8.89\pm0.94$ & $145.91\pm3.04$\\
         3 & $0.217\pm0.0548$ & $2.02\pm0.03$ & $14.22\pm0.94$ & $150.23\pm1.9$\\
         & & & & \\
         \hline
         & & & & \\
         simultaneous & $0.174\pm 0.031$& $2.04\pm 0.01$ &$11.8\pm 0.54$ & $148.17\pm1.32$ \\
         & & & & \\
         \hline
    \end{tabular}
    \caption{Results from Spectro-polarimetric fitting of the Stokes $I,\,Q,\,U$ spectra from each DU with $1\sigma$ statistical uncertainties. The last row corresponds to a simultaneous fit of all the spectra from all DUs which is reflected in the smaller statistical uncertainties.}
    \label{tab:xspec-results}
\end{table}
\paragraph{}
The polarization vectors estimated from the Stokes $Q$ and $U$ maps are overplotted on Stokes $I$ maps from \plk observations at 217 GHz, IXPE data from this work and IRAM data at 90 GHz \cite{Aumont2009} in Figure \ref{fig:crab-stokesi-polvec}. We find that the beam of \plk is such that any features of the underlying magnetic field are smeared. However, in the IRAM and IXPE maps, signatures of what could be a toroidal magnetic field are observed \cite{Bucciantini:2017gdz}.
\section{Discussion}
\label{sec:x-ray-discussion}
We analyzed IXPE data from one observation cycle of the Crab nebula and created Stokes $I$, $Q$, $U$ maps from the Level-2 event lists. We also extracted Stokes $I$, $Q$, $U$ spectra from an aperture of interest. We find the position angle of polarization of $148.17^\circ\pm 1.32^\circ$ ($-89.46^\circ\pm 1.32^\circ$) in Equatorial (Galactic) coordinates in IAU convention within an aperture of $0.5^\prime$ in the nebular region of Crab using Spectro-polarimetry. It is apparent that the measured polarization properties are inconsistent within the detector units when fit separately i.e. when the parameters defining the fitting model are free and independent for each DU. This indicates the possibility of low photon statistics and unknown systematics that require further investigation. 
\paragraph{}
The inclusion of data from the first two observation cycles would improve the photon statistics. Removal of particle and instrumental background events \cite{DiMarco:2023xli}, aspect corrections \cite{Baldini:2022nve} and leakage effects due to the PSF \cite{Dinsmore:2024chc} are some of the systematics that need to be accounted for in future work. We also need to consider a selection of different apertures across the field of view to study the variation of polarization properties and understand if the isolation of the off-pulse events was truly achieved. There is a possibility of \textit{leakage} of pulsed events due to the PSF when considering relatively large PSF. However choosing smaller pixel sizes could result in low photon statistics per pixel.
\paragraph{}
The position angle of polarization measured at 90 GHz by \cite{Aumont2009} at the region of peak polarization flux was found to be $149^\circ\pm1.4^\circ$ (from maps with pixel size of $27^{\prime\prime}$). This value is in agreement with our preliminary measurements, from maps with pixel size of $5.2^{\prime\prime}$, in the X-rays. In Chapter \ref{ch:tauA-pol-angle}, we find an inverse-noise-weighted mean value of $150.71^\circ\pm0.06^\circ$ measured from \npp data in the microwave regime. A direct comparison of the position angle of polarization measured from X-ray data with measurements from the microwave data should be made with caution as the beam size of \plk instruments is in the order of a few arcminutes. However, these preliminary results are promising.
\paragraph{}
In the interest of using these measurements for absolute calibration of CMB experiments, one would need to perform the calibration at the TOD level where systematic effects are better accounted for. Absolute calibration by selecting TOD from close to the apertures considered for X-ray measurements, with the inclusion of modelling of the beam asymmetries, is plausible in the near-future as the position angle of polarization from the nebular part of Crab appears to be stable across a large range of wavelengths (microwave to X-rays) and over time.

%% file: chapters/chp08.tex
\chapter{Conclusions and Outlook}
\label{ch:conclusion}
The CMB has been the cornerstone of Modern Cosmology. It offers an observational window into probing new physics such as the inflationary paradigm and parity-violating phenomena occurring due to the coupling of pseudo-scalar fields to the electromagnetic field. The measurement of the primordial $B-$modes are limited by the calibration of the orientation of the polarization sensitivity of the detectors, secondary anisotropies such as tSZ and gravitational lensing, and Galactic foregrounds. The measurement of the cosmic birefringence angle is degenerate with the instrument miscalibration angle. Some of these challenges can be tackled by synergies with other experiments, multi-wavelength polarization-sensitive observations, component separation methods, modelling of the secondary anisotropies, and independent methods of calibration of the instruments.
\paragraph{}
We introduced the goals of this thesis in Chapter \ref{ch:introduction} followed by a discussion of the astrophysical objects of interest, namely radio halos and the Crab nebula, in Chapter \ref{ch:astro-objects}. The experiments, whose observations and simulations are used in this thesis, are presented in Chapter \ref{ch:observatories}. The physical processes describing the phenomena discussed in the subsequent chapters are presented in Chapter \ref{ch:physical-processes}.
\paragraph{}
In Chapter \ref{ch:ntSZ}, we explored the feasibility of measuring the ntSZ effect to constrain the non-thermal energy budget in radio halos from \plk data and further constrain the magnetic field strength in conjunction with synchrotron emission observations at 1.4 GHz. We find that an improvement in the constraints on magnetic field strength by a factor of two with upcoming CMB experiments can be achieved.
\paragraph{}
We study the properties of $1/f$ noise at TOD and map levels resulting from detector and atmospheric noise in the context of an upcoming ground-based experiment in Chapter \ref{ch:TOAST-TOD}. In Chapters \ref{ch:tauA-pol-angle} and \ref{ch:x-ray}, polarization properties of the Crab nebula in the microwave and X-ray regimes are inferred, respectively, with the intention of performing relative and absolute calibration of the orientation of polarization sensitivity of \plk detectors. We also explore a novel method of absolute calibration of the measured position angle of linear polarization of the Crab nebula in the microwave regime in Chapter \ref{ch:tauA-pol-angle} by assuming the cosmic birefringence angle constrained from an analysis of the CMB and the Galactic foregrounds by \cite{Eskilt:2022cff}.
\paragraph{}
In this work, we have demonstrated the possibility of constraining the non-thermal energy budget and the magnetic field strength within galaxy clusters with known radio halos with observations in the microwave regime using the non-thermal SZ effect in conjunction with the measurements of synchrotron emission at 1.4 GHz. We also looked at the constraining power of upcoming ground-based telescopes with higher sensitivity and frequency coverage on the future prospects of a similar study. While constraints from \plk data on the amplitude of the SZ effect were weak, the constraints from upcoming experiments are very promising. The constraints on the magnetic field strength with upcoming surveys are especially promising as such a synergy would already enable to rule out certain models of the non-thermal electrons in the ICM. Measurements of the ntSZ effect provide an estimate of the non-thermal electron number density which could be used as a prior for normalization of the spectra of Inverse-Compton effect observed in the X$-$rays. The proposed observations of synchrotron emission in galaxy clusters to aid similar studies with Faraday rotation measure by other collaborations would provide much needed synergies to enable a better understanding of the complex nature of non-thermal components within the ICM, the origin of cosmic rays and merger histories of galaxy clusters.
\paragraph{}
We have analyzed \plk data and IXPE data of the Crab nebula in the microwave and X-ray regimes, respectively, to measure its polarization properties. We find that the measured position angle of linear polarization of the nebula is consistent across such a wide range of wavelengths, justifying our assumptions of the same populations of non-thermal electrons emitting polarized synchrotron emission across the different wavelengths and the polarization angle remaining stable across wavelengths and time. There is a possibility to make more robust measurements from X-ray data by the inclusion of data from all observing cycles of the IXPE, accounting for the systematic effects and considering different regions of the nebula within the FoV. We have also demonstrated the need for dedicated ground-based observations of the Crab nebula to enable accurate modelling of systematic effects and minimize the uncertainty of the measured polarization properties in order to have an independent method of absolute polarization angle calibration of upcoming CMB experiments.
\paragraph{}
Finally, we have developed a time-ordered data simulation pipeline for FYST and demonstrated how such simulations aid in the study of the effects of various noise components at the TOD level and at map level. We also explored the characteristics of different scanning strategies and map-making methods which influence the mitigation of the $1/f$ noise due to detectors and atmosphere. We find that the correlated noise residuals in destriped temperature and polarization (Stokes $Q$ and $U$) maps have different characteristics due to the differing PSD of atmospheric and instrument noise. This informs the choice of baseline length and a characterization of the resulting noise in the estimated angular power spectra for destriping method of map reconstruction. Such a study has the potential to inform (i) the development of data reduction and analysis pipelines of observed data and (ii) optimal map-making methods for a given science case and observing field.
\paragraph{}
The CMB contains a wealth of information that is yet to be extracted. While the technology of instruments have progressed significantly, the limiting factors remain systematic effects introduced by data reduction techniques and effects of the instruments that must be well understood for efficient extraction of information on foregrounds and the CMB. We have demonstrated some of the challenges encountered in the measurement of polarization properties and spectral distortions of the CMB. Independent methods of calibration are necessary to characterize foreground emission and systematic effects better in order to minimize the uncertainties on the measured polarization and secondary anisotropies of the CMB.

%% file: chapters/app01.tex
\chapter{\plk Bandpass}
\textit{Appendix \ref{app:SZ-paper-section} is associated with Chapter \ref{ch:ntSZ} that constitutes as the paper that has been submitted.}
\section{Modeling relativistic tSZ and kSZ}
\label{app:SZ-paper-section}
\label{sec:modelsz}
\subsection{Relativistic tSZ}
\label{app:rSZ}
The momentum distribution of scattering electrons which give rise to the $tSZ_{rel}$ effect are modelled using the Maxwell-J\"{u}ttner distribution. Here, a distribution of electron momenta can be described in terms of the normalized thermal energy-parameter, $\Theta = \frac{k_\mathrm{B}T_\mathrm{e}}{m_\mathrm{e} c^2}$, as
\begin{equation}
    f_\mathrm{e,th}(p;\Theta) = \frac{1}{\Theta K_2(1/\Theta)}p^2 \mathrm{exp}(-\frac{\sqrt{1+p^2}}{\Theta}) ,
    \label{eq:maxjutt}
\end{equation}
where $K_v(x)$ denotes the modified Bessel function of the second kind which is introduced for appropriate normalization of the distribution. 
The total IC spectrum for a Planck distribution of photons with  specific intensity of CMB is computed as,
\begin{equation}
    \delta i(x) = \Bigg(\int_0^\infty\int_{-\infty}^\infty f_\mathrm{e,th}(p;\Theta)\: K(e^\mathrm{s};p) e^\mathrm{s} \: i(x/e^\mathrm{s})\: ds\: dp - i(x)\Bigg) I_0\tau_\mathrm{e,th} .
    \label{eq:thermal1}
\end{equation}
Eq. (\ref{eq:thermal1}) is numerically integrated employing the following limits on the integrands,
\begin{equation}
    \delta i(x) = \Bigg(\int_{p_1}^{p_2}\int_{-s_{m}(p_1)}^{s_{m}(p_2)} f_\mathrm{e,th}(p;\Theta) K(e^\mathrm{s};p) e^\mathrm{s}\: \Bigg[\frac{(x/e^\mathrm{s})^3}{(e^{x/e^\mathrm{s}}-1)}\Bigg]\: ds\: dp - \Bigg[\frac{x^3}{(e^x-1)}\Bigg] \Bigg) I_0\frac{m_\mathrm{e} c^2}{k_\mathrm{B}T_\mathrm{e}} y_\mathrm{th} ,
    \label{eq:thermal2}
\end{equation}
where $s_m(p) = 2\mathrm{arcsinh}(p),\, K(e^\mathrm{s};p)$ is described in Eq. (\ref{eq:ksp}) and we have used $\tau_\mathrm{e,th}=\frac{m_\mathrm{e} c^2}{k_\mathrm{B}T_\mathrm{e}} y_\mathrm{th}$, with $T_\mathrm{e}$ representing the temperature of the scattering electrons.
Eq. (\ref{eq:thermal2}) provides the correct estimation of the tSZ effect with relativistic corrections ($\mathrm{tSZ_{rel}}$) for electron energies $> 1\,$keV, which is the case with the ICM.
\paragraph{} In order to test that the implementation of the photon redistribution function to estimate the SZ effect for a given distribution of electron momenta is correct, $\mathrm{tSZ_{rel}}$ is computed using Eq. (\ref{eq:thermal2}) and compared with the methods provided in \cite{chlubaszpack} and \cite{Itoh1998}.
\subsection{kSZ}
\label{app:ksz}
The kSZ effect is the distortion in the specific intensity or temperature of the CMB due to scattering of the CMB photons by free electrons undergoing bulk motion. The distortion in specific intensity of CMB due to the kSZ effect is written as
\begin{equation}
    \Delta I_\mathrm{kSZ} =  -I_0\, \tau_\mathrm{e}\,\Big(\frac{v_\mathrm{pec}}{c}\Big)\frac{x^4e^x}{(e^x-1)^2},\quad x=\frac{h\nu}{k_\mathrm{B}T_\mathrm{CMB}},
    \label{eq:iksz}
\end{equation}
where $I_0$ is the specific intensity of the CMB, $v_\mathrm{pec}$ is the peculiar velocity associated with the cluster along line-of-sight and $\tau_\mathrm{e}$ is the optical depth due to the free electrons. The optical depth can be expressed in terms of the Compton-y parameter ($y_\mathrm{th}$) as
\begin{equation}
    \tau_\mathrm{e} = \int\sigma_T\,n_\mathrm{e}\,dl = \frac{m_\mathrm{e}c^2}{k_\mathrm{B}T_\mathrm{e}}y_\mathrm{th},
\end{equation}
and a parameter analogous to $y_\mathrm{th}$ for the kSZ effect is defined as
\begin{equation}
    y_\mathrm{kSZ} = \tau_\mathrm{e}\,\Big(\frac{v_\mathrm{pec}}{c}\Big) = \frac{m_\mathrm{e}c^2}{k_\mathrm{B}T_\mathrm{e}}\,\Big(\frac{v_\mathrm{pec}}{c}\Big)y_\mathrm{th}.
    \label{eq:yksz}
\end{equation}

\begin{figure}
    \centering
    \includegraphics[width=0.7\textwidth]{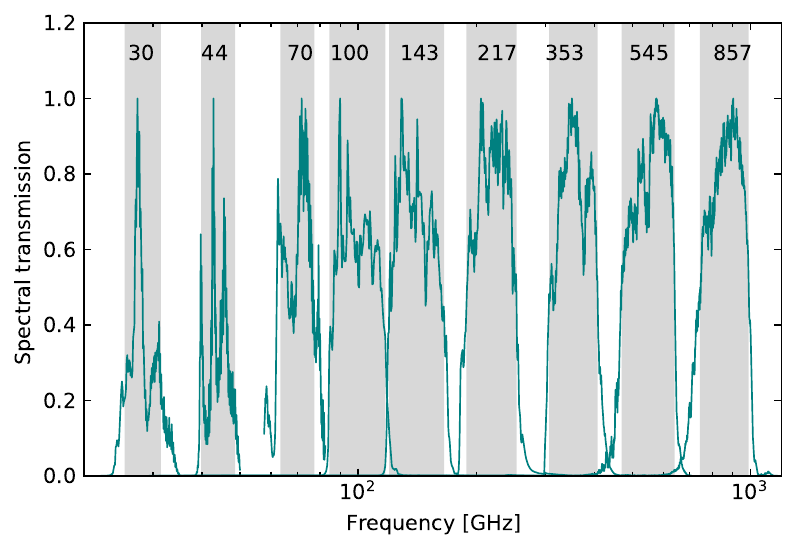}
    \caption{\plk LFI and HFI spectral transmission (teal curves) as a function of frequency. The grey shaded regions correspond to the bandwidths of respective frequency channels mentioned in text (in GHz).}
    \label{fig:planck-bandpass}
\end{figure}
\subsection{Bandpass corrections: SZ effect spectra}
\label{sec:bandpass}
Application of bandpass corrections is necessary to reduce errors due to  systematic effects introduced by the variation in spectral response of each detector in a frequency channel. The spectral response of the detectors was measured through ground-based tests \cite{LFIresponse, PlanckHFIresponse}.  Following the formalism presented in \cite{PlanckHFIresponse}, the bandpass corrected SZ-spectra are computed as
\begin{equation}
    \Delta \tilde{I}_\mathrm{tSZ_{rel}}(x) = y_\mathrm{th}\:I_0\:\frac{\int d\nu\: \tau(\nu)\:g(x,T_\mathrm{e})}{\int d\nu\: \tau(\nu)\: \big(\frac{\nu_c}{\nu}\big)},
    \label{eq:bandpasstsz}
\end{equation}
for the $\mathrm{tSZ_{rel}}$ effect spectrum for scattering electron temperature $T_\mathrm{e}$ and
\begin{equation}
    \Delta \tilde{I}_\mathrm{ntSZ}(x) = y_\mathrm{nth}\:I_0\:\frac{\int d\nu\: \tau(\nu)\:\tilde{g}(x)}{\int d\nu\: \tau(\nu)\: \big(\frac{\nu_c}{\nu}\big)},
    \label{eq:bandpassntsz}
\end{equation}
for the ntSZ effect spectra where $\Delta \tilde{I}_\mathrm{ntSZ}(x)$ is computed separately for each of the non-thermal electron distributions considered in this work. In the equations, $\nu_c$ denotes the central frequency of the frequency bands, and $\tau(\nu)$ is the spectral transmission\footnote{2018 release of filter bandpass transmissions used in this work are available at: \url{https://wiki.cosmos.esa.int/planck-legacy-archive/index.php/The_RIMO}} at frequency $\nu$.

\section{Colour Correction: Crab Nebula SED}
\label{subsec:colour-correction}
The observation of an astrophysical source with a known SED is used to spectrally calibrate broad-band photometric instruments to account for the variable response of detectors within a frequency band. There are two methods one can implement to perform photometric calibration: (i) keeping the reference frequency fixed, one can aim to measure the intensity of the source at the reference frequency with the assumed SED. (ii) keeping the reference intensity fixed, one can determine an \textit{effective} frequency with the assumed SED. The local motion of the \plk satellite with respect to the CMB, known as the CMB dipole, is used to calibrate the LFI and HFI. The CMB dipole signal can be described by the temperature derivative of the Planck function, $B_\nu(T,\nu)$, as
\begin{align}
    b_\nu' &= \frac{\partial B_\nu(T_\mathrm{CMB},\nu)}{\partial T}\\
    &=\Bigg[\frac{2h\,\nu^3}{c^2(e^x-1)}\Bigg]\Bigg(\frac{e^x}{e^x-1}\Bigg)\times\frac{h\nu}{k_BT_\mathrm{CMB}^2}\,\Bigg[\frac{\mathrm{W}}{\mathrm{m^2\,sr\,Hz\,K}}\Bigg]
\end{align}
where $x=(h\nu/(k_BT_\mathrm{CMB}))$ with $T_\mathrm{CMB}=2.7255\,$K. Colour correction is the description of data with respect to an SED  different from the SED used for calibration of the instrument but at the same reference frequency \cite{Griffin:2013yva}. In the case of the source SED being described by a power-law of index $\alpha$ over the field of interest,
\begin{equation}
    \mathrm{d}I_{\nu,\alpha} = \Bigg(\frac{\nu}{\nu_c}\Bigg)^\alpha\mathrm{d}I_{c,\alpha}\,\Bigg[\frac{\mathrm{W}}{\mathrm{m^2\,sr\,Hz\,K}}\Bigg],
\end{equation}
where $\nu_c$ is the reference frequency and $I_{c,\alpha}$ is the associated intensity.
\paragraph{}
Thus, the unit conversion and colour correction are estimated from the \plk instrument model\footnote{The instrument models are available in the \plk legacy archive at \url{https://pla.esac.esa.int}.} as
\begin{equation}
    C = \frac{\int d\nu\: \tau(\nu)\:\big(\frac{\nu_c}{\nu}\big)}{\int d\nu\: \tau(\nu)\: \big(\frac{\nu}{\nu_c}\big)^\alpha} \: \Bigg[\frac{\mathrm{Hz}}{\mathrm{Hz}}\Bigg].
\end{equation}
When we consider an array of detectors, the colour corrections for each are appended into a vector $\mathbf{C}$ which is used in eq.\ \eqref{eq:ch4-tod-def} in Chapter \ref{ch:tauA-pol-angle}.

%% file: chapters/app02.tex
\chapter{X-ray data analysis}
\label{ch:app-x-ray-analysis}

\section{Gas Pixel Detector (GPD)}
\label{app:gpd}
\begin{figure}[ht]
    \centering
    \begin{subfigure}[b]{0.49\textwidth}
        \includegraphics[width=\textwidth]{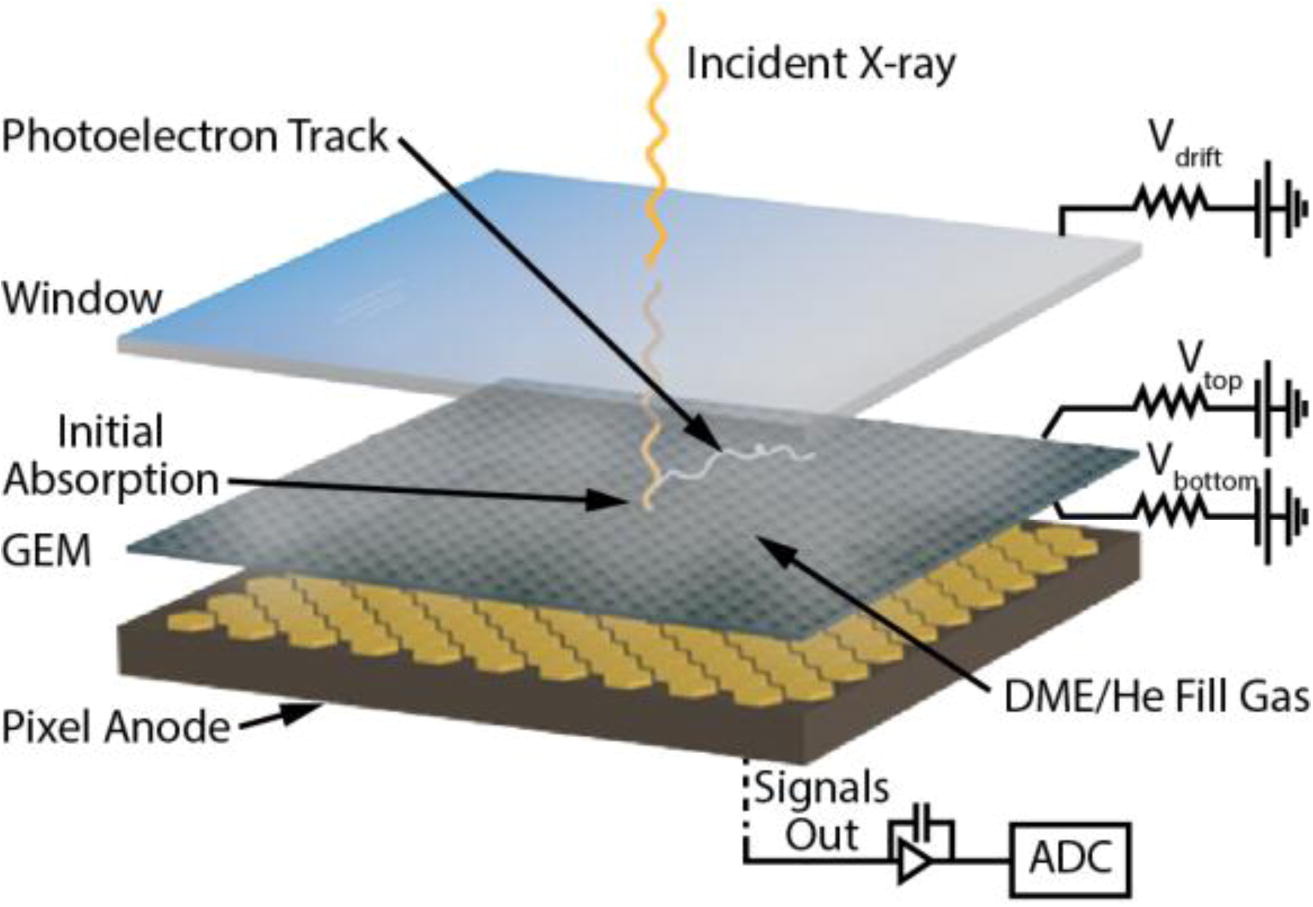}
    \end{subfigure}
    \begin{subfigure}[b]{0.49\textwidth}
        \includegraphics[width=\textwidth]{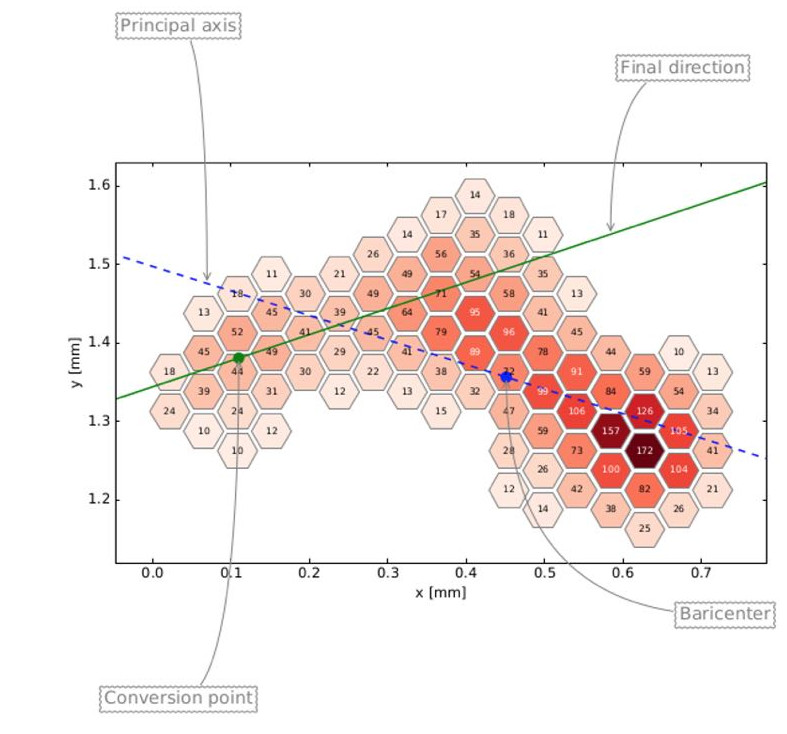}
    \end{subfigure}
   \caption{\textit{Left}: A graphic describing some of the components of the GPD and an incident X-ray photon's interaction with the suspended gas in the GPD demonstrating the basic principle of detection. Image adapted from \cite{Baldini:2021}. \textit{Right}: A photoelectron track resulting from the interaction of a photon of energy 5.9 keV with gas with the track axis and the direction of emission highlighted in the image. The colour scale and numbers on the pixels indicate the charge measured in the pixelized anode of the ASIC. Image adapted from \cite{DiMarco:2022}.}
    \label{fig:app-gpd-graphic}
\end{figure}

The GPD technology was first developed in the 2000s and became advanced enough for in-flight operations recently. It was implemented for the first time onboard PolarLight without the imaging capabilities. The GPD technology with its full capabilities have been implemented onboard IXPE. A Beryllium window seals the components of the detector and is transparent to X-rays. Incident photons enter the gas volume through the Beryllium window and interact with the suspended gas which then emits a photoelectron. The photoelectron further ionizes the gas leaving behind a photoelectron track of ionized electrons. The electric field applied parallel to the optical axis \textit{guide} the primary ionization electrons towards the GEM. The GEM consists of a polymer foil chemically pierced with holes through which the ionization electrons pass to be amplified into an avalanche of charge by an application of a differential potential that are then collected at the Application Specific Integrated Circuit (ASIC) \cite{Bellazzini:2006} which acts as a pixelized anode. The ASIC enables the photoelectron track reconstruction. The GEM preserves the photoelectron track information while providing the necessary gas gain for charge amplification. The holes in the GEM and the pixelization on the anode follow a hexagonal pattern. The \textit{left} panel of Figure \ref{fig:app-gpd-graphic} shows a schematic of the major components of the GPD that an incident X-ray photon encounters.
\paragraph{}
The polarization information from the azimuthal distribution of the photoelectron directions of emission projected onto the readout plane is recovered through a moment analysis \cite{DiMarco:2022} which can be divided into two steps: (i) the identification of the charge barycenter and the main track axis which enables the recognition of the absorption point of the incident photon; (ii) a second moment analysis weights the pixel charges according to the distance from the absorption point to estimate better the direction of the photoelectron tracks. Thus, such a photoelectron track reconstruction allows for simultaneous estimation of the energy, time, incoming direction, and polarization state of the incident photons. The \textit{right} panel of Figure \ref{fig:app-gpd-graphic} shows an example of a photoelectron track observed by a GPD due to an incident 5.9 keV photon. The charge content of each pixel in the anode indicates the energy of the incident photon, the trigger output of the ASIC indicates the time of incidence. The point of absorption and the direction of photoelectron emission indicate the incoming direction and polarization state of the incident photon, respectively. 
\section{Minimum Detectable Polarization (MDP)}
\label{app:mdp}
\begin{figure}
    \centering
    \includegraphics[width=0.7\textwidth]{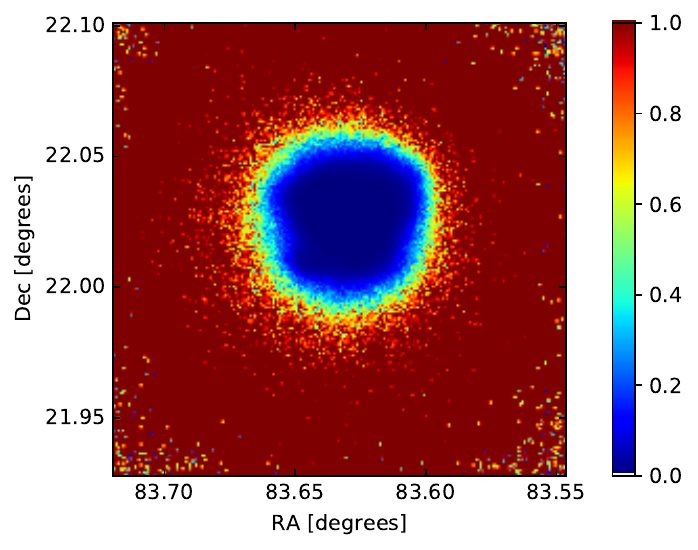}
    \caption{MDP at 99\% confidence level of the Crab nebula field corresponding to the dataset observed by IXPE considered in this work (Chapter \ref{ch:x-ray}). Smaller the MDP, larger is the significance of the measured polarization properties.}
    \label{fig:app-mdp-99}
\end{figure}

The Minimum Detectable Polarization (MDP) is defined as the fractional polarization which will be measured in the absence of any true polarization  with a chance probability of 1 \cite{Krawczynski:2011fm}. The MDP on 99\% confidence level is expressed in terms of the number of photon events $N$ and the modulation factor of the instrument $\mu$ as \cite{Krawczynski:2011fm, Kislat:2014sdf}
\begin{equation}
    \mathrm{MDP} = \frac{4.29}{\mu\sqrt{N}}.
    \label{eq:app-mdp-1}
\end{equation}
It is apparent from eq.\ \eqref{eq:app-mdp-1} that since $\mathrm{MDP}\sim \mu^{-1}$, it is important to design instruments with optimal modulation factor response. A map of the MDP at 99\% confidence level estimated for the dataset observed by IXPE as part of the polarimetric analysis in Chapter \ref{ch:x-ray} is shown in Figure \ref{fig:app-mdp-99}. The polarization properties measured at lower MDP have higher significance.
\section{Data products}
\label{app:xray-data-products}
The IXPE data products used for the analysis in Chapter \ref{ch:x-ray} are described in this section.
\subsection{Redistribution Matrix File (RMF)}
The \textit{detector response} gives a measure of the probability that a photon with energy $E$ is detected in a channel. The level-2 event files do not contain information about the energy of the incident photons but rather the channels or pulse intervals (PIs). The RMF file is then used to convert the PIs to incident photon energies.
\paragraph{}The RMF file is used to convert PIs of the events in the level-2 event files to energy. Photons in the range $2.0\,\mathrm{keV}\leq E\leq 8.0\,$keV are selected.
\subsection{Ancillary Response File (ARF)}
The sensitivity of a detector to a photon of a given energy is dependent on the off-axis angle i.e. how far from the optical axis a photon is incident on a detector. \textit{Effective area} gives a measure of the sensitivity of a detector by taking into account the effects of mirror vignetting. The ARF contains information on effective area, window transmission and detector efficiency components as a function of photon energy. The Effective area of each DU is plotted as a function of energy in the \textit{left} panel of Figure \ref{fig:effarea-modf}.
\subsection{Modulation Response File (MRF)}
\begin{figure}[ht]
    \centering
    \begin{subfigure}{0.49\textwidth}
        \includegraphics[width=\textwidth]{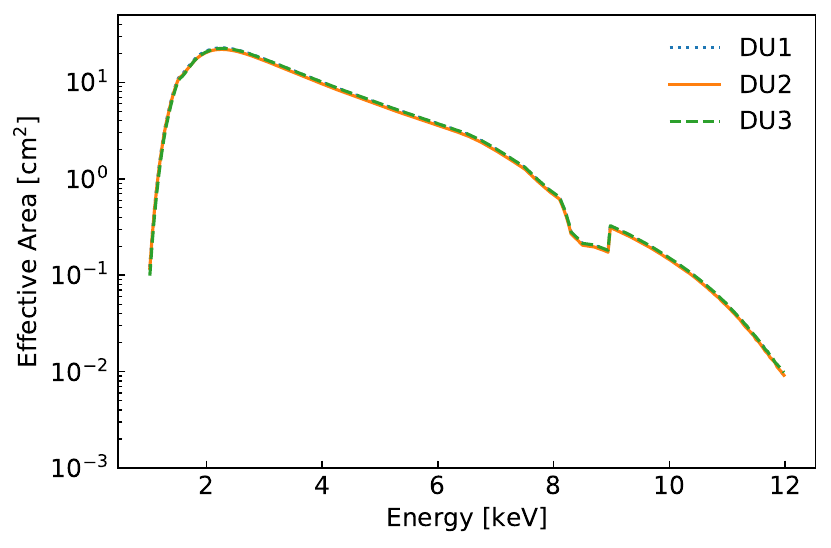}
    \end{subfigure}
        \begin{subfigure}{0.49\textwidth}
       \includegraphics[width=\textwidth]{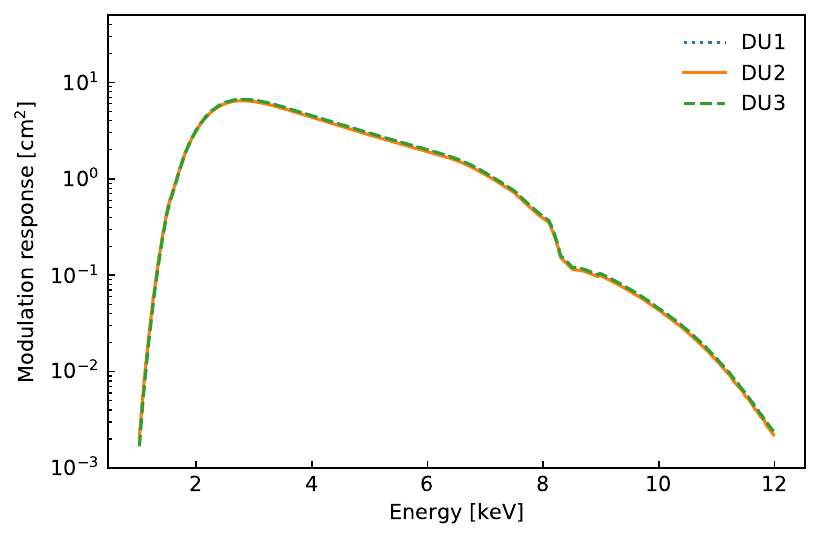}
    \end{subfigure}
    \caption{\textit{Left}: Effective area as a function of energy is plotted for the three detector units (DUs), constructed from the ARF files. \textit{Right}: Modulation factor, $\mu_E$, is plotted as a function of energy for each DU constructed from the MRF files.}
    \label{fig:effarea-modf}
\end{figure}

The Modulation Response File essentially contains the cosine modulation response as a function of energy of the detectors multiplied by the effective area of the instrument. The modulation response of each DU is plotted as a function of energy in the \textit{right} panel of Figure \ref{fig:effarea-modf}. 
\paragraph{}
As is evident in Figure \ref{fig:effarea-modf}, the effective area and modulation response of the DUs fall with increasing energies and is found to be optimal in the $2-8\,$keV range which is the reason for our choice of the energy range considered in the analysis of IXPE data in Chapter \ref{ch:x-ray}.

%% file: danksagung.tex
\chapter*{Acknowledgements}\addcontentsline{toc}{chapter}{\protect Acknowledgements}
I would first like to express my gratitude towards my supervisor and mentor, Prof. Dr. Eiichiro Komatsu, for giving me the opportunity to work on the projects that eventually became this doctoral thesis, and for his extensive patience that was necessary in his supervision of me. His valuable insights, feedback and guidance over the course of my work were absolutely essential. I am grateful to Dr. Fabian Schmidt and Dr. Benedetta Ciardi for being part of my thesis committee and providing me guidance over the course of my PhD. I am grateful to Dr. Kaustuv Basu for having worked on the ntSZ project with me. I would like to thank the International Max-Planck Research School (IMPRS) on Astrophysics for accepting me into its doctoral programme which enabled my work contract with the Max-Planck-Institut f\"{u}r Astrophysik (MPA), and for the provision of one-week lectures that became useful in building my knowledge in Astrophysics and Cosmology. I am grateful for the wonderful and hardworking secretaries of MPA, especially Frau Sonja Gr\"{u}ndl, Frau Gabriela Kratschmann and Frau Maria Depner, who helped navigate the bureaucracy within the institute and the city of Munich. I thank Frau Annette Hilbert for her assistance with everything related to IMPRS.
\paragraph{}
This work would not have been possible without my parents, Muralidhara V. and Geetha Muralidhara, who inflicted me onto this world and have cared for me ever since despite my many flaws. My brother Varun Muralidhar, my sister-in-law Sowmya Sathyanarayana and my niece Aarya have given me crucial support over the course of my education. I thank Vasanth Balakrishna Subramani for his companionship over the course of my PhD. I am grateful for my friends Manali Jeste, Devina Misra, Sanya Munjal, Vinee Chauhan and Vishnu Balakrishnan for all the virtual (and sometimes in-person) hangouts that gave me some semblance of familiarity and sanity during the pandemic (and after). I thank Melitta Pelger and Lisa Schuster for being like my second family once I moved to Munich.
\paragraph{}
I appreciate Martin Reinecke for the problem-solving that became necessary with the programming aspects of my work. I also appreciate Patricia Diego-Palazuelos for our discussions on \npp simulations, miscalibration angles and foregrounds. I am grateful to have had Elisa Ferreira, Julia Stadler and Luisa Lucie-Smith as my mentors. I would also like to thank Reijo Keskitalo for providing access to \npp data and answering my many questions regarding TOD processing (within \npp and \texttt{TOAST}). I thank Dr. Duncan Watts for certain clarifications about systematics within Cosmoglobe DR1. I am grateful to Prof. Hua Feng, Dr. Yu Zhou, Dr. Tsunefomi Mizuno, Prof. Giorgio Matt, Dr. Eugene Churazov and Dr. Ildar Khabibullin for useful discussions on the components of the Crab nebula and X-ray data analysis.
\paragraph{}
I have had a very enriching experience being part of the CCAT collaboration. I am grateful to have collaborated with Ankur Dev on TOD simulations for FYST. I am especially fond of all the memories I made with Ankur, Maude Charmetant and Chris Karoumpis in New York city after the fourth CCAT collaboration meeting.
\paragraph{}
I am especially grateful for Marta Monelli and Laura Herold for their friendship which became absolutely essential. I find myself very fortunate to have worked with them and seen them grow into incredible scientists. I am grateful for everytime Marta welcomed me into her home for all the amazing dinners and much-needed conversations. Aleksandra Grudskaia, whom I initially only knew as my office-mate, became my close friend who made my time in Munich and at MPA more lively. I am grateful to Lennart van Laake for his proofreading of this thesis and for his invaluable feedback. I am very fortunate to have had Swathi Karanth and Lennart in my life without whom this thesis would not have been written to completion.